\documentstyle[preprint,aps]{revtex}
\begin{document}
\draft

\title
{Dirac's Observables for the SU(3)xSU(2)xU(1) Standard Model.}

\author{Luca Lusanna}

\address
{Sezione INFN di Firenze\\
L.go E.Fermi 2 (Arcetri)\\
50125 Firenze, Italy\\
E-mail LUSANNA@FI.INFN.IT}

\author{and}

\author{Paolo Valtancoli}

\address
{Dipartimento di Fisica\\
Universita' di Firenze\\
L.go E.Fermi 2 (Arcetri)\\
50125 Firenze, Italy\\
E-mail VALTANCOLI@FI.INFN.IT}

\maketitle

\begin{abstract}

The complete, missing, 
Hamiltonian treatment of the standard SU(3)xSU(2)xU(1) model with
Grassmann-valued fermion fields in the Higgs phase is given. We bypass the
complications of the Hamiltonian theory in the Higgs phase,  
resulting from the spontaneous symmetry breaking with the Higgs mechanism,
by studying the Hamiltonian formulation of the Higgs phase for the gauge 
equivalent Lagrangian in the unitary gauge. A canonical basis of Dirac's
observables is found and the reduced physical Hamiltonian is evaluated. Its
self-energy part is nonlocal for the electromagnetic and strong interactions,
but local for the weak ones. Therefore, the Fermi 4-fermion interaction
reappears at the nonperturbative level.

\vskip 1truecm
\noindent June 1996
\vskip 1truecm
\noindent This work has been partially supported by the network ``Constrained 
Dynamical Systems" of the E.U. Programme ``Human Capital and Mobility".

\end{abstract}
\vfill\eject

\section
{Introduction}

In two previous papers\cite{lv1,lv2} (referred to as I and II) we made the
complete canonical reduction of the Higgs model with fermions and with
spontaneous symmetry breaking in the Abelian (I) and non-Abelian SU(2) (II)
cases. In both cases there is an ambiguity in solving the Gauss law first
class constraints, which reflects the existence of disjoint sectors of
solutions of the Euler-Lagrange equations. While in the Abelian case there are 
two sectors of solutions, the electromagnetic and the Higgs phases, in the
non-Abelian SU(2) case the sectors correspond to six phases, one of which is
the Higgs phase and one to the SU(2)-symmetric phase [the remaining four 
phases have partially broken SU(2) symmetry and are not SU(2) covariant]. The
Dirac observables and the physical Hamiltonians and Lagrangians of the Higgs
phase have been found in both cases. In the Hamiltonian, the self-energy term
turns out to be local, but not polynomial, and contains a local
four-fermion interaction. Therefore, the nonrenormalizability of the unitary
gauge (our method of canonical reduction is similar to it, but without the
introduction of gauge-fixings) is confirmed.

In this paper we will give a complete Hamiltonian formulation of the Higgs
sector of the standard SU(3)xSU(2)xU(1) model of elementary particles with
Grassmann-valued fermion fields together with its canonical reduction. Namely,
using the results of Refs.\cite{lv1,lv2} and those of Ref.\cite{lusa}, we will
find a complete set of canonical Dirac's observables and the reduced physical
Hamiltonian. This will be done in the case of a trivial principal 
SU(3)xSU(2)xU(1)-bundle (so that there are no monopole solutions)
over a fixed $x^o$, $R^3$ slice of a 3+1 decomposition 
of Minkowski spacetime, without never going to Euclidean space. 
Since the reduction is non covariant, the next step
will be to covariantize the results by reformulating the theory on spacelike
hypersurfaces foliating Minkowski spacetime and, then, by restricting the 
description to the Wigner hyperplanes orthogonal to the total 4-momentum of
the system (assumed timelike). In this way the standard model will be described
in the ``covariant relativistic rest-frame instant form" of the dynamics,
which was defined in Ref.\cite{lus1,albad} for the system of N charged scalar
particles (with Grassmann-valued electric charges)
plus the electromagnetic field [for this system one found the Dirac's 
observables, the physical Hamiltonian with the Coulomb potential extracted from
field theory (with the Coulomb self-energies regularized by the property 
$Q^2_i=0$ of the Grassmann-valued electric charges), the second order equations 
of motion for the field and the particles and the Lienard-Wiechert potentials]. 
In this form of the dynamics there is a
universal breaking of Lorentz covariance connected with the description of the
center of mass of the isolated system, but all the other variables have
Wigner covariance. This implies that the relative dynamics with respect to the 
center of mass on the Wigner hyperplane is naturally ``Euclidean": under a
Lorentz transformation the hyperplane is rotated in Minkowski spacetime (and
the canonical center of mass transforms noncovariantly like the Newton-Wigner
position operator, i.e. it has only the rotational covariance implied by the
little group of massive Poincar\'e representations), but the relative
Wigner-covariant variables inside it only feel induced Wigner SO(3) rotations.
The Wigner hyperplane seems to be the natural candidate to solve the Lorentz
covariance problem of lattice gauge theory. It is also possible to formulate
covariant 1-time relativistic statistical mechanics on this hyperplane
\cite{lus1}.

Moreover, the noncovariance of the center of mass
identifies a classical unit of lenght (the M$\o$ller radius $\rho =\sqrt{-W^2}
/cP^2=|\vec S| / c\sqrt{P^2}$) to be used as a ultraviolet cutoff in a future
attempt to quantize these nonlocal and nonpolynomial reduced field theories. 
In Ref.\cite{alba} the results of Ref.\cite{lus1} were extended to N scalar 
particles with Grassmann-valued color charges plus the SU(3) color Yang-Mills 
field (pseudoclassical relativistic scalar-quark model). The Dirac observables,
the physical Hamiltonian with the interquark potential
and the second order equations of motion for both the
field and the particles have been found. In the N=2 (meson) 
case, a form of the requirement of having only color singlets, suitable for a
field-independent quark model, implies a ``pseudoclassical asymptotic freedom"
and a regularization of the quark self-energies.
To reformulate the standard model in this way, one  needs the completion of the
description of Dirac and chiral fields and of spinning particles on spacelike
hypersurfaces\cite{dep} by adapting the method 
of Refs.\cite{hen} for
the canonical description of fermion fields in curved spacetimes to
spacelike hypersurfaces in Minkowski spacetime.

See Refs.\cite{re} for a review of the full research program and of its 
achievements till now.

To apply the results of Ref.\cite{lusa}, we must assume 
that the SU(3) gauge potentials and gauge transformations belong to a suitable 
weigthed Sobolev space\cite{mon,can}, so that any form of Gribov ambiguity is
absent. Instead, it is not necessary that the SU(2)xU(1) gauge potentials and
gauge transformations belong to the same special spaces, because the 
Hamiltonian reduction associated with the Gauss laws is purely algebraic and 
does not require to solve elliptic equations as in the case of the Gauss laws 
of SU(3). However, if one wishes to have homogeneous Hamiltonian boundary
conditions for the various fields [and also to have the possibility to try
to make the reduction also of the other 
non-Higgs phases], one has to work in those special spaces for the whole 
SU(3)xSU(2)xU(1).

In Section II a review of the standard SU(3)xSU(2)xU(1) model is given to fix 
the notations.

In Section III we give the Lagrangian density in the unitary gauge and we 
introduce the mass eigenstates for the fermions.

In Section IV we give the Euler-Lagrange equations, the Hamiltonian and the
primary and secondary constraints. Also the energy-momentum tensor and the
Hamiltonian boundary conditions for the standard model are given. At the end of 
the Section we show that, if we try to reformulate the Hamiltonian theory in 
terms of the vector boson fields rather than in terms of the original gauge
fields, the constraints change nature and the theory becomes extremely 
complicated.

Therefore, in Section V we study the Hamiltonian formulation of the
Lagrangian in the unitary gauge. Now we get only primary and secondary 
constraints, with those referring to the vector boson being of second class.

In Section VI we find the electromagnetic and color Dirac observables.

In Section VII a canonical basis of Dirac observables of the standard model
is found and the physical reduced Hamiltonian is given. Its self-energy part
is nonlocal for the electromagnetic and strong interactions, but local for the
weak ones. Therefore, the Fermi 4-fermions interaction reappears at the 
nonperturbative level.

Finally, in Section VIII we evaluate the physical Hamilton equations.

In the Conclusions some remarks are made.

\section
{The Lagrangian of the standard model}

In this Section we shall make a brief review of the standard SU(3)xSU(2)xU(1)
model to fix the notations.

The standard model is described by the following Lagrangian density\cite{sm}
[see also Refs.\cite{dgh,daw,her,pich}]

\begin{eqnarray}
{\cal L}(x)&=&-{1\over {4g^2_s}}G_A^{\mu\nu}(x)G_{A\mu\nu}(x)-{1\over {4g^2_w}}
W_a^{\mu\nu}(x)W_{a\mu\nu}(x)-{1\over {4g^2_y}}V^{\mu\nu}(x)V_{\mu\nu}(x)+
\nonumber \\
&+&[D^{(W,V)}_{\mu}\phi (x)]^{\dagger}D^{(W,V)\mu}\phi (x)-\lambda 
(\phi^{\dagger}(x)\phi (x) -\phi_o^2)^2+\nonumber \\
&+&{\bar \psi}^{(l)}_{Li}(x)i\gamma^{\mu}(\partial_{\mu}+W_{a\mu}(x)T^a_w+
V_{\mu}(x)Y_w)\psi^{(l)}_{Li}(x)+\nonumber \\
&+&{\bar \psi}^{(l)}_{Ri}(x)i\gamma^{\mu}(\partial_{\mu}+
V_{\mu}(x)Y_w)\psi^{(l)}_{Ri}(x)+\nonumber \\
&+&{\bar \psi}^{(l)}_{Li}(x)\cdot {{\phi (x)}\over {\phi_o}}M^{(l)}_{ij}\psi
^{(l)}_{Rj}(x)+{\bar \psi}^{(l)}_{Ri}(x)
M_{ij}^{(l)\dagger}{{\phi^{\dagger}(x)}\over
{\phi_o}}\cdot \psi^{(l)}_{Lj}(x)+\nonumber \\
&+&{\bar \psi}^{(q)}_{Li}(x)i\gamma^{\mu}(\partial_{\mu}+W_{a\mu}(x)T^a_w+
V_{\mu}(x)Y_w+G_{A\mu}(x)T^A_s)\psi^{(q)}_{Li}(x)+\nonumber \\
&+&{\bar \psi}^{(q)}_{Ri}(x)i\gamma^{\mu}(\partial_{\mu}+V_{\mu}(x)Y_w+
G_{A\mu}(x)T^A_s)\psi^{(q)}_{Ri}(x)+\nonumber \\
&+&{\bar {\tilde \psi}}^{(q)}_{Ri}(x)i\gamma^{\mu}(\partial_{\mu}+V_{\mu}(x)
Y_w+G_{A\mu}(x)T_s^A){\tilde \psi}^{(q)}_{Ri}(x)+\nonumber \\
&+&{\bar \psi}^{(q)}_{Li}(x)\cdot {{\phi (x)}\over {\phi_o}}M^{(q)}_{ij}\psi
^{(q)}_{Rj}(x)+{\bar \psi}^{(q)}_{Ri}(x)M_{ij}^{(q)\dagger}{{\phi^{\dagger}(x)}
\over {\phi_o}}\cdot \psi^{(q)}_{Lj}(x)+\nonumber \\
&+&{\bar \psi}^{(q)}_{Li}(x)\cdot {{\tilde \phi (x)}\over {\phi_o}}{\tilde M}
^{(q)}_{ij}{\tilde \psi}^{(q)}_{Rj}(x)+{\bar {\tilde \psi}}^{(q)}_{Ri}(x)
{\tilde M}_{ij}^{(q)\dagger}{{{\tilde \phi}^{\dagger}(x)}\over {\phi_o}}\cdot 
\psi^{(q)}_{Lj}(x)+\nonumber \\
&+&\theta {{g_s}\over {32\pi^2}}G_{A\mu\nu}(x)\, {*}G_A^{\mu\nu}(x)=\nonumber \\
&=&.....+G_{A\mu}(x)J^{\mu}_{sA}(x)+V_{\mu}J^{\mu}_{Y_w}(x)+W_{a\mu}(x)
J^{\mu}_{wa}(x),\nonumber \\
&&{}\nonumber \\
J^{\mu}_{sA}(x)&=&{\bar \psi}^{(q)}_{Li}(x)\gamma^{\mu}iT^A_s\psi^{(q)}_{Li}(x)+
{\bar \psi}^{(q)}_{Ri}(x)\gamma^{\mu}iT^A_s\psi^{(q)}_{Ri}(x)+{\bar {\tilde
\psi}}^{(q)}_{Ri}(x)\gamma^{\mu}iT^A_s{\tilde \psi}^{(q)}_{Ri}(x),\nonumber \\
J^{\mu}_{Y_w}(x)&=&{\bar \psi}^{(l)}_{Li}(x)\gamma^{\mu}iY_w\psi^{(l)}_{Li}(x)
+{\bar \psi}^{(l)}_{Ri}(x)\gamma^{\mu}iY_w\psi^{(l)}_{Ri}(x)+{\bar \psi}^{(q)}
_{Li}(x)\gamma^{\mu}iY_w\psi^{(q)}_{Li}(x)+\nonumber \\
&+&{\bar \psi}^{(q)}_{Ri}(x)\gamma^{\mu}iY_w\psi^{(q)}_{Ri}(x)+{\bar {\tilde
\psi}}^{(q)}_{Ri}(x)\gamma^{\mu}iY_w{\tilde \psi}^{(q)}_{Ri}(x),\nonumber \\
J^{\mu}_{wa}(x)&=&{\bar \psi}^{(l)}_{Li}(x)\gamma^{\mu}iT^a_w\psi^{(l)}_{Li}(x)
+{\bar \psi}^{(q)}_{Li}(x)\gamma^{\mu}iT^a_w\psi^{(q)}_{Li}(x).
\label{1}
\end{eqnarray}

The fields $G_{A\mu}(x)=g_s{\tilde G}_{A\mu}(x)$ [A=1,..,8], $W_{a\mu}(x)=g_w
{\tilde W}_{a\mu}(x)$ [a=1,2,3] and $V_{\mu}(x)=g_y{\tilde V}_{\mu}(x)$ are 
the SU(3), SU(2) and U(1) gauge potentials respectively; here $g_s, g_w$ and 
$g_y$ are the associated strong (color), weak isospin and weak hypercharge 
coupling constants. The generators of the Lie algebras su(3) of color [${\hat G}_{\mu}=
G_{A\mu}{\hat T}^A_s$] and su(2) of weak isospin [${\hat W}_{\mu}=W_{a\mu}
{\hat T}^a_w$] in the adjoint representation [8-dimensional for SU(3) and
3-dimensional for SU(2)] and of the Lie algebra u(1) of 
weak hypercharge are [$c_{ABC}$ are the SU(3) totally antisymmetric structure
constants]

\begin{eqnarray}
&&{\hat T}^A_s=-{\hat T}^{A\dagger}_s,\quad\quad ({\hat T}^A_s)_{BC}=c_{ABC},
\quad\quad [{\hat T}^A_s,{\hat T}_s^B]=c_{ABC}{\hat  T}^C_s,\nonumber \\
&&{\hat T}^a_w=-{\hat T}^{a\dagger}_w,\quad\quad ({\hat T}^a_w)_{bc}=\epsilon
_{abc},\quad\quad [{\hat T}^a_w,{\hat T}^b_w]=\epsilon_{abc}{\hat T}^c_w,
\nonumber \\
&&Y_w=-{i\over 2}y=-iY.
\label{2}
\end{eqnarray}

The field strengths and the covariant derivatives associated with $G_{A\mu}, 
W_{a\mu}, V_{\mu}$ are

\begin{eqnarray}
G_{A\mu\nu}(x)&=&\partial_{\mu}G_{A\nu}(x)-\partial_{\nu}G_{A\mu}(x)+c_{ABC}
G_{B\mu}(x)G_{C\nu}(x),\nonumber \\
W_{a\mu\nu}(x)&=&\partial_{\mu}W_{a\nu}(x)-\partial_{\nu}W_{a\mu}(x)+
\epsilon_{abc}W_{b\mu}(x)W_{c\nu}(x), \nonumber \\
V_{\mu\nu}(x)&=&\partial_{\mu}V_{\nu}(x)-\partial_{\nu}V_{\mu}(x),\nonumber \\
&&{}\nonumber \\
({\hat D}^{(G)}_{\mu})_{AC}&=&\delta_{AC}\partial_{\mu}+c_{ABC}G_{B\mu}=
(\partial_{\mu}-{\hat G}_{\mu})_{AC},\nonumber \\
({\hat D}^{(W)}_{\mu})_{ac}&=&\delta_{ac}\partial_{\mu}+\epsilon_{abc}W_{b\mu}=
(\partial_{\mu}-{\hat W}_{\mu})_{ac},\nonumber \\
{\hat D}^{(V)}_{\mu}&=&\partial_{\mu}+V_{\mu}Y_w,
\label{3}
\end{eqnarray}

\noindent and the gauge transformations are defined as [${\hat G}_{\mu\nu}=
G_{A\mu\nu}{\hat T}^A_s$, ${\hat W}_{\mu\nu}=W_{a\mu\nu}{\hat T}^a_w$]

\begin{eqnarray}
{\hat G}_{\mu}(x)&\mapsto& {\hat G}^U_{\mu}(x)=U_s^{-1}(x){\hat G}_{\mu}(x)U
_s(x)+U_s^{-1}(x)\partial_{\mu}U_s(x)=\nonumber \\
&=&{\hat G}_{\mu}(x)+U^{-1}_s(x)\, (\partial_{\mu}U_s(x)+[{\hat G}_{\mu}(x),U
_s(x)]),\nonumber \\
{\hat W}_{\mu}(x)&\mapsto& {\hat W}^U_{\mu}(x)=U_w^{-1}(x){\hat W}_{\mu}(x)U
_w(x)+U_w^{-1}(x)\partial_{\mu}U_w(x)=\nonumber \\
&=&{\hat W}_{\mu}(x)+U_w^{-1}(x)\, (\partial_{\mu}U_w(x)+[{\hat W}_{\mu}(x),U
_w(x)]),\nonumber \\
V_{\mu}(x)Y_w&\mapsto& V^U_{\mu}(x)Y_w=V_{\mu}(x)Y_w+U^{-1}_y(x)\partial_{\mu}
U_y(x)=[V_{\mu}(x)+\partial_{\mu}\Lambda_y(x)]Y_w,\nonumber \\
&&{}\nonumber \\
{\hat G}_{\mu\nu}(x)&\mapsto& {\hat G}^U_{\mu\nu}(x)=U^{-1}_s(x){\hat G}
_{\mu\nu}(x)U_s(x)=
{\hat G}_{\mu\nu}(x)+U^{-1}_s(x)\, [{\hat G}_{\mu\nu}(x),U_s(x)],\nonumber \\
{\hat W}_{\mu\nu}(x)&\mapsto& {\hat W}^U_{\mu\nu}(x)=U^{-1}_w(x){\hat W}
_{\mu\nu}(x)U_w(x)=
{\hat W}_{\mu\nu}(x)+U^{-1}_w(x)\, [{\hat W}_{\mu\nu}(x),U_w(x)],\nonumber \\
V_{\mu\nu}(x)&\mapsto& V^U_{\mu\nu}(x)=V_{\mu\nu}(x).
\label{4}
\end{eqnarray}

\noindent Here $U_s, U_w, U_y=e^{\Lambda_yY_w}$ are the realizations in the
adjoint representation of the SU(3), SU(2) and U(1) gauge transformations 
respectively.

The last term in Eq.(\ref{1}) is the topological $\theta$-term [it is the
source of strong CP-violation, whose experimental absence requires
$\theta \leq 2\cdot 10^{-10}$]; in it ${*}G_{A}^{\mu\nu}=
{1\over 2}\epsilon^{\mu\nu\alpha\beta}G_{A\alpha\beta}$ is the dual field
strength (${*}$ is the Hodge star operator) and one has $G_{A\mu\nu}{*}G_A
^{\mu\nu}=-{1\over 4}{\vec E}_A\cdot {\vec B}_A=\partial_{\mu}\epsilon
^{\mu\nu\alpha\beta}(G_{A\nu}G_{A\alpha\beta}-{1\over 6}c_{ABC}G_{A\nu}
G_{B\alpha}G_{C\beta})$.

The field $\phi (x)$ is a complex Higgs field in the fundamental doublet
representation of the weak isospin SU(2) 

\begin{eqnarray}
&&\phi(x)=\left( \begin{array} {l} \phi_{+}(x) \\ \phi_0(x) \end{array} \right)
=e^{\theta_a(x)T^a_w} \left( \begin{array}{c} 0 \\ \phi_o+{1\over {\sqrt{2}}}
H(x)\end{array} \right)={1\over {\sqrt{2}}}e^{\theta_a(x)T^a_w} \left( 
\begin{array}{c} 0 \\ v + H(x)\end{array} \right) ,\nonumber \\
&&\tilde \phi (x)=\phi^c(x)=2iT^2_w\phi^{*}(x)= i\tau^2 \phi^{*}(x)=
\left( \begin{array}{c} \phi^{*}
_0(x)\\ -\phi_{-}(x)=-\phi_{+}^{*}(x) \end{array} \right) ,
\label{5}
\end{eqnarray}

\noindent where the lower subscript of the components denotes the electric 
charge. The field $\phi^c$ is the charge conjugate of the Higgs field. The
su(2) generators in the doublet representation are [$\tau^a$ are the Pauli
matrices]

\begin{eqnarray}
T^a_w&=&-i{{\tau^a}\over 2},\nonumber \\
&&[T^a_w,T^b_w]=\epsilon_{abc}T_w^c, \quad\quad T^a_wT^b_w+T^b_wT^a_w=
-{1\over 2}\delta^{ab}.
\label{6}
\end{eqnarray}

If ${\tilde U}_w$ is the realization of the SU(2) gauge transformations in the
doublet representation and $W_{\mu}=W_{a\mu}T^a_w$, $W_{\mu\nu}=W_{a\mu\nu}
T^a_w$, then in analogy with Eqs.(\ref{4}) one has $W_{\mu}\mapsto W^U_{\mu}=
{\tilde U}^{-1}_wW_{\mu}{\tilde U}_w+{\tilde U}^{-1}_w\partial_{\mu}{\tilde
U}_w$, $W_{\mu\nu} \mapsto W^U_{\mu\nu}={\tilde U}^{-1}_wW_{\mu\nu}{\tilde
U}_w$.

The constant $\phi_o=\phi_o^{*}$ appearing in the Higgs potential $V(\phi )=-
\lambda (\phi^{\dagger}\phi -\phi_o^2)^2$ is real; the three phases $\theta_a
(x)$ parametrize the absolute minima $\phi^{\dagger}\phi =\phi_o^2$ of
$V(\phi )$. At the quantum level $< \phi > =\phi_o \not= 0$ is the gauge
not-invariant formulation of symmetry breaking. The covariant derivative of 
the Higgs field and their gauge transformations are

\begin{eqnarray}
&&D^{(W,V)}_{\mu \, ab}\phi_b(x)=[\delta_{ab}(\partial_{\mu}+V_{\mu}(x)Y_w)+
(W_{c\mu}(x)T_w^c)_{ab}]\phi_b(x),\nonumber \\
&&\phi (x) \mapsto \phi^U(x)={\tilde U}^{-1}_w(x)U^{-1}_y(x)\phi (x).
\label{7}
\end{eqnarray}

The relation between the fields $V_{\mu}$, $W_{3\mu}$, and the electromagnetic,
$A_{\mu}$, and neutral vector boson, $Z_{\mu}$, fields is

\begin{eqnarray}
&&V_{\mu}=g_y{\tilde V}_{\mu}=g_y [-sin\, \theta_w {\tilde Z}_{\mu}+cos\, 
\theta_w {\tilde A}_{\mu}] =A_{\mu}-tg\, \theta_w Z_{\mu},\nonumber \\
&&W_{3\mu}=g_w{\tilde W}_{3\mu}=g_w [cos\, \theta_w {\tilde Z}_{\mu}+sin\, 
\theta_w {\tilde A}_{\mu}] = A_{\mu}+ cot\, \theta_w Z_{\mu},\nonumber \\
&&{}\nonumber \\
&&A_{\mu}=e{\tilde A}_{\mu}= e[cos\, \theta_w {\tilde V}_{\mu}+sin\, \theta_w
{\tilde W}_{3\mu}]= {1\over {g^2_w+g^2_y}} [g^2_wV_{\mu}+g^2_y W_{3\mu}],
\nonumber \\
&&Z_{\mu}=e{\tilde Z}_{\mu}= e [-sin\, \theta_w {\tilde V}_{\mu}+cos\, \theta_w 
{\tilde W}_{3\mu}] = {{g_wg_y}\over {g^2_w+g^2_y}} (W_{3\mu} - V_{\mu}).
\label{8}
\end{eqnarray}

\noindent Here $\theta_w$ is the Weinberg angle and $e$ is the unit of electric
charge; their relation to the original coupling constants $g_w, g_y$ is

\begin{eqnarray}
&&tg\, \theta_w = {{g_y}\over {g_w}},\quad\quad e = {{g_wg_y}\over 
{g^2_w+g^2_y}},\nonumber \\
&&{}\nonumber \\
&&g_y = {e\over {cos\, \theta_w}},\quad\quad g_w = {e\over {sin\, \theta_w}}.
\label{9}
\end{eqnarray}

The charged vector boson fields are

\begin{eqnarray}
&&W_{\pm \mu}={1\over {\sqrt{2}}} (W_{1 \mu} \mp i W_{2 \mu}),\quad\quad
T_w^{\pm}={1\over {\sqrt{2}}} (T_w^1\mp i T_w^2),\nonumber \\
&&W_{1 \mu} = {1\over {\sqrt{2}}} (W_{+ \mu} + W_{- \mu}),\quad\quad
W_{2 \mu} = {i\over {\sqrt{2}}} (W_{+ \mu} - W_{- \mu}),\nonumber \\
&&{}\nonumber \\
&&W_{a \mu} T_w^a + V_{\mu} Y_w = W_{+ \mu} T_w^{-} + W_{- \mu} T_w^{+} + 
A_{\mu} (T^3_w + Y_w) + Z_{\mu} (cot\, \theta_w T^3_w - tg\, \theta_w Y_w)=
\nonumber \\
&&= W_{+ \mu} T^{-}_w + W_{- \mu} T_w^{+} -i Q_{em} A_{\mu} -i Q_Z Z_{\mu}=
\nonumber \\
&&=-i\left( \begin{array}{cc} {1\over 2}(1+y) A_{\mu}+{1\over 2}(cot\, \theta_w
-y tg\, \theta_w) Z_{\mu} & {1\over {\sqrt{2}}} W_{-\mu} \\
{1\over {\sqrt{2}}} W_{+\mu} & {1\over 2}(-1+y) A_{\mu}-{1\over 2}(cot\, 
\theta_w +y tg\, \theta_w) Z_{\mu} \end{array} \right).
\label{10}
\end{eqnarray}

The last line of this equation defines the electric $Q_{em}=i(T^3_w+Y_w)={1\over
2} \left( \begin{array}{cc} 1+y & 0\\ 0 & -1+y \end{array} \right)$ and neutral
$Q_Z=i(cot\, \theta_w T^3_w -tg\, \theta_w Y_w)={1\over 2} \left( \begin{array}
{cc} cot\, \theta_w -y tg\, \theta_w & 0 \\ 0 & -cot\, \theta_w -y tg\,
\theta_w \end{array} \right)$ charge generators for the doublet SU(2) 
representation [in the singlet SU(2) representation one has $Q_{em}=iY_w=
{1\over 2}y=Y$ and $Q_Z=-itg\, \theta_w Y_w=-{1\over 2} tg\, \theta_w y$
and $V_{\mu} Y_w=-i Q_{em} A_{\mu} -i Q_Z Z_{\mu}=-{i\over 2} y (A_{\mu}-
tg\, \theta_w Z_{\mu})$]

\begin{eqnarray}
&&Q_{em} = i(T^3_w+Y_w) = {1\over 2}(\tau^3 + y) = {1\over 2} \tau^3 + Y,
\nonumber \\
&&Q_Z = i(cot\, \theta_w T^3_w - tg\, \theta_w Y_w) = {1\over 2}(cot\,
\theta_w \tau^3 -tg\, \theta_w y),\nonumber \\
&&{}\nonumber \\
&&iT^3_w = sin\, \theta_w (sin\, \theta_w Q_{em} + cos\, \theta_w Q_Z),
\nonumber \\
&&iY_w = cos\, \theta_w ( cos\, \theta_w Q_{em} - sin\, \theta_w Q_Z).
\label{11}
\end{eqnarray}

For the Higgs field $\phi = \left( \begin{array}{l} \phi_{+}\\ \phi_o 
\end{array} \right)$ one has the assignements $Y={1\over 2}$ [y=1] and
$Q_{em}\left( \begin{array}{l} \phi_{+}\\ \phi_o 
\end{array} \right) = \left( \begin{array}{l} \phi_{+}\\ 0
\end{array} \right)$, $Q_Z\left( \begin{array}{l} \phi_{+}\\ \phi_o 
\end{array} \right)={1\over {sin\, 2\theta_w}}\left( \begin{array}{c} 
(1-2sin^2\, \theta_w) \phi_{+}\\ -\phi_o \end{array} \right)$.

The Grassmann-valued
fermion fields $\psi^{(l)}_{.\, i\alpha}(x), \psi^{(q)}_{.\, i\alpha}(x)$
represent leptons and quarks respectively; $\alpha$ is a spinor index, while 
the index i=1,2,3, denotes the families. 
The fields $\psi^{(.)}_{Lia\alpha}(x)$, $\psi
^{(.)}_{Ri\alpha}(x)$ denote left and right fields [$\psi_L={1\over 2}(1-
\gamma_5)\psi, \psi_R={1\over 2}(1+\gamma_5)\psi, {\bar \psi}_L={1\over 2}
\bar \psi (1+\gamma_5), {\bar \psi}_R={1\over 2}\bar \psi (1-\gamma_5),
\bar \psi \psi ={\bar \psi}_L\psi_R+{\bar \psi}_R\psi_L, \bar \psi \gamma^{\mu}
\psi ={\bar \psi}_L\gamma^{\mu}\psi_L+{\bar \psi}_R\gamma^{\mu}\psi_R$]:
the left fields belong
to the doublet representation of the weak isospin SU(2), while the right ones
are SU(2) singlets. The quark fields $\psi^{(q)}_{LiAa\alpha}(x), \psi^{(q)}
_{RiA\alpha}(x)$ also belong to the fundamental triplet representation of the
color SU(3), whose generators are [$\lambda^A$ are the $3\times 3$ Gell-Mann
matrices; $d_{ABC}$ are totally symmetric coefficients]

\begin{eqnarray}
T^A_s&=&-{i\over 2}\lambda^A,\quad\quad 
Tr\, \lambda^A\lambda^B=2\delta_{AB},\quad\quad
c_{ABC}T^B_sT^C_s={3\over 2}T^A_s,\nonumber \\
&&[T_s^A,T_s^B]=c_{ABC}T_s^C,\quad\quad T_s^AT_s^B+T_s^BT_s^A=-{1\over 3}
\delta_{AB}-id_{ABC}T_s^C,\nonumber \\
&&\sum_A(T^A_s)_{ab}(T^A_s)_{cd}={1\over 6}\delta_{ab}\delta_{cd}-{1\over 2}
\delta_{ad}\delta_{bc} \quad\quad (a,b,c,d=1,2,3).
\label{12}
\end{eqnarray}

\noindent The SU(3) Casimirs for a representation R, namely $C_2(R)=\sum_A
[T^A_s(R)]^2$ and $C_3(R)=\sum_{ABC}T^A_s(R)T^B_s(R)T^C_s(R)$, have the
values $C_2(3)={4\over 3}$, $C_3(3)={10\over 9}$ for the triplet (R=3)
and $C_2(8)=3$, $C_3(8)=0$, for  the adjoint (R=8) representations
respectively.

The covariant derivatives and gauge transformations [${\tilde U}_w, {\tilde U}
_s$ are their realizations in the SU(2) doublet and SU(3) triplet 
representations respectively] of the fermion fields are [$G_{\mu}=G_{A\mu}T^A_s,
W_{\mu}=W_{a\mu}T^a_w$]

\begin{eqnarray}
&&D^{(W,V)}_{\mu \, ab}\psi^{(l)}_{Lib}=[\delta_{ab}(\partial_{\mu}+V_{\mu}Y_w)
+(W_{\mu})_{ab}]\, \psi^{(l)}_{Lib},\nonumber \\
&&D^{(V)}_{\mu}\psi^{(l)}_{Ri}=(\partial_{\mu}+V_{\mu}Y_w)\psi^{(l)}_{Ri},
\nonumber \\
&&D^{(W,V,G)}_{\mu \, AB\, ab}\psi^{(q)}_{LiBb}=[\delta_{ab}\delta_{AB}(\partial
_{\mu}+V_{\mu}Y_w)+\delta_{AB}(W_{\mu})_{ab}+\delta_{ab}(G_{\mu})_{AB}]\psi
^{(q)}_{LiBb},\nonumber \\
&&D^{(V,G)}_{\mu \, AB}\psi^{(q)}_{RiB}=[\delta_{AB}(\partial_{\mu}+V_{\mu}Y_w)
+(G_{\mu})_{AB}]\psi^{(q)}_{RiB}\quad\quad same\, \, for\, \, {\tilde \psi}
^{(q)}_{RiA},\nonumber \\
&&{}\nonumber \\     
&&\psi^{(l)}_{Li} \mapsto \psi^{(l)U}_{Li}={\tilde U}^{-1}_wU^{-1}_y\psi^{(l)}
_{Li},\nonumber \\
&&\psi^{(l)}_{Ri} \mapsto \psi^{(l)U}_{Ri}=U^{-1}_y\psi^{(l)}_{Ri},\nonumber \\
&&\psi^{(q)}_{Li} \mapsto \psi^{(q)U}_{Li}={\tilde U}^{-1}_wU^{-1}_y{\tilde U}
^{-1}_s\psi^{(q)}_{Li},\nonumber \\
&&\psi^{(q)}_{Ri} \mapsto \psi^{(q)U}_{Ri}=U^{-1}_y{\tilde U}^{-1}_s\psi^{(q)}
_{Ri},\quad\quad same\, \, for\, \, {\tilde \psi}^{(q)}_{Ri}.
\label{13}
\end{eqnarray}

The known leptons [electron, muon, tau and the associated massless left
neutrinos (right ones are absent)] and quarks [six flavours: up, down, charme, 
strange, top, bottom, each one with three color components] are described by the
following fermion fields (weak interaction or gauge eigenstates)

\begin{eqnarray}
&&\psi^{(l)}_{L1}(x)=\left( \begin{array}{l} \nu_{eL}(x)\\ e_L(x) \end{array} 
\right) ,\quad\quad \psi^{(l)}_{L2}(x)=\left( \begin{array}{l} \nu_{\mu L}(x)\\
\mu_L(x) \end{array} \right) ,\quad\quad \psi^{(l)}_{L3}(x)=\left( 
\begin{array}{l} \nu_{\tau L}(x)\\ \tau_L(x) \end{array} \right) \nonumber \\
&&\psi^{(l)}_{R1}(x)=e_R(x) \quad\quad \psi^{(l)}_{R2}(x)=\mu_R(x)
\quad\quad \psi^{(l)}_{R3}(x)=\tau_R(x) ,\nonumber \\
&&\psi^{(q)}_{L1}(x)=\left( \begin{array}{l} u_L(x)\\ d_L(x) 
\end{array} \right) ,\quad\quad \psi^{(q)}_{L2}(x)=\left( \begin{array}{l} 
c_L(x)\\ s_L(x) \end{array} \right) ,\quad\quad \psi^{(q)}_{L3}(x)=
\left( \begin{array}{l} t_L(x)\\ b_L(x) \end{array} \right) , 
\nonumber \\
&&{\tilde \psi}^{(q)}_{R1}(x)= u_R(x) ,\quad\quad {\tilde \psi}^{(q)}
_{R2}(x)= c_R(x) ,\quad\quad {\tilde \psi}^{(q)}_{R3}(x)= t_R(x),
\nonumber \\
&&\psi^{(q)}_{R1}(x)=d_R(x) ,\quad\quad \psi^{(q)}_{R2}(x)=s_R(x), \quad\quad
\psi^{(q)}_{R3}(x)=b_R(x).
\label{14}
\end{eqnarray}

The charge assignements for quarks are

\begin{eqnarray}
&&\left( \begin{array}{lll} \nu_{eL}& \nu_{\mu L}& \nu_{\tau L}\end{array} 
\right) :\quad\quad Q_{em}=0,\quad\quad Y=-{1\over 2}\quad [y=-1],\quad\quad
Q_Z={1\over {sin\, 2\theta_w}},\nonumber \\
&&\left( \begin{array}{lll}e_L& \mu_L& \tau_L\end{array} \right) :\quad\quad
Q_{em}=-1,\quad\quad Y=-{1\over 2}\quad [y=-1],\quad\quad Q_Z={{-1+2sin^2\,
\theta_w}\over {sin\, 2\theta_w}},\nonumber \\
&&\left( \begin{array}{lll} e_R& \mu_R& \tau_R\end{array} \right) :\quad\quad
Q_{em}=-1,\quad\quad Y=-1\quad [y=-2],\quad\quad Q_Z={{2sin^2\, \theta_w}\over
{sin\, 2\theta_w}},\nonumber \\
&&\left( \begin{array}{lll} u_L&  c_L& t_L\end{array}
\right) :\quad\quad Q_{em}={2\over 3},\quad\quad Y={1\over 6}\quad [y={1\over 
3}],\quad\quad Q_Z={{1-{4\over 3}sin^2\, \theta_w}\over {sin\, 2\theta_w}},
\nonumber \\
&&\left( \begin{array}{lll} d_L& s_L& b_L\end{array} \right) :\quad\quad
Q_{em}=-{1\over 3},\quad\quad Y={1\over 6}\quad [y={1\over 3}],\quad\quad
Q_Z={{-1+{2\over 3}sin^2\, \theta_w}\over {sin\, 2\theta_w}},\nonumber \\
&&\left( \begin{array}{lll}  u_R&  c_R&  t_R\end{array}
\right) :\quad\quad Q_{em}={2\over 3},\quad\quad Y={2\over 3}\quad [y={4\over 
3}],\quad\quad Q_Z={{-4sin^2\, \theta_w}\over {3sin\, 2\theta_w}},\nonumber \\
&&\left( \begin{array}{lll} d_R& s_R& b_R\end{array} \right) :\quad\quad
Q_{em}=-{1\over 3},\quad\quad Y=-{1\over 3}\quad [y=-{2\over 3}],\quad\quad
Q_Z={{2sin^2\, \theta_w}\over {3sin\, 2\theta_w}}.
\label{15}
\end{eqnarray}

Due to Eq.(\ref{10}), Eqs.(\ref{3}), (\ref{7}) and (\ref{13}) imply

\begin{eqnarray}
&&({\hat D}^{(W)}_{\mu})_{ab}=[\partial_{\mu}-W_{+ \mu}{\hat T}^{-}_w-
W_{- \mu}{\hat T}^{+}_w-(A_{\mu}+cot\, \theta_w Z_{\mu}){\hat T}^3_w]_{ab},
\nonumber \\
&&{\hat D}^{(V)}_{\mu}=D_{\mu}^{(V)}=\partial_{\mu}+(A_{\mu}-tg\, \theta_w
Z_{\mu})Y_w,\nonumber \\
&&D^{(W,V)}_{\mu ab}=[\partial_{\mu}+W_{+ \mu}T^{-}_w+W_{- \mu}T^{+}_w-
iQ_{em}A_{\mu}-iQ_Z Z_{\mu}]_{ab},\nonumber \\
&&D^{(W,V,G)}_{\mu AB ab}=\delta_{AB}D^{(W,V)}_{\mu ab}+\delta_{ab}(G_{\mu})
_{AB},\nonumber \\
&&D^{(V,G)}_{\mu AB}=\delta_{AB}D^{(V)}_{\mu}+(G_{\mu})_{AB}.
\label{16}
\end{eqnarray}

The Lagrangian density (\ref{1}) has the following form in terms of the fields
$A_{\mu}=e{\tilde A}_{\mu}$, $Z_{\mu}=e{\tilde Z}_{\mu}$, $W_{\pm \mu}={e\over 
{sin\, \theta_w}}{\tilde W}_{\pm \mu}$ [we define the following ``Abelian"
field strengths: $A_{\mu\nu}=\partial_{\mu}A_{\nu}-
\partial_{\nu}A_{\mu}$, $Z_{\mu\nu}=\partial_{\mu}
Z_{\nu}-\partial_{\nu}Z_{\mu}$, ${\cal W}_{\pm \, \mu\nu}=
\partial_{\mu}W_{\pm \, \nu}-\partial_{\nu}W_{\pm \, \mu}$]

\begin{eqnarray}
{\hat {\cal L}}(x)&=&-{1\over {4g_s^2}}G^{\mu\nu}_A(x)G_{A\mu\nu}(x)-{1\over 
{4e^2}}A^{\mu\nu}(x)A_{\mu\nu}(x)-{1\over {4e^2}}Z^{\mu\nu}(x)Z
_{\mu\nu}(x)-\nonumber \\
&-&{{sin^2\, \theta_w}\over {2e^2}}{\cal W}_{+}^{\mu\nu}(x){\cal W}
_{-\, \mu\nu}(x)+ie[A^{\mu\nu}(x)+cot\, \theta_wZ^{\mu\nu}(x)]W_{+\, \mu}(x)
W_{-\, \nu}(x)+\nonumber \\
&+&ie[{\cal W}_{+}^{\mu\nu}(x)W_{-\, \mu}(x)-
{\cal W}^{\mu\nu}_{-}(x)W_{+\, \mu}(x)][A_{\nu}(x)+cot\, 
\theta_wZ_{\nu}(x)]+\nonumber \\
&+&{{sin^2\, \theta_w}\over {2e^2}}[W^2_{+}(x)W^2_{-}(x)-
(W_{+}(x)\cdot W_{-}(x))^2]-\nonumber \\
&-&{{sin^2\, \theta_w}\over {e^2}}[A(x)+cot\, \theta_w
Z(x)]^2 W_{+}(x)\cdot W_{-}(x)+\nonumber \\
&+&{{sin^2\, \theta_w}\over {e^2}}W_{+}(x)\cdot [A(x)+cot\, \theta_wZ(x)]
W_{-}(x)\cdot [A(x)+cot\, \theta_wZ(x)]+\nonumber \\
&+&[(\partial_{\mu}+(W_{+\, \mu}(x)T^{-}_w+W_{-\, \mu}(x)
T^{+}_w)-iQ_{em}A_{\mu}(x)-iQ_Z\, Z_{\mu}(x))\phi (x)]
^{\dagger}\cdot \nonumber \\
&\cdot& [(\partial^{\mu}+(W^{\mu}_{+}(x)T^{-}_w+W^{\mu}
_{-}(x)T^{+}_w)-iQ_{em}A^{\mu}(x)-iQ_ZZ^{\mu}(x))\phi (x)]-
\nonumber \\
&-&\lambda (\phi^{\dagger}(x)\phi (x)-\phi_o^2)^2+\nonumber \\
&+&{\bar \psi}^{(l)}_{Li}(x)i\gamma^{\mu}[\partial_{\mu}+(W_{+\mu}
(x)T^{-}_w+W_{-\mu}(x)T^{+}_w)-\nonumber \\
&-&iQ_{em}A_{\mu}(x)-
iQ_ZZ_{\mu}(x)]\psi^{(l)}_{Li}(x)+\nonumber \\
&+&{\bar \psi}^{(l)}_{Ri}(x)i\gamma^{\mu}[\partial_{\mu}+(A_{\mu}(x)
-tg\, \theta_wZ_{\mu}(x))Y_w]\psi^{(l)}_{Ri}(x)+\nonumber \\
&+&{\bar \psi}^{(l)}_{Li}(x)\cdot {{\phi (x)}\over {\phi_o}}M^{(l)}_{ij}\psi
^{(l)}_{Rj}(x)+{\bar \psi}^{(l)}_{Ri}(x)
M_{ij}^{(l)\dagger}{{\phi^{\dagger}(x)}\over
{\phi_o}}\cdot \psi^{(l)}_{Lj}(x)+\nonumber \\
&+&{\bar \psi}^{(q)}_{Li}(x)i\gamma^{\mu}[\partial_{\mu}+(W^{\mu}
_{+}(x)T^{-}_w+W^{\mu}_{-}(x)T^{+}_w)-\nonumber \\
&-&iQ_{em}A_{\mu}(x)-
iQ_ZZ_{\mu}(x)+G_{A\mu}(x)T^A_s]\psi^{(q)}_{Li}(x)+\nonumber \\
&+&{\bar \psi}^{(q)}_{Ri}(x)i\gamma^{\mu}[\partial_{\mu}+(A_{\mu}(x)
-tg\, \theta_wZ_{\mu}(x))Y_w+
G_{A\mu}(x)T^A_s]\psi^{(q)}_{Ri}(x)+\nonumber \\
&+&{\bar {\tilde \psi}}^{(q)}_{Ri}(x)i\gamma^{\mu}[\partial_{\mu}+
(A_{\mu}(x)-tg\, \theta_wZ_{\mu}(x))Y_w
+G_{A\mu}(x)T_s^A]{\tilde \psi}^{(q)}_{Ri}(x)+\nonumber \\
&+&{\bar \psi}^{(q)}_{Li}(x)\cdot {{\phi (x)}\over {\phi_o}}M^{(q)}_{ij}\psi
^{(q)}_{Rj}(x)+{\bar \psi}^{(q)}_{Ri}(x)M_{ij}^{(q)\dagger}{{\phi^{\dagger}(x)}
\over {\phi_o}}\cdot \psi^{(q)}_{Lj}(x)+\nonumber \\
&+&{\bar \psi}^{(q)}_{Li}(x)\cdot {{\tilde \phi (x)}\over {\phi_o}}{\tilde M}
^{(q)}_{ij}{\tilde \psi}^{(q)}_{Rj}(x)+{\bar {\tilde \psi}}^{(q)}_{Ri}(x)
{\tilde M}_{ij}^{(q)\dagger}{{{\tilde \phi}^{\dagger}(x)}\over {\phi_o}}\cdot 
\psi^{(q)}_{Lj}(x)+\nonumber \\
&+&\theta {{g_s}\over {32\pi^2}}G_{A\mu\nu}(x)\, {*}G_A^{\mu\nu}(x)=\nonumber \\
&=&......+G_{A\mu}(x)J^{\mu}_{sA}(x)+A_{\mu}(x) {\tilde j}^{\mu}
_{(em)}(x)+Z_{\mu}(x){\tilde j}^{\mu}_{(NC)}(x) +\nonumber \\
&+&[W_{+\mu}(x){\tilde j}^{\mu}_{(CC)-}(x)+W_{-\, \mu}(x){\tilde j}^{\mu}
_{(CC)+}],\nonumber \\
&&{}\nonumber \\
J^{\mu}_{sA}(x)&=&{\bar \psi}^{(q)}_{Li}(x)\gamma^{\mu}iT^A_s\psi^{(q)}_{Li}(x)+
{\bar \psi}^{(q)}_{Ri}(x)\gamma^{\mu}iT^A_s\psi^{(q)}_{Ri}(x)+{\bar {\tilde
\psi}}^{(q)}_{Ri}(x)\gamma^{\mu}iT^A_s{\tilde \psi}^{(q)}_{Ri}(x),\nonumber \\
{\tilde j}^{\mu}_{(em)}(x)&=&J^{\mu}_{w3}(x)+J^{\mu}_{Y_w}(x)=\nonumber \\
&=&{\bar \psi}^{(l)}_{Li}(x)\gamma^{\mu}Q_{em}\psi^{(l)}_{Li}(x)
+{\bar \psi}^{(l)}_{Ri}(x)\gamma^{\mu}iY_w\psi^{(l)}_{Ri}(x)+{\bar \psi}^{(q)}
_{Li}(x)\gamma^{\mu}Q_{em}\psi^{(q)}_{Li}(x)+\nonumber \\
&+&{\bar \psi}^{(q)}_{Ri}(x)\gamma^{\mu}iY_w\psi^{(q)}_{Ri}(x)+{\bar {\tilde
\psi}}^{(q)}_{Ri}(x)\gamma^{\mu}iY_w{\tilde \psi}^{(q)}_{Ri}(x),\nonumber \\
{\tilde j}^{\mu}_{(NC)}(x)&=&cot\, \theta_w\, J^{\mu}_{w3}(x)-tg\, \theta_w\,
J^{\mu}_{Y_w}(x)=\nonumber \\
&=&{\bar \psi}^{(l)}_{Li}(x)\gamma^{\mu}Q_Z\psi^{(l)}_{Li}(x)-tg\, \theta_w
{\bar \psi}^{(l)}_{Ri}(x)\gamma^{\mu}iY_w\psi^{(l)}_{Ri}(x)+{\bar \psi}^{(q)}
_{Li}(x)\gamma^{\mu}Q_Z\psi^{(q)}_{Li}(x)-\nonumber \\
&-&tg\, \theta_w {\bar \psi}^{(q)}_{Ri}(x)\gamma^{\mu}iY_w\psi^{(q)}_{Ri}(x)-
tg\, \theta_w {\bar {\tilde \psi}}^{(q)}_{Ri}(x)\gamma^{\mu}iY_w{\tilde \psi}
^{(q)}_{Ri}(x),\nonumber \\
{\tilde j}^{\mu}_{(CC)\mp}(x)&=&{1\over {\sqrt{2}}}[J^{\mu}_{w1}(x)\pm i J^{\mu}
_{w2}(x)]=\nonumber \\
&=&{\bar \psi}^{(l)}_{Li}(x)\gamma^{\mu}iT^{\mp}_w\psi^{(l)}_{Li}(x)
+{\bar \psi}^{(q)}_{Li}(x)\gamma^{\mu}iT^{\mp}_w\psi^{(q)}_{Li}(x),\nonumber \\
&&[alternative\, notation\, {\tilde j}^{\mu}_{(CH)}={\tilde j}^{\mu}_{(CC)-},
\quad {\tilde j}^{{*}\mu}_{(CH)}={\tilde j}^{\mu}_{(CC)+}].
\label{17}
\end{eqnarray}

As a consequence of the spontaneous symmetry breaking with the Higgs
mechanism, one can make a field-dependent SU(2) gauge transformation
$U^{(\theta )}_w(x)=e^{\theta_a(x){\hat T}^a_w}$, ${\tilde U}^{(\theta )}_w(x)=
e^{\theta_a(x)T^a_w}$, to the (not renormalizable) ``unitary gauge" where the
original $SU(3)\times SU(2)\times U(1)$ gauge symmetry is broken to $SU(3)
\times U_{em}(1)$ describing the remaining massless color and electromagnetic
interactions. The  explicit action of this gauge transformation is

\begin{eqnarray}
&&\phi (x) \mapsto \phi^{'}(x)={\tilde U}^{(\theta )\, -1}_w(x)\phi (x)=
\left( \begin{array}{c} 0 \\ \phi_o+{1\over {\sqrt{2}}} H(x) \end{array}
\right) = {1\over {\sqrt{2}}} \left( \begin{array}{c} 0 \\ v+H(x) \end{array}
\right) ,\nonumber \\
&&W_{a \mu}(x)T^a_w \mapsto W^{'}_{a \mu}(x)T^a_w=
e^{-\theta_a(x)T^a_w}(W_{b\mu}(x)T^b_w+\partial^{\mu})
e^{\theta_c(x)T^c_w},\nonumber \\
&&\psi^{(l)}_{Li}(x) \mapsto \psi ^{(l) {'}}_{Li}(x)={\tilde U}^{(\theta ) -1}
_w(x) \psi^{(l)}_{Li}(x),\nonumber \\
&&\psi^{(q)}_{Li}(x) \mapsto \psi^{(q) {'}}_{Li}(x)= {\tilde U}^{(\theta ) -1}
_w(x) \psi^{(q)}_{Li}(x).
\label{18}
\end{eqnarray}

\section
{The Lagrangian density in the unitary gauge}

Since the Lagrangian density (\ref{1}) is invariant under this field-dependent
SU(2) gauge transformation, we can rewrite it in terms of the new fields
$\phi^{'}$ (or H), $W^{'}_{a \mu}$, $\psi^{(l) {'}}_{Li}$, $\psi^{(q) {'}}
_{Li}$, and of the not trasformed ones $G_{A\mu}$, $V_{\mu}$, $\psi^{(l)}_{Ri}$, 
$\psi^{(q)}_{Ri}$, ${\tilde \psi}^{(q)}_{Ri}$. The new Lagrangian density
does not depend on the three would-be Goldstone bosons $\theta_a(x)$, which
are absorbed to generate the mass terms for the vector bosons $W^{'}_{\pm \mu}$,
$Z^{'}_{\mu}$, but not for the electromagnetic field $A^{'}_{\mu}$ [$Z^{'}
_{\mu}$ and $A^{'}_{\mu}$ are obtained from Eq.(\ref{8}) with $W^{'}_{3 \mu}$].

One gets

\begin{eqnarray}
&&D^{(W^{'},V)}_{\mu ab} \phi^{'}_b= \left( \begin{array}{c} -{i\over 2}
(v+H)W^{'}_{+ \mu} \\ {1\over {\sqrt{2}}}[\partial_{\mu}H+{{i(v+H)}\over 
{sin^2\, 2\theta_w}} Z^{'}_{\mu}] \end{array} \right) ,\nonumber \\
&&{}\nonumber \\
&&-\lambda [\phi^{{'} \dagger}\phi^{'} - \phi_o^2 ]^2=\lambda v^2 H^2
(1+{H\over {2v}})^2=-{1\over 2} m_H^2 H^2 (1+{e\over {2sin\, 2\theta_w\, m_Z}}
H)^2,\nonumber \\
&&{}\nonumber \\
&&[D^{(W^{'},V)}_{\mu} \phi^{'}]^{\dagger} D^{(W^{'},V) \mu} \phi^{'}=
\nonumber \\
&&={1\over 2}\partial_{\mu}H \partial^{\mu}H + {1\over 4}(v+H)^2 [W^{'}_{+ \mu}
W^{{'} \mu}_{-} +{2\over {sin^2\, 2\theta_w}} Z^{'}_{\mu} Z^{{'} \mu}]=
\nonumber \\
&&={1\over 2}\partial_{\mu}H \partial^{\mu}H +{1\over 4} g^2_wv^2[{\tilde W}
^{'}_{+ \mu} {\tilde W}^{{'} \mu}_{-} +{1\over {2 cos^2\, \theta_w}} {\tilde Z}
^{'}_{\mu}{\tilde Z}^{{'} \mu} ] (1+{H\over v})^2=\nonumber \\
&&={1\over 2}\partial_{\mu}H \partial^{\mu}H + [ m_W^2 {\tilde W}^{'}_{+ \mu} 
{\tilde W}^{{'} \mu}_{-} +{1\over 2} m^2_Z {\tilde Z}^{'}_{\mu} {\tilde Z}
^{{'} \mu} ] (1+{e\over {sin\, 2\theta_w\, m_Z}} H)^2,
\label{19}
\end{eqnarray}

\noindent where ($v=\sqrt{2} \phi_o$)

\begin{eqnarray}
&&m_H=v\sqrt{2\lambda}=2\phi_o \sqrt{\lambda},\quad\quad m_W={1\over 2}v g_w=
{1\over {\sqrt{2}}} \phi_o g_w,\quad \quad m_Z={{m_W}\over {cos\, \theta_w}}=
{1\over {\sqrt{2}}} \phi_o \sqrt{g^2_w+g^2_y},\nonumber \\
&&\Rightarrow v=\sqrt{2} \phi_o={{sin\, 2\theta_w\, m_Z}\over e}=
{1\over {\sqrt{\sqrt{2} G_F}}},\quad\quad
\lambda ={{e^2 m^2_H}\over {8 sin^2\, \theta_w\, m_W^2}}.={{e^2m^2_H}\over
{2sin^2\, 2\theta_wm^2_Z}}.
\label{20}
\end{eqnarray}

\noindent where $G_F$ is the Fermi constant. Therefore $g_w, g_y, \phi_o, 
\lambda$ are replaced by $e, \theta_w$ [or $G_F$], $m_Z, m_H$,
while $m_W=m_Zcos\, \theta_w$ is a derived quantity ($m_Z$ is known with a
better accuracy); experimentally one has $\alpha^{-1}=(e^2/4\pi )^{-1}=
137.0359895 \pm 0.0000061$ [in Heaviside-Lorentz units and with $\hbar = c
= 1$, so that $e$ is adimensional], 
$G_F={{g^2_w}\over {4\sqrt{2} m_W^2}}={1\over 
{2\sqrt{2}\phi_o^2}}=(1.16639 \pm 0.00002) \times 10^{-5} GeV^{-2}\approx 1/
(293 GeV)^2$, $sin^2\, \theta_w =0.23$, $m_Z=(91.1884 \pm 0.0022) GeV$, $m_H
> 65.1 GeV (95CL)$, so that  $m_W=(80.26 \pm 0.16) GeV$ [$m_W=m_Zcos\, \theta_w$
only at the tree level; radiative corrections give a 
six percent contribution],
$< \phi > =\phi_o ={v\over {\sqrt{2}}} = {1\over {2^{3/4}\sqrt{G_F}}}= 
246.221 GeV$, and, if $\rho =m_W^2/m_Z^2 cos^2\, \theta_w =1$, $\rho - 1 =
\triangle \rho =(4.1 \pm 1.55)\times 10^{-3}$.

Let us remark that the range of the electromagnetic force is infinite since
the electromagnetic field remains massless; at the quantum level the
renormalized electromagnetic coupling constant is $\alpha (r)=\alpha /
(1-{{\alpha}\over {3\pi}} log (1+{{\hbar}\over {m_erc}}))$ ($m_e$ is the
electron mass), so that $\alpha (r)\approx \alpha$ if $r$ is much higher of
the electron Compton wavelength ($r >> \hbar /m_ec$) and $\alpha (r)
\rightarrow \infty$ if one probes distances $r \approx {{\hbar}\over {m_ec}}
e^{-3\pi /\alpha} \approx 10^{-300} m$. Instead for the strong color force,
where $\alpha_s=g_s^2/4\pi$ is the coupling constant, the QCD renormalization 
gives $\alpha_s(r)=\alpha_s/ {{\alpha_s}\over {4\pi}}(11-{2\over 3}N_f)
log {{\hbar c}\over {\Lambda_s r}}$ ($N_f=6$ and $N_c=3$ are the number of
quark flavours and colors respectively; $\Lambda_s\approx 0.2-0.3 Gev$ is the
hadronic color energy scale); for$N_f=6 < 33/2$ the sign in the
denominator is opposite to the electromagnetic one, so that for $r\rightarrow 
0$ one has the ``asymptotic freedom" of quarks $\alpha_s(r)\rightarrow 0$,
while $\alpha_s(r)\rightarrow \infty$ (breakdown of QCD perturbative
expansion) for $r\rightarrow R_s={{\hbar c}\over {\Lambda_s}} \approx
10^{-15} m$ (the range of strong color interactions) signalling the
confinement of quarks and gluons. The range of weak interactions is
determined by the Compton wavelength of the W vector boson, $R_w=
{{\hbar }\over {m_Wc}}\approx 2.5\times 10^{-18}m.$. For comparison the distance
at which the standard description of the (infinite range) gravitational 
interaction is supposed to break down is the Planck length $R_p=\sqrt{ {{\hbar
G_N}\over {c^3}} }= 1.616\, \, 10^{-33}\, cm.$, where $G_N$ is the Newton 
constant.

After the field-dependent SU(2) gauge transformation $U^{(\theta )}_w(x)$,
the gauge invariant Lagrangian density (\ref{1}) becomes [remember that
$A^{'}_{\mu}=e{\tilde A}^{'}_{\mu}$, $Z^{'}_{\mu}=e{\tilde Z}^{'}_{\mu}$,
$W^{'}_{a\mu}={e\over {sin\, \theta_w}} {\tilde W}^{'}_{a\mu}$, $V_{\mu}= 
{e\over {cos\, \theta_w}} {\tilde V}_{\mu}$, $m_W=
m_Z cos\, \theta_w$, and that ${\tilde \phi}^{'}=i\tau^2\phi^{'}={1\over
{\sqrt{2}}} \left( \begin{array}{c} v+H \\ 0 \end{array} \right)$; in the
terms $-{1\over 4}{\tilde W}^{{'}\mu\nu}_a{\tilde W}^{'}_{a\mu\nu}-{1\over 4}
{\tilde V}^{\mu\nu}{\tilde V}_{\mu\nu}$ Eqs.(\ref{8}), (\ref{10}) have been
used]

\begin{eqnarray}
{\cal L}^{'}(x)&=&-{1\over {4g^2_s}}G_A^{\mu\nu}(x)G_{A\mu\nu}(x)-{1\over {4}}
{\tilde A}^{{'}\mu\nu}(x){\tilde A}^{'}_{\mu\nu}(x) -{1\over 4} {\tilde Z}
^{{'}\mu\nu}(x){\tilde Z}^{'}_{\mu\nu}(x)-\nonumber \\
&-&{1\over 2}{\tilde W}^{{'}\mu\nu}_{+}
(x){\tilde W}^{'}_{-\, \mu\nu}(x)+\nonumber \\
&+&[ m_W^2{\tilde W}^{'}_{+\mu}(x){\tilde W}^{{'} \mu}_{-}(x)+{1\over 2} m_Z^2
{\tilde Z}^{'}_{\mu}(x){\tilde Z}^{{'} \mu}(x)] (1+{e\over {sin\, 2\theta_w 
m_Z}}H(x))^2+\nonumber \\
&+& ie ({\tilde A}^{{'}\mu\nu}(x)+cot\, \theta_w {\tilde Z}^{{'}\mu\nu
}(x))
{\tilde W}^{'}_{+\, \mu}(x){\tilde W}^{'}_{-\, \nu}(x)+\nonumber \\
&+& ie [{\tilde W}_{+}^{{'}\mu\nu}(x) {\tilde W}^{'}_{-\, \mu}(x)- 
{\tilde W}^{{'}\mu\nu}_{-}(x) {\tilde W}^{'}_{+\, \mu}(x)]
({\tilde A}^{'}_{\nu}(x)+cot\, \theta_w {\tilde Z}^{'}_{\nu}(x))+\nonumber \\
&+& {{e^2}\over {2sin^2\, \theta_w}} 
[{\tilde W}^{{'}\, 2}_{+}(x){\tilde W}^{{'}\,
2}_{-}(x) -({\tilde W}^{'}_{+}(x)\cdot {\tilde W}^{'}_{-}(x))^2]-\nonumber \\
&-& e^2({\tilde A}^{{'}\mu}(x)+cot\, \theta_w {\tilde Z}^{{'}\mu}(x))({\tilde A}
^{'}_{\mu}(x)+cot\, \theta_w {\tilde Z}^{'}_{\mu}(x)) {\tilde W}^{'}_{+}(x) 
\cdot  {\tilde W}^{'}_{-}(x)+\nonumber \\
&+&e^2{\tilde W}^{{'}\mu}_{+}(x) ({\tilde A}^{'}_{\mu}(x)+cot\, \theta_w
{\tilde Z}^{'}_{\mu}(x))\, {\tilde W}^{{'}\nu}_{-}(x) ({\tilde A}^{'}_{\nu}(x)+
cot\, \theta_w{\tilde Z}^{'}_{\nu}(x))+\nonumber \\
&+&{1\over 2} \partial_{\mu} H(x)\partial^{\mu} H(x)- {1\over 2} m_H^2 H^2(x) 
(1+{e\over {2sin\, 2\theta_w m_Z}}H(x))^2+\nonumber \\
&+&{\bar \psi}^{(l){'}}_{Li}(x)i\gamma^{\mu}[\partial_{\mu}+{e\over {sin\,
\theta_w}}({\tilde W}^{'}_{+\mu}(x)T^{-}_w+{\tilde W}^{'}_{-\, \mu}(x)T^{+}_w)
-\nonumber \\
&-&ieQ_{em}{\tilde A}^{'}_{\mu}(x)-ieQ_Z{\tilde Z}^{'}_{\mu}(x)]
\psi^{(l){'}}_{Li}(x)+\nonumber \\
&+&{\bar \psi}^{(l)}_{Ri}(x)i\gamma^{\mu}[\partial_{\mu}+e({\tilde A}^{'}
_{\mu}(x)-tg\, \theta_w{\tilde Z}^{'}_{\mu}(x))Y_w]\psi^{(l)}_{Ri}(x)+
\nonumber \\
&+&(1+{e\over {sin\, 2\theta_wm_Z}} H(x)) \nonumber \\
&&[{\bar \psi}^{(l){'}}_{Li}(x)\cdot 
\left( \begin{array}{l} 0\\ 1 \end{array} \right)  M^{(l)}_{ij}\psi
^{(l)}_{Rj}(x)+{\bar \psi}^{(l)}_{Ri}(x)M_{ij}^{(l)\dagger} 
\left( \begin{array}{ll}
0 & 1 \end{array} \right)\cdot \psi^{(l){'}}_{Lj}(x)]+\nonumber \\
&+&{\bar \psi}^{(q){'}}_{Li}(x)i\gamma^{\mu}[\partial_{\mu}+{e\over {sin\, 
\theta_w}}({\tilde W}^{'}_{+\mu}(x)T^{-}_w+{\tilde W}^{'}_{-\, \mu}(x)T^{+}_w)
-\nonumber \\
&-&ieQ_{em}{\tilde A}^{'}_{\mu}(x)-ieQ_Z{\tilde Z}^{'}_{\mu}(x)
+G_{A\mu}(x)T^A_s]\psi^{(q){'}}_{Li}(x)+\nonumber \\
&+&{\bar \psi}^{(q)}_{Ri}(x)i\gamma^{\mu}[\partial_{\mu}+e({\tilde A}^{'}
_{\mu}(x)-tg\, \theta_w{\tilde Z}^{'}_{\mu}(x))Y_w
+ G_{A\mu}(x)T^A_s]\psi^{(q)}_{Ri}(x)+\nonumber \\
&+&{\bar {\tilde \psi}}^{(q)}_{Ri}(x)i\gamma^{\mu}[\partial_{\mu}+
e({\tilde A}^{'}_{\mu}(x)-tg\, \theta_w{\tilde Z}^{'}_{\mu}(x))Y_w
+G_{A\mu}(x)T_s^A]{\tilde \psi}^{(q)}_{Ri}(x)+\nonumber \\
&+&(1+{e\over {sin\, 2\theta_wm_Z}} H(x))\nonumber \\
&& [{\bar \psi}^{(q){'}}_{Li}(x)\cdot 
\left( \begin{array}{l} 0 \\ 1 \end{array} \right) M^{(q)}_{ij}\psi
^{(q)}_{Rj}(x)+{\bar \psi}^{(q)}_{Ri}(x)M_{ij}^{(q)\dagger} 
\left( \begin{array}{ll}
0 & 1 \end{array} \right)\cdot \psi^{(q){'}}_{Lj}(x)+\nonumber \\
&+&{\bar \psi}^{(q){'}}_{Li}(x)\cdot \left( \begin{array}{l} 1 \\ 0 \end{array}
\right){\tilde M}^{(q)}_{ij}{\tilde \psi}^{(q)}_{Rj}(x)+{\bar {\tilde \psi}}
^{(q)}_{Ri}(x){\tilde M}_{ij}^{(q)\dagger} \left( \begin{array}{ll} 1 & 0 
\end{array} \right) \cdot \psi^{(q){'}}_{Lj}(x)]+\nonumber \\
&+&\theta {{g_s}\over {32\pi^2}}G_{A\mu\nu}(x)\, {*}G_A^{\mu\nu}(x)=\nonumber \\
&=&....+G_{A\mu}(x)J^{\mu}_{sA}(x)+e{\tilde A}^{'}_{\mu}(x){\tilde j}^{{'}\mu}
_{(em)}(x)+e{\tilde Z}^{'}_{\mu}(x){\tilde j}^{{'}\mu}_{(NC)}(x)+\nonumber \\
&+&{e\over
{sin\, \theta_w}} [{\tilde W}^{'}_{+\, \mu}(x){\tilde j}^{{'}\mu}_{(CC)-}(x)+
{\tilde W}^{'}_{-\, \mu}(x){\tilde j}^{{'}\mu}_{(CC)+}(x)],\nonumber \\
&&{}\nonumber \\
J^{\mu}_{sA}(x)&=&{\bar \psi}^{(q)}_{Li}(x)\gamma^{\mu}iT^A_s\psi^{(q)}_{Li}(x)+
{\bar \psi}^{(q)}_{Ri}(x)\gamma^{\mu}iT^A_s\psi^{(q)}_{Ri}(x)+{\bar {\tilde
\psi}}^{(q)}_{Ri}(x)\gamma^{\mu}iT^A_s{\tilde \psi}^{(q)}_{Ri}(x),\nonumber \\
{\tilde j}^{{'}\mu}_{(em)}(x)&=&
{\bar \psi}^{(l){'}}_{Li}(x)\gamma^{\mu}Q_{em}\psi^{(l){'}}_{Li}(x)
+{\bar \psi}^{(l)}_{Ri}(x)\gamma^{\mu}iY_w\psi^{(l)}_{Ri}(x)+{\bar \psi}^{(q)
{'}}_{Li}(x)\gamma^{\mu}Q_{em}\psi^{(q){'}}_{Li}(x)+\nonumber \\
&+&{\bar \psi}^{(q)}_{Ri}(x)\gamma^{\mu}iY_w\psi^{(q)}_{Ri}(x)+{\bar {\tilde
\psi}}^{(q)}_{Ri}(x)\gamma^{\mu}iY_w{\tilde \psi}^{(q)}_{Ri}(x),\nonumber \\
{\tilde j}^{{'}\mu}_{(NC)}(x)&=&
{\bar \psi}^{(l){'}}_{Li}(x)\gamma^{\mu}Q_Z\psi^{(l){'}}_{Li}(x)-tg\, \theta_w
{\bar \psi}^{(l)}_{Ri}(x)\gamma^{\mu}iY_w\psi^{(l)}_{Ri}(x)+{\bar \psi}^{(q){'}}
_{Li}(x)\gamma^{\mu}Q_Z\psi^{(q){'}}_{Li}(x)-\nonumber \\
&-&tg\, \theta_w {\bar \psi}^{(q)}_{Ri}(x)\gamma^{\mu}iY_w\psi^{(q)}_{Ri}(x)-
tg\, \theta_w {\bar {\tilde \psi}}^{(q)}_{Ri}(x)\gamma^{\mu}iY_w{\tilde \psi}
^{(q)}_{Ri}(x),\nonumber \\
{\tilde j}^{{'}\mu}_{(CC)\mp}(x)&=&
{\bar \psi}^{(l){'}}_{Li}(x)\gamma^{\mu}iT^{\mp}_w\psi^{(l){'}}_{Li}(x)
+{\bar \psi}^{(q){'}}_{Li}(x)\gamma^{\mu}iT^{\mp}_w\psi^{(q){'}}_{Li}(x).
\label{21}
\end{eqnarray}

The complete set of fermionic currents is 
[these equations define the currents $J^{{'}\mu}_{wa}$ and $J^{{'}\mu}_{Y_w}$]

\begin{eqnarray}
{\tilde j}^{{'}\mu}_{w\, a}(x)&=&
{{g_w}\over e}J^{{'}\mu}_{wa}(x)={1\over {sin\, \theta_w}}
J^{{'}\mu}_{wa}(x)={\bar \psi}^{(l){'}}_{Li}(x)\gamma^{\mu}iT^a_w\psi^{(l){'}}
_{Li}(x)+{\bar \psi}^{(q){'}}_{Li}(x)\gamma^{\mu}iT^a_w\psi^{(q){'}}_{Li}(x),
\nonumber \\
&&{\tilde j}^{{'}\mu}_{w\, 1}(x)={1\over {\sqrt{2}}}[{\tilde j}^{{'}\mu}
_{(CC)+}(x)+{\tilde j}^{{'}\mu}_{(CC)-}(x)],\quad\quad {\tilde j}^{{'}\mu}
_{w\, 2}(x)={i\over {\sqrt{2}}}[{\tilde j}^{{'}\mu}_{(CC)+}(x)-
{\tilde j}^{{'}\mu}_{(CC)-}(x)],\nonumber \\
&&{\tilde j}^{{'}\mu}_{w\, 3}(x)={1\over {sin\, \theta_w}}J^{{'}\mu}_{w3}(x)=
sin\, \theta_w [sin\, \theta_w {\tilde j}^{{'}\mu}_{(em)}(x)-cos\, \theta_w 
{\tilde j}^{{'}\mu}_{(NC)}(x)]=\nonumber \\
&=&{\bar \psi}^{(l){'}}_{Li}(x)\gamma^{\mu}iT^3_w\psi^{(l){'}}_{Li}(x)+{\bar 
\psi}^{(q){'}}_{Li}(x)\gamma^{\mu}iT^3_w\psi^{(q){'}}_{Li}(x),\nonumber \\
&&{}\nonumber \\
&&{\tilde j}^{{'}\mu}_{Y_w}(x)={1\over {cos\, \theta_w}}J^{{'}\mu}_{Y_w}(x)=
cos\, \theta_w [cos\, \theta_w {\tilde j}^{{'}
\mu}_{(em)}(x)-sin\, \theta_w {\tilde j}^{{'}\mu}_{(NC)}(x)]=\nonumber \\
&=&{\bar \psi}^{(l){'}}_{Li}(x)\gamma^{\mu}iY_w\psi^{(l){'}}_{Li}(x)+{\bar 
\psi}^{(l)}_{Ri}(x)\gamma^{\mu}iY_w\psi^{(l)}_{Ri}(x)+\nonumber \\
&&+{\bar \psi}^{(q){'}}_{Li}(x)\gamma^{\mu}iY_w\psi^{(q){'}}_{Li}(x)+{\bar 
\psi}^{(q)}_{Ri}(x)\gamma^{\mu}iY_w\psi^{(q)}_{Ri}(x)+{\bar {\tilde \psi}}
^{(q)}_{Ri}(x)\gamma^{\mu}iY_w{\tilde \psi}^{(q)}_{Ri}(x),\nonumber \\
&&{}\nonumber \\
{\tilde j}^{{'}\mu}_{(em)}(x)&=& cos\, \theta_w {\tilde j}^{{'}\mu}_{Y_w}(x)+
sin\, \theta_w {\tilde j}^{{'}\mu}_{w3}(x),\nonumber \\
{\tilde j}^{{'}\mu}_{(NC)}(x)&=& -sin\, \theta_w {\tilde j}^{{'}\mu}_{Y_w}(x) 
+cos\, \theta_w {\tilde j}^{{'}\mu}_{w3}(x),\nonumber \\
{\tilde j}^{{'}\mu}_{(CC)\pm}(x)&=&{1\over {\sqrt{2}}} [{\tilde j}^{{'}\mu}_{w1}
(x)\mp i {\tilde j}^{{'}\mu}_{w2}(x)].
\label{29}
\end{eqnarray}

One can also present the following quartic terms in a different way: 

$-e^2({\tilde A}^{{'}\mu}(x)+cot\, \theta_w {\tilde Z}^{{'}\mu}(x))({\tilde A}
^{'}_{\mu}(x)+cot\, \theta_w {\tilde Z}^{'}_{\mu}(x)) {\tilde W}^{'}_{+}(x) 
\cdot  {\tilde W}^{'}_{-}(x)+e^2{\tilde W}^{{'}\mu}_{+}(x) ({\tilde A}^{'}
_{\mu}(x)+cot\, \theta_w{\tilde Z}^{'}_{\mu}(x))\, {\tilde W}^{{'}\nu}_{-}(x) 
({\tilde A}^{'}_{\nu}(x)+cot\, \theta_w{\tilde Z}^{'}_{\nu}(x))=-e^2\lbrace
{\tilde A}^{{'} 2}(x) {\tilde W}_{+}^{'}(x)\cdot {\tilde W}_{-}^{'}(x)-{\tilde
W}_{+}^{'}(x)\cdot {\tilde A}^{'}(x) {\tilde W}_{-}^{'}(x)\cdot {\tilde A}^{'}
(x) +cot^2\, \theta_w ({\tilde Z}^{{'} 2}(x){\tilde W}_{+}^{'}(x)\cdot {\tilde 
W}_{-}^{'}(x)-{\tilde W}_{+}^{'}(x)\cdot {\tilde Z}^{'}(x) {\tilde W}_{-}^{'}
(x)\cdot {\tilde Z}^{'}(x)) +cot\, \theta_w (2{\tilde A}^{'}(x)\cdot {\tilde Z}
^{'}(x)-{\tilde W}_{+}^{'}(x)\cdot {\tilde A}^{'}(x) {\tilde W}_{-}^{'}
(x)\cdot {\tilde Z}^{'}(x))-{\tilde W}_{+}^{'}(x)\cdot {\tilde Z}^{'}(x) 
{\tilde W}_{-}^{'}(x)\cdot {\tilde A}^{'}(x))\rbrace $.

It can be shown that in the unitary gauge the complex mass matrices containing 
the Yukawa couplings (replacing the not-gauge-invariant Dirac mass terms) can be
diagonalized by means of unitary left and right matrices [$S^{(l)-1}_{L,R}=
S^{(l)\dagger}_{L,R}$, $S^{(q)-1}_{L,R}=S^{(q)\dagger}_{L,R}$, ${\tilde S}
^{(q)-1}_{L,R}={\tilde S}^{(q)\dagger}_{L,R}$]

\begin{eqnarray}
&&(1+{e\over {sin\, 2\theta_wm_Z}} H(x)) [{\bar \psi}^{(l){'}}_{Li}(x)\cdot 
\left( \begin{array}{l} 0\\ 1 \end{array} \right)  M^{(l)}_{ij}\psi
^{(l)}_{Rj}(x)+{\bar \psi}^{(l)}_{Ri}(x)M_{ij}^{(l)\dagger} 
\left( \begin{array}{ll}
0 & 1 \end{array} \right)\cdot \psi^{(l){'}}_{Lj}(x)]=\nonumber \\
&&=(1+{e\over {sin\, 2\theta_wm_Z}} H(x)) [\left( \begin{array}{lll} {\bar e}_L
^{(m)}(x) & {\bar \mu}_L^{(m)}(x) & {\bar \tau}_L^{(m)}(x) \end{array} \right) 
\left( \begin{array}{ccc} m_e & 0 & 0 \\ 0 & m_{\mu} & 0 \\ 0 & 0 & m_{\tau}
\end{array} \right) \left( \begin{array}{c} e_R^{(m)}(x) \\ \mu^{(m)}_R(x) \\
\tau_R^{(m)}(x)\end{array} \right) + \nonumber \\
&&+\left( \begin{array}{lll} {\bar e}_R^{(m)}(x) & {\bar \mu}_R^{(m)}(x) & 
{\bar \tau}_R^{(m)}(x) \end{array} \right) 
\left( \begin{array}{ccc} m_e & 0 & 0 \\ 0 & m_{\mu} & 0 \\ 0 & 0 & m_{\tau}
\end{array} \right) \left( \begin{array}{c} e_L^{(m)}(x) \\ \mu^{(m)}_L(x) \\
\tau_L^{(m)}(x)\end{array} \right) ]=\nonumber \\
&&=(1+{e\over {sin\, 2\theta_wm_Z}} H(x)) [m_e {\bar e}^{(m)}(x) e^{(m)}(x) + 
m_{\mu} {\bar \mu}^{(m)}(x) \mu^{(m)}(x) + m_{\tau} {\bar \tau}^{(m)}(x) 
\tau^{(m)}(x)]\nonumber \\
&&{}\nonumber \\
&&(1+{e\over {sin\, 2\theta_wm_Z}} H(x)) [{\bar \psi}^{(q){'}}_{Li}(x)\cdot 
\left( \begin{array}{l} 0 \\ 1 \end{array} \right) M^{(q)}_{ij}\psi
^{(q)}_{Rj}(x)+{\bar \psi}^{(q)}_{Ri}(x)M_{ij}^{(q)\dagger} 
\left( \begin{array}{ll}
0 & 1 \end{array} \right)\cdot \psi^{(q){'}}_{Lj}(x)+\nonumber \\
&&+{\bar \psi}^{(q){'}}_{Li}(x)\cdot \left( \begin{array}{l} 1 \\ 0 \end{array}
\right){\tilde M}^{(q)}_{ij}{\tilde \psi}^{(q)}_{Rj}(x)+{\bar {\tilde \psi}}
^{(q)}_{Ri}(x){\tilde M}_{ij}^{(q)\dagger} \left( \begin{array}{ll} 1 & 0 
\end{array} \right)\cdot \psi^{(q){'}}_{Lj}(x)]=\nonumber \\
&&=(1+{e\over {sin\, 2\theta_wm_Z}} H(x)) [\left( \begin{array}{ccc} {\bar d}
^{(m)}_L(x) & {\bar s}^{(m)}_L(x) & {\bar b}^{(m)}_L(x) \end{array} \right)  
\left( \begin{array}{ccc} m_d & 0 & 0 \\ 0 & m_s & 0 \\ 0 & 0 & m_b \end{array}
\right) \left( \begin{array}{c} d_R^{(m)}(x) \\ s_R^{(m)}(x) \\ b_R^{(m)}(x) 
\end{array} \right) + \nonumber \\
&&+\left( \begin{array}{ccc} {\bar u}_L^{(m)}(x) & {\bar c}_L^{(m)}(x) &
{\bar t}_L^{(m)}(x) \end{array} \right) \left( \begin{array}{ccc} m_u & 0 & 0 \\
0 & m_c & 0 \\ 0 & 0 & m_t \end{array} \right) \left( \begin{array}{c} u_R
^{(m)}(x) \\ c_R^{(m)}(x) \\ t_R^{(m)}(x) \end{array} \right) +\nonumber \\
&&+\left( \begin{array}{ccc} {\bar d}^{(m)}_R(x) & {\bar s}^{(m)}_R(x) & 
{\bar b}^{(m)}_R(x) \end{array} \right)  \left(
\begin{array}{ccc} m_d & 0 & 0 \\ 0 & m_s & 0 \\ 0 & 0 & m_b \end{array}
\right) \left( \begin{array}{c} d_L^{(m)}(x) \\ s_L^{(m)}(x) \\ b_L^{(m)}(x) 
\end{array} \right) +\nonumber \\
&&+ \left( \begin{array}{ccc} {\bar u}_R^{(m)}(x) & {\bar c}_R^{(m)}(x) &
{\bar t}_R^{(m)}(x) \end{array} \right) \left( \begin{array}{ccc} m_u & 0 & 0 \\
0 & m_c & 0 \\ 0 & 0 & m_t \end{array} \right) \left( \begin{array}{c} u_L
^{(m)}(x) \\ c_L^{(m)}(x) \\ t_L^{(m)}(x) \end{array} \right)]=\nonumber \\
&&=(1+{e\over {sin\, 2\theta_wm_Z}} H(x)) [m_d {\bar d}^{(m)}(x) d^{(m)}(x) + m_s
{\bar s}^{(m)}(x) s^{(m)}(x) + m_b {\bar b}^{(m)}(x) b^{(m)}(x) + \nonumber \\
&&+m_u {\bar u}^{(m)}(x)u^{(m)}(x) + m_c {\bar c}^{(m)}(x) c^{(m)}(x) + 
m_t {\bar t}^{(m)}(x) t^{(m)}(x)],
\label{22}
\end{eqnarray}

\noindent where the mass eigenstates of leptons and quarks are defined by

\begin{eqnarray}
&&\left( \begin{array}{c} e_L^{(m)} \\ \mu_L^{(m)} \\ \tau_L^{(m)} \end{array}
\right) = S^{(l)\dagger}_L \left( \begin{array}{c} e^{'}_L \\ \mu^{'}_L \\
\tau^{'}_L \end{array} \right) ,\quad\quad
\left( \begin{array}{c} e_R^{(m)} \\ \mu_R^{(m)} \\ \tau_R^{(m)} \end{array}
\right) = S^{(l)\dagger}_R \left( \begin{array}{c} e_R \\ \mu_R \\
\tau_R \end{array} \right) ,\quad\quad
\left( \begin{array}{c} \nu_{eL}^{(m)} \\ \nu_{\mu L}^{(m)} \\ \nu_{\tau L}
^{(m)} \end{array} \right) = S^{(l)\dagger}_L \left( \begin{array}{c} \nu_{eL}
^{'}\\ \nu_{\mu L}^{'} \\ \nu_{\tau L}^{'} \end{array} \right) ,\nonumber \\
&&{}\nonumber \\
&&\left( \begin{array}{c} d_L^{(m)} \\ s_L^{(m)} \\ b_L
^{(m)} \end{array} \right) = S^{(q)\dagger}_L \left( \begin{array}{c} d_L
^{'}\\ s_L^{'} \\ b_L^{'} \end{array} \right) ,\quad\quad
\left( \begin{array}{c} d_R^{(m)} \\ s_R^{(m)} \\ b_R
^{(m)} \end{array} \right) = S^{(q)\dagger}_R \left( \begin{array}{c} d_R
\\ s_R \\ b_R \end{array} \right) ,\nonumber \\
&&\left( \begin{array}{c} u_L^{(m)} \\ c_L^{(m)} \\ t_L
^{(m)} \end{array} \right) = {\tilde S}^{(q)\dagger}_L \left( \begin{array}{c} 
u_L^{'}\\ c_L^{'} \\ t_L^{'} \end{array} \right) ,\quad\quad
\left( \begin{array}{c} u_R^{(m)} \\ c_R^{(m)} \\ t_R
^{(m)} \end{array} \right) = {\tilde S}^{(q)\dagger}_L \left( \begin{array}{c} 
u_R\\ c_R \\ t_R \end{array} \right) .
\label{23}
\end{eqnarray}

The parameters $m_e, m_{\mu}, m_{\tau}, m_d, m_s, m_b, m_u, m_c, m_t$
[$m_{\nu_e}=m_{\nu_{\mu}}=m_{\nu_{\tau}}=0$], are called lepton and ``current"
quark masses. For leptons they coincide with the asymptotic free (on-shell)
states, which however do not exist for quarks according to the confinement
hypothesis. For quarks, at the quantum level, these parameters are thought
to be running with the renormalization scale, $m_q(\mu )$, usually in the 
$\overline{MS}$ renormalization scheme; for the light quarks u, d, s, one
chooses $\mu = 1 Gev$, while for c and b one can choose $m_q=m_q(\mu =m_q)$ due
to perturbative QCD ($m_t$ is still a preliminary result). The chiral symmetry
properties of u, d, s, allow to fix in a scale independent way (QCD does not
feel flavour) the ratios $2m_s/(m_d+m_u)=22.6\pm 3.3$, $(m_d-m_u)/(m_d+m_u)=
0.25\pm 0.04$. For heavy quarks one can define the mass $m^{pole}_q$ associated
with a perturbative quark propagator (a kinematical on-shell mass like for
leptons), $m_q^{pole}=m_q(\mu =m_q^{pole})[1+{4\over 3\pi} \alpha_s(m_q^{pole})
+)(\alpha_s^2)]$ (note that $m_t^{pole}-m_t(\mu =m_t)= 7 Gev$). For the study
of light hadrons (bound states of quarks) one uses also the ``constituent"
quark masses, $m^{const}_q=m_q+ \Lambda_s/c^2$, since $\Lambda_s$ gives the
order of magnitude of the quark kinetic energy; in this way, even if one
sends to zero the current mass of u, d, s, quarks, one still has for the
proton and the neutron $m_p\approx 2m_u^{const}+m_d^{const}\approx m_n$.
The experimental values of the lepton and current quark masses are

\begin{eqnarray}
&&m_e=(0.51099906\pm 0.00000015) MeV,\quad\quad m_{\mu}=(105.658389\pm 0.000034)
MeV, \nonumber \\
&&m_{\tau}=(1777.0\pm 0.3) MeV,\nonumber \\
&&m_{\nu_e} < 7.0 eV (95CL),\quad\quad m_{\nu_{\mu}} < 0.27 MeV (90CL),
\quad\quad m_{\nu_{\tau}} < 24 MeV (95CL),\nonumber \\ 
&&m_d(1 GeV)=(8.5\pm 2.5) MeV,\quad\quad m_s(1 GeV)=(180\pm 25)MeV,\nonumber \\
&&m_b=(4.25\pm 0.10) GeV, \quad\quad m_u(1 Gev)=(5.0\pm 2.5) MeV, \nonumber \\ 
&&m_c=(1.25\pm 0.05) GeV,\quad\quad m_t=(175\pm 6) GeV.
\label{24}
\end{eqnarray}

Finally, by using the mass eigenstates, the unitary gauge Lagrangian density
(\ref{21}) becomes

\begin{eqnarray}
{\tilde {\cal L}}^{'}(x)&=&
-{1\over {4g^2_s}}G_A^{\mu\nu}(x)G_{A\mu\nu}(x)-{1\over {4}}
{\tilde A}^{{'}\mu\nu}(x){\tilde A}^{'}_{\mu\nu}(x) -{1\over 4} {\tilde Z}
^{{'}\mu\nu}(x){\tilde Z}^{'}_{\mu\nu}(x)-\nonumber \\
&-&{1\over 2} {\tilde W}^{{'}\mu\nu}_{+}(x){\tilde W}^{'}_{-\, \mu\nu}(x)+\nonumber \\
&+&[ m_W^2{\tilde W}^{'}_{+\mu}(x){\tilde W}^{{'} \mu}_{-}(x)+{1\over 2} m_Z^2
{\tilde Z}^{'}_{\mu}(x){\tilde Z}^{{'} \mu}(x)] (1+{e\over {sin\, 2\theta_w 
m_Z}}H(x))^2+\nonumber \\
&+& ie ({\tilde A}^{{'}\mu\nu}(x)+cot\, \theta_w {\tilde Z}^{{'}\mu\nu
}(x)){\tilde W}^{'}_{+\, \mu}(x){\tilde W}^{'}_{-\, \nu}(x)+\nonumber \\
&+& ie [{\tilde W}_{+}^{{'}\mu\nu}(x) {\tilde W}^{'}_{-\, \mu}(x)
{\tilde W}^{{'}\mu\nu}_{-}(x) {\tilde W}^{'}_{+\, \mu}
(x)] ({\tilde A}^{'}_{\nu}(x)+cot\, \theta_w {\tilde Z}^{'}_{\nu}(x))+
\nonumber \\
&+& {{e^2}\over {2sin^2\, \theta_w}} [{\tilde W}^{{'}\, 2}_{+}(x){\tilde W}^{{'}\,
2}_{-}(x) -({\tilde W}^{'}_{+}(x)\cdot {\tilde W}^{'}_{-}(x))^2]-\nonumber \\
&-& e^2({\tilde A}^{{'}\mu}(x)+cot\, \theta_w {\tilde Z}^{{'}\mu}(x))({\tilde A}
^{'}_{\mu}(x)+cot\, \theta_w {\tilde Z}^{'}_{\mu}(x)) {\tilde W}^{'}_{+}(x) 
\cdot  {\tilde W}^{'}_{-}(x)+\nonumber \\
&+&e^2{\tilde W}^{{'}\mu}_{+}(x) ({\tilde A}^{'}_{\mu}(x)+cot\, \theta_w
{\tilde Z}^{'}_{\mu}(x))\, {\tilde W}^{{'}\nu}_{-}(x) ({\tilde A}^{'}_{\nu}(x)+
cot\, \theta_w{\tilde Z}^{'}_{\nu}(x))+\nonumber \\
&+&{1\over 2} \partial_{\mu} H(x)\partial^{\mu} H(x)- {1\over 2} m_H^2 H^2(x) 
(1+{e\over {2sin\, 2\theta_w m_Z}}H(x))^2+\nonumber \\
&+& \left( \begin{array}{ccc} {\bar \nu}^{(m)}_e(x) & {\bar \nu}^{(m)}_{\mu}(x)
& {\bar \nu}^{(m)}_{\tau}(x) \end{array} \right) i\gamma^{\mu}\partial_{\mu}
{1\over 2}(1-\gamma_5)
\left( \begin{array}{c} \nu^{(m)}_e(x)\\ \nu^{(m)}_{\mu}(x)\\ \nu^{(m)}_{\tau}
(x) \end{array} \right) +\nonumber \\
&+& \left( \begin{array}{ccc} {\bar e}^{(m)}(x) & {\bar \mu}^{(m)}(x) & {\bar
\tau}^{(m)}(x) \end{array} \right) \nonumber \\
&&[i\gamma^{\mu}\partial_{\mu} + (1+ {e\over
{sin\, 2\theta_w m_Z}} H(x)) \left( \begin{array}{ccc} 
m_e & 0 & 0\\ 0 & m_{\mu}
& 0\\ 0 & 0 & m_{\tau} \end{array} \right) ] \left( \begin{array}{c} e^{(m)}
(x) \\ \mu^{(m)}(x) \\ \tau^{(m)}(x) \end{array} \right) +\nonumber \\
&+&\left( \begin{array}{ccc} {\bar u}^{(m)}(x) & {\bar c}^{(m)}(x) & {\bar
t}^{(m)}(x) \end{array} \right) \nonumber \\
&&[i\gamma^{\mu}\partial_{\mu} + (1+ {e\over
{sin\, 2\theta_w m_Z}} H(x)) \left( \begin{array}{ccc} m_u & 0 & 0\\ 0 & m_c
& 0\\ 0 & 0 & m_t \end{array} \right) ] \left( \begin{array}{c} u^{(m)}
(x) \\ c^{(m)}(x) \\ t^{(m)}(x) \end{array} \right) +\nonumber \\
&+&\left( \begin{array}{ccc} {\bar d}^{(m)}(x) & {\bar s}^{(m)}(x) & {\bar
b}^{(m)}(x) \end{array} \right) \nonumber \\
&&[i\gamma^{\mu}\partial_{\mu} + (1+ {e\over
{sin\, 2\theta_w m_Z}} H(x)) \left( \begin{array}{ccc} m_d & 0 & 0\\ 0 & m_s
& 0\\ 0 & 0 & m_b \end{array} \right) ] \left( \begin{array}{c} d^{(m)}
(x) \\ s^{(m)}(x) \\ b^{(m)}(x) \end{array} \right) +\nonumber \\
&+&G_{A\mu}(x) J^{\mu}_{sA}(x)+
e{\tilde A}^{'}_{\mu}(x){\tilde j}^{{'}\mu}_{(em)}(x)+e{\tilde Z}^{'}_{\mu}
(x) {\tilde j}^{{'}\mu}_{(NC)}(x)+\nonumber \\
&+&{e\over {sin\, \theta_w}} ({\tilde W}^{'}
_{+\, \mu}(x){\tilde j}^{{'}\mu}_{(CC)\, -}(x)+ {\tilde W}^{'}_{-\, \mu}(x) 
{\tilde j}^{{'}\mu}_{(CC)\, +}(x))+\nonumber \\ 
&+&\theta {{g_s}\over {32\pi^2}}G_{A\mu\nu}(x)\, {*}G_A^{\mu\nu}(x),
\label{25}
\end{eqnarray}

\noindent with the electromagnetic, neutral, charge changing and strong
currents defined by the following equations

\begin{eqnarray}
{\tilde j}^{{'}\mu}_{(em)}(x)&=& 
\left( \begin{array}{ccc} {\bar e}^{(m)}(x) & {\bar \mu}
^{(m)}(x) & {\bar \tau}^{(m)}(x) \end{array} \right) \gamma^{\mu} \left(
\begin{array}{c} e^{(m)}(x)\\ \mu^{(m)}(x)\\ \tau^{(m)}(x) \end{array}
\right) +\nonumber \\
&&+{1\over 3} \left( \begin{array}{ccc} {\bar d}^{(m)}(x) & {\bar s}
^{(m)}(x) & {\bar b}^{(m)}(x) \end{array} \right) \gamma^{\mu} \left(
\begin{array}{c} d^{(m)}(x)\\ s^{(m)}(x)\\ b^{(m)}(x) \end{array}
\right) -\nonumber \\
&-&{2\over 3}\left( \begin{array}{ccc} {\bar u}^{(m)}(x) & {\bar c}
^{(m)}(x) & {\bar t}^{(m)}(x) \end{array} \right) \gamma^{\mu} \left(
\begin{array}{c} u^{(m)}(x)\\ c^{(m)}(x)\\ t^{(m)}(x) \end{array}
\right) ,\nonumber \\
&&{}\nonumber \\
{\tilde j}^{{'}\mu}_{(NC)}(x)&=&
\left( \begin{array}{ccc} {\bar \nu}_e^{(m)}(x) & {\bar \nu}
^{(m)}_{\mu}(x) & {\bar \nu}^{(m)}_{\tau}(x) \end{array} \right) {{\gamma^{\mu}
{1\over 2}(1-\gamma_5)}\over {sin\, 2\theta_w}} \left( \begin{array}{c}
\nu_e^{(m)}(x) \\ \nu_{\mu}^{(m)}(x) \\ \nu_{\tau}^{(m)}(x) \end{array}
\right)+\nonumber \\
&&+\left( \begin{array}{ccc} {\bar e}^{(m)}(x) & {\bar \mu}
^{(m)}(x) & {\bar \tau}^{(m)}(x) \end{array} \right) {{(2sin^2\, \theta_w-
{1\over 2})\gamma^{\mu}+{1\over 2}\gamma^{\mu}\gamma_5}\over {sin\, 2\theta_w}} 
\left( \begin{array}{c} e^{(m)}(x) \\ \mu^{(m)}(x) \\ \tau^{(m)}(x) \end{array}
\right)+\nonumber \\
&&+\left( \begin{array}{ccc} {\bar u}^{(m)}(x) & {\bar c}
^{(m)}(x) & {\bar t}^{(m)}(x) \end{array} \right) {{({1\over 2}-{4\over 3}
sin^2\, \theta_w)\gamma^{\mu}-{1\over 2} \gamma^{\mu}\gamma_5}\over {sin\, 
2\theta_w}} \left( \begin{array}{c} u^{(m)}(x) \\ c^{(m)}(x) \\ t^{(m)}(x) 
\end{array} \right)+\nonumber \\
&&+\left( \begin{array}{ccc} {\bar d}^{(m)}(x) & {\bar s}
^{(m)}(x) & {\bar b}^{(m)}(x) \end{array} \right) {{({2\over 3}sin^2\,
\theta_w-{1\over 2})\gamma^{\mu}-{1\over 2}\gamma^{\mu}\gamma_5}\over {sin\, 
2\theta_w}} \left( \begin{array}{c} d^{(m)}(x) \\ s^{(m)}(x) \\ b^{(m)}(x) 
\end{array} \right) ,\nonumber \\
&&{}\nonumber \\
{\tilde j}^{{'}\mu}_{(CC)\, -}(x)&=&
\left( \begin{array}{ccc} {\bar e}^{(m)}(x) & {\bar \mu}
^{(m)}(x) & {\bar \tau}^{(m)}(x) \end{array} \right) \gamma^{\mu} {1\over 2}
(1-\gamma_5) \left( \begin{array}{c} \nu_e^{(m)}(x)\\ \nu_{\mu}^{(m)}(x)\\
\nu_{\tau}^{(m)}(x) \end{array} \right) +\nonumber \\
&&+\left( \begin{array}{ccc} {\bar u}^{(m)}(x) & {\bar c}^{(m)}(x) & {\bar t}
^{(m)}(x) \end{array} \right) \gamma^{\mu} {1\over 2}(1-\gamma_5) V_{CKM}
\left( \begin{array}{c} d^{(m)}(x)\\ s^{(m)}(x)\\ b^{(m)}(x) \end{array}
\right) ,\nonumber \\
{\tilde j}^{{'}\mu}_{(CC)\, +}(x)&=&
\left( \begin{array}{ccc} {\bar \nu}_e^{(m)}(x) & {\bar 
\nu}_{\mu}^{(m)}(x) & {\bar \nu}_{\tau}^{(m)}(x) \end{array} \right) \gamma
^{\mu} {1\over 2}(1-\gamma_5) \left( \begin{array}{c} e^{(m)}(x)\\ \mu^{(m)}
(x)\\ \tau^{(m)}(x) \end{array} \right) +\nonumber \\
&&+\left( \begin{array}{ccc} {\bar d}^{(m)}(x) & {\bar s}^{(m)}(x) & {\bar b}
^{(m)}(x) \end{array} \right) \gamma^{\mu} {1\over 2}(1-\gamma_5) V_{CKM}
^{\dagger} \left( \begin{array}{c} u^{(m)}(x)\\ c^{(m)}(x)\\ t^{(m)}(x) 
\end{array} \right) ,\nonumber \\
&&{}\nonumber \\
J^{\mu}_{sA}(x)&=&\left( \begin{array}{ccc} {\bar u}^{(m)}(x) & 
{\bar c}^{(m)}(x) & {\bar t}^{(m)}(x) \end{array} \right) \gamma^{\mu}iT_s^A 
\left( \begin{array}{c} u^{(m)}
(x)\\ c^{(m)}(x)\\ t^{(m)}(x) \end{array} \right) +\nonumber \\
&&+\left( \begin{array}{ccc} {\bar d}^{(m)}(x) & {\bar s}^{(m)}(x)
& {\bar b}^{(m)}(x) \end{array} \right) \gamma^{\mu}iT_s^A \left( 
\begin{array}{c} d^{(m)}(x)\\ s^{(m)}(x)\\ b^{(m)}(x) \end{array} \right) .
\label{26}
\end{eqnarray}

The neutral current ${\tilde j}^{\mu}_{(NC)}(x)$ is also written in the
alternative forms ($\sum_f$is the sum over all fermions): ${\tilde j}^{\mu}
_{(NC)}(x)={1\over {sin\, 2\theta_w}} \sum_f{\bar \psi}^{(m)}_f(x)(g_v^{f)}
\gamma^{\mu}-g_a^{(f)}\gamma^{\mu}\gamma_5)\psi^{(m)}_f(x)=\sum_f{\bar \psi}
^{(m)}_f(x)(v_f\gamma^{\mu}-a_f\gamma^{\mu}\gamma_5)\psi^{(m)}_f(x)$, where
$g_v=sin\, 2\theta_w\, v =i(T^3_w-2sin^2\, \theta_w Q_{em})$, $g_a=sin\, 
2\theta_w\, a =iT^3_w$ [the fermion assignements are: $g_v^{(\nu_e,\nu_{\mu},
\nu_{\tau})} =g_a^{(\nu_e,\nu_{\mu},\nu_{\tau})}={1\over 2}$; $g_v^{(e,\mu ,
\tau )} ={1\over 2}(4sin^2\, \theta_w-1)$, $g_a^{(e,\mu ,\tau )}=-{1\over 2}$;
$g_v^{(u,c,t)}={1\over 2}(1-{8\over 3}sin^2\, \theta_w)$, $g_a^{(u,c,t)}=
{1\over 2}$; $g_v^{(d,s,b)}={1\over 2}({4\over 3}sin^2\, \theta_w-1)$, $g_a^{(d,
s,b)}=-{1\over 2}$].

In the charge-changing currents of the V-A type (V=vector $\gamma^{\mu}$, 
A=axial-vector $\gamma^{\mu}\gamma_5$), the Cabibbo-Kobayashi-Maskawa matrix
$V_{CKM}={\tilde S}^{(q)\dagger}_LS_L^{(q)}$ appears; it can be shown that it
depends on three angles $\theta_{12}=\theta_C, \theta_{13}, \theta_{23}$
giving the mixing of the quarks d, s, b, of the three families [$c_{ij}=
cos\, \theta_{ij} \geq 0, s_{ij}=sin\, \theta_{ij} \geq 0$] and a complex
phase $e^{i\delta_{13}}$ [$0 \leq \delta_{13} \leq 2\pi$], 
unique source of the weak CP-violation observed in
the K system. With only two families, only the Cabibbo angle $\theta_C$
remains, which is enough to explain the GIM mechanism (absence of flavour
changing neutral currents since $\bar d d+\bar s s={\bar d}_Cd_C+{\bar s}_C
s_C$ with $d_C=cos\, \theta_C d+sin\, \theta_C s$, $s_C=-sin\, \theta_C d+
cos\, \theta_C s$) and the different strength of hadronic $\triangle S=0$
and $\triangle S =\triangle Q =1$ processes (S is the strong strangeness).
One has

\begin{equation}
V_{CKM}=\left( \begin{array}{ccc} c_{12}c_{13} & s_{12}c_{13} & s_{13}e^{-i
\delta_{13}}\\ -s_{12}c_{23}-c_{12}s_{23}s_{13}e^{i\delta_{13}}& c_{12}c_{23}-
s_{12}s_{23}s_{13}e^{i\delta_{13}}& s_{23}c_{13}\\ s_{12}s_{23}-c_{12}c_{23}
s_{13}e^{i\delta_{13}}& -c_{12}s_{23}-s_{12}c_{23}s_{13}e^{i\delta_{13}}&
c_{23}c_{13} \end{array} \right) ;
\label{27}
\end{equation}

\noindent the matrix of the moduli has the following form and the moduli
have the following experimental range of values

\begin{eqnarray}
| V_{CKM} |&=& \left( \begin{array}{ccc} |V_{ud}|& |V_{us}|& |V_{ub}|\\
|V_{cd}|& |V_{cs}|& |V_{cb}|\\ |V_{td}|& |V_{ts}|& |V_{tb}|\end{array}
\right) =\nonumber \\
&=&\left( \begin{array}{ccc} 1-{{\lambda^2}\over 2}& \lambda & A\lambda^3
(\rho -i\eta )\\ -\lambda & 1-{{\lambda^2}\over 2}& A\lambda^2\\ A\lambda^3
(1-\rho -i\eta )& -A\lambda^2& 1\end{array} \right) +O(\lambda^4)=\nonumber \\
&=&\left( \begin{array}{ccc} 0.9745\, to\, 0.9757& 0.219\, to\, 0.224& 0.002\,
to\, 0.005\\ 0.218\, to\, 0.224& 0.9736\, to\, 0.9750& 0036\, to\, 0.046\\
0.004\, to\, 0.014& 0.034\, to\, 0.046& 0.9989\, to\, 0.9993 \end{array}
\right) ,
\label{28}
\end{eqnarray}

\noindent where $\lambda = |V_{us}| =0.2205\pm 0.0018$, $A=|V_{cb}|/\lambda^2
=0.80\pm 0.04$, $\sqrt{\rho^2+\eta^2}=|V_{ub}|/\lambda |V_{cb}|=0.36\pm 0.10$;
one has $s_{12}=0.219\, to\, 0.223$, $s_{23}=0.036\, to\, 0.046$, $s_{13}=
0.002\, to\, 0.005$.

The total number of free parameters of the standard model is 19: the nine masses
(or Yukawa couplings) $m_e, m_{\mu}, m_{\tau}, m_d, m_s, m_b, m_u, m_c, m_t$;
the three mixing angles $\theta_{12}=\theta_C, \theta_{23}, \theta_{13}$;
the phase $\delta_{13}$ [weak CP-violation]; the electromagnetic coupling
$\alpha$; the Weinberg angle $\theta_w$; the vector boson mass $m_Z$ [or
$m_W$]; the Higgs mass $m_H$; the strong coupling $\alpha_s(m_Z^2)$ or the
QCD scale $\Lambda_s$; the $\theta$-angle [strong CP-violation].

The unitary gauge Lagrangian density has the following exact global (1st
Noether theorem) and local (2nd Noether theorem) symmetries;

1) The global groups $U^{(l)}_i(1)$: $\psi^{(l)}_i(x)\mapsto e^{i\alpha^{(l)}_i}
\psi^{(l)}_i(x)$, whose conserved quantities are the lepton numbers $N_i$ of 
the three lepton families. The associated conserved currents are $J^{\mu}_{Ni}
(x)= {\bar \psi}^{(l)}_{Li}(x) \gamma^{\mu} \psi^{(l)}_{Li}(x)+{\bar \psi}^{(l)}
_{Ri}(x) \gamma^{\mu} \psi^{(l)}_{Ri}(x)$ [$J^{\mu}_{N1}(x)={\bar e}^{(m)}(x) 
\gamma^{\mu} e^{(m)}(x)+{\bar \nu}^{(m)}_{eL}\gamma^{\mu} \nu^{(m)}_{eL}(x)$ 
and so on], $\partial_{\mu}J^{\mu}_{Ni}(x) {\buildrel \circ \over =} 0$, where 
``${\buildrel \circ \over =}$" means evaluated on the equations of motion.

2) The $U_{s\, V}(1)$ global group [the matrix $V_{CKM}$ mixes the quark 
families]: $\psi^{(q)}_{Li}(x)\mapsto e^{i\alpha^{(q)}}\psi^{(q)}_{Li}(x)$,
$\psi^{(q)}_{Ri}(x)\mapsto e^{i\alpha^{(q)}}\psi^{(q)}_{Ri}(x)$, ${\tilde 
\psi}^{(q)}_{Ri}(x)\mapsto e^{i\alpha^{(q)}}{\tilde \psi}^{(q)}_{Ri}(x)$,
whose conserved quantity is the baryon number B. The associated conserved 
current is $J^{\mu}_B(x)=\sum_{i=1}^3 [{\bar \psi}^{(q)}_{Li}(x) \gamma^{\mu}
\psi^{(q)}_{Li}(x)+{\bar \psi}^{(q)}_{Ri}(x)\gamma^{\mu} \psi^{(q)}_{Ri}(x)+
{\bar {\tilde \psi}}^{(q)}_{Ri}(x)\gamma^{\mu}{\tilde \psi}^{(q)}_{Ri}(x)] =
{\bar d}^{(m)}(x)\gamma^{\mu}d^{(m)}(x)+{\bar s}^{(m)}(x)\gamma^{\mu}s^{(m)}(x)
+{\bar b}^{(m)}(x)\gamma^{\mu}b^{(m)}(x)+{\bar u}^{(m)}(x)\gamma^{\mu}u^{(m)}
(x)+{\bar c}^{(m)}(x)\gamma^{\mu}c^{(m)}(x)+{\bar t}^{(m)}(x)\gamma^{\mu}
t^{(m)}(x)$

3) The local strong color group SU(3), $G_{A\mu}(x){\hat T}^A\mapsto U^{-1}_s
(x)G_{A\mu}(x){\hat T}^A_sU_s(x)+U_s^{-1}(x)\partial_{\mu}U_s(x)$, $\psi^{(q)}
_{Li}(x)\mapsto U^{-1}_s(x)\psi^{(q)}_{Li}(x)$, $\psi^{(q)}_{Ri}(x)\mapsto
U^{-1}_s(x)\psi^{(q)}_{Ri}(x)$, ${\tilde \psi}^{(q)}_{Ri}(x)\mapsto U^{-1}_s(x)
{\tilde \psi}^{(q)}_{Ri}(x)$, giving the conservation of the non-Abelian
SU(3) charges $Q_A$ (improper conservation law from the 2nd Noether theorem
and Gauss theorem). The associated conserved current is $J_{sA}^{\mu}(x)$ of
Eqs.(\ref{26}).

4) The local electromagnetic gauge group $U_{em}(1)$ giving the conservation
of the electric charge (improper conservation law from the 2nd Noether
theorem and Gauss theorem). It is called the custodial symmetry.
The associated gauge transformations are
$A^{'}_{\mu}(x)\mapsto A^{'}_{\mu}(x)+U^{-1}_{em}(x)\partial_{\mu}U_{em}(x)$,
$\psi^{(l)}_{Li}(x)\mapsto U^{-1}_{em}(x)\psi^{(l)}_{Li}(x)$, $\psi^{(l)}_{Ri}
(x)\mapsto U^{-1}_{em}(x)\psi^{(l)}_{Ri}(x)$, $\psi^{(q)}_{Li}(x)\mapsto
U^{-1}_{em}(x)\psi^{(q)}_{Li}(x)$, $\psi^{(q)}_{Ri}(x)\mapsto U^{-1}_{em}(x)
\psi^{(q)}_{Ri}(x)$, ${\tilde \psi}^{(q)}_{Ri}(x)\mapsto U^{-1}_{em}(x){\tilde
\psi}^{(q)}_{Ri}(x)$. As we shall see in Section VI, in the Higgs sector at each
instant there is a su(2)xu(1) algebra of non conserved charges in the
electroweak sector.

Moreover, the standard model has approximate global symmetries

1) Strong chiral symmetry

1a) If we put $m_u=m_d=m_s=0$, $\theta_{13}=\theta_{23}=\delta_{13}=0$
($\theta_{12}=\theta_c$), and rearrange the $u^{(m)}(x)$, $d^{(m)}(x)$,
$s^{(m)}(x)$, quark fields in the triplet form $q(x)= \left( \begin{array}{l}
q_1(x)\\ q_2(x)\\ q_3(x)\end{array} \right) =\left( \begin{array}{l}
u^{(m)}(x)\\ d^{(m)}(x)\\ s^{(m)}(x)\end{array} \right)$, the Lagrangian
density (\ref{25}) is invariant under the (strong interactions) global
Noether transformations associated with an $U_{sV}(1)\times U_{sA}(1)\times
SU_{sV}(3)\times SU_{sA}(3)$ group, whose infinitesimal form is
[$\alpha_V$, $\alpha_A$, $\alpha_{V,\bar A}$, $\alpha_{A,\bar A}$ are the
constant parameters; $\lambda^{\bar A}$ are the SU(3) Gell-Mann matrices in the
fundamental triplet representation]

\begin{eqnarray}
&&q_i(x) \mapsto q_i(x)+i \alpha_V q_i(x),\quad\quad i=1,2,3,\nonumber \\
&&q_i(x) \mapsto q_i(x) +i \alpha_{V,\bar A} ({{\lambda^{\bar A}}\over 2})_{ij}
q_j(x),\quad\quad q_i(x)\mapsto e^{i[\alpha_V+\alpha_{V,\bar A}
{{\lambda^{\bar A}}\over 2}]}q_i(x),\nonumber \\
&&q_i(x) \mapsto q_i(x) +i \alpha_A \gamma_5 q_i(x),\quad\quad q_i(x)\mapsto
e^{i\gamma_5[\alpha_A+\alpha_{A,\bar A}{{\lambda^{\bar A}}\over 2}]}q_i(x),
\nonumber \\
&&q_i(x) \mapsto q_i(x) + i\alpha_{A,\bar A} ({{\lambda^{\bar A}}\over 2})
_{ij} \gamma_5 q_j(x),
\label{a1}
\end{eqnarray}

\noindent whose associated conserved Noether vector and axial-vector currents
and charges are

\begin{eqnarray}
{\cal V}^{\mu}(x) &=&i\bar q(x) \gamma^{\mu} q(x),\quad\quad Q_V=\int d^3x\, 
{\cal V}^o(\vec x,x^o),\nonumber \\
{\cal A}^{\mu}(x) &=& i\bar q(x) \gamma^{\mu}\gamma_5 q(x),\quad\quad Q_A=
\int d^3x\, {\cal A}^o(\vec x,x^o),\nonumber \\
{\cal V}^{\mu}_{\bar A}(x)&=&i\bar q(x) \gamma^{\mu}{{\lambda^{\bar A}}\over 2}
q(x),\quad\quad Q_{V,\bar A} =\int d^3x\, {\cal V}^o_{\bar A}(\vec x,x^o),
\nonumber \\
{\cal A}^{\mu}_{\bar A}(x)&=&i\bar q(x) \gamma^{\mu}\gamma_5 {{\lambda^{\bar A}}
\over 2} q(x),\quad\quad Q_{A,\bar A}= \int d^3x\, {\cal A}^o_{\bar A}(\vec x,
x^o).
\label{a2}
\end{eqnarray}

$Q_V$ is the part of baryon number $B=\int d^3x\, J^o_B(\vec x,x^o)$ containing
the $u^{(m)}$, $d^{(m)}$, $s^{(m)}$, quarks, while $Q_{V,\bar A}$ are the
global approximatively conserved (the scale of the breaking is given by $m_s$)
Gell-Mann flavour charges of the standard quark model of hadrons.
They are: strong
isospin $T_s^a={1\over 2} \lambda^a$, a=1,2,3; strong hypercharge $Y_s={1\over
{\sqrt{3}}} \lambda^8$, strangeness $S_s=Y_s-B$, electric charge $Q_{em}=
T_s^3+{1\over 2}Y_s$; U-spin $U_s^1={1\over 2}\lambda^6$, $U_s^2={1\over 2}
\lambda^7$, $U_s^3={1\over 4}(\sqrt{3} \lambda^8-\lambda^3)={3\over 4}Y_s-
{1\over 2}T^3_s$; V-spin $V_s^1={1\over 2}\lambda^4$, $V_s^2={1\over 2}
\lambda^5$, $V_s^3={1\over 4}(\sqrt{3}\lambda^8+\lambda^3)={3\over 4}Y_s+
{1\over 2}T_s^3$; the quark assignements are 

$\begin{array}{ccccccccccc} {}& B& T_s& T^3_s& Q_{em}& Y_s& S_s& U_s& U_s^3&
V_s& V_s^3\\
u^{(m)}&1/3& 1/2& 1/2& 2/3& 1/3& 0& 0& 0& 1/2& 1/2\\
d^{(m)}& 1/3& 1/2& -1/2& -1/3& 1/3& 0& 1/2& 1/2& 0& 0\\
s^{(m)}& 1/3& 0& 0& -1/3& 4/3& -1& 1/2& -1/2& 1/2& -1/2. \end{array}$

Since we have $\lbrace Q_{V,\bar A}, Q_{V,\bar B}\rbrace =c_{\bar A\bar B\bar C}
Q_{V,\bar C}$, $\lbrace Q_{V,\bar A}, Q_{A,\bar B}\rbrace =c_{\bar A\bar B\bar
C} Q_{A,\bar C}$, $\lbrace Q_{A,\bar A}, Q_{A,\bar B}\rbrace =c_{\bar A\bar B
\bar C} Q_{V,\bar C}$, we can define the left and right charges

\begin{eqnarray}
Q_R&=&{1\over 2}(Q_V-Q_A),\quad\quad Q_V=Q_L+Q_R,\nonumber \\
Q_L&=&{1\over 2}(Q_V+Q_A),\quad\quad Q_A=Q_L-Q_R,\nonumber \\
Q_{R,\bar A}&=&{1\over 2}(A_{V,\bar A}-Q_{A,\bar A}),\quad\quad
Q_{V,\bar A}=Q_{L,\bar A}+Q_{R,\bar A},\nonumber \\
Q_{L,\bar A}&=&{1\over 2}(Q_{V,\bar A}+Q_{A,\bar A}),\quad\quad Q_{A,\bar A}=
Q_{L,\bar A}-Q_{R,\bar A},\nonumber \\
&&{}\nonumber \\
&&\lbrace Q_{R,\bar A}, Q_{R,\bar B}\rbrace =c_{\bar A\bar B\bar C}Q_{R,\bar
C},\nonumber \\
&&\lbrace Q_{R,\bar A}, Q_{L,\bar B}\rbrace =0,\nonumber \\
&&\lbrace Q_{L,\bar A}, Q_{L,\bar B}\rbrace =c_{\bar A\bar B\bar C}Q_{L,\bar C}.
\label{a3}
\end{eqnarray}

In this form the group $U_{sV}(1)\times U_{sA}(1)\times SU_{sV}(3)\times SU_{sA}
(3)$ is replaced by the global strong chiral group $U_{sR}(1)\times U_{sL}(1)
\times SU_{sR}(3)\times SU_{sL}(3)$. 

\noindent At the quantum level one has:

A) The vector current ${\cal V}^{\mu}(x)$ is still conserved.

B) The axial-vector current ${\cal A}^{\mu}(x)$ is no more conserved due to the
global chiral anomaly [$U_{sA}(1)$-anomaly]. On one side, this phenomenon
explains the otherwise forbidden decay $\pi^0\rightarrow 2\gamma$, but on the
other side it constitutes the $U_{sA}(1)$-problem, because one cannot
invoke a dynamical spontaneous symmetry breaking mechanism, since the
associated Goldstone boson should be the $\eta^{'}$ pseudoscalar boson,
which has too big a mass. The way out seems to be topological, i.e.
connected with the $\theta$ vacuum and its strong CP problem, for which
there are various interpretations (existence of the axion,...).

Let us remark that with 3 colors, $N_c=3$, there is no local 
$SU(3)\times SU(2)\times U(1)$ chiral anomaly, which would spoil the 
renormalizability of the standard model.

C) $SU_{sR}(3)\times SU_{sL}(3)$ is supposed to be 
dynamically spontaneously broken to the
diagonal $SU_{sL+R}(3)=SU_{sV}(3)$ approximate flavour Gell-Mann symmetry
group (valid also for $m_u=m_d=m_s\not=0$) by the formation of a quark
condensate $< \bar q(x)q(x) > \not= 0$ [instead a gluon condensate ${{g_s}
\over {4\pi}} < 0| F_{A\mu\nu}(x)F_A^{\mu\nu}(x) |0 > ={{g_s}\over {2\pi}}
< 0| \sum_A({\vec B}^2_A(x)-{\vec E}_A^2(x)) |0 >$ should correspond to a 
magnetic color configuration of the vacuum, responsible for the confinement
of the electric flux between quarks and for the string tension $k\approx (450\,
Mev)^2$
(the coefficient of the linear confining potential)]. This condensate of
quarks pairs breaks chirality [$< 0| {\bar q}_i(x)q_i(x) |0 > \approx -
(220\, Mev)^2$ for each i] with a nonperturbative dynamical mechanism (for
instance Nambu-Jona Lasinio). The quark condensate dynamically generates the
``constituent" mass for the quarks, much larger than the ``current" mass
[$m^{const}_u\approx m^{const}_d\approx 300 Mev$, $m^{const}_s\approx 450
Mev$], to be used in the quark model as an effective mass.
In the limit of exact $SU_{sV}(3)$,
the $SU_{sV}(3)$ octet of pseudoscalar mesons $\pi , K, \eta,$ would be
massless and would correspond to the eight Goldstone bosons associated with
the spontaneous symmetry breaking.

1b) One could also put $m_u=m_d=m_s=m_c=m_b=m_t=0$ and study the approximate
$SU_{sL}(6)\times SU_{sR}(6)$ global symmetry, but it is much less interesting
due to the big breaking of this symmetry measured by the value of $m_t$.

2) Weak chiral symmetry

If we put $m_e=m_{\mu}=m_{\tau}=0$ and $m_u=m_d=m_s=m_c=m_b=m_t=0$ (i.e. all
the leptons and quarks are massless), one has the global Noether symmetry
$SU_{wL}(2)\times SU_{wR}(2)$ which should be spontaneously broken to
the weak isospin $SU_{wL+R}(2)=SU_w(2)$ global custodial symmetry. See Ref.
\cite{dob}.

3) Heavy quark symmetry: for this approximate symmetry see Ref.\cite{heavy}.

\section
{Euler-Lagrange equations and constraints from ${\cal L}(x)$}

The Euler-Lagrange equations deriving from the Lagrangian density (\ref{1})
are [$V(\phi )=\lambda (\phi^{\dagger}\phi -\phi_o^2)^2$ is the Higgs
potential]

\begin{eqnarray}
L_A^{(G)\, \mu}&=&g_s^2({{\partial {\cal L}}\over {\partial G_{A\mu}}}-\partial
_{\nu}{{\partial {\cal L}}\over {\partial \partial_{\nu}G_{A\mu}}})={\hat D}
^{(G)}_{\nu AB}G^{\nu\mu}_B+g^2_sJ^{\mu}_{sA} {\buildrel \circ \over =} 0,
\nonumber \\
&&{}\nonumber \\
J&^{\mu}&_{sA}=i{\bar \psi}^{(q)}_{Li} \gamma^{\mu}T^A_s\psi^{(q)}_{Li}+
i{\bar \psi}^{(q)}_{Ri}\gamma^{\mu}T^A_s\psi^{(q)}_{Ri}+i{\bar {\tilde \psi}}
^{(q)}_{Ri}\gamma^{\mu}T^A_s{\tilde \psi}^{(q)}_{Ri},\nonumber \\
L_a^{(W)\, \mu}&=&g_w^2({{\partial {\cal L}}\over {\partial W_{a\mu}}}-\partial
_{\nu}{{\partial {\cal L}}\over {\partial \partial_{\nu}W_{a\mu}}})={\hat D}
^{(W)}_{\nu ab}W^{\nu\mu}_b+g^2_w{\hat J}^{\mu}_{wa} {\buildrel \circ \over =} 0
,\nonumber \\
&&{}\nonumber \\
{\hat J}&^{\mu}&_{wa}=i{\bar \psi}^{(l)}_{Li} \gamma^{\mu}T^a_w\psi^{(l)}_{Li}+
i{\bar \psi}^{(q)}_{Li}\gamma^{\mu}T^a_w\psi^{(q)}_{Li}-\nonumber \\
&-&\phi^{\dagger}[T_w^aD^{(W,V)\mu}-
\loarrow{D^{(W,V)\mu {\dagger}}}T_w^a]\phi= \nonumber \\
&=&J^{\mu}_{wa}-\phi^{\dagger}[T_w^aD^{(W,V)\mu}-
\loarrow{D^{(W,V)\mu {\dagger}}}T_w^a]\phi ,\nonumber \\
L^{(V)\, \mu}&=&g_y^2({{\partial {\cal L}}\over {\partial V_{\mu}}}-\partial
_{\nu}{{\partial {\cal L}}\over {\partial \partial_{\nu}V_{\mu}}})=\partial
_{\nu}V^{\nu\mu}+g^2_y{\hat J}^{\mu}_{Y_w} {\buildrel \circ \over =} 0,
\nonumber \\
&&{}\nonumber \\
{\hat J}&^{\mu}&_{Y_w}=i{\bar \psi}^{(l)}_{Li} \gamma^{\mu}Y_w\psi^{(l)}_{Li}+
i{\bar \psi}^{(l)}_{Ri}\gamma^{\mu}Y_w\psi^{(l)}_{Ri}+i{\bar \psi}^{(q)}_{Li}
\gamma^{\mu}Y_w\psi^{(q)}_{Li}+i{\bar \psi}^{(q)}_{Ri}\gamma^{\mu}Y_w\psi^{(q)}
_{Ri}+\nonumber \\
&+&i{\bar {\tilde \psi}}^{(q)}_{Ri}\gamma^{\mu}Y_w{\tilde \psi}^{(q)}_{Ri}-
\phi^{\dagger}[Y_wD^{(W,V)\mu}-\loarrow{D^{(W,V)\mu {\dagger}}}Y_w]\phi =
\nonumber \\
&=&J^{\mu}_{Y_w}-\phi^{\dagger}[Y_wD^{(W,V)\mu}-\loarrow{D^{(W,V)\mu 
{\dagger}}}Y_w]\phi ,\nonumber \\
&&{}\nonumber \\
L_{\phi a}&=&{{\partial {\cal L}}\over {\partial \phi_a}}-\partial_{\mu}
{{\partial {\cal L}}\over {\partial \partial_{\mu}\phi_a}}=-
{[D^{(W,V)\mu}D^{(W,V)}_{\mu}\phi ]}_a^{\dagger}-{{\partial V(\phi )}\over
{\partial \phi_a}} +\nonumber \\
&+&{1\over {\phi_o}}[{\bar \psi}^{(l)}_{Lia} M^{(l)}_{ij}\psi^{(l)}_{Rj} +
{\bar \psi}^{(q)}_{Lia} M^{(q)}_{ij}\psi^{(q)}_{Rj}+{\bar {\tilde \psi}}^{(q)}
_{Ri}{\tilde M}^{(q)\dagger}_{ij}(2iT^3_w)_{ab}\psi^{(q)}_{Ljb}]
{\buildrel \circ \over =} 0,\nonumber \\
L_{\phi^{*}a}&=&{{\partial {\cal L}}\over {\partial \phi^{*}_a}}-\partial_{\mu}
{{\partial {\cal L}}\over {\partial \partial_{\mu}\phi^{*}_a}}=-
[D^{(W,V)\mu}D^{(W,V)}_{\mu}\phi ]_a-{{\partial V(\phi )}\over {\partial 
\phi^{*}_a}}+\nonumber \\
&+&{1\over {\phi_o}}[{\bar \psi}^{(l)}_{Ri}M^{(l)\dagger}_{ij}\psi^{(l)}_{Lia}+
{\bar \psi}^{(q)}_{Ri}M^{(q)\dagger}_{ij}\psi^{(q)}_{Lja}+ {\bar \psi}^{(q)}
_{Lib}(2iT^3_w)_{ba}{\tilde M}^{(q)}_{ij}{\tilde \psi}^{(q)}_{Rj}]
{\buildrel \circ \over =} 0,\nonumber \\
&&{}\nonumber \\
L_{\psi Li}^{(l)}&=&{{\partial {\cal L}}\over {\partial \psi^{(l)}_{Li}}}-
\partial_{\mu}{{\partial {\cal L}}\over {\partial \partial_{\mu}\psi^{(l)}
_{Li}}}=-{\bar \psi}^{(l)}_{Li} [\loarrow{i(\partial_{\mu}-W_{a\mu}T_w^a-
V_{\mu}Y_w)}\gamma^{\mu}]-{\bar \psi}^{(l)}_{Rj}M_{ji}^{(l)\dagger} {{\phi
^{\dagger}}\over {\phi_o}} {\buildrel \circ \over =} 0,\nonumber \\
L_{\bar \psi Li}^{(l)}&=&{{\partial {\cal L}}\over {\partial {\bar \psi}^{(l)}
_{Li}}}-\partial_{\mu}{{\partial {\cal L}}\over {\partial \partial_{\mu}{\bar 
\psi}^{(l)}_{Li}}}=[\gamma^{\mu}i(\partial_{\mu}+W_{a\mu}T_w^a)+V_{\mu}Y_w]
\psi^{(l)}_{Li}-{{\phi}\over {\phi_o}} M^{(l)}_{ij}\psi^{(l)}_{Rj} 
{\buildrel \circ \over =} 0,\nonumber \\
L^{(l)}_{\psi Ri}&=&{{\partial {\cal L}}\over {\partial \psi^{(l)}_{Ri}}}-
\partial_{\mu}{{\partial {\cal L}}\over {\partial \partial_{\mu}\psi^{(l)}
_{Ri}}}=-{\bar \psi}^{(l)}_{Ri} [\loarrow{i(\partial_{\mu}-V_{\mu}Y_w)}
\gamma^{\mu}]-{\bar \psi}^{(l)}_{Lj}\cdot {{\phi}\over {\phi_o}} M^{(l)}_{ji}
{\buildrel \circ \over =} 0,\nonumber \\
L^{(l)}_{\bar \psi Ri}&=&{{\partial {\cal L}}\over {\partial {\bar \psi}^{(l)}
_{Ri}}}-\partial_{\mu}{{\partial {\cal L}}\over {\partial \partial_{\mu}{\bar 
\psi}^{(l)}_{Ri}}}=[\gamma^{\mu}i(\partial_{\mu}+V_{\mu}Y_w)]\psi^{(l)}_{Ri}-
M^{(l)\dagger}_{ij} {{\phi^{\dagger}}\over {\phi_o}}\cdot \psi^{(l)}_{Lj} 
{\buildrel \circ \over =} 0,\nonumber \\
&&{}\nonumber \\
L^{(q)}_{\psi Li}&=&{{\partial {\cal L}}\over {\partial \psi^{(q)}_{Li}}}-
\partial_{\mu}{{\partial {\cal L}}\over {\partial \partial_{\mu}\psi^{(q)}
_{Li}}}=-{\bar \psi}^{(q)}_{Li} [\loarrow{i(\partial_{\mu}-W_{a\mu}T_w^a-
V_{\mu}Y_w-G_{A\mu}T^A_s)}\gamma^{\mu}]-\nonumber \\
&-&{\bar \psi}^{(q)}_{Rj}M^{(q)\dagger}_{ji} {{\phi^{\dagger}}\over {\phi_o}}
-{\bar {\tilde \psi}}^{(q)}_{Rj}{\tilde M}^{(q)\dagger}_{ji}{{{\tilde \phi}
^{\dagger}}\over {\phi_o}} {\buildrel \circ \over =} 0,\nonumber \\
L^{(q)}_{\bar \psi Li}&=&{{\partial {\cal L}}\over {\partial {\bar \psi}^{(q)}
_{Li}}}-\partial_{\mu}{{\partial {\cal L}}\over {\partial \partial_{\mu}{\bar 
\psi}^{(q)}_{Li}}}=[\gamma^{\mu}i(\partial_{\mu}+W_{a\mu}T_w^a+V_{\mu}Y_w+
G_{A\mu}T^A_s)]\psi^{(q)}_{Li}-\nonumber \\
&-& {{\phi}\over {\phi_o}}M^{(q)}_{ij}\psi^{(q)}_{Rj}-{{\tilde \phi}\over
{\phi_o}}{\tilde M}^{(q)}_{ij}{\tilde \psi}^{(q)}_{Rj}
{\buildrel \circ \over =} 0,\nonumber \\
L^{(q)}_{\psi Ri}&=&{{\partial {\cal L}}\over {\partial \psi^{(q)}_{Ri}}}-
\partial_{\mu}{{\partial {\cal L}}\over {\partial \partial_{\mu}\psi^{(q)}
_{Ri}}}=-{\bar \psi}^{(q)}{Ri} [\loarrow{i(\partial_{\mu}-V_{\mu}Y_w-G_{A\mu}
T^A_s)}\gamma^{\mu}]-{\bar \psi}^{(q)}_{Lj}\cdot {{\phi}\over {\phi_o}}
M^{(q)}_{ji} {\buildrel \circ \over =} 0,\nonumber \\
L^{(q)}_{\bar \psi Ri}&=&{{\partial {\cal L}}\over {\partial {\bar \psi}^{(q)}
_{Ri}}}-\partial_{\mu}{{\partial {\cal L}}\over {\partial \partial_{\mu}{\bar 
\psi}^{(q)}_{Ri}}}=[\gamma^{\mu}i(\partial_{\mu}+V_{\mu}Y_w+G_{A\mu}T^A_s)]
\psi^{(q)}_{Ri} -M_{ij}^{(q)\dagger}{{\phi^{\dagger}}\over {\phi_o}}\psi^{(q)}
_{Lj} {\buildrel \circ \over =} 0,\nonumber \\
L^{(q)}_{\tilde \psi Ri}&=&{{\partial {\cal L}}\over {\partial {\tilde \psi}
^{(q)}_{Ri}}}-\partial_{\mu}{{\partial {\cal L}}\over {\partial \partial_{\mu}
{\tilde \psi}^{(q)}_{Ri}}}=-{\bar {\tilde \psi}}^{(q)}_{Ri}
[\loarrow{i(\partial_{\mu}-V_{\mu}Y_w-G_{A\mu}T^A_s)}\gamma^{\mu}]-{\bar \psi}
^{(q)}_{Lj}\cdot {{\tilde \phi}\over {\phi_o}}{\tilde M}^{(q)}_{ji}
{\buildrel \circ \over =} 0,\nonumber \\
L^{(q)}_{{\bar {\tilde \psi}} Ri}&=&{{\partial {\cal L}}\over {\partial {\bar 
{\tilde \psi}}^{(q)}_{Ri}}}-\partial_{\mu}{{\partial {\cal L}}\over {\partial 
\partial_{\mu}{\bar {\tilde \psi}}^{(q)}_{Ri}}}=[\gamma^{\mu}
i(\partial_{\mu}+V_{\mu}Y_w+G_{A\mu}T^A_s)]{\tilde \psi}^{(q)}_{Ri}-
{\tilde M}^{(q)\dagger}_{ij}{{{\tilde \phi}^{\dagger}}\over {\phi_o}}\cdot
\psi^{(q)}_{Lj} {\buildrel \circ \over =} 0,\nonumber \\
&&{}\nonumber \\
{\tilde L}^{(\tilde A) \mu}&=&{{\partial {\cal L}}\over {\partial {\tilde A}
_{\mu}}}-\partial_{\nu}{{\partial {\cal L}}\over {\partial \partial_{\nu}
{\tilde A}_{\mu}}}=\partial_{\nu}{\tilde A}^{\nu\mu}-2ie\partial_{\nu}({\tilde 
W}^{\mu}_{+}
{\tilde W}_{-}^{\nu}-{\tilde W}^{\nu}_{+}{\tilde W}^{\mu}_{-})+\nonumber \\
&+&ie({\tilde W}_{-\, \nu}{\tilde W}^{\nu\mu}_{+}-{\tilde W}_{+\, 
\nu}{\tilde W}^{\nu\mu}_{-})-2e^2{\tilde W}_{+}\cdot {\tilde W}_{-} 
({\tilde A}^{\mu}+cot\, \theta_w {\tilde Z}^{\mu})+\nonumber \\
&+&e^2({\tilde W}^{\mu}_{+}{\tilde W}^{\nu}_{-}+{\tilde W}^{\mu}_{-}
{\tilde W}^{\nu}_{+})({\tilde A}_{\nu}+cot\, \theta_w {\tilde Z}
_{\nu})+\nonumber \\
&+&ie\lbrace [Q_{em}\phi (x)]^{\dagger}\cdot
[(\partial^{\mu}+g_w({\tilde W}^{\mu}_{+}(x)T^{-}_w+{\tilde W}^{\mu}
_{-}(x)T^{+}_w)-\nonumber \\
&-&ieQ_{em}{\tilde A}^{\mu}(x)-ieQ_Z{\tilde Z}^{\mu}(x))\phi (x)]-
\nonumber \\
&-&[(\partial^{\mu}+g_w({\tilde W}_{+}^{\mu}(x)T^{-}_w+{\tilde W}_{-}^{\mu}(x)
T^{+}_w)-\nonumber \\
&-&ieQ_{em}{\tilde A}^{\mu}(x)-ieQ_Z{\tilde Z}^{\mu}(x))\phi (x)]
^{\dagger}\cdot [Q_{em}\phi (x)]\rbrace +\nonumber \\
&+&e{\tilde j}^{\mu}_{(em)}{\buildrel \circ \over =} 0,\nonumber \\
&&{}\nonumber \\
{\tilde j}&^{\mu}&_{(em)}={\bar \psi}^{(l)}_{Li} \gamma^{\mu}Q_{em}
\psi^{(l)}_{Li}+{\bar \psi}^{(l)}_{Ri}\gamma^{\mu}iY_w\psi^{(l)}_{Ri}+
{\bar \psi}^{(q)}_{Li}\gamma^{\mu}Q_{em}\psi^{(q)}_{Li}+\nonumber \\
&+&{\bar \psi}^{(q)}_{Ri}\gamma^{\mu} iY_w\psi^{(q)}_{Ri}+{\bar {\tilde \psi}}
^{(q)}_{Ri}\gamma^{\mu} iY_w{\tilde \psi}^{(q)}_{Ri},\nonumber \\
{\tilde L}^{(\tilde Z) \mu}&=&{{\partial {\cal L}}\over {\partial {\tilde Z}
_{\mu}}}-\partial_{\nu}{{\partial {\cal L}}\over {\partial \partial_{\nu}
{\tilde Z}_{\mu}}}=\partial_{\nu}{\tilde Z}^{{'}
\nu\mu}-2iecot\, \theta_w \partial_{\nu}({\tilde W}^{{'}\mu}_{+}{\tilde W}_{-}
^{{'}\nu}-{\tilde W}^{{'}\nu}_{+}{\tilde W}^{{'}\mu}_{-})+\nonumber \\
&+&ie\lbrace [Q_Z\phi (x)]^{\dagger}\cdot
[(\partial^{\mu}+g_w({\tilde W}^{\mu}_{+}(x)T^{-}_w+{\tilde W}^{\mu}
_{-}(x)T^{+}_w)-\nonumber \\
&-&ieQ_{em}{\tilde A}^{\mu}(x)-ieQ_Z{\tilde Z}^{\mu}(x))\phi (x)]-
\nonumber \\
&-&[(\partial^{\mu}+g_w({\tilde W}_{+}^{\mu}(x)T^{-}_w+{\tilde W}_{-}^{\mu}(x)
T^{+}_w)-ieQ_{em}{\tilde A}^{\mu}(x)-\nonumber \\
&-&ieQ_Z{\tilde Z}^{\mu}(x))\phi (x)]
^{\dagger}\cdot [Q_Z\phi (x)]\rbrace +\nonumber \\
&+&iecot\, \theta_w ({\tilde W}^{'}_{-\, \nu}{\tilde W}^{{'}\nu\mu}_{+}-
{\tilde W}^{'}_{+\, \nu}{\tilde W}^{{'}\nu\mu}_{-})-2e^2cot\, \theta_w
{\tilde W}^{'}_{+}\cdot {\tilde W}^{'}_{-} ({\tilde A}^{{'}\mu}+cot\, \theta_w 
{\tilde Z}^{{'}\mu})+\nonumber \\
&+&e^2cot\, \theta_w ({\tilde W}^{{'}\mu}_{+}{\tilde W}^{{'}\nu}_{-}+{\tilde W}
^{{'}\mu}_{-}{\tilde W}^{{'}\nu}_{+})({\tilde A}^{'}_{\nu}+cot\, \theta_w 
{\tilde Z}^{'}_{\nu})+ e{\tilde j}^{\mu}_{(NC)}{\buildrel \circ \over =} 0,
\nonumber \\
&&{}\nonumber \\
{\tilde j}&^{\mu}&_{(NC)}={\bar \psi}^{(l)}_{Li} \gamma^{\mu}Q_Z
\psi^{(l)}_{Li}-tg\, \theta_w {\bar \psi}^{(l)}_{Ri}\gamma^{\mu}iY_w\psi
^{(l)}_{Ri}+{\bar \psi}^{(q)}_{Li}\gamma^{\mu}Q_Z\psi^{(q)}_{Li}-
\nonumber \\
&-&tg\, \theta_w {\bar \psi}^{(q)}_{Ri}\gamma^{\mu} iY_w\psi^{(q)}_{Ri}-
tg\, \theta_w {\bar {\tilde \psi}}^{(q)}_{Ri}\gamma^{\mu} iY_w{\tilde \psi}
^{(q)}_{Ri},\nonumber \\
{\tilde L}^{({\tilde W}_{\pm}) \mu}&=&{{\partial {\cal L}}\over {\partial 
{\tilde W}_{\pm \mu}}}-\partial_{\nu}{{\partial {\cal L}}\over {\partial 
\partial_{\nu}{\tilde W}_{\pm \mu}}}=
\partial_{\nu}{\tilde W}^{\nu\mu}_{\mp}+\nonumber \\
&+&g_w\lbrace [T^{\mp}_w\phi (x)]^{\dagger}\cdot
[(\partial^{\mu}+g_w({\tilde W}^{\mu}_{+}(x)T^{-}_w+{\tilde W}^{\mu}
_{-}(x)T^{+}_w)-\nonumber \\
&-&ieQ_{em}{\tilde A}^{\mu}(x)-ieQ_Z{\tilde Z}^{\mu}(x))\phi (x)]+
\nonumber \\
&+&[(\partial^{\mu}+g_w({\tilde W}_{+}^{\mu}(x)T^{-}_w+{\tilde W}_{-}^{\mu}(x)
T^{+}_w)-\nonumber \\
&-&ieQ_{em}{\tilde A}^{\mu}(x)-ieQ_Z{\tilde Z}^{\mu}(x))\phi (x)]
^{\dagger}\cdot [T^{\mp}_w\phi (x)]\rbrace +\nonumber \\
&+&ie\partial_{\nu} [{\tilde W}^{\nu}_{\mp}({\tilde A}^{\mu}+cot\,
\theta_w {\tilde Z}^{\mu})-{\tilde W}^{\mu}_{\mp}({\tilde A}^{\nu}
+cot\, \theta_w {\tilde Z}^{\nu})]-\nonumber \\
&-&ie[({\tilde A}^{\mu\nu}+cot\, \theta_w
{\tilde Z}^{\mu\nu}){\tilde W}_{\mp \, \nu}-{\tilde W}^{\mu\nu}_{\mp}
({\tilde A}_{\nu}+cot\, \theta_w {\tilde Z}_{\nu})]+\nonumber \\
&+&{{e^2}\over {sin^2\, \theta_w}}[{\tilde W}^{\mu}_{\pm}{\tilde W}^{2}
_{\mp}-{\tilde W}^{\mu}_{\mp}{\tilde W}_{+}\cdot {\tilde W}_{-}]-e^2
{\tilde W}^{\mu}_{\mp} ({\tilde A}+cot\, \theta_w {\tilde Z})^2+
\nonumber \\
&+&e^2({\tilde A}^{\mu}+cot\, \theta_w {\tilde Z}^{\mu}) {\tilde W}
_{\mp}\cdot ({\tilde A}+cot\, \theta_w {\tilde Z})+ {e\over {
sin\, \theta_w}} {\tilde j}^{\mu}_{(CC)\, \mp}
{\buildrel \circ \over =} 0,\nonumber \\
&&{}\nonumber \\
{\tilde j}&^{\mu}&_{(CC)\, \mp}={\bar \psi}^{(l)}_{Li} \gamma^{\mu}iT^{\mp}_w
\psi^{(l)}_{Li}+{\bar \psi}^{(q)}_{Li}\gamma^{\mu}iT^{\mp}_w\psi^{(q)}
_{Li}.
\label{30}
\end{eqnarray}

\noindent In the last lines we added the Euler-Lagrange equations for ${\tilde 
A}_{\mu}=e^{-1}A_{\mu}$, ${\tilde Z}_{\mu}=e^{-1}Z_{\mu}$, ${\tilde W}
_{\pm \mu}=e^{-1}sin\, \theta_w W_{\pm \mu}$, obtained from Eq.
(\ref{17}).

The canonical momenta implied by the Lagrangian density (\ref{1}) are

\begin{eqnarray}
&&\pi^{(G) o}_A(x)={{\partial {\cal L}(x)}\over {\partial \partial_oG_{Ao}
(x)}}=0,\nonumber \\
&&\pi^{(G) k}_A(x)={{\partial {\cal L}(x)}\over {\partial \partial_oG_{Ak}(x)}}=
-g_s^{-2}G_A^{ok}(x)=g_s^{-2}E^{(G) k}_A(x),\nonumber \\
&&\pi^{(W) o}_a(x)={{\partial {\cal L}(x)}\over {\partial \partial_oW_{ao}(x)}}
=0,\nonumber \\
&&\pi^{(W) k}_a(x)={{\partial {\cal L}(x)}\over {\partial \partial_oW_{ak}(x)}}=
-g_w^{-2}W_a^{ok}(x)=g_w^{-2}E^{(W) k}_a(x),\nonumber \\
&&\pi^{(V) o}(x)={{\partial {\cal L}(x)}\over {\partial \partial_oV_o(x)}}=0,
\nonumber \\
&&\pi^{(V) k}(x)={{\partial {\cal L}(x)}\over {\partial \partial_oV_k(x)}}=
-g_y^{-2}V^{ok}(x)=g_y^{-2}E^{(V) k}(x),\nonumber \\
&&{}\nonumber \\
&&\pi_{\phi \, a}(x)={{\partial {\cal L}(x)}\over {\partial \partial_o
\phi_a(x)}}={[D^{(W,V)}_o\phi (x)]}_a^{\dagger},\nonumber \\
&&\pi_{\phi^{\dagger}\, a}(x)={{\partial {\cal L}(x)}\over {\partial \partial_o
\phi_a^{*}(x)}}={[D^{(W,V)}_o\phi (x)]}_a,\nonumber \\
&&{}\nonumber \\
&&\pi^{(l)}_{\psi Lia\alpha}(x)={{\partial {\cal L}(x)}\over {\partial 
\partial_o\psi^{(l)}_{Lia\alpha}(x)}}=-{i\over 2}{({\bar \psi}^{(l)}_{Li}(x)
\gamma_o)}_{a\alpha},\nonumber \\
&&\pi^{(l)}_{\bar \psi Lia\alpha}(x)={{\partial {\cal L}(x)}\over {\partial 
\partial_o{\bar \psi}^{(l)}_{Lia\alpha}(x)}}=-{i\over 2}{(\gamma_o\psi^{(l)}
_{Li}(x))}_{a\alpha},\nonumber \\
&&\pi^{(l)}_{\psi Ri\alpha}(x)={{\partial {\cal L}(x)}\over {\partial \partial_o
\psi^{(l)}_{Ri\alpha}(x)}}=-{i\over 2}{({\bar \psi}^{(l)}_{Ri}(x)\gamma_o)}
_{\alpha},\nonumber \\
&&\pi^{(l)}_{\bar \psi Ri\alpha}(x)={{\partial {\cal L}(x)}\over {\partial 
\partial_o{\bar \psi}^{(l)}_{Ri\alpha}(x)}}=-{i\over 2}{(\gamma_o\psi^{(l)}
_{Ri}(x))}_{\alpha},\nonumber \\
&&{}\nonumber \\
&&\pi^{(q)}_{\psi LiAa\alpha}(x)={{\partial {\cal L}(x)}\over {\partial 
\partial_o\psi^{(q)}_{LiAa\alpha}(x)}}=-{i\over 2}{({\bar \psi}^{(q)}_{Li}(x)
\gamma_o)}_{Aa\alpha},\nonumber \\
&&\pi^{(q)}_{\bar \psi LiAa\alpha}(x)={{\partial {\cal L}(x)}\over {\partial 
\partial_o{\bar \psi}^{(q)}_{LiAa\alpha}(x)}}=-{i\over 2}{(\gamma_o\psi^{(q)}
_{Li}(x))}_{Aa\alpha},\nonumber \\
&&\pi^{(q)}_{\psi RiA\alpha}(x)={{\partial {\cal L}(x)}\over {\partial 
\partial_o\psi^{(q)}_{RiA\alpha}(x)}}=-{i\over 2}{({\bar \psi}^{(q)}_{Ri}(x)
\gamma_o)}_{A\alpha},\nonumber \\
&&\pi^{(q)}_{\bar \psi RiA\alpha}(x)={{\partial {\cal L}(x)}\over {\partial 
\partial_o{\bar \psi}^{(q)}_{RiA\alpha}(x)}}=-{i\over 2}{(\gamma_o\psi^{(q)}
_{Ri}(x))}_{A\alpha},\nonumber \\
&&\pi^{(q)}_{{\bar {\tilde \psi}} RiA\alpha}(x)={{\partial {\cal L}(x)}\over 
{\partial \partial_o{\tilde \psi}^{(q)}_{RiA\alpha}(x)}}=-{i\over 2}{({\bar 
{\tilde \psi}}^{(q)}_{Ri}(x)\gamma_o)}_{A\alpha},\nonumber \\
&&\pi^{(q)}_{{\bar {\tilde \psi}} RiA\alpha}(x)={{\partial {\cal L}(x)}\over 
{\partial \partial_o{\bar {\tilde \psi}}^{(q)}_{RiA\alpha}(x)}}=-{i\over 2}
{(\gamma_o{\tilde \psi}^{(q)}_{Ri}(x))}_{A\alpha}.
\label{31}
\end{eqnarray}

They satisfy the standard Poisson brackets

\begin{eqnarray}
&&\lbrace G_{A\mu}(\vec x,x^o), \pi^{(G)\nu}_B(\vec y,x^o)\rbrace =\delta_{AB}
\delta^{\nu}_{\mu}\delta^3(\vec x-\vec y),\nonumber \\
&&\lbrace W_{a\mu}(\vec x,x^o),\pi_b^{(W)\nu}(\vec y,x^o)\rbrace =\delta_{ab}
\delta^{\nu}_{\mu}\delta^3(\vec x-\vec y),\nonumber \\
&&\lbrace V_{\mu}(\vec x,x^o), \pi^{(V)\nu}(\vec y,x^o)\rbrace =\delta^{\nu}
_{\mu}\delta^3(\vec x-\vec y),\nonumber \\
&&{}\nonumber \\
&&\lbrace \phi_a(\vec x,x^o),\pi_{\phi \, b}(\vec y,x^o)\rbrace =
\lbrace \phi^{*}_a(\vec x,x^o),\pi_{\phi^{\dagger}\, b}(\vec y,x^o)\rbrace =
\delta_{ab}\delta^3(\vec x-\vec y),\nonumber \\
&&\lbrace H(\vec x,x^o), \pi_H(\vec y,x^o)\rbrace =\delta^3(\vec x-\vec y),
\nonumber \\
&&\lbrace \theta_a(\vec x,x^o), \pi_{\theta \, b}(\vec y,x^o)\rbrace =\delta
_{ab}\delta^3(\vec x-\vec y),\nonumber \\
&&{}\nonumber \\
&&\lbrace \psi^{(l)}_{Lia\alpha}(\vec x,x^o),\pi^{(l)}_{\psi Ljb\beta}(\vec y,
x^o)\rbrace =\lbrace {\bar \psi}^{(l)}_{Lia\alpha}(\vec x,x^o),\pi^{(l)}
_{\bar \psi Ljb\beta}(\vec y,x^o)\rbrace =\nonumber \\
&&=-\delta_{ij}\delta_{ab}\delta_{\alpha\beta}\delta^3(\vec x-\vec y),
\nonumber \\
&&\lbrace \psi^{(l)}_{Ri\alpha}(\vec x,x^o),\pi^{(l)}_{\psi Rj\beta}(\vec y,x^o)
\rbrace =\lbrace {\bar \psi}^{(l)}_{Ri\alpha}(\vec x,x^o),\pi^{(l)}
_{\bar \psi Rj\beta}(\vec y,x^o)\rbrace =\nonumber \\
&&=-\delta_{ij}\delta_{\alpha\beta}\delta^3(\vec x-\vec y),\nonumber \\
&&\lbrace \psi^{(q)}_{LiAa\alpha}(\vec x,x^o),\pi^{(q)}_{\psi LjBb\beta}(\vec 
y,x^o)\rbrace =\lbrace {\bar \psi}^{(q)}_{LiAa\alpha}(\vec x,x^o),\pi
^{(q)}_{\bar \psi LjBb\beta}(\vec y,x^o)\rbrace =\nonumber \\
&&=\delta_{ij}\delta_{AB}
\delta_{ab}\delta_{\alpha\beta}\delta^3(\vec x-\vec y),\nonumber \\
&&\lbrace \psi^{(q)}_{RiA\alpha}(\vec x,x^o),\pi^{(q)}_{\psi RjB\beta}(\vec y,
x^o)\rbrace =\lbrace {\bar \psi}^{(q)}_{RiA\alpha}(\vec x,x^o),\pi^{(q)}
_{\bar \psi RjB\beta}(\vec y,x^o)\rbrace =\nonumber \\
&&=\delta_{ij}\delta_{\alpha\beta}\delta^3(\vec x-\vec y),\nonumber \\
&&\lbrace {\tilde \psi}^{(q)}_{RiA\alpha}(\vec x,x^o),\pi^{(q)}_{\tilde \psi 
RjB\beta}(\vec y,x^o)\rbrace =\lbrace {\bar {\tilde \psi}}^{(q)}_{RiA\alpha}
(\vec x,x^o),\pi^{(q)}_{{\bar {\tilde \psi}}RjB\beta}(\vec y,x^o)\rbrace 
=\nonumber \\
&&=\delta_{ij}\delta_{\alpha\beta}\delta^3(\vec x-\vec y).
\label{32}
\end{eqnarray}

All the fermionic momenta generate second class constraints of the type
$\pi_{\psi}(x)+{i \over 2}(\bar \psi (x)\gamma_o)\approx 0$, $\pi_{\bar \psi}
(x)+{i\over 2}\gamma_o\psi (x)\approx 0$, which are eliminated \cite{lusa}
by going to Dirac brackets; then the surviving variables $\psi (x), \bar \psi
(x)$ satisfy (for the sake of simplicity we still use the notation
$\lbrace .,.\rbrace$ for the Dirac brackets)

\begin{eqnarray}
&&\lbrace \psi^{(l)}_{Lia\alpha}(\vec x,x^o),{\bar \psi}^{(l)}_{Ljb\beta}(\vec 
y,x^o)\rbrace =-i\delta_{ij}\delta_{ab}{(\gamma^o)}_{\alpha\beta}\delta^3(\vec 
x-\vec y),\nonumber \\
&&\lbrace \psi^{(l)}_{Ri\alpha}(\vec x,x^o),{\bar \psi}^{(l)}_{Rj\beta}(\vec y,
x^o)\rbrace =-i\delta_{ij}{(\gamma^o)}_{\alpha\beta}\delta^3(\vec x-
\vec y),\nonumber \\
&&\lbrace \psi^{(q)}_{LiAa\alpha}(\vec x,x^o),{\bar \psi}^{(q)}_{LjBb\beta}
(\vec y,x^o)\rbrace =-i\delta_{ij}\delta_{AB}\delta_{ab}{(\gamma^o)}
_{\alpha\beta}\delta^3(\vec x-\vec y),\nonumber \\
&&\lbrace \psi^{(q)}_{RiA\alpha}(\vec x,x^o),{\bar \psi}^{(q)}_{RjB\beta}(\vec 
y,x^o)\rbrace =-i\delta_{ij}\delta_{AB}{(\gamma^o)}_{\alpha\beta}\delta^3(\vec 
x-\vec y),\nonumber \\
&&\lbrace {\tilde \psi}^{(q)}_{RiA\alpha}(\vec x,x^o),{\bar {\tilde \psi}}
^{(q)}_{RjB\beta}(\vec y,x^o)\rbrace =-i\delta_{ij}\delta_{AB}{(\gamma^o)}
_{\alpha\beta}\delta^3(\vec x-\vec y).
\label{33}
\end{eqnarray}

The resulting Dirac Hamiltonian density is [after  allowed integrations by 
parts; $\lambda^{(G)}_{Ao}(x), \lambda^{(W)}_{ao}(x), \lambda^{(V)}_o(x)$,
are Dirac multipliers; $B^{(G)k}_A(x)=-{1\over 2}\epsilon^{kij}G^{ij}_A(x)$,
$B^{(W)k}_a(x)=-{1\over 2}\epsilon^{kij}W^{ij}_a(x)$, $B^{(V)k}(x)=-{1\over 2}
\epsilon^{kij}V^{ij}(x)$ are the magnetic fields for the corresponding
interactions; $\vec \alpha =\gamma^o\vec \gamma$, $\beta =\gamma^o$]

\begin{eqnarray}
{\cal H}_D(x)&=&{1\over 2}\sum_A [g_s^2{\vec \pi}^{(G) 2}_A(x)+g^{-2}_s{\vec
B}^{(G)\, 2}_A(x)]+\nonumber \\
&+&{1\over 2}\sum_a [g_w^2{\vec \pi}^{(W)\, 2}_a(x)+g_w^{-2}{\vec B}^{(W)\, 2}
_a(x)] + {1\over 2}[g_y^2{\vec \pi}^{(V)\, 2}(x)+{\vec B}^{(V)\, 2}(x)]+
\nonumber \\
&+&\pi_{\phi}(x) \pi_{\phi^{\dagger}}(x)+ [{\vec D}^{(W,V)}\phi (x)]^{\dagger}
\cdot [{\vec D}^{(W,V)}\phi (x)] + \lambda (\phi^{\dagger}(x)\phi (x)
-\phi_o^2)^2+\nonumber \\
&+&{1\over 2} \psi^{(l)\dagger}_{Li}(x) [\vec \alpha \cdot (\vec \partial +
{\vec W}_a(x)T^a_w + \vec V(x) Y_w)-\loarrow{ (\vec \partial -{\vec W}_a(x)T^a_w
-\vec V(x)Y_w)}\cdot \vec \alpha ]\psi^{(l)}_{Li}(x)+\nonumber \\
&+&{i\over 2}\psi^{(l)\dagger}_{Ri}(x) [\vec \alpha \cdot (\vec \partial +\vec 
V(x) Y_w)-\loarrow{(\vec \partial -\vec V(x)Y_w)}\cdot \vec \alpha ]\psi^{(l)}
_{Ri}(x)-\nonumber \\
&-&{\bar \psi}^{(l)}_{Li}(x)\cdot {{\phi (x)}\over {\phi_o}} M^{(l)}_{ij}
\psi^{(l)}_{Rj}(x)- {\bar \psi}^{(l)}_{Ri}(x) M^{(l)\dagger}_{ij} {{\phi
^{\dagger}}\over {\phi_o}}\cdot \psi^{(l)}_{Lj}(x)+\nonumber \\
&+&{i\over 2}\psi^{(q)\dagger}_{Li}(x) [\vec \alpha \cdot (\vec \partial +
{\vec W}_a(x)T^a_w+\vec V(x)Y_w+{\vec G}_A(x)T^a_s)-\nonumber \\
&-&\loarrow{(\vec \partial -{\vec W}_a(x)T^a_w-\vec V(x)Y_w-{\vec G}_A(x)T^a_s)
}\cdot \vec \alpha ]\psi^{(q)}_{Li}(x)+\nonumber \\
&+&{i\over 2} \psi^{(q)\dagger}_{Ri}(x) [\vec \alpha \cdot (\vec \partial +
\vec V(x)Y_w+{\vec G}_A(x)T^A_s)-\loarrow{(\vec \partial -\vec V(x)Y_w-{\vec 
G}_A(x)T^A_s)}\cdot \vec \alpha ] \psi^{(q)}_{Ri}(x)+\nonumber \\
&+&{i\over 2} {\tilde \psi}^{(q)\dagger}_{Ri}(x) [\vec \alpha \cdot (\vec 
\partial +\vec V(x)Y_w+{\vec G}_A(x)T^A_s)-\loarrow{(\vec \partial -\vec V(x)
Y_w-{\vec G}_A(x)T^A_s)}\cdot \vec \alpha ] {\tilde \psi}^{(q)}_{Ri}(x)-
\nonumber \\
&-&{\bar \psi}^{(q)}_{Li}(x)\cdot {{\phi (x)}\over {\phi_o}} M^{(q)}_{ij}
\psi^{(q)}_{Rj}(x) - {\bar \psi}^{(q)}_{Ri}(x) M^{(q)\dagger}_{ij} {{\phi
^{\dagger}(x)}\over {\phi_o}}\cdot \psi^{(q)}_{Lj}(x)-\nonumber \\
&-&{\bar \psi}^{(q)}_{Li}(x)\cdot {{\tilde \phi (x)}\over {\phi_o}} {\tilde
M}^{(q)}_{ij} {\tilde \psi}^{(q)}_{Rj}(x) - {\bar {\tilde \psi}}^{(q)}_{Ri}(x)
{\tilde M}^{(q)\dagger}_{ij} {{{\tilde \phi}^{\dagger}(x)}\over {\phi_o}}\cdot
\psi^{(q)}_{Lj}(x)-\nonumber \\
&-& G_{Ao}(x) [-{\hat {\vec D}}^{(G)}_{AB}\cdot {\vec \pi}^{(G)}_B(x)+i\psi
^{(q)\dagger}_{Li}(x)T^A_s\psi^{(q)}_{Li}(x)+\nonumber \\
&+&i\psi^{(q)\dagger}_{Ri}(x)T^A_s
\psi^{(q)}_{Ri}(x)+i{\tilde \psi}^{(q)\dagger}_{Ri}(x)T^A_s{\tilde \psi}^{(q)}
_{Ri}(x)]-\nonumber \\
&-&W_{ao}(x) [-{\hat {\vec D}}^{(W)}_{ab}\cdot {\vec \pi}^{(W)}_b(x)+i\psi^{(l)
\dagger}_{Li}(x)T^a_w\psi^{(l)}_{Li}(x)+i\psi^{(q)\dagger}_{Li}(x)T^a_w\psi
^{(q)}_{Li}(x)-\nonumber \\
&-& (\pi_{\phi}(x)T^a_w \phi (x)- \phi^{\dagger}(x) T^a_w \pi_{\phi^{\dagger}}
(x))]-\nonumber \\
&-& V_o(x) [-\vec \partial \cdot {\vec \pi}^{(V)}(x)+i\psi^{(l)\dagger}_{Li}(x)
Y_w\psi^{(l)}_{Li}(x)+i\psi^{(l)\dagger}_{Ri}(x)Y_w\psi^{(l)}_{Ri}(x)+
\nonumber \\
&+&i\psi^{(q)\dagger}_{Li}(x)Y_w\psi^{(q)}_{Li}(x)+i\psi^{(q)\dagger}_{Ri}(x)Y_w
\psi^{(q)}_{Ri}(x)+i{\tilde \psi}^{(q)\dagger}_{Ri}(x)Y_w{\tilde \psi}^{(q)}
_{Ri}(x)-\nonumber \\
&-&(\pi_{\phi}(x)Y_w\phi (x)-\phi^{\dagger}(x)Y_w\pi_{\phi^{\dagger}}(x))]-
\nonumber \\
&-&\theta {{g^3_s}\over {8\pi^2}} {\vec \pi}^{(G)}_A(x) \cdot {\vec B}^{(G)}_A
(x)+\nonumber \\
&+&\lambda_{Ao}(x) \pi^{(G) o}_A(x) +\lambda_{ao}(x) \pi^{(W) o}_a(x)+
\lambda^{(V)}_o(x) \pi^{(V) o}(x).
\label{34}
\end{eqnarray}

\noindent We get $\lambda_{Ao}(x){\buildrel \circ \over =} {{\partial}\over
{\partial x^o}} G_{Ao}(x), \lambda_{a)}(x){\buildrel \circ \over =} {{\partial}
\over {\partial x^o}} W_{ao}(x), \lambda_o(x){\buildrel \circ \over =} 
{{\partial}\over {\partial x^o}} V_o(x)$.

The time constancy of the primary constraints $\pi^{(G) o}_A(x)\approx 0$,
$\pi^{(W) o}_a(x)\approx 0, \pi^{(V) o}(x)\approx 0$, yields the Gauss law
secondary constraints

\begin{eqnarray}
\Gamma^{(G)}_A(x)&=&g^{-2}_s L^{(G) o}_A(x) = -\vec \partial \cdot {\vec \pi}
^{(G)}_A(x)-c_{ABC} {\vec G}_B(x)\cdot {\vec \pi}^{(G)}_C(x)+\nonumber \\
&+&i\psi^{(q)\dagger}_{Li}(x)T^A_s\psi^{(q)}_{Li}(x)+i\psi^{(q)\dagger}_{Ri}(x)
T^A_s\psi^{(q)}_{Ri}(x)+i{\tilde \psi}^{(q)\dagger}_{Ri}(x)T^A_s{\tilde \psi}
^{(q)}_{Ri}(x) =\nonumber \\
&=&-{\hat {\vec D}}^{(G)}_{AB}(x)\cdot {\vec \pi}^{(G)}_B(x)+J^o_{sA}(x)
\approx 0,\nonumber \\
\Gamma^{(W)}_a(x)&=&g_w^{-2} L^{(W) o}_a(x)=-\vec \partial \cdot {\vec \pi}_a
^{(W)}(x)-\epsilon_{abc}{\vec W}_b(x)\cdot {\vec \pi}_c^{(W)}(x)+i\psi
^{(l)\dagger}
_{Li}(x)T^a_w\psi^{(l)}_{Li}(x)+\nonumber \\
&+&i\psi^{(q)\dagger}_{Li}(x)T^a_w\psi^{(q)}_{Li}(x)-(\pi_{\phi}(x)T^a_w\phi (x)
-\phi^{\dagger}(x)T^a_w\pi_{\phi^{\dagger}}(x))=\nonumber \\
&=&-{\hat {\vec D}}^{(W)}_{ab}(x)\cdot {\vec \pi}^{(W)}_b(x)-(\pi_{\phi}(x)
T^a_w\phi (x)-\phi^{\dagger}(x)T^a_w\pi_{\phi^{\dagger}}(x))+J^o_{wa}(x) 
\approx 0,\nonumber \\
\Gamma^{(V)}(x)&=&g_y^{-2} L^{(V) o}(x)=-\vec \partial \cdot {\vec \pi}^{(V)}
(x)+i\psi^{(l)\dagger}_{Li}(x)
Y_w\psi^{(l)}_{Li}(x)+i\psi^{(l)\dagger}_{Ri}(x)Y_w\psi^{(l)}_{Ri}(x)+
\nonumber \\
&+&i\psi^{(q)\dagger}_{Li}(x)Y_w\psi^{(q)}_{Li}(x)+i\psi^{(q)\dagger}_{Ri}(x)Y_w
\psi^{(q)}_{Ri}(x)+i{\tilde \psi}^{(q)\dagger}_{Ri}(x)Y_w{\tilde \psi}^{(q)}
_{Ri}(x)-\nonumber \\
&-&(\pi_{\phi}(x)Y_w\phi (x)-\phi^{\dagger}(x)Y_w\pi_{\phi^{\dagger}}(x))=
\nonumber \\
&=&-\vec \partial \cdot {\vec \pi}^{(V)}(x) -(\pi_{\phi}(x)Y_w\phi (x)-
\phi^{\dagger}(x)Y_w\pi_{\phi^{\dagger}}(x))+J^o_{Y_w}(x) \approx 0.
\label{35}
\end{eqnarray}

The secondary constraints are constants of the motion and the 16+6+2=24 primary 
and secondary constraints are all first class with the only nonvanishing
Poisson brackets

\begin{eqnarray}
&&\lbrace \Gamma^{(G)}_A(\vec x,x^o), \Gamma^{(G)}_B(\vec y,x^o)\rbrace =
c_{ABC} \Gamma^{(G)}_C(\vec x,x^o) \delta^3(\vec x-\vec y),\nonumber \\
&&\lbrace \Gamma^{(W)}_a(\vec x,x^o), \Gamma^{(W)}_b(\vec y,x^o) \rbrace =
\epsilon_{abc} \Gamma^{(W)}_c(\vec x,x^o) \delta^3(\vec x-\vec y).
\label{36}
\end{eqnarray}

Let us make a digression on the choice of the boundary conditions on the
various fields. The conserved energy-momentum and angular momentum tensor 
densities  and Poincar\'e generators are [$\sigma^{\mu\nu}={i\over 2}[\gamma
^{\mu},\gamma^{\nu}]$, $\sigma^i={1\over 2}\epsilon^{ijk}\sigma^{jk}$; 
$D^{(.)\dagger}_{\mu}$ are defined as in Eq.(\ref{13}) with a change of sign 
in the fields]

\begin{eqnarray}
\Theta^{\mu\nu}(x)&=&\Theta^{\nu\mu}(x)=
g^{-2}_s(G_A^{\mu\alpha}(x)G_{A\alpha}{}^{\nu}(x)+{1\over 4}
\eta^{\mu\nu}G_A^{\alpha\beta}(x)G_{A\alpha\beta}(x))+\nonumber \\
&+&g^{-2}_w(W_a^{\mu\alpha}(x)W_{a\alpha}{}^{\nu}(x)+{1\over 4}
\eta^{\mu\nu}W_a^{\alpha\beta}(x)W_{a\alpha\beta}(x))+\nonumber \\
&+&g^{-2}_y(V^{\mu\alpha}(x)V_{\alpha}{}^{\nu}(x)+{1\over 4}
\eta^{\mu\nu}V^{\alpha\beta}(x)V_{\alpha\beta}(x))+\nonumber \\
&+&{[D^{(W,V)\mu}\phi (x)]}^{\dagger}\, [D^{(W,V)\nu}\phi (x)]+{[D^{(W,V)\nu}
\phi (x)]}^{\dagger} [D^{(W,V)\mu}\phi (x)]-\nonumber \\
&-&\eta^{\mu\nu}[\, {[D^{(W,V)\alpha}\phi (x)]}^{\dagger} [D^{(W,V)}{}_{\alpha}
\phi (x)]-V(\phi )]+\nonumber \\
&+&{i\over 2}{\bar \psi}^{(l)}_{Li}(x)[\gamma^{\mu}D^{(W,V)\nu}-
\loarrow{D^{(W,V){\dagger}\nu}}\gamma^{\mu}]\psi^{(l)}_{Li}(x)+
\nonumber \\
&+&{i\over 2}{\bar \psi}^{(l)}_{Ri}(x)[\gamma^{\mu}D^{(V)\nu}-
\loarrow{D^{(V){\dagger}\nu}}\gamma^{\mu}]\psi^{(l)}_{Ri}(x)+
\nonumber \\
&-&\eta^{\mu\nu} [{\bar \psi}^{(l)}_{Li}(x)\cdot 
{{\phi (x)}\over {\phi_o}}M^{(l)}_{ij}\psi
^{(l)}_{Rj}(x)+{\bar \psi}^{(l)}_{Ri}(x)
M_{ij}^{(l)\dagger}{{\phi^{\dagger}(x)}\over
{\phi_o}}\cdot \psi^{(l)}_{Lj}(x)]+\nonumber \\
&+&{i\over 2}{\bar \psi}^{(q)}_{Li}(x)[\gamma^{\mu}D^{(W,V,G)\nu}-
\loarrow{D^{(W,V,G){\dagger}\nu}}\gamma^{\mu}]\psi^{(q)}_{Li}(x)+
\nonumber \\
&+&{i\over 2}{\bar \psi}^{(q)}_{Ri}(x)[\gamma^{\mu}D^{(V,G)\nu}-
\loarrow{D^{(V,G){\dagger}\nu}}\gamma^{\mu}]\psi^{(q)}_{Ri}(x)+
\nonumber \\
&+&{i\over 2}{\bar {\tilde \psi}}^{(q)}_{Ri}(x)[\gamma^{\mu}D^{(V,G)\nu}-
\loarrow{D^{(V,G){\dagger}\nu}}\gamma^{\mu}]{\tilde \psi}^{(q)}_{Ri}(x)+
\nonumber \\
&-&\eta^{\mu\nu} [
{\bar \psi}^{(q)}_{Li}(x)\cdot {{\phi (x)}\over {\phi_o}}M^{(q)}_{ij}\psi
^{(q)}_{Rj}(x)+{\bar \psi}^{(q)}_{Ri}(x)M_{ij}^{(q)\dagger}{{\phi^{\dagger}(x)}
\over {\phi_o}}\cdot \psi^{(q)}_{Lj}(x)] +\nonumber \\
&-&\eta^{\mu\nu} [
{\bar \psi}^{(q)}_{Li}(x)\cdot {{\tilde \phi (x)}\over {\phi_o}}{\tilde M}
^{(q)}_{ij}{\tilde \psi}^{(q)}_{Rj}(x)+{\bar {\tilde \psi}}^{(q)}_{Ri}(x)
{\tilde M}_{ij}^{(q)\dagger}{{{\tilde \phi}^{\dagger}(x)}\over {\phi_o}}\cdot 
\psi^{(q)}_{Lj}(x)],
\nonumber \\
{{\cal M}}^{\mu\alpha\beta}(x)&=&x^{\alpha}\Theta^{\mu\beta}(x)-x^{\beta}
\Theta^{\mu \alpha}(x)+{1\over 4}{\bar \psi}^{(l)}_{Li}(x)(\gamma^{\mu}\sigma
^{\alpha\beta}+\sigma^{\alpha\beta}\gamma^{\mu})\psi^{(l)}_{Li}(x)+\nonumber \\
&+&{1\over 4}{\bar \psi}^{(l)}_{Ri}(x)(\gamma^{\mu}\sigma
^{\alpha\beta}+\sigma^{\alpha\beta}\gamma^{\mu})\psi^{(l)}_{Ri}(x)+
{1\over 4}{\bar \psi}^{(q)}_{Li}(x)(\gamma^{\mu}\sigma
^{\alpha\beta}+\sigma^{\alpha\beta}\gamma^{\mu})\psi^{(q)}_{Li}(x)+\nonumber \\
&+&{1\over 4}{\bar \psi}^{(q)}_{Ri}(x)(\gamma^{\mu}\sigma
^{\alpha\beta}+\sigma^{\alpha\beta}\gamma^{\mu})\psi^{(q)}_{Ri}(x)+
{1\over 4}{\bar {\tilde \psi}}^{(q)}_{Ri}(x)(\gamma^{\mu}\sigma
^{\alpha\beta}+\sigma^{\alpha\beta}\gamma^{\mu}){\tilde \psi}^{(q)}_{Ri}(x),
\nonumber \\
&&{}\nonumber \\
&&\partial_{\nu}\Theta^{\nu\mu}(x){\buildrel \circ \over =} 0,\quad\quad
\partial_{\mu}{\cal M}^{\mu\alpha\beta}(x){\buildrel \circ \over =} 0,
\nonumber \\
&&{}\nonumber \\
P^{\mu}&=&\int d^3x \Theta^{o\mu}(\vec x,x^o),\nonumber \\
J^{\mu\nu}&=&\int d^3x {\cal M}^{o\mu\nu}(\vec x,x^o),\nonumber \\
&&{}\nonumber \\
P^o&=&\int d^3x\, \lbrace {1\over 2}\sum_A[g_s^2{\vec \pi}_A^{(G)2}(\vec x,x^o)+
g^{-2}_s{\vec B}_A^{(G)2}(\vec x,x^o)]+\nonumber \\
&+&{1\over 2}\sum_a[g_w^2{\vec \pi}_a^{(W)2}(\vec x,x^o)+
g^{-2}_w{\vec B}_a^{(W)2}(\vec x,x^o)]+\nonumber \\
&+&{1\over 2}[g_y^2{\vec \pi}^{(V)2}(\vec x,x^o)+
g^{-2}_y{\vec B}^{(V)2}(\vec x,x^o)]+\nonumber \\
&+&\pi_{\phi}(\vec x,x^o)\cdot \pi_{\phi^{\dagger}}(\vec x,x^o)+
{[{\vec D}^{(W,V)}\phi
(\vec x,x^o)]}^{\dagger}\cdot [{\vec D}^{(W,V)}\phi (\vec x,x^o)]+V(\phi (\vec 
x,x^o))+\nonumber \\
&+&{i\over 2}\psi^{(l)\dagger}_{Li}(\vec x,x^o)[D^{(W,V) o}-\loarrow{D^{(W,V)
{\dagger}\, o}}]\psi^{(l)}_{Li}(\vec x,x^o)+ \nonumber \\
&+&{i\over 2}\psi^{(l)\dagger}_{Ri}(\vec x,x^o)[D^{(V) o}-\loarrow{D^{(V)
{\dagger}\, o}}]\psi^{(l)}_{Ri}(\vec x,x^o)+ \nonumber \\
&+&{\bar \psi}^{(l)}_{Li}(x)\cdot 
{{\phi (x)}\over {\phi_o}}M^{(l)}_{ij}\psi
^{(l)}_{Rj}(x)+{\bar \psi}^{(l)}_{Ri}(x)
M_{ij}^{(l)\dagger}{{\phi^{\dagger}(x)}\over
{\phi_o}}\cdot \psi^{(l)}_{Lj}(x)+\nonumber \\
&+&{i\over 2}\psi^{(q)\dagger}_{Li}(\vec x,x^o)[D^{(W,V,G) o}-\loarrow{D^{(W,
V,G){\dagger}\, o}}]\psi^{(q)}_{Li}(\vec x,x^o)+ \nonumber \\
&+&{i\over 2}\psi^{(q)\dagger}_{Ri}(\vec x,x^o)[D^{(V,G) o}-\loarrow{D^{(V,G)
{\dagger}\, o}}]\psi^{(q)}_{Ri}(\vec x,x^o)+ \nonumber \\
&+&{i\over 2}{\tilde \psi}^{(q)\dagger}_{Ri}(\vec x,x^o)[D^{(V,G) o}-
\loarrow{D^{(V,G){\dagger}\, o}}]{\tilde \psi}^{(q)}_{Ri}(\vec x,x^o)+
\nonumber \\
&+& {\bar \psi}^{(q)}_{Li}(x)\cdot {{\phi (x)}\over {\phi_o}}M^{(q)}_{ij}\psi
^{(q)}_{Rj}(x)+{\bar \psi}^{(q)}_{Ri}(x)M_{ij}^{(q)\dagger}{{\phi^{\dagger}(x)}
\over {\phi_o}}\cdot \psi^{(q)}_{Lj}(x)] +\nonumber \\
&+&{\bar \psi}^{(q)}_{Li}(x)\cdot {{\tilde \phi (x)}\over {\phi_o}}{\tilde M}
^{(q)}_{ij}{\tilde \psi}^{(q)}_{Rj}(x)+{\bar {\tilde \psi}}^{(q)}_{Ri}(x)
{\tilde M}_{ij}^{(q)\dagger}{{{\tilde \phi}^{\dagger}(x)}\over {\phi_o}}\cdot 
\psi^{(q)}_{Lj}(x) \rbrace , \nonumber \\
P^i&=&\int d^3x\, \lbrace {({\vec \pi}^{(G)}_A(\vec x,x^o)\times {\vec B}^{(G)}
_A(\vec x,x^o))}^i+{({\vec \pi}^{(W)}_a(\vec x,x^o)\times {\vec B}^{(W)}
_a(\vec x,x^o))}^i+\nonumber \\
&+&{({\vec \pi}^{(V)}(\vec x,x^o)\times {\vec B}^{(V)}(\vec x,x^o))}^i+
\nonumber \\
&+&\pi_{\phi}(\vec x,x^o)D^{(W,V)i}\phi (\vec x,x^o)+{[D^{(W,V)i}\phi
(\vec x,x^o)]}^{\dagger}\pi_{\phi^{\dagger}}(\vec x,x^o)+\nonumber \\
&+&{i\over 2}\psi^{(l)\dagger}_{Li}(\vec x,x^o)[D^{(W,V)i}
+\loarrow{D^{(W,V){\dagger}i}}]\psi^{(l)}_{Li}(\vec x,x^o) \nonumber \\
&+&{i\over 2}\psi^{(l)\dagger}_{Ri}(\vec x,x^o)[D^{(V)i}
+\loarrow{D^{(V){\dagger}i}}]\psi^{(l)}_{Ri}(\vec x,x^o) \nonumber \\
&+&{i\over 2}\psi^{(q)\dagger}_{Li}(\vec x,x^o)[D^{(W,V,G)i}
+\loarrow{D^{(W,V,G){\dagger}i}}]\psi^{(q)}_{Li}(\vec x,x^o) \nonumber \\
&+&{i\over 2}\psi^{(q)\dagger}_{Ri}(\vec x,x^o)[D^{(V,G)i}
+\loarrow{D^{(V,G){\dagger}i}}]\psi^{(q)}_{Ri}(\vec x,x^o) \nonumber \\
&+&{i\over 2}{\tilde \psi}^{(q)\dagger}_{Ri}(\vec x,x^o)[D^{(V,G)i}
+\loarrow{D^{(V,G){\dagger}i}}]{\tilde \psi}^{(q)}_{Ri}(\vec x,x^o)\rbrace ,
\nonumber \\
J^i&=&{1\over 2}\epsilon^{ijk}J^{jk}=\int d^3x\, \lbrace {[\vec x\times({\vec
\pi}_A^{(G)}(\vec x,x^o)\times {\vec B}_A^{(G)}(\vec x,x^o))]}^i+\nonumber \\
&+&{[\vec x\times({\vec \pi}_a^{(W)}(\vec x,x^o)\times {\vec B}_a^{(W)}(\vec x,
x^o))]}^i+\nonumber \\
&+&{[\vec x\times({\vec \pi}^{(V)}(\vec x,x^o)\times {\vec B}^{(V)}(\vec x,x^o)
)]}^i-\nonumber \\
&+&{[\vec x\times (\pi_{\phi}(\vec x,x^o){\vec D}^{(W,V)}\phi (\vec x,x^o)+
({\vec D}^{(W,V)}\phi (\vec x,x^o))^{\dagger}\pi_{\phi^{\dagger}}
(\vec x,x^o))]}^i+\nonumber \\
&+&{i\over 2}\psi^{(l)\dagger}_{Li}(\vec x,x^o)[\vec x\times(D^{(W,V)i}
+\loarrow{D^{(W,V){\dagger}i}})]\psi^{(l)}_{Li}(\vec x,x^o) \nonumber \\
&+&{i\over 2}\psi^{(l)\dagger}_{Ri}(\vec x,x^o)[\vec x\times(D^{(V)i}
+\loarrow{D^{(V){\dagger}i}})]\psi^{(l)}_{Ri}(\vec x,x^o) \nonumber \\
&+&{i\over 2}\psi^{(q)\dagger}_{Li}(\vec x,x^o)[\vec x\times(D^{(W,V,G)i}
+\loarrow{D^{(W,V,G){\dagger}i}})]\psi^{(q)}_{Li}(\vec x,x^o) \nonumber \\
&+&{i\over 2}\psi^{(q)\dagger}_{Ri}(\vec x,x^o)[\vec x\times(D^{(V,G)i}
+\loarrow{D^{(V,G){\dagger}i}})]\psi^{(q)}_{Ri}(\vec x,x^o) \nonumber \\
&+&{i\over 2}{\tilde \psi}^{(q)\dagger}_{Ri}(\vec x,x^o)[\vec x\times(D^{(V,G)i}
+\loarrow{D^{(V,G){\dagger}i}})]{\tilde \psi}^{(q)}_{Ri}(\vec x,x^o)\rbrace 
\nonumber \\
&+&{1\over 2}\psi^{(l)\dagger}_{Li}(\vec x,x^o)\sigma^i\psi^{(l)}_{Li}
(\vec x,x^o)+{1\over 2}
\psi^{(l)\dagger}_{Ri}(\vec x,x^o)\sigma^i\psi^{(l)}_{Ri}(\vec x,x^o)+
\nonumber \\
&+&{1\over 2}
\psi^{(q)\dagger}_{Li}(\vec x,x^o)\sigma^i\psi^{(q)}_{Li}(\vec x,x^o)+
\nonumber \\
&+&{1\over 2}
\psi^{(q)\dagger}_{Ri}(\vec x,x^o)\sigma^i\psi^{(q)}_{Ri}(\vec x,x^o)+
{1\over 2}
{\tilde \psi}^{(q)\dagger}_{Ri}(\vec x,x^o)\sigma^i{\tilde \psi}^{(q)}_{Ri}
(\vec x,x^o)\rbrace , \nonumber \\
K^i&=&J^{oi}=x^oP^i-\int d^3x\, x^i \Theta^{oo}(\vec x,x^o).
\label{37}
\end{eqnarray}

\noindent Now, following Ref.\cite{lusa}, we will assume the following
not-Lorentz-invariant phase-space oriented boundary conditions for $r=|\,
\vec x\, | \rightarrow \infty$, implying that the ten Poincar\'e generators
are finite [these boundary conditions are natural for the covariantization of
the theory by reformulating it on spacelike hypersurfaces\cite{lusa,lus1,lv1};
see the comments in the Introduction]

\begin{eqnarray}
&&G_{A0}(\vec x,x^o),\quad W_{ao}(\vec x,x^o),\quad V_o(\vec x,x^o)\, 
\rightarrow {{const.}\over {r^{1+\epsilon}}} +O(r^{-2}),\nonumber \\
&&\pi^{(G)\, o}(\vec x,x^o),\quad \pi^{(W)\, o}(\vec x,x^o),\quad \pi^{(V)\, o}
(\vec x,x^o)\, \rightarrow {{const.}\over {r^{1+\epsilon}}} +O(r^{-2}),
\nonumber \\
&&{\vec G}_A(\vec x,x^o),\quad {\vec W}_a(\vec x,x^o),\quad \vec V(\vec x,x^o)
\, \rightarrow {{const.}\over {r^{2+\epsilon}}}+O(r^{-3}),\nonumber \\
&&\Rightarrow {\vec B}^{(G)}_A(\vec x,x^o),\quad {\vec B}^{(W)}_a(\vec x,x^o),
\quad {\vec B}^{(V)}(\vec x,x^o)\, \rightarrow {{const.}\over {r^{3+\epsilon}}}
+O(r^{-4}),\nonumber \\
&&{\vec \pi}^{(G)}_A(\vec x,x^o),\quad {\vec \pi}^{(W)}_a(\vec x,x^o),\quad
{\vec \pi}^{(V)}(\vec x,x^o)\, \rightarrow {{const.}\over {r^{2+\epsilon}}}
+O(r^{-3}),\nonumber \\
&&\lambda_{Ao}(\vec x,x^o),\quad \lambda_{ao}(\vec x,x^o),\quad \lambda_o
(\vec x,x^o)\, \rightarrow {{const.}\over {r^{1+\epsilon}}}+O(r^{-2}),
\nonumber \\
&&\pi^{(G)\, o}_A(\vec x,x^o),\quad \pi^{(W)\, o}_a(\vec x,x^o),\quad
\pi^{(V)\, o}(\vec x,x^o)\, \rightarrow {{const.}\over {r^{1+\epsilon}}}
+O(r^{-2}),\nonumber \\
&&U_s(\vec x,x^o),\quad U_w(\vec x,x^o),\quad U_y(\vec x,x^o)\, \rightarrow
U_{s,w,y,\, \infty}+O(r^{-1}),\quad\quad U_{s,w,y,\, \infty}=const.,\nonumber \\
&&\vec \partial U_s(\vec x,x^o),\quad \vec \partial U_w(\vec x,x^o),\quad
\vec \partial U_y(\vec x,x^o)\, \rightarrow {{const.}\over {r^{2+\epsilon}}}
+O(r^{-3}),\nonumber \\
&&\partial^oU_s(\vec x,x^o),\quad \partial^oU_w(\vec x,x^o),\quad \partial^o
U_y(\vec x,x^o)\, \rightarrow {{const.}\over {r^{1+\epsilon}}}+O(r^{-2}),
\nonumber \\
&&\phi (\vec x,x^o)\, \rightarrow const. +{{const.}\over {r^{2+\epsilon}}}+
O(r^{-3}),\quad \Rightarrow \phi (x)=\phi_o\, allowed,\nonumber \\
&&\pi_{\phi}(\vec x,x^o)\, \rightarrow {{const.}\over {r^{2+\epsilon}}}+
O(r^{-3}),\nonumber \\
&&\psi (\vec x,x^o)\, \rightarrow {{const.}\over {r^{3/2+\epsilon}}}+
O(r^{-2}),\quad \psi =\psi^{(l)}_{Li}, \psi^{(l)}_{Ri}, \psi^{(q)}_{Li},
\psi^{(q)}_{Ri}, {\tilde \psi}^{(q)}_{Ri},\nonumber \\
&&\Rightarrow \Gamma_A^{(G)}(\vec x,x^o),\quad \Gamma^{(W)}_a(\vec x,x^o),
\quad \Gamma^{(V)}(\vec x,x^o)\, \rightarrow {{const.}\over {r^{3+\epsilon}}}
+O(r^{-4}),\nonumber \\
&&\Rightarrow \alpha^{(G)}_A(\vec x,x^o),\quad \alpha^{(W)}_a(\vec x,x^o),
\quad \alpha^{(V)}(\vec x,x^o)\, \rightarrow {{const.}\over {r^{3+\epsilon}}}
+O(r^{-4}).
\label{38}
\end{eqnarray}

In the last line of the previous equations we have denoted with $\alpha^{(G)}
_A(x), \alpha^{(W)}_a(x), \alpha^{(V)}(x)$, the parameters of the infinitesimal
gauge transformations generated by the first class constraints $\Gamma^{(G)}_A
(x)\approx 0, \Gamma^{(W)}_a(x)\approx 0, \Gamma^{(V)}(x)\approx 0$; as shown in
Ref.\cite{lusa}, they must have the same boundary conditions as the Gauss laws.

With these angle-independent limits for $r\rightarrow \infty$, the non-Abelian
color charges (see Ref.\cite{lusa}) transform covariantly under the gauge
transformations, which in turn preserve the boundary conditions on the
fields. We have assumed the same boundary conditions of the strong
interactions also for the electroweak ones, so that also the (unbroken)
non-abelian $SU(2)\times U(1)$ charges transform covariantly under gauge
transformations. Moreover, with some refining of these boundary conditions
(see Ref.\cite{lusa}), one can avoid any form of Gribov ambiguity for all
the interactions.

Let us now reformulate the Hamiltonian theory in terms of the fields H(x),
$\theta_a(x)$, $A_{\mu}(x)$, $Z_{\mu}(x)$, $W_{\pm \mu}(x)$ [see Eq.(\ref{17})]
, rather than in terms of the fields $\phi_a(x)$, $V_{\mu}(x)$, $W_{a\mu}(x)$.

By using Eqs.(\ref{5}), (\ref{7}), (\ref{8}), 
(\ref{10}), (\ref{18}), the formula
$e^{-\theta_bT^b_w} T^a_w e^{\theta_cT^c_w}= T^c_w (e^{\theta_b{\hat T}^b_w})
_{ca}$ [$({\hat T}^c_w)_{ab}=\epsilon_{abc}$; see Ref.\cite{lusa}], the
identities $\tau^a\tau^b=\delta_{ab}+i\epsilon_{abc}\tau^c$, $\tau^c\tau^a
\tau^d=i\epsilon_{adc}+\delta_{ad}\tau^c+\delta_{ac}\tau^d-\delta_{cd}\tau^a$,
$\tau^c\tau^a\tau^b-\tau^a\tau^b\tau^c=2(i\epsilon_{abc}+\delta_{ac}\tau^b-
\delta_{bc}\tau^a)$, $\tau^c\tau^a\tau^b\tau^d=\delta_{ab}\delta_{cd}+\delta
_{ac}\delta_{bd}+i\epsilon_{abc}\tau^d+i(\delta_{ab}\epsilon_{cde}+\delta_{ac}
\epsilon_{bde}-\delta_{bc}\epsilon_{ade})\tau^e$, 
$\left( \begin{array}{ll} 0 & 1\end{array} \right) \tau^a
\left( \begin{array}{l} 0\\ 1\end{array} \right) =-\delta_{a3}$,
$\left( \begin{array}{ll} 0 & 1\end{array} \right) \tau^a\tau^b\left( 
\begin{array}{l} 0\\ 1\end{array} \right) =\delta_{ab}$, and $e^{\theta_bT^b
_w}=e^{-{i\over 2}\theta {\hat n}_b\tau^b}=cos\, {{\theta}\over 2}-isin\,
{{\theta}\over 2} {\hat n}_b\tau^b$ [with $\sum_b{\hat n}^2_b=1$, $\theta =
\sqrt{\theta^2_1+\theta^2_2+\theta^2_3}$, ${\hat n}_a=\theta_a/\theta$, 
$\partial \theta /\partial \theta_a={\hat n}_a$, $\partial {\hat n}_a/
\partial \theta_b={1\over {\theta}}(\delta_{ab}-{\hat n}_a{\hat n}_b)$],
we get (see also II) with $Y_w\phi (x)=-{i\over 2}\phi (x)$

\begin{eqnarray}
&&{}\nonumber \\
W^{'}_{a\mu}(x)T^a_w&=&e^{-\theta_b(x)T^b_w}[W_{a\mu}(x)T^a_w+\partial_{\mu}]
e^{\theta_c(x)T^c_w}=\nonumber \\
&=&(\, [cos\theta(x)\delta_{ab} + 2 sin^2{\theta(x)\over 2} 
{\hat n}_a(x) {\hat n}_b(x)
+ sin\theta(x) \epsilon_{abc} {\hat n}_c(x) ] W_{b\mu}(x) + \nonumber \\
&+& [ {\hat n}_a(x){\hat n}_b(x) + {{sin\theta(x)}\over {\theta (x)}}(\delta
_{ab}-{\hat n}_a(x){\hat n}_b(x))+\nonumber \\
&+&{2\over {\theta (x)}} sin^2 {\theta(x)\over 2}
\epsilon_{abc} {\hat n}_c(x)] \partial_\mu \theta_b(x)\, )T^a_w = \nonumber \\
&=&(\, [cos\, \theta (x) \delta_{a3}+2sin^2\, {{\theta (x)}\over 2}{\hat n}_a
{\hat n}_3+sin\, \theta (x) \epsilon_{a3c}{\hat n}_c(x)](A_{\mu}(x)+cot\,
\theta_w Z_{\mu}(x))+\nonumber \\
&+&[cos\, \theta (x) {{\delta_{a1}+i\delta_{a2}}\over {\sqrt{2}}}+2sin^2\,
{{\theta (x)}\over 2}{\hat n}_a(x){\hat n}_{-}(x)+\nonumber \\
&+&{{sin\, \theta (x)}\over
{\sqrt{2}}}(\epsilon_{a1c}+i\epsilon_{a2c}){\hat n}_c(x)]W_{+\mu}(x)+
\nonumber \\
&+&[cos\, \theta (x) {{\delta_{a1}-i\delta_{a2}}\over {\sqrt{2}}}+2sin^2\,
{{\theta (x)}\over 2}{\hat n}_a(x){\hat n}_{+}(x)+\nonumber \\
&+&{{sin\, \theta (x)}\over
{\sqrt{2}}}(\epsilon_{a1c}-i\epsilon_{a2c}){\hat n}_c(x)]W_{-\mu}(x)+
\nonumber \\
&+&[{\hat n}_a(x){\hat n}_b(x)+{{sin\, \theta (x)}\over {\theta (x)}}(\delta
_{ab}-{\hat n}_a(x){\hat n}_b(x))+\nonumber \\
&+&{2\over {\theta^2(x)}}sin^2\, {{\theta (x)}
\over 2} \epsilon_{abc}{\hat n}_c(x)] \partial_{\mu}\theta_b(x)\, )T^a_w,
\nonumber \\
&&{}\nonumber \\
&&D^{(W,V)}_{\mu}\phi (x)=\nonumber \\
&=&e^{\theta_d(x)T^d_w}\lbrace {1\over {\sqrt{2}}}\partial_{\mu}H(x)-{i\over 2}
(\phi_o+{1\over {\sqrt{2}}}H(x))[V_{\mu}(x)+W^{'}_{a\mu}(x)\tau^a]\rbrace 
\left( \begin{array}{l} 0\\ 1 \end{array} \right) ,\nonumber \\
&&{}\nonumber \\
&&[D^{(W,V)}_{\mu}\phi (x)]^{\dagger}=\nonumber \\
&=&\left( \begin{array}{ll} 0 & 1\end{array}
\right) \lbrace {1\over {\sqrt{2}}}\partial_{\mu}H(x)+{i\over 2}(\phi_o+{1\over 
{\sqrt{2}}}H(x))[V_{\mu}(x)+W^{'}_{a\mu}(x)\tau^a]\rbrace e^{-\theta_d(x)
T^d_w},\nonumber \\
&&{}\nonumber \\
&&[D^{(W,V)}_{\mu}\phi (x)]^{\dagger}[D^{(W,V)\mu}\phi (x)]-\lambda (\phi
^{\dagger}(x)\phi (x) -\phi_o^2)^2=\nonumber \\
&=&{1\over 2}\partial_{\mu}H(x)\partial^{\mu}H(x)-{1\over 2}m_H^2 H^2(x) 
(1+{e\over {2sin\, 2\theta_wm_Z}}H(x))^2+\nonumber \\
&+&{1\over 4} (\phi_o+{1\over {\sqrt{2}}}H(x))^2 
[V^{\mu}(x)V_{\mu}(x)-2V^{\mu}(x)W^{'}_{3\mu}(x)+\sum_aW^{'}_{a\mu}(x)W^{{'}\,
\mu}_a(x)]=\nonumber \\
&=&{1\over 2}\partial_{\mu}H(x)\partial^{\mu}H(x)-{1\over 2}m_H^2 H^2(x) 
(1+{e\over {2sin\, 2\theta_wm_Z}}H(x))^2+\nonumber \\
&+&{1\over 4} (\phi_o+{1\over {\sqrt{2}}}H(x))^2 \lbrace (A(x)-tg\, \theta_w 
Z(x))^2+(A(x)+cot\, \theta_w Z(x))^2\nonumber \\
&-&2(cos\, \theta (x)+2sin^2\, {{\theta (x)}\over 
2}{\hat n}_3(x))(A^2(x)-Z^2(x)+2cot\, \theta_wA(x)\cdot Z(x))-\nonumber \\
&-&2(A^{\mu}(x)-tg\, \theta_wZ^{\mu}(x))[i sin\, \theta (x) ({\hat n}_{+}(x)
W_{- \mu}(x)-{\hat n}_{-}(x)W_{+ \mu}(x))+\nonumber \\
&+&2sin^2{{\theta (x)}\over 2} {\hat n}_3(x)({\hat n}_{+}(x)W_{- \mu}(x)+
{\hat n}_{-}(x)W_{+ \mu}(x))]+2 W_{+}(x)\cdot W_{-}(x)+\nonumber \\
&+&[{\hat n}_a(x){\hat n}_b(x)+{4\over {\theta^2(x)}}sin^2{{\theta (x)}\over 2}
(\delta_{ab}-{\hat n}_a(x){\hat n}_b(x))]\partial_{\mu}\theta_a(x) \partial
^{\mu}\theta_b(x)+\nonumber \\
&+&4 ({{Z^{\mu}(x)}\over {sin\, 2\theta_w}} [{\hat n}_3(x){\hat n}_a(x)+
{{sin\, \theta (x)}\over {\theta (x)}}(\delta_{3a}-{\hat n}_3(x){\hat n}_a
(x))]-\nonumber \\
&-&{2\over {\theta (x)}} sin^2{{\theta (x)}\over 2} \epsilon_{3ab}{\hat n}_b(x)
(A^{\mu}(x)+cot\, 2\theta_w Z^{\mu}(x))\, )\partial_{\mu}\theta_a(x)
\rbrace , \nonumber \\
&&{}\nonumber \\
&&\theta_{\pm}(x)={1\over {\sqrt{2}}}(\theta_1(x)\mp i\theta_2(x)),\quad\quad
{\hat n}_{\pm}(x)={1\over {\sqrt{2}}} ({\hat n}_1(x)\mp i{\hat n}_2(x)).
\label{39}
\end{eqnarray}

The new canonical momenta associated with the Lagrangian density (\ref{17})
are [see Eq.(\ref{17}) for the expression of the
kinetic term $-{1\over {4g^2_w}}W^{\mu\nu}_aW_{a\mu\nu}-{1\over {4g^2_y}}
V^{\mu\nu}V_{\mu\nu}$ in terms of $A_{\mu}, Z_{\mu}, W_{\pm \mu}$;
${\cal W}^{oi}_{\pm}=\partial^oW^i_{\pm}-\partial^iW^o_{\pm}$]

\begin{eqnarray}
\pi_H(x)&=&{{\partial {\cal L}(x)}\over {\partial \partial_oH(x)}}=\partial_o
H(x),\nonumber \\
\pi_{\theta a}(x)&=&{{\partial {\cal L}(x)}\over {\partial \partial_o\theta
_a (x)}}={1\over 2}(\phi_o+{1\over {\sqrt{2}}}H(x))^2\lbrace [{\hat n}_a(x)
{\hat n}_b(x)+{4\over {\theta^2(x)}}sin^2\, {{\theta (x)}\over 2}\cdot
\nonumber \\
&\cdot& (\delta_{ab}-{\hat n}_a(x){\hat n}_b(x))] \partial^o\theta_b(x)+2
[{{Z^o(x)}\over {sin\, 2\theta_w}}({\hat n}_3(x){\hat n}_a(x)+\nonumber \\
&+&{{sin\, \theta 
(x)}\over {\theta (x)}}(\delta_{3a}-{\hat n}_3(x){\hat n}_a(x))-{2\over {\theta
(x)}}sin^2\, {{\theta (x)}\over 2}\epsilon_{3ab}{\hat n}_b(x)(A^o(x)+cot\,
2\theta_w Z^o(x))]\rbrace ,\nonumber \\
&&{}\nonumber \\
&&\Rightarrow  \partial^o\theta_a(x)={2\over {(\phi_o+{1\over {\sqrt{2}}}H(x))^2
}}[{\hat n}_a(x){\hat n}_b(x)+\nonumber \\
&+&{{\theta^2(x)}\over {4sin^2\, {{\theta (x)}\over
2} }}(\delta_{ab}-{\hat n}_a(x){\hat n}_b(x))]\, \pi_{\theta_b}(x)-\nonumber \\
&&-{{2Z^o(x)}\over {sin\, 2\theta_w}}[{\hat n}_a(x){\hat n}_3(x)+{{\theta (x)
sin\, \theta (x)}\over {4sin^2\, {{\theta (x)}\over 2} }}(\delta_{a3}-{\hat n}
_a(x){\hat n}_3(x))]-\nonumber \\
&-&\theta (x)\epsilon_{3ab}{\hat n}_b(x)(A^o(x)+cot\, 
2\theta_w Z^o(x)),\nonumber \\
&&{}\nonumber \\
\pi^{(A) o}(x)&=&{{\partial {\cal L}(x)}\over {\partial \partial_oA_o(x)}}=
0,\nonumber \\
\pi^{(A) i}(x)&=&{{\partial {\cal L}(x)}\over {\partial \partial_oA_i(x)}}=
-e^{-2}A^{oi}(x)+ie^{-2}sin^2\, \theta_w[W^o_{+}(x)W^i_{-}(x)-W^i_{+}(x)W^o
_{-}(x)],\nonumber \\
&&{}\nonumber \\
&&\Rightarrow \partial^oA^i(x)=\partial^iA^o(x)-e^2\pi^{(A)i}(x)+i sin^2\,
\theta_w (W^o_{+}(x)W^i_{-}(x)-W^i_{+}(x)W^o_{-}(x)),\nonumber \\
\pi^{(Z) o}(x)&=&{{\partial {\cal L}(x)}\over {\partial \partial_oZ_o(x)}}=
0,\nonumber \\
\pi^{(Z) i}(x)&=&{{\partial {\cal L}(x)}\over {\partial \partial_oZ_i(x)}}=
=-e^{-2}Z^{oi}(x)+{i\over 2}e^{-2}sin^2\, \theta_w[W^o_{+}(x)W^i_{-}(x)-W^i
_{+}(x)W^o_{-}(x)],\nonumber \\
&&{}\nonumber \\
&&\Rightarrow \partial^oZ^i(x)=\partial^iZ^o(x)-e^2\pi^{(Z)i}(x)+{i\over 2}
sin\, 2\theta_w (W^o_{+}(x)W^i_{-}(x)-W^i_{+}(x)W^o_{-}(x)),\nonumber \\
\pi^{(W_{\pm}) o}(x)&=&{{\partial {\cal L}(x)}\over {\partial \partial_oW_{\pm
o}(x)}}=0,\nonumber \\
\pi^{(W_{\pm}) i}(x)&=&{{\partial {\cal L}(x)}\over {\partial \partial_oW_{\pm
i}(x)}}=-e^{-2}sin^2\, \theta_w {\cal W}^{oi}_{\mp}(x)\pm \nonumber \\
&\pm& ie^{-2}sin^2\, \theta_w [W^o_{\mp}(x) (A^i(x)+cot\, \theta_w Z^i(x))- 
W^i_{\mp}(x) (A^o(x)+cot\, \theta_w Z^o(x))],\nonumber \\
&&{}\nonumber \\
&&\Rightarrow \partial^oW^i_{\mp}(x)=\partial^iW^o_{\mp}(x)-{{e^2}\over
{sin^2\, \theta_w}}\pi^{(W_{\pm})i}(x)\pm \nonumber \\
&\pm& i [W^o_{\mp}(x)(A^i(x)+cot\, \theta_w Z^i(x))-W^i_{\mp}(x)(A^o(x)+cot\,
\theta_w Z^o(x))];
\label{40}
\end{eqnarray}

\noindent they satisfy standard Poisson brackets.

These new momenta are related to $\pi_{\phi}(x), \pi_{\phi^{\dagger}}(x), 
\pi^{(V)\mu}(x), \pi_a^{(W)\mu}(x)$, by the equations 

\begin{eqnarray}
\pi_{\phi}(x)&=&[D^{(W,V) o}\phi (x)]^{\dagger}=\nonumber \\
&=&\left( \begin{array}{ll} 0 & 1\end{array} \right) 
\lbrace {1\over {\sqrt{2}}} \pi_H(x)+{i\over{2}}(\phi_o+
{1\over {\sqrt{2}}}H(x))(V^o(x)+W^{{'} o}_a\tau^a)]e^{-\theta_d(x)T^d_w},
\nonumber \\
&&{}\nonumber \\
\pi_{\phi^{\dagger}}(x)&=&D^{(W,V) o}\phi (x)=\nonumber \\
&=&e^{\theta_d(x)T^d_w}\lbrace {1\over {\sqrt{2}}} \pi_H(x)-{i\over 2}
(\phi_o+{1\over {\sqrt{2}}}H(x))(V^o(x)+W^{{'} o}_a\tau^a)]
\left( \begin{array}{l} 0\\ 1 \end{array} \right) ,\nonumber \\
&&{}\nonumber \\
V^o(x)&+&W^{{'} o}_a(x)\tau^a=A^o(x)-tg\, \theta_w Z^o(x)+\nonumber \\
&+&\lbrace [cos\, \theta (x)\delta_{a3}+2sin^2{{\theta (x)}\over 2}{\hat n}_a
(x){\hat n}_3(x)+sin\, \theta (x)\epsilon_{a3b}{\hat n}_b(x)](A^o(x)+cot\,
\theta_w Z^o(x))+\nonumber \\
&+&[cos\, \theta (x) {{\delta_{a1}+i\delta_{a2}}\over {\sqrt{2}}}+2 sin^2\,
{{\theta (x)}\over 2}{\hat n}_a(x){\hat n}_{-}(x)+{{sin\, \theta (x)}\over
{\sqrt{2}}}(\epsilon_{a1b}+i\epsilon_{a2b}){\hat n}_b(x)]W^o_{+}(x)+\nonumber \\
&+&[cos\, \theta (x){{\delta_{a1}-i\delta_{a2}}\over {\sqrt{2}}}+2 sin^2\,
{{\theta (x)}\over 2}{\hat n}_a(x){\hat n}_{+}(x)+{{sin\, \theta (x)}\over
{\sqrt{2}}}(\epsilon_{a1b}-i\epsilon_{a2b}){\hat n}_b(x)]W^o_{-}(x)+
\nonumber \\
&+&{2\over {(\phi_o+{1\over {\sqrt{2}}}H(x))^2}}[{\hat n}_a(x){\hat n}_b(x)+
{{\theta (x)sin\, \theta (x)}\over {4 sin^2\, {{\theta (x)}\over 2} }}
(\delta_{ab}-{\hat n}_a(x){\hat n}_b(x))+\nonumber \\
&+&{1\over 2}\epsilon_{abc}{\hat n}_c(x)]
\pi_{\theta_b}(x)-\nonumber \\
&-&{{2 Z^o(x)}\over {sin\, 2\theta_w}} [cos^2{{\theta (x)}\over 2}\delta_{a3}+
sin^2\, {{\theta (x)}\over 2}{\hat n}_a(x){\hat n}_3(x)+{{sin\, \theta (x)}
\over {2\theta (x)}}\epsilon_{a3b}{\hat n}_b(x)]+\nonumber \\
&+&(A^o(x)+cot\, 2\theta_w Z^o(x))[sin\, \theta (x) \epsilon_{3ab}{\hat n}_b
(x)+{2\over {\theta (x)}}sin^2\, {{\theta (x)}\over 2}(\delta_{a3}-{\hat n}_3
(x){\hat n}_a(x))] \rbrace \tau^a ,\nonumber \\
&&{}\nonumber \\
&&{\vec \pi}^{(W)}_1(x) = {1\over {\sqrt{2}}} ({\vec \pi}^{(W_{-})}(x)+{\vec
\pi}^{(W_{+})}(x)),\nonumber \\
&&{\vec \pi}^{(W)}_2(x) = {i\over {\sqrt{2}}} ({\vec \pi}^{(W_{-})}(x)-{\vec
\pi}^{(W_{+})}(x)),\nonumber \\
&&{\vec \pi}^{(W)}_3(x) = sin\, \theta_w (sin\, \theta_w {\vec \pi}^{(A)}(x)+
cos\, \theta_w {\vec \pi}^{(Z)}(x)),\nonumber \\
&&{\vec \pi}^{(V)}(x) = cos\, \theta_w (cos\, \theta_w {\vec \pi}^{(A)} -sin\,
\theta_w {\vec \pi}^{(Z)}(x)),\nonumber \\
&&{}\nonumber \\
&&{\vec \pi}^{(A)}(x) = {\vec \pi}^{(W)}_3(x)+{\vec \pi}^{(V)}(x), \nonumber \\
&&{\vec \pi}^{(Z)}(x) = cot\, \theta_w {\vec \pi}^{(W)}_3(x)-tg\, \theta_w 
{\vec \pi}^{(V)}(x), \nonumber \\
&&{\vec \pi}^{(W_{\mp})}(x) = {1\over {\sqrt{2}}} ({\vec \pi}^{(W)}_1(x)\mp i 
{\vec \pi}^{(W)}_2(x)).
\label{41}
\end{eqnarray}

The secondary constraints (\ref{38}) may be rewritten in terms of the new
momenta in the following form:

\begin{eqnarray}
\Gamma^{(G)}_A(x)&=&-\vec \partial \cdot {\vec \pi}^{(G)}_A(x)-c_{ABC}{\vec
G}_B(x)\cdot {\vec \pi}^{(G)}_C(x)+J^o_{sA}(x)\approx 0,\nonumber \\
\Gamma^{(A)}(x)&=&{1\over {\sqrt{2}}} (\Gamma^{(W)}_3(x) + \Gamma^{(V)}(x)) =
\nonumber \\
&=&-\vec \partial \cdot {\vec \pi}^{(A)}(x) +i({\vec W}_{+}(x)
\cdot {\vec \pi}^{(W_{+})}(x) - {\vec W}_{-}(x)\cdot {\vec \pi}^{(W_{+})}(x))
+\nonumber \\
&+&{i\over 2}[\pi_{\phi}(x)(1+\tau^3)\phi (x)-\phi^{\dagger}(x)(1+\tau^3)\pi
_{\phi^{\dagger}}(x)]+{\tilde j}^o_{(em)}(x)\approx 0,\nonumber \\
\Gamma^{(Z)}(x)&=&cot\, \theta_w \Gamma^{(W)}_3(x) -tg\, \theta_w \Gamma^{(V)}
(x)=\nonumber \\
&=& -\vec \partial \cdot {\vec \pi}^{(Z)}(x) +icot\, \theta_w ({\vec W}_{+}(x)
\cdot {\vec \pi}^{(W_{+})}(x) - {\vec W}_{-}(x)\cdot {\vec \pi}^{(W_{+})}(x))
+\nonumber \\
&+&{i\over 2}[\pi_{\phi}(x)(cot\, \theta_w\tau^3-tg\, \theta_w)\phi (x)-\phi
^{\dagger}(x)(cot\, \theta_w\tau^3-tg\, \theta_w)\pi
_{\phi^{\dagger}}(x)]+{\tilde j}^o_{(NC)}(x)\approx 0,\nonumber \\
\Gamma^{(W_{\pm})}(x)&=&{1\over {\sqrt{2}}} (\Gamma^{(W)}_1(x) \mp i \Gamma
^{(W)}_2(x))=\nonumber \\
&=& -\vec \partial \cdot {\vec \pi}^{(W_{\mp})}(x)\mp i ({\vec W}_{\pm}(x)\cdot
sin\, \theta_w (sin\, \theta_w {\vec \pi}^{(A)}(x)+cos\, \theta_w {\vec \pi}
^{(Z)}(x)) - \nonumber \\
&-&(\vec A(x)+cot\, \theta_w \vec Z(x))\cdot {\vec \pi}^{(W_{\mp})}
(x)) +\nonumber \\
&+&{i\over 2}[\pi_{\phi}(x)\tau^{\mp}\phi (x)-\phi^{\dagger}(x)\tau^{\mp}\pi
_{\phi^{\dagger}}(x)]+{\tilde j}^o_{(CC)\mp}(x)\approx 0,\nonumber \\
&&{}\nonumber \\
\pi_{\phi}\phi&& -\phi^{\dagger}\pi_{\phi^{\dagger}}=i(\phi_o+{1\over {\sqrt{2}
}}H)^2(V^o-W^{{'}o}_3),\nonumber \\
\pi_{\phi}\tau^3\phi&&-\phi^{\dagger}\tau^3\pi_{\phi^{\dagger}}=
-i(\phi_o+{1\over {\sqrt{2}}}H)^2[(cos\, \theta +2sin^2\, {{\theta}\over 2}
{\hat n}^2_3)(V^o-W^{{'}o}_3)+\nonumber \\
&+&(sin\, \theta {\hat n}_2-2 sin^2\,{{\theta}\over 2}{\hat n}_1{\hat n}_3)
W^{{'} o}_1-(sin\, \theta {\hat n}_1+2 sin^2{{\theta}\over 2}{\hat n}_2{\hat
n}_3)W^{{'} o}_2],\nonumber \\
\pi_{\phi}\tau^1\phi&&-\phi^{\dagger}\tau^1\pi_{\phi^{\dagger}}=
-i(\phi_o+{1\over {\sqrt{2}}}H)^2[(sin\, \theta {\hat n}_2+2 sin^2\, {{\theta}
\over 2}{\hat n}_1{\hat n}_3)(V^o-W^{{'} o}_3)-\nonumber \\
&-&(cos\, \theta +2 sin^2\, {{\theta}\over 2}{\hat n}^2_1)W^{{'}o}_1+(sin\,
\theta {\hat n}_3-2 sin^2\, {{\theta}\over 2}{\hat n}_1{\hat n}_2)W^{{'} o}_2],
\nonumber \\
\pi_{\phi}\tau^2\phi&&-\phi^{\dagger}\tau^2\pi_{\phi^{\dagger}}=
-i(\phi_o+{1\over {\sqrt{2}}}H)^2[-(sin\, \theta {\hat n}_1-2 sin^2\, {{\theta}
\over 2}{\hat n}_2{\hat n}_3)(V^o-W^{{'}o}_3)-\nonumber \\
&-&(sin\, \theta {\hat n}_3+2 sin^2\, {{\theta}\over 2}{\hat n}_1{\hat n}_2)
W^{{'}o}_1-(cos\, \theta +2 sin^2\, {{\theta}\over 2}{\hat n}^2_2)W^{{'}o}_2],
\nonumber \\
&&{}\nonumber \\
V^o&&-W^{{'}o}_3=-{2\over {(\phi_o+{1\over {\sqrt{2}}}H)^2}}
[{\hat n}_3{\hat n}_b+{{\theta sin\, \theta}\over {4sin^2\, {{\theta}
\over } }}(\delta_{3b}-{\hat n}_3{\hat n}_b)+{{\theta}\over 2}\epsilon_{3bc}
{\hat n}_c]\pi_{\theta_b}-\nonumber \\
&-& [{{2sin^2{{\theta}\over 2} }\over {sin\,
2\theta_w}}(1-{\hat n}_3) Z^o+\nonumber \\
&+&2sin^2{{\theta}\over 2} {\hat n}_3 (W^o_{+} {\hat n}_{-}+W^o_{-} {\hat n}
_{+})-i sin\, \theta (W^o_{+} {\hat n}_{-}-W^o_{-} {\hat n}_{+})],\nonumber \\
W^{{'}o}_1&&={2\over {(\phi_o+{1\over {\sqrt{2}}}H)^2}}[{\hat n}_1{\hat n}_b+
{{\theta sin\, \theta}\over {4 sin^2\, {{\theta}\over 2} }}(\delta_{1b}-
{\hat n}_1{\hat n}_b)+{{\theta}\over 2}\epsilon_{1bc}{\hat n}_c]\pi_{\theta_b}
+\nonumber \\
&+&{{Z^o}\over {sin\, 2\theta_w}}(sin\, \theta {\hat n}_2-2 sin^2\, {{\theta}
\over 2}{\hat n}_1{\hat n}_3)+2 sin^2\, {{\theta}\over 2}{\hat n}_1
(W^o_{+}{\hat n}_{-}+W^o_{-}{\hat n}_{+})+\nonumber \\
&+&{1\over {\sqrt{2}}}[(cos\, \theta +i sin\, \theta {\hat n}_3)W^o_{+}+(cos\,
\theta -i sin\, \theta {\hat n}_3)W^o_{-}],\nonumber \\
&&W^{{'}o}_2={2\over {(\phi_o+{1\over {\sqrt{2}}}H)^2}}[{\hat n}_2{\hat n}_b+
{{\theta sin\, \theta}\over {4 sin^2\, {{\theta}\over 2} }}(\delta_{2b}-{\hat n}
_2{\hat n}_b)+{{\theta}\over 2}\epsilon_{2bc}{\hat n}_c] \pi_{\theta_b}-
\nonumber \\
&-&{{Z^o}\over {sin\, 2\theta_w}} (sin\, \theta {\hat n}_1+2 sin^2\, {{\theta}
\over 2}{\hat n}_2{\hat n}_3)+2 sin^2\, {{\theta}\over 2}{\hat n}
_2 (W^o_{+}{\hat n}_{-}+W^o_{-}{\hat n}_{+})-\nonumber \\
&-&{1\over {\sqrt{2}}}[(sin\, \theta {\hat n}_3-i cos\, \theta )W^o_{+}+(sin\,
\theta {\hat n}_3+i cos\, \theta )W^o_{-}].
\label{43}
\end{eqnarray}

The main point is that $\Gamma^{(A)}$, $\Gamma^{(Z)}$, $\Gamma^{(W_{\pm})}$,
are independent from $\pi^H(x)$ and $A^o(x)$. However they depend on $Z^o(x)$,
$W^o_{\pm}(x)$, and this implies the existence of 3 tertiary constraints and
of 3 quaternary ones [see Ref.\cite{chai} for the general patterns of second
class constraints]. Namely, in the new variables the constraints change nature
and number with respect to the Hamiltonian formulation associated with the 
Lagrangian (\ref{1}) [in contrast to papers I and II, this is due to the mixing
in Eqs.(\ref{8}) together with the non-Abelian nature of $W^{\mu}_a$].

While the color and electromagnetic Gauss laws $\Gamma^{(G)}_A(x)\approx 0$,
$\Gamma^{(A)}(x)\approx 0$, are elliptic equations in the momenta ${\vec \pi}
^{(G)}_A(x)$, ${\vec \pi}^{(A)}(x)$, respectively, the weak ones $\Gamma^{(Z)}
(x)\approx 0$, $\Gamma^{(W_{\pm})}(x)\approx 0$, are ambiguous: each one of them
(in analogy to I, II, and momentarily forgetting their dependence on $Z^o$
and $W^o_{\pm}$)
can be thought either as an elliptic equation in one of the momenta ${\vec \pi}
^{(Z)}(x)$, ${\vec \pi}^{(W_{\pm})}(x)$, or as an algebraic equation in the 
Higgs momenta (the would-be Goldstone bosons) $\pi_{\theta \pm}(x), \pi_{\theta
_3}(x)$. Since the Gauss laws are the subset of the Euler-Lagrange equations 
independent from the accelerations, it turns out that the space of solutions of 
the Euler-Lagrange equations of the standard model is the disjoint union of 8 
sectors (for 6 of them there may be many inequivalent copies corresponding to 
different choices of which combinations of the Higgs momenta have to be
determined):

i) the $SU(2)\times U(1)$ symmetric phase (0 broken and 4 unbroken generators),
in which all the fields $A_{\mu}, Z_{\mu}, W_{\pm \mu}$ (or $V_{\mu}, W_{a\mu}$)
are massless.

ii) 3 phases with 1 broken and 3 unbroken generators, in which $SU(2)\times
U(1)$ is broken to three non-commuting U(1)'s. The three phases are: a)
$A_{\mu}, W_{\pm \mu}$ massless and $Z_{\mu}$ massive; b) $A_{\mu}, Z_{\mu},
W_{+ \mu}$ massless and $W_{- \mu}$ massive; c) $A_{\mu}, Z_{\mu}, W_{- \mu}$
massless and $W_{+ \mu}$ massive. However, in general there are phases
corresponding to all the possible choices of which combination of $V_{\mu}, 
W_{a\mu}$, becomes massive.

iii) 3 phases with 2 broken and 2 unbroken generators, in which $SU(2)\times
U(1)$ is broken to two (in general non-commuting) U(1)'s. The three phases
are: a) $A_{\mu}, Z_{\mu}$ massless and $W_{\pm \mu}$ massive; b) $A_{\mu},
W_{+ \mu}$, massless and $Z_{\mu}, W_{- \mu}$, massive; c) $A_{\mu},
W_{- \mu}$, massless and $Z_{\mu}, W_{+ \mu}$, massive. Again there may be
many other copies of these phases.

iv) The Higgs phase with 3 broken and 1 unbroken generators, in which
$SU(2)\times U(1)$ is broken to $U_{(em)}(1)$: $A_{\mu}$ massless and $Z_{\mu},
W_{\pm \mu}$ massive.

However, in contrast to I and II, here the situation is much more complicated
due to the presence of $Z^o$ and $W^o_{\pm}$ in the constraints. The expected
3 tertiary and 3 quaternary constraints should be such that at the end one gets
the following situation in the Higgs phase: i) $A^o(x)$ is a gauge variable
conjugate to the 1st class constraint $\pi^{(A)o}(x)\approx 0$; ii) $\theta_a
(x)$ and $\pi_{\theta_a}(x)$ are determined by 3 pairs of 2nd class constraints
(containing $\Gamma^{(Z)}(x)\approx 0$, $\Gamma^{(W_{\pm})}(x)\approx 0$
and the 3 tertiary constraints); iii) $\pi^{(Z)o}(x)\, [\approx 0]$, $\pi
^{(W_{\pm})o}(x)\, [\approx 0]$ and the 3 quaternary constraints determining
$Z^o(x)$, $W^o_{\pm}(x)$, form 3 pairs of 2nd class constraints; iv)
$\pi^{(G)o}_A(x)\approx 0$, $\Gamma^{(G)}_A(x)\approx 0$ are 1st class
constraints.

Due to this extremely complicated situation, we will not study the direct
canonical reduction in the Higgs sector of the constrained phase space 
associated with the Lagrangian density (\ref{17}) [one should evaluate the 
Hamiltonian associated with the Lagrangian (\ref{17}) and make a complete 
analysis of the constraints], but, following I and II, we will study the easier
canonical reduction of the first and second class constraints deriving from 
the  Lagrangian density ${\cal L}^{'}(x)$ of Eq.(\ref{21}) in the unitary gauge.

\section
{Euler-Lagrange equations and  constraints from ${\cal L}^{'}(x)$}

The Euler-Lagrange equations
associated with the Lagrangian density (\ref{21}), which 
does not depend on the (would-be Goldstone bosons) Higgs fields $\theta_a(x)$,
but only on $H(x), G_{A\mu}(x), {\tilde A}^{'}_{\mu}(x), {\tilde Z}^{'}_{\mu}
(x), {\tilde W}^{'}_{\pm \mu}(x), \psi^{(l)\, {'}}_{Li}(x), \psi^{(l)}_{Ri}(x),
\psi^{(q)\, {'}}_{Li}(x), \psi^{(q)}_{Ri}(x), {\tilde \psi}^{(q)}_{Ri}(x)$, are

\begin{eqnarray}
L_A^{(G)\, \mu}&=&g_s^2({{\partial {\cal L}^{'}}\over 
{\partial G_{A\mu}}}-\partial
_{\nu}{{\partial {\cal L}}\over {\partial \partial_{\nu}G_{A\mu}}}={\hat D}
^{(G)}_{\nu AB}G^{\nu\mu}_B+g^2_sJ^{\mu}_{sA} {\buildrel \circ \over =} 0,
\nonumber \\
&&{}\nonumber \\
&J&^{\mu}_{sA}=i{\bar \psi}^{(q)}_{Li} \gamma^{\mu}T^A_s\psi^{(q)}_{Li}+
i{\bar \psi}^{(q)}_{Ri}\gamma^{\mu}T^A_s\psi^{(q)}_{Ri}+i{\bar {\tilde \psi}}
^{(q)}_{Ri}\gamma^{\mu}T^A_s{\tilde \psi}^{(q)}_{Ri},\nonumber \\
&&{}\nonumber \\
{\tilde L}^{(\tilde A){'}\, \mu}&=&{{\partial {\cal L}^{'}}\over 
{\partial {\tilde A}^{'}_{\mu}}}-\partial_{\nu}{{\partial {\cal L}^{'}}\over 
{\partial \partial_{\nu}{\tilde A}^{'}_{\mu}}}=\partial_{\nu}{\tilde A}^{{'}
\nu\mu}-2ie\partial_{\nu}({\tilde W}^{{'}\mu}_{+}{\tilde W}_{-}^{{'}\nu}-
{\tilde W}^{{'}\nu}_{+}{\tilde W}^{{'}\mu}_{-})+\nonumber \\
&+&ie({\tilde W}^{'}_{-\, \nu}{\tilde W}^{{'}\nu\mu}_{+}-{\tilde W}^{'}_{+\, 
\nu}{\tilde W}^{{'}\nu\mu}_{-})-2e^2{\tilde W}^{'}_{+}\cdot {\tilde W}^{'}_{-} 
({\tilde A}^{{'}\mu}+cot\, \theta_w {\tilde Z}^{{'}\mu})+\nonumber \\
&+&e^2({\tilde W}^{{'}\mu}_{+}{\tilde W}^{{'}\nu}_{-}+{\tilde W}^{{'}\mu}_{-}
{\tilde W}^{{'}\nu}_{+})({\tilde A}^{'}_{\nu}+cot\, \theta_w {\tilde Z}^{'}
_{\nu})+ e{\tilde j}^{{'}\mu}_{(em)}{\buildrel \circ \over =} 0,\nonumber \\
&&{}\nonumber \\
{\tilde j}&^{{'}\mu}&_{(em)}={\bar \psi}^{(l){'}}_{Li} \gamma^{\mu}Q_{em}
\psi^{(l){'}}_{Li}+{\bar \psi}^{(l)}_{Ri}\gamma^{\mu}iY_w\psi^{(l)}_{Ri}+
{\bar \psi}^{(q){'}}_{Li}\gamma^{\mu}Q_{em}\psi^{(q){'}}_{Li}+\nonumber \\
&+&{\bar \psi}^{(q)}_{Ri}\gamma^{\mu} iY_w\psi^{(q)}_{Ri}+{\bar {\tilde \psi}}
^{(q)}_{Ri}\gamma^{\mu} iY_w{\tilde \psi}^{(q)}_{Ri}=\nonumber \\
&=&{\bar \psi}^{(l){'}}_{Li} \gamma^{\mu}Q_{em}\psi^{(l){'}}_{Li}+
{\bar \psi}^{(q){'}}_{Li}\gamma^{\mu}Q_{em}\psi^{(q){'}}_{Li}-{\bar \psi}^{(l)}
_{Ri}\gamma^{\mu}\psi^{(l)}_{Ri}-{1\over 3}{\bar \psi}^{(q)}_{Ri}\gamma^{\mu}
\psi^{(q)}_{Ri}+{2\over 3}{\bar {\tilde \psi}}^{(q)}_{Ri}\gamma^{\mu}{\tilde 
\psi}^{(q)}_{Ri},\nonumber \\
&&{}\nonumber \\
{\tilde L}^{(\tilde Z){'}\, \mu}&=&{{\partial {\cal L}^{'}}\over 
{\partial {\tilde Z}^{'}_{\mu}}}-\partial_{\nu}{{\partial {\cal L}^{'}}\over 
{\partial \partial_{\nu}{\tilde Z}^{'}_{\mu}}}=\partial_{\nu}{\tilde Z}^{{'}
\nu\mu}-2iecot\, \theta_w \partial_{\nu}({\tilde W}^{{'}\mu}_{+}{\tilde W}_{-}
^{{'}\nu}-{\tilde W}^{{'}\nu}_{+}{\tilde W}^{{'}\mu}_{-})+\nonumber \\
&+&m_Z^2 (1+{e\over {sin\, 2\theta_w m_Z}}H)^2 {\tilde Z}^{{'}\mu}+\nonumber \\
&+&iecot\, \theta_w ({\tilde W}^{'}_{-\, \nu}{\tilde W}^{{'}\nu\mu}_{+}-
{\tilde W}^{'}_{+\, \nu}{\tilde W}^{{'}\nu\mu}_{-})-2e^2cot\, \theta_w
{\tilde W}^{'}_{+}\cdot {\tilde W}^{'}_{-} ({\tilde A}^{{'}\mu}+cot\, \theta_w 
{\tilde Z}^{{'}\mu})+\nonumber \\
&+&e^2cot\, \theta_w ({\tilde W}^{{'}\mu}_{+}{\tilde W}^{{'}\nu}_{-}+{\tilde W}
^{{'}\mu}_{-}{\tilde W}^{{'}\nu}_{+})({\tilde A}^{'}_{\nu}+cot\, \theta_w 
{\tilde Z}^{'}_{\nu})+ e{\tilde j}^{{'}\mu}_{(NC)}{\buildrel \circ \over =} 0,
\nonumber \\
&&{}\nonumber \\
{\tilde j}&^{{'}\mu}&_{(NC)}={\bar \psi}^{(l){'}}_{Li} \gamma^{\mu}Q_Z
\psi^{(l){'}}_{Li}-tg\, \theta_w {\bar \psi}^{(l)}_{Ri}\gamma^{\mu}iY_w\psi
^{(l)}_{Ri}+{\bar \psi}^{(q){'}}_{Li}\gamma^{\mu}Q_Z\psi^{(q){'}}_{Li}-
\nonumber \\
&-&tg\, \theta_w {\bar \psi}^{(q)}_{Ri}\gamma^{\mu} iY_w\psi^{(q)}_{Ri}-
tg\, \theta_w {\bar {\tilde \psi}}^{(q)}_{Ri}\gamma^{\mu} iY_w{\tilde \psi}
^{(q)}_{Ri}=\nonumber \\
&=&{\bar \psi}^{(l){'}}_{Li} \gamma^{\mu}Q_Z\psi^{(l){'}}_{Li}+
{\bar \psi}^{(q){'}}_{Li}\gamma^{\mu}Q_Z\psi^{(q){'}}_{Li}-tg\, \theta_w [-
{\bar \psi}^{(l)}_{Ri}\gamma^{\mu}\psi^{(l)}_{Ri}-{1\over 3}{\bar \psi}^{(q)}
_{Ri}\gamma^{\mu}\psi^{(q)}_{Ri}+{2\over 3}{\bar {\tilde \psi}}^{(q)}_{Ri}
\gamma^{\mu}{\tilde \psi}^{(q)}_{Ri}],\nonumber \\
&&{}\nonumber \\
{\tilde L}^{({\tilde W}_{\pm}){'}\, \mu}&=&{{\partial {\cal L}^{'}}\over 
{\partial {\tilde W}^{'}_{\pm \, \mu}}}-\partial_{\nu}{{\partial {\cal L}^{'}}
\over {\partial \partial_{\nu} {\tilde W}^{'}_{\pm \, \mu}}}=\partial
_{\nu}{\tilde W}^{{'}\nu\mu}_{\mp}+m_W^2(1+{e\over {sin\, 2\theta_w m_Z}}H)^2
{\tilde W}^{{'}\mu}_{\mp}+\nonumber \\
&+&ie\partial_{\nu} [{\tilde W}^{{'}\nu}_{\mp}({\tilde A}^{{'}\mu}+cot\,
\theta_w {\tilde Z}^{{'}\mu})-{\tilde W}^{{'}\mu}_{\mp}({\tilde A}^{{'}\nu}
+cot\, \theta_w {\tilde Z}^{{'}\nu})]-\nonumber \\
&-&ie[({\tilde A}^{{'}\mu\nu}+cot\, \theta_w
{\tilde Z}^{{'}\mu\nu}){\tilde W}^{'}_{\mp \, \nu}-{\tilde W}^{{'}\mu\nu}_{\mp}
({\tilde A}^{'}_{\nu}+cot\, \theta_w {\tilde Z}^{'}_{\nu})]+\nonumber \\
&+&{{e^2}\over {sin^2\, \theta_w}}[{\tilde W}^{{'}\mu}_{\pm}{\tilde W}^{{'}2}
_{\mp}-{\tilde W}^{{'}\mu}_{\mp}{\tilde W}^{'}_{+}\cdot {\tilde W}^{'}_{-}]-e^2
{\tilde W}^{{'}\mu}_{\mp} ({\tilde A}^{'}+cot\, \theta_w {\tilde Z}^{'})^2+
\nonumber \\
&+&e^2({\tilde A}^{{'}\mu}+cot\, \theta_w {\tilde Z}^{{'}\mu}) {\tilde W}^{'}
_{\mp}\cdot ({\tilde A}^{'}+cot\, \theta_w {\tilde Z}^{'})+ {e\over { 
sin\, \theta_w}} {\tilde j}^{{'}\mu}_{(CC)\, \mp}
{\buildrel \circ \over =} 0,\nonumber \\
&&{}\nonumber \\
{\tilde j}&^{{'}\mu}&_{(CC)\, \mp}={\bar \psi}^{(l){'}}_{Li} \gamma^{\mu}iT^{\mp}_w
\psi^{(l){'}}_{Li}+{\bar \psi}^{(q){'}}_{Li}\gamma^{\mu}iT^{\mp}_w\psi^{(q){'}}
_{Li},\nonumber \\
&&{}\nonumber \\
L_H&=&{{\partial {\cal L}^{'}}\over {\partial H}}-\partial_{\nu}{{\partial 
{\cal L}^{'}}\over {\partial \partial_{\nu} H}}=-\Box H -m_H^2 H (1+{e\over
{2sin\, 2\theta_w m_Z}} H) (1+{e\over {sin\, 2\theta_w m_Z}} H)+\nonumber \\
&+&{{2e}\over {sin\, 2\theta_w m_Z}} (1+{e\over {sin\, 2\theta_w m_Z}} H)
[m^2_W {\tilde W}^{'}_{+} \cdot {\tilde W}^{'}_{-} +{1\over 2} m_Z^2 {\tilde
Z}^{{'} 2}]+\nonumber \\
&+&{e\over {sin\, 2\theta_wm_Z}} [{\bar \psi}^{(l){'}}_{Li}\cdot \left(
\begin{array}{l} 0\\ 1\end{array} \right) M^{(l)}_{ij}\psi^{(l)}_{Rj}+
{\bar \psi}^{(l)}_{Ri}M^{(l)\dagger}_{ij} \left( \begin{array}{ll} 0& 1
\end{array} \right) \cdot \psi^{(l){'}}_{Lj}+\nonumber \\
&+&{\bar \psi}^{(q){'}}_{Li}\cdot \left( \begin{array}{l} 0\\ 1\end{array}
\right) M^{(q)}_{ij}\psi^{(q)}_{Rj}+{\bar \psi}^{(q)}_{Ri}M^{(q)\dagger}_{ij}
\left( \begin{array}{ll} 0& 1\end{array} \right) \cdot \psi^{(q){'}}_{Lj}+
\nonumber \\
&+&{\bar \psi}^{(q){'}}_{Li}\cdot \left( \begin{array}{l} 1\\ 0\end{array}
\right) {\tilde M}^{(q)}_{ij}{\tilde \psi}^{(q)}_{Rj}+{\bar {\tilde \psi}}^{(q)}
_{Ri}{\tilde M}^{(q)\dagger}_{ij} \left( \begin{array}{ll} 1& 0\end{array}
\right) \cdot \psi^{(q){'}}_{Lj}]{\buildrel \circ \over =} 0,\nonumber \\
&&{}\nonumber \\
L_{\psi^{'} Li}^{(l)}&=&{{\partial {\cal L}^{'}}\over {\partial \psi^{(l){'}}
_{Li}}}-\partial_{\mu}{{\partial {\cal L}^{'}}\over {\partial \partial_{\mu}
\psi^{(l){'}}_{Li}}}=-{\bar \psi}^{(l){'}}_{Li} [\loarrow{i(\partial_{\mu}-
{e\over {sin\, \theta_w}}({\tilde W}^{'}_{+\, \mu}T^{-}_w+{\tilde W}^{'}_{-\, 
\mu}T^{+}_w)+}\nonumber \\
&+&ieQ_{em}{\tilde A}^{'}_{\mu}+ieQ_Z{\tilde Z}^{'}_{\mu})\gamma
^{\mu}]-\nonumber \\
&-&(1+{e\over {sin\, 2\theta_wm_Z}} H){\bar \psi}^{(l)}_{Rj} M^{(l)\dagger}_{ji}
\left( \begin{array}{ll} 0& 1\end{array} \right)
{\buildrel \circ \over =} 0,\nonumber \\
L_{{\bar \psi}^{'} Li}^{(l)}&=&{{\partial {\cal L}^{'}}\over {\partial {\bar 
\psi}^{(l){'}}_{Li}}}-\partial_{\mu}{{\partial {\cal L}^{'}}\over {\partial 
\partial_{\mu}{\bar \psi}^{(l){'}}_{Li}}}=[\gamma^{\mu}i(\partial_{\mu}+
{e\over {sin\, \theta_w}}({\tilde W}^{'}_{+\, \mu}T^{-}_w+{\tilde W}^{'}_{-\, 
\mu}T^{+}_w)-\nonumber \\
&-&ieQ_{em}{\tilde A}^{'}_{\mu}-ieQ_Z{\tilde Z}^{'}_{\mu})]
\psi^{(l){'}}_{Li}+\nonumber \\
&+&(1+{e\over {sin\, 2\theta_wm_Z}}H)\left( \begin{array}{l} 0\\ 1\end{array}
\right) M^{(l)}_{ij} \psi^{(l)}_{Rj}{\buildrel \circ \over =} 0,\nonumber \\
L^{(l)}_{\psi Ri}&=&{{\partial {\cal L}^{'}}\over {\partial \psi^{(l)}_{Ri}}}-
\partial_{\mu}{{\partial {\cal L}^{'}}\over {\partial \partial_{\mu}\psi^{(l)}
_{Ri}}}=-{\bar \psi}^{(l)}_{Ri} [\loarrow{i(\partial_{\mu}-e({\tilde A}^{'}
_{\mu}-tg\, \theta_w{\tilde Z}^{'}_{\mu})}\gamma^{\mu}]-\nonumber \\
&-&(1+{e\over {sin\, 2\theta_wm_Z}} H){\bar \psi}^{(l){'}}_{Lj}\cdot \left(
\begin{array}{l} 0\\ 1\end{array} \right)  M^{(l)}_{ji}
{\buildrel \circ \over =} 0,\nonumber \\
L^{(l)}_{\bar \psi Ri}&=&{{\partial {\cal L}^{'}}\over {\partial {\bar \psi}
^{(l)}_{Ri}}}-\partial_{\mu}{{\partial {\cal L}^{'}}\over {\partial \partial
_{\mu}{\bar \psi}^{(l)}_{Ri}}}=[\gamma^{\mu}i(\partial_{\mu}+e({\tilde A}^{'}
_{\mu}-tg\, \theta_w{\tilde Z}^{'}_{\mu})]\psi^{(l)}_{Ri}+\nonumber \\
&+&(1+{e\over {sin\, 2\theta_wm_Z}} H) M^{(l)\dagger}_{ij} \left(
\begin{array}{ll} 0& 1\end{array} \right) \cdot \psi^{(l){'}}_{Lj} 
{\buildrel \circ \over =} 0,\nonumber \\
&&{}\nonumber \\
L^{(q)}_{\psi^{'} Li}&=&{{\partial {\cal L}^{'}}\over {\partial \psi^{(q){'}}
_{Li}}}-\partial_{\mu}{{\partial {\cal L}^{'}}\over {\partial \partial_{\mu}
\psi^{(q){'}}_{Li}}}=-{\bar \psi}^{(q){'}}_{Li} [\loarrow{i(\partial_{\mu}
-{e\over {sin\, \theta_w}}({\tilde W}^{'}_{+\, \mu}T^{-}_w+{\tilde W}^{'}_{-\,
\mu}T^{+}_w)+}\nonumber \\
&+&ieQ_{em}{\tilde A}^{'}_{\mu}+ieQ_Z{\tilde Z}^{'}_{\mu}
-G_{A\mu}T^A_s)\gamma^{\mu}]-\nonumber \\
&-&(1+{e\over {sin\, 2\theta_wm_Z}} H) [{\bar \psi}^{(q)}_{Rj}M^{(q)\dagger}
_{ji} \left( \begin{array}{ll} 0& 1\end{array} \right) +
{\bar {\tilde \psi}}^{(q)}_{Rj}{\tilde M}^{(q)\dagger}_{ji} \left(
\begin{array}{ll} 1& 0\end{array} \right) ] {\buildrel \circ \over =} 0,
\nonumber \\
L^{(q)}_{{\bar \psi}^{'} Li}&=&{{\partial {\cal L}^{'}}\over {\partial {\bar 
\psi}^{(q){'}}_{Li}}}-\partial_{\mu}{{\partial {\cal L}^{'}}\over {\partial 
\partial_{\mu}{\bar \psi}^{(q){'}}_{Li}}}=[\gamma^{\mu}i(\partial_{\mu}+
{e\over {sin\, \theta_w}}({\tilde W}^{'}_{+\, \mu}T^{-}_w+{\tilde W}^{'}_{-\, 
\mu}T^{+}_w)-\nonumber \\
&-&ieQ_{em}{\tilde A}^{'}_{\mu}-ieQ_Z{\tilde Z}^{'}_{\mu}
+G_{A\mu}T^A_s)]\psi^{(q){'}}_{Li}+\nonumber \\
&+& (1+{e\over {sin\, 2\theta_wm_Z}} H) [\left( \begin{array}{l} 0\\ 1
\end{array} \right) M^{(q)}_{ij}\psi^{(q)}_{Rj} +\left( \begin{array}{l}
1\\ 0\end{array} \right) {\tilde M}^{(q)}_{ij}{\tilde \psi}^{(q)}_{Rj}
{\buildrel \circ \over =} 0,\nonumber \\
L^{(q)}_{\psi Ri}&=&{{\partial {\cal L}^{'}}\over {\partial \psi^{(q)}_{Ri}}}-
\partial_{\mu}{{\partial {\cal L}^{'}}\over {\partial \partial_{\mu}\psi^{(q)}
_{Ri}}}=-{\bar \psi}^{(q)}{Ri} [\loarrow{i(\partial_{\mu}-e({\tilde A}^{'}_{\mu}
-tg\, \theta_w {\tilde Z}^{'}_{\mu})-G_{A\mu}T^A_s)}\gamma^{\mu}]-\nonumber \\
&-&(1+{e\over {sin\, 2\theta_wm_Z}} H){\bar \psi}^{(q){'}}_{Lj}\cdot \left(
\begin{array}{l} 0\\ 1\end{array} \right) 
M^{(q)}_{ji} {\buildrel \circ \over =} 0,\nonumber \\
L^{(q)}_{\bar \psi Ri}&=&{{\partial {\cal L}^{'}}
\over {\partial {\bar \psi}^{(q)}_{Ri}}}-\partial_{\mu}
{{\partial {\cal L}^{'}}\over {\partial \partial_{\mu}{\bar 
\psi}^{(q)}_{Ri}}}=[\gamma^{\mu}i(\partial_{\mu}+e({\tilde A}^{'}_{\mu}
-tg\, \theta_w {\tilde Z}^{'}_{\mu})+G_{A\mu}T^A_s)]\psi^{(q)}_{Ri} +
\nonumber \\
&+&(1+{e\over {sin\, 2\theta_wm_Z}} H) M_{ij}^{(q)\dagger} \left(
\begin{array}{ll} 0& 1\end{array} \right) \cdot
\psi^{(q){'}}_{Lj} {\buildrel \circ \over =} 0,\nonumber \\
L^{(q)}_{\tilde \psi Ri}&=&{{\partial {\cal L}^{'}}\over {\partial {\tilde \psi}
^{(q)}_{Ri}}}-\partial_{\mu}{{\partial {\cal L}^{'}}
\over {\partial \partial_{\mu}
{\tilde \psi}^{(q)}_{Ri}}}=-{\bar {\tilde \psi}}^{(q)}_{Ri}
[\loarrow{i(\partial_{\mu}-e({\tilde A}^{'}_{\mu}
-tg\, \theta_w {\tilde Z}^{'}_{\mu})-G_{A\mu}T^A_s)}\gamma^{\mu}]-\nonumber \\
&-&(1+{e\over {sin\, 2\theta_wm_Z}} H) {\bar \psi}^{(q){'}}_{Lj}\cdot \left(
\begin{array}{l} 1\\ 0\end{array} \right) {\tilde M}^{(q)}_{ji}
{\buildrel \circ \over =} 0,\nonumber \\
L^{(q)}_{{\bar {\tilde \psi}} Ri}&=&{{\partial {\cal L}^{'}}\over 
{\partial {\bar {\tilde \psi}}^{(q)}_{Ri} }}-
\partial_{\mu}{{\partial {\cal L}^{'}}\over 
{\partial \partial_{\mu}{\bar {\tilde \psi}}^{(q)}_{Ri}}}=[\gamma^{\mu}
i(\partial_{\mu}+e({\tilde A}^{'}_{\mu}-tg\, \theta_w {\tilde Z}^{'}_{\mu})
+G_{A\mu}T^A_s)]{\tilde \psi}^{(q)}_{Ri}+\nonumber \\
&+&(1+{e\over {sin\, 2\theta_wm_Z}} H) {\tilde M}^{(q)\dagger}_{ij} \left(
\begin{array}{ll} 1& 0\end{array} \right) \cdot
\psi^{(q){'}}_{Lj} {\buildrel \circ \over =} 0.
\label{44}
\end{eqnarray}

The canonical momenta associated with the Lagrangian density (\ref{21})
[${\tilde E}^{(\tilde A){'} i}=-{\tilde A}^{{'} oi}$, ${\tilde E}^{(\tilde Z)
{'} i}=-{\tilde Z}^{{'} oi}$, ${\tilde E}^{({\tilde W}_{\pm}){'} i}=-(\partial^o
{\tilde W}^{{'} i}_{\mp} -\partial^i{\tilde W}^{{'} o}_{\mp})=-{\tilde {\cal 
W}}^{{'}
oi}_{\mp}$ are natural definitions of ``Abelian" electric fields; the last one
must not be confused with the non-Abelian field strength ${\tilde W}^{'}
_{a\mu\nu}=\partial_{\mu}{\tilde W}^{'}_{a\nu}-\partial_{\nu}{\tilde W}^{'}
_{a\mu}+{e\over {sin\, \theta_w}} \epsilon_{abc}
{\tilde W}^{'}_{b\mu}{\tilde W}^{'}_{c\nu}$]

\begin{eqnarray}
\pi^{(G) o}_A(x)&=& 0,\nonumber \\
\pi^{(G) i}_A(x)&=&-g^{-2}_sG^{oi}_A(x)=g^{-2}_s E^{(G) i}_A(x),\nonumber \\
\pi_H(x)&=&\partial^oH(x),\nonumber \\
{\tilde \pi}^{(\tilde A){'} o}(x)&=&0,\nonumber \\
{\tilde \pi}^{(\tilde A){'} i}(x)&=&-{\tilde A}^{{'} oi}(x)+ie[{\tilde W}^{{'} 
o}_{+}(x){\tilde W}^{{'} i}_{-}(x)-{\tilde W}^{{'} i}_{+}(x){\tilde W}^{{'} o}
_{-}(x)]=\nonumber \\
&=&{\tilde E}^{(\tilde A){'} i}(x)+ie[{\tilde W}^{{'} 
o}_{+}(x){\tilde W}^{{'} i}_{-}(x)-{\tilde W}^{{'} i}_{+}(x){\tilde W}^{{'} o}
_{-}(x)],\nonumber \\
{\tilde \pi}^{(\tilde Z){'} o}(x)&=&0,\nonumber \\
{\tilde \pi}^{(\tilde Z){'} i}(x)&=&-{\tilde Z}^{{'} oi}(x)+ie cot\, \theta_w
[{\tilde W}^{{'} o}_{+}(x) {\tilde W}^{{'} i}_{-}(x)-{\tilde W}^{{'} i}_{+}(x)
{\tilde W}^{{'} o}_{-}(x)]=\nonumber \\
&=&{\tilde E}^{(\tilde Z){'} i}+ie cot\, \theta_w
[{\tilde W}^{{'} o}_{+}(x) {\tilde W}^{{'} i}_{-}(x)-{\tilde W}^{{'} i}_{+}(x)
{\tilde W}^{{'} o}_{-}(x)],\nonumber \\
{\tilde \pi}^{({\tilde W}_{\pm}){'} o}(x)&=&0,\nonumber \\
{\tilde \pi}^{({\tilde W}_{\pm}){'} i}(x)&=&-{\tilde {\cal W}}^{{'} oi}
_{\mp}(x)\pm \nonumber \\
&\pm& ie [{\tilde W}^{{'} o}_{\mp}(x) ({\tilde A}^{{'} i}(x)+cot\, \theta_w
{\tilde Z}^{{'} i}(x))-{\tilde W}^{{'} i}_{\mp}(x) ({\tilde A}^{{'} o}(x)+
cot\, \theta_w {\tilde Z}^{{'} o}(x))]=\nonumber \\
&=&{\tilde E}^{({\tilde W}_{\pm}){'} i}(x)\pm ie [{\tilde W}^{{'} o}_{\mp}(x) 
({\tilde A}^{{'} i}(x)+cot\, \theta_w
{\tilde Z}^{{'} i}(x))-\nonumber \\
&-&{\tilde W}^{{'} i}_{\mp}(x) ({\tilde A}^{{'} o}(x)+
cot\, \theta_w {\tilde Z}^{{'} o}(x))],\nonumber \\
&&{}\nonumber \\
&&\{G_{A\mu}(\vec x,x^o),\pi_B^{(G)\nu}(\vec y,x^o)\}=\delta_{AB}\delta^{\nu}
_{\mu}\delta^3(\vec x-\vec y),\nonumber \\
&&\{ {\tilde A}_{\mu}^{'}(\vec x,x^o),{\tilde \pi}^{(\tilde A){'}\nu}(\vec y,
x^o) \} = \{ {\tilde Z}^{'}_{\mu}(\vec x,x^o),{\tilde \pi}^{(\tilde Z){'}\nu}
(\vec y,x^o) \} =\delta^{\nu}_{\mu}\delta^3(\vec x-\vec y).\nonumber \\
&&\{{\tilde W}^{'}_{\pm \mu}(\vec x,x^o),{\tilde \pi}^{({\tilde W}_{\pm}){'}
\nu}(\vec y,x^o) \} =\delta^{\nu}_{\mu} \delta^3(\vec x-\vec y),
\label{45}
\end{eqnarray}

\noindent plus the fermion momenta $\pi^{(l){'}}_{\psi Lia\alpha}(x)$, $\pi
^{(l){'}}_{\bar\psi Lia\alpha}(x)$, $\pi^{(l)}_{\psi Ri\alpha}(x)$, $\pi^{(l)}
_{\bar \psi Ri\alpha}(x)$, $\pi^{(q){'}}_{\psi LiAa\alpha}(x)$, $\pi^{(q){'}}
_{\bar \psi LiAa\alpha}(x)$, $\pi^{(q)}_{\psi RiA\alpha}(x)$, $\pi^{(q)}_{\bar
\psi RiA\alpha}(x)$, $\pi^{(q)}_{\tilde \psi RiA\alpha}(x)$, $\pi^{(q)}_{ {\bar
{\tilde \psi}} RiA\alpha}(x)$. The fermion momenta generate second class
constraints, which can be eliminated with the Dirac brackets (\ref{33}).

The resulting Dirac Hamiltonian density is [the 
``Abelian" magnetic fields are defined as
 $B^{(\tilde A){'} i}=-{1\over 2}
\epsilon^{ijk}{\tilde A}^{{'} jk}$, $B^{(\tilde Z){'} i}=-{1\over 2}\epsilon
^{ijk}{\tilde Z}^{{'} jk}$, $B^{({\tilde W}_{\pm}){'} i}=-{1\over 2}\epsilon
^{ijk}\partial^j{\tilde W}^{{'} k}_{\pm}=\epsilon^{ijk}{\tilde {\cal W}}_{\pm}
^{{'} jk}$; instead $B^{(G)i}_A=-{1\over 2}\epsilon^{ijk}G^{jk}_A$ are
non-Abelian magnetic fields; $m_W=m_Z cos\, \theta_w$]

\begin{eqnarray}
{\cal H}^{'}_D(x)&=&{1\over 2}\sum_A [g^2_s{\vec \pi}^{(G) 2}_A(x)+ g^{-2}_s 
{\vec B}^{(G) 2}_A(x)]+\nonumber \\
&+&{1\over 2} [{\vec {\tilde \pi}}^{(\tilde A){'} 2}(x)+{\vec B}^{(\tilde A){'}
2}(x)]+{1\over 2} [{\vec {\tilde \pi}}^{(\tilde Z){'} 2}(x) +{\vec B}
^{(\tilde Z){'} 2}(x)]+\nonumber \\
&+&{\vec {\tilde \pi}}^{({\tilde W}_{+}){'}}(x) \cdot {\vec {\tilde \pi}}
^{({\tilde W}_{-}){'}}(x)+ {\vec B}^{({\tilde W}_{+}){'}}(x) \cdot {\vec B}
^{({\tilde W}_{-}){'}}(x)+\nonumber \\
&+& (1+{e\over {sin\, 2\theta_w m_Z}} H(x))^2 [m_W^2 {\vec {\tilde W}}^{'}_{+}
(x) \cdot {\vec {\tilde W}}^{'}_{-}(x) + {1\over 2} m^2_Z {\vec {\tilde Z}}^{'}
(x)]+\nonumber \\
&+&{1\over 2} [\pi^2_H(x) + (\vec \partial H(x))^2] + {1\over 2} m_H^2 H^2(x)
(1+{e\over {2sin\, 2\theta_w m_Z}} H(x))^2+\nonumber \\
&+&ie [{\vec {\tilde W}}^{'}_{+}(x) \times {\vec {\tilde W}}^{'}_{-}(x) \cdot
({\vec B}^{(\tilde A){'}}(x)+ cot\, \theta_w {\vec B}^{(\tilde Z){'}}(x))]+
\nonumber \\
&+&ie({\vec {\tilde A}}^{'}(x)+cot\, \theta_w {\vec {\tilde Z}}^{'}(x)) [\times
{\vec {\tilde W}}^{'}_{-}(x) \cdot {\vec B}^{({\tilde W}_{+}){'}}(x)-
\times {\vec {\tilde W}}^{'}_{+}(x) \cdot {\vec B}^{({\tilde W}_{-}){'}}(x)) ]
-\nonumber \\
&-&{{e^2}\over {2sin^2\, \theta_w}} [ {\vec {\tilde W}}^{{'} 2}_{+}(x) {\vec 
{\tilde W}}^{{'} 2}_{-}(x) -( {\vec {\tilde W}}^{'}_{+}(x) \cdot {\vec {\tilde 
W}}^{'}_{-}(x) )^2]+\nonumber \\
&+& e^2 [{\vec {\tilde W}}^{'}_{+}(x) \cdot {\vec {\tilde W}}^{'}_{-}(x)
({\vec {\tilde A}}^{'}(x)+cot\, \theta_w {\vec {\tilde Z}}^{'}(x))^2 -
\nonumber \\
&-&{\vec {\tilde W}}^{'}_{+}(x) \cdot ({\vec {\tilde A}}^{'}(x)+cot\, \theta_w
{\vec {\tilde Z}}^{'}(x)) {\vec {\tilde W}}^{'}_{-}(x) \cdot ({\vec {\tilde 
A}}^{'}(x)+cot\, \theta_w {\vec {\tilde Z}}^{'}(x))]+\nonumber \\
&+&{i\over 2}\psi^{(l){'} \dagger}_{Li}(x) [\vec \alpha \cdot (\vec \partial +
{e\over {sin\, \theta_w}}({\vec {\tilde W}}^{'}_{+}(x)T^{-}_w+{\vec {\tilde 
W}}^{'}_{-}(x)T^{+}_w)-ieQ_{em}{\vec {\tilde A}}^{'}(x) -\nonumber \\
&-&ieQ_Z{\vec {\tilde Z}}^{'}(x) )-\nonumber \\
&-&\loarrow{ (\vec \partial -
{e\over {sin\, \theta_w}}({\vec {\tilde W}}^{'}_{+}(x)T^{-}_w+{\vec {\tilde 
W}}^{'}_{-}(x)T^{+}_w)+ieQ_{em}{\vec {\tilde A}}^{'}(x) +}\nonumber \\
&+&ieQ_Z{\vec {\tilde Z}}^{'}(x) )\cdot \vec \alpha ] \psi^{(l){'}}_{Li}(x)+
\nonumber \\
&+&{i\over 2}\psi^{(l)\dagger}_{Ri}(x) [\vec \alpha \cdot (\vec \partial +
{e\over {cos\, \theta_w}}({\vec {\tilde A}}^{'}(x)-tg\, \theta_w {\vec {\tilde 
Z}}^{'}(x)) Y_w)-\nonumber \\
&-&\loarrow{(\vec \partial -{e\over {cos\, \theta_w}}({\vec {\tilde A}}^{'}(x)
-tg\, \theta_w {\vec {\tilde Z}}^{'}(x)) Y_w)}\cdot \vec \alpha ] \psi^{(l)}
_{Ri}(x)-\nonumber \\
&-& (1+{e\over {sin\, 2\theta_wm_Z}} H(x)) \nonumber \\
&&[{\bar \psi}^{(l){'}}_{Li}(x)\cdot
\left( \begin{array}{l} 0\\ 1\end{array} \right) M^{(l)}_{ij} \psi^{(l)}_{Rj}
(x)+ {\bar \psi}^{(l)}_{Ri}(x) M^{(l)\dagger}_{ij} \left( \begin{array}{ll}
0 & 1\end{array} \right) \cdot \psi^{(l){'}}_{Lj}(x)]+\nonumber \\
&+&{i\over 2}\psi^{(q){'}\dagger}_{Li}(x) [\vec \alpha \cdot (\vec \partial +
{e\over {sin\, \theta_w}}({\vec {\tilde W}}^{'}_{+}(x)T^{-}_w+{\vec {\tilde 
W}}^{'}_{-}(x)T^{+}_w)-ieQ_{em}{\vec {\tilde A}}^{'}(x)-\nonumber \\
&-&ieQ_Z{\vec {\tilde Z}}^{'}(x)+{\vec G}_A(x)T^A_s)-\nonumber \\
&-&\loarrow{(\vec \partial -{e\over {sin\, \theta_w}}({\vec {\tilde W}}^{'}_{+}
(x)T^{-}_w+{\vec {\tilde W}}^{'}_{-}(x)T^{+}_w)+ieQ_{em}{\vec {\tilde A}}^{'}
(x)+}\nonumber \\
&+&ieQ_Z{\vec {\tilde Z}}^{'}(x)-{\vec G}_A(x)T^A_s)\cdot \vec \alpha ]
\psi^{(l){'}}_{Li}(x)+\nonumber \\
&+&{i\over 2}\psi^{(q)\dagger}_{Ri}(x)[\vec \alpha \cdot (\vec \partial +e
({\vec {\tilde A}}^{'}(x)-tg\, \theta_w {\vec {\tilde 
Z}}^{'}(x))Y_w +{\vec G}_A(x)T^A_s)-\nonumber \\
&-&\loarrow{(\vec \partial -e
({\vec {\tilde A}}^{'}(x)-tg\, \theta_w {\vec {\tilde 
Z}}^{'}(x))Y_w -{\vec G}_A(x)T^A_s)}\cdot \vec \alpha ] \psi^{(q)}_{Ri}(x)+
\nonumber \\
&+&{i\over 2}{\tilde \psi}^{(q)\dagger}_{Ri}(x)[\vec \alpha \cdot (\vec 
\partial +e({\vec {\tilde A}}^{'}(x)-tg\, \theta_w 
{\vec {\tilde Z}}^{'}(x))Y_w +{\vec G}_A(x)T^A_s)-\nonumber \\
&-&\loarrow{(\vec \partial -e({\vec {\tilde A}}^{'}(x)
-tg\, \theta_w {\vec {\tilde Z}}^{'}(x))Y_w -{\vec G}_A(x)T^A_s)}\cdot \vec 
\alpha ] {\tilde \psi}^{(q)}_{Ri}(x)-\nonumber \\
&-&(1+{e\over {sin\, 2\theta_wm_Z}} H(x)) \nonumber \\
&&[{\bar \psi}^{(q){'}}_{Li}(x)\cdot 
\left( \begin{array}{l} 0 \\ 1 \end{array} \right) M^{(q)}_{ij}\psi
^{(q)}_{Rj}(x)+{\bar \psi}^{(q)}_{Ri}(x)M_{ij}^{(q)\dagger} 
\left( \begin{array}{ll}
0 & 1 \end{array} \right) \cdot \psi^{(q){'}}_{Lj}(x)+\nonumber \\
&+&{\bar \psi}^{(q){'}}_{Li}(x)\cdot \left( \begin{array}{l} 1 \\ 0 \end{array}
\right) {\tilde M}^{(q)}_{ij}{\tilde \psi}^{(q)}_{Rj}(x)+{\bar {\tilde \psi}}
^{(q)}_{Ri}(x){\tilde M}_{ij}^{(q)\dagger} \left( \begin{array}{ll} 1 & 0 
\end{array} \right) \cdot \psi^{(q){'}}_{Lj}(x)]-\nonumber \\
&-&\theta {{g^3_s}\over {8\pi^2}}{\vec \pi}^{(G)}_A(x)\cdot {\vec B}^{(G)}_A(x)
-\nonumber \\
&-&{\tilde A}^{{'} o}(x) [-\vec \partial \cdot {\vec {\tilde \pi}}^{(\tilde A)
{'}}(x)+ie({\vec {\tilde W}}^{'}_{+}(x)\cdot {\vec {\tilde \pi}}^{({\tilde
W}_{+}){'}}(x)-{\vec {\tilde W}}^{'}_{-}(x)\cdot {\vec {\tilde \pi}}
^{({\tilde W}_{-}){'}}(x))+\nonumber \\
&+&e \psi^{(l){'}\dagger}_{Li}(x)Q_{em}\psi^{(l){'}}_{Li}(x)+e\psi^{(l)\dagger}
_{Ri}(x)iY_w\psi^{(l)}_{Ri}(x)+\nonumber \\
&+&e\psi^{(q){'}\dagger}_{Li}(x)Q_{em}\psi^{(q){'}}_{Li}(x)+\nonumber \\
&+&e\psi^{(q)\dagger}_{Ri}(x)iY_w\psi^{(q)}_{Ri}(x)+e{\tilde \psi}^{(q)\dagger}
_{Ri}(x)iY_w{\tilde \psi}^{(q)}_{Ri}(x)]-\nonumber \\
&-&{\tilde Z}^{{'} o}(x) [-\vec \partial \cdot {\vec {\tilde \pi}}^{(\tilde
Z){'}}(x)+ie cot\, \theta_w ({\vec {\tilde W}}^{'}_{+}(x)\cdot {\vec {\tilde 
\pi}}^{({\tilde W}_{+}){'}}(x)-{\vec {\tilde W}}^{'}_{-}(x)\cdot {\vec {\tilde 
\pi}}^{({\tilde W}_{-}){'}}(x))+\nonumber \\
&+&e \psi^{(l){'}\dagger}_{Li}(x)Q_Z\psi^{(l){'}}_{Li}(x)-etg\, \theta_w
\psi^{(l)\dagger}_{Ri}(x)iY_w\psi^{(l)}_{Ri}(x)+\nonumber \\
&+&e\psi^{(q){'}\dagger}_{Li}(x)Q_Z\psi^{(q){'}}_{Li}(x)-etg\, \theta_w
\psi^{(q)\dagger}_{Ri}(x)iY_w\psi^{(q)}_{Ri}(x)-\nonumber \\
&-&etg\, \theta_w{\tilde \psi}^{(q)\dagger}_{Ri}(x)iY_w
{\tilde \psi}^{(q)}_{Ri}(x)]-\nonumber \\
&-&{\tilde W}^{{'} o}_{+}(x) [-\vec \partial \cdot 
{\vec {\tilde \pi}}^{({\tilde W}_{-}){'}}(x)+ie
({\vec {\tilde W}}^{'}_{-}(x)\cdot ({\vec {\tilde \pi}}^{(\tilde A){'}}(x)+
\nonumber \\
&+&cot\, \theta_w {\vec {\tilde \pi}}^{(\tilde
Z){'}}(x))-({\vec {\tilde A}}^{'}(x)+cot\, \theta_w {\vec {\tilde Z}}^{'}(x))
\cdot {\vec {\tilde \pi}}^{({\tilde W}_{+}){'}}(x))+\nonumber \\
&+&{{ie}\over {sin\, \theta_w}}(\psi^{(l){'}\dagger}_{Li}(x)T^{-}_w\psi^{(l)
{'}}_{Li}(x)+\psi^{(q){'}\dagger}_{Li}(x)T^{-}_w\psi^{(q){'}}_{Li}(x))]-
\nonumber \\
&-&{\tilde W}^{{'} o}_{-}(x) [-\vec \partial \cdot {\vec {\tilde \pi}}
^{({\tilde W}_{+}){'}}(x)-ie
({\vec {\tilde W}}^{'}_{+}(x)\cdot ({\vec {\tilde \pi}}^{(\tilde A){'}}(x)+
\nonumber \\
&+&cot\, \theta_w {\vec {\tilde \pi}}^{(\tilde
Z){'}}(x))-({\vec {\tilde A}}^{'}(x)+cot\, \theta_w {\vec {\tilde Z}}^{'}(x))
\cdot {\vec {\tilde \pi}}^{({\tilde W}_{-}){'}}(x))+\nonumber \\
&+&{{ie}\over {sin\, \theta_w}}(\psi^{(l){'}\dagger}_{Li}(x)T^{+}_w\psi^{(l)
{'}}_{Li}(x)+\psi^{(q){'}\dagger}_{Li}(x)T^{+}_w\psi^{(q){'}}_{Li}(x))]-
\nonumber \\
&-&(1+{e\over {sin\, 2\theta_w\, m_Z}} H(x))^2 [m_W^2 {\tilde W}
^{{'} o}_{+}(x) {\tilde W}^{{'} o}_{-}(x)+{1\over 2} m_Z^2 {\tilde Z}^{{'} o 2}
(x)]-\nonumber \\
&-&G^o_A(x) [-{\hat {\vec D}}^{(G)}_{AB}\cdot {\vec \pi}^{(G)}_B(x)+
i\psi^{(q){'}\dagger}_{Li}(x)T^A_s\psi^{(q){'}}_{Li}(x)+\nonumber \\
&+&i\psi^{(q)\dagger}
_{Ri}(x)T^A_s\psi^{(q)}_{Ri}(x)+{\tilde \psi}^{(q)\dagger}_{Ri}(x)T^A_s
{\tilde \psi}^{(q)}_{Ri}(x)]+\nonumber \\
&+&\lambda_{Ao}(x) \pi^{(G) o}_A(x)+\lambda_o^{(\tilde A)}(x) {\tilde \pi}
^{(\tilde A){'} o}(x)+\lambda_o^{(\tilde  Z)}(x) {\tilde \pi}^{(\tilde Z){"}
o}(x)+\nonumber \\
&+&\lambda_o^{(-)}(x){\tilde \pi}^{({\tilde W}_{+}){"} o}(x)+
\lambda_o^{(+)}(x) {\tilde \pi}^{({\tilde W}_{-}){'} o}(x)=\nonumber \\
&&{}\nonumber \\
&=&{1\over 2}\sum_A [g^2_s{\vec \pi}^{(G) 2}_A(x)+ g^{-2}_s 
{\vec B}^{(G) 2}_A(x)]+\nonumber \\
&+&{1\over 2} [{\vec {\tilde \pi}}^{(\tilde A){'} 2}(x)+{\vec B}^{(\tilde A){'}
2}(x)]+{1\over 2} [{\vec {\tilde \pi}}^{(\tilde Z){'} 2}(x) +{\vec B}
^{(\tilde Z){'} 2}(x)]+\nonumber \\
&+&{\vec {\tilde \pi}}^{({\tilde W}_{+}){'}}(x) \cdot {\vec {\tilde \pi}}
^{({\tilde W}_{-}){'}}(x)+ {\vec B}^{({\tilde W}_{+}){'}}(x) \cdot {\vec B}
^{({\tilde W}_{-}){'}}(x)+\nonumber \\
&+& (1+{e\over {sin\, 2\theta_w m_Z}} H(x))^2 [m_W^2 {\vec {\tilde W}}^{'}_{+}
(x) \cdot {\vec {\tilde W}}^{'}_{-}(x) + {1\over 2} m^2_Z {\vec {\tilde Z}}^{'}
(x)]+\nonumber \\
&+&{1\over 2} [\pi^2_H(x) + (\vec \partial H(x))^2] + {1\over 2} m_H^2 H^2(x)
(1+{e\over {2sin\, 2\theta_w m_Z}} H(x))^2+\nonumber \\
&+&ie [{\vec {\tilde W}}^{'}_{+}(x) \times {\vec {\tilde W}}^{'}_{-}(x) \cdot
({\vec B}^{(\tilde A){'}}(x)+ cot\, \theta_w {\vec B}^{(\tilde Z){'}}(x))]+
\nonumber \\
&+& ({\vec {\tilde A}}^{'}(x)+cot\, \theta_w {\vec {\tilde Z}}^{'}(x)) [\times
{\vec {\tilde W}}^{'}_{-}(x) \cdot {\vec B}^{({\tilde W}_{+}){'}}(x)-
\times {\vec {\tilde W}}^{'}_{+}(x) \cdot {\vec B}^{({\tilde W}_{-}){'}}(x)) ]
-\nonumber \\
&-&{{e^2}\over {2sin^2\, \theta_w}} [ {\vec {\tilde W}}^{{'} 2}_{+}(x) {\vec 
{\tilde W}}^{{'} 2}_{-}(x) -( {\vec {\tilde W}}^{'}_{+}(x) \cdot {\vec {\tilde 
W}}^{'}_{-}(x) )^2]+\nonumber \\
&+& e^2 [{\vec {\tilde W}}^{'}_{+}(x) \cdot {\vec {\tilde W}}^{'}_{-}(x)
({\vec {\tilde A}}^{'}(x)+cot\, \theta_w {\vec {\tilde Z}}^{'}(x))^2 -
\nonumber \\
&-&{\vec {\tilde W}}^{'}_{+}(x) \cdot ({\vec {\tilde A}}^{'}(x)+cot\, \theta_w
{\vec {\tilde Z}}^{'}(x)) {\vec {\tilde W}}^{'}_{-}(x) \cdot ({\vec {\tilde 
A}}^{'}(x)+cot\, \theta_w {\vec {\tilde Z}}^{'}(x))]+\nonumber \\
&+& \left( \begin{array}{ccc} {\bar \nu}^{(m)}_e(x) & {\bar \nu}^{(m)}_{\mu}(x)
& {\bar \nu}^{(m)}_{\tau}(x) \end{array} \right) i\vec \alpha \cdot \vec 
\partial {1\over 2}(1-\gamma_5)
\left( \begin{array}{c} \nu^{(m)}_e(x)\\ \nu^{(m)}_{\mu}(x)\\ \nu^{(m)}_{\tau}
(x) \end{array} \right) +\nonumber \\
&+& \left( \begin{array}{ccc} {\bar e}^{(m)}(x) & {\bar \mu}^{(m)}(x) & {\bar
\tau}^{(m)}(x) \end{array} \right) \nonumber \\
&&[i\vec \alpha \cdot \vec \partial +\beta (1+ {e\over
{sin\, 2\theta_w m_Z}} H(x)) \left( \begin{array}{ccc} 
m_e & 0 & 0\\ 0 & m_{\mu}
& 0\\ 0 & 0 & m_{\tau} \end{array} \right) ] \left( \begin{array}{c} e^{(m)}
(x) \\ \mu^{(m)}(x) \\ \tau^{(m)}(x) \end{array} \right) +\nonumber \\
&+&\left( \begin{array}{ccc} {\bar u}^{(m)}(x) & {\bar c}^{(m)}(x) & {\bar
t}^{(m)}(x) \end{array} \right) \nonumber \\
&&[i\vec \alpha \cdot \vec \partial +\beta (1+ {e\over
{sin\, 2\theta_w m_Z}} H(x)) \left( \begin{array}{ccc} m_u & 0 & 0\\ 0 & m_c
& 0\\ 0 & 0 & m_t \end{array} \right) ] \left( \begin{array}{c} u^{(m)}
(x) \\ c^{(m)}(x) \\ t^{(m)}(x) \end{array} \right) +\nonumber \\
&+&\left( \begin{array}{ccc} {\bar d}^{(m)}(x) & {\bar s}^{(m)}(x) & {\bar
b}^{(m)}(x) \end{array} \right) \nonumber \\
&&[i\vec \alpha \cdot \vec \partial +\beta (1+ {e\over
{sin\, 2\theta_w m_Z}} H(x)) \left( \begin{array}{ccc} m_d & 0 & 0\\ 0 & m_s
& 0\\ 0 & 0 & m_b \end{array} \right) ] \left( \begin{array}{c} d^{(m)}
(x) \\ s^{(m)}(x) \\ b^{(m)}(x) \end{array} \right) +\nonumber \\
&+&{\vec G}_{A}(x)\cdot {\vec J}_{sA}(x)+
e{\vec {\tilde A}}^{'}(x)\cdot {\vec {\tilde j}}^{'}_{(em)}(x)+e{\vec {\tilde 
Z}}^{'}(x) {\vec {\tilde j}}^{'}_{(NC)}(x)+\nonumber \\
&+&{e\over {sin\, \theta_w}} ({\vec {\tilde W}}^{'}
_{+}(x)\cdot {\vec{\tilde j}}^{'}_{(CC)\, -}(x)+ {\vec {\tilde W}}^{'}_{-}(x) 
\cdot {\vec {\tilde j}}^{'}_{(CC)\, +}(x))-\nonumber \\ 
&-&\theta {{g^3_s}\over {8\pi^2}}{\vec \pi}^{(G)}_A(x)\cdot {\vec B}^{(G)}_A(x)
-\nonumber \\
&-&G^o_A(x) \Gamma^{(G)}_A(x)-
{\tilde A}^{{'} o}(x) \Gamma^{(\tilde A)}(x) -\nonumber \\
&-&{\tilde Z}^{{'}o}(x)[{\tilde \zeta}
^{(\tilde Z)}(x)-m^2_Z(1+{e\over {sin\, 2\theta_w
m_Z}} H(x))^2 {\tilde Z}^{{'}o}(x)]-\nonumber \\
&-&{\tilde W}^{{'}o}_{+}(x) [{\tilde \zeta}
^{({\tilde W}_{+})}(x)-m_W^2(1+{e\over 
{sin\, 2\theta_wm_Z}} H(x))^2 {\tilde  W}^{{'}0}_{-}(x)] -\nonumber \\
&-&{\tilde W}^{{'}o}_{-}(x) [{\tilde \zeta}
^{({\tilde W}_{-})}(x)-m_W^2(1+{e\over {sin\, 
2\theta_wm_Z}} H(x))^2 {\tilde W}^{{'}o}_{+}(x)]-\nonumber \\
&-&(1+{e\over {sin\, 2\theta_wm_Z}} H(x))^2 [m_W^2 {\tilde  W}^{{'}o}_{+}(x)
{\tilde W}^{{'}o}_{-}(x)+{1\over 2}m_Z^2 {\tilde Z}^{{'}o\, 2}(x)]+\nonumber \\
&+&\lambda_{Ao}(x) \pi^{(G) o}_A(x)+\lambda_o^{(\tilde A)}(x) {\tilde \pi}
^{(\tilde A){'} o}(x)+\lambda_o^{(\tilde  Z)}(x) {\tilde \pi}^{(\tilde Z){"}
o}(x)+\nonumber \\
&+&\lambda_o^{(-)}(x){\tilde \pi}^{({\tilde W}_{+}){"} o}(x)+
\lambda_o^{(+)}(x) {\tilde \pi}^{({\tilde W}_{-}){'} o}(x),
\label{46}
\end{eqnarray}

\noindent with $\Gamma^{(G)}_A$, $\Gamma^{(\tilde A)}$, ${\tilde \zeta}
^{(\tilde Z)}$,
${\tilde \zeta}^{({\tilde W}_{\pm})}$, defined in the next Eq.(\ref{47}).

The time constancy of the primary constraints generates the secondary ones

\begin{eqnarray}
\Gamma^{(G)}_A(x)&=&-{\hat {\vec D}}^{(G)}_{AB}\cdot {\vec \pi}^{(G)}_B(x)+
i\psi^{(q){'}\dagger}_{Li}(x)T^A_s\psi^{(q){'}}_{Li}(x)+\nonumber \\
&+&i\psi^{(q)\dagger}
_{Ri}(x)T^A_s\psi^{(q)}_{Ri}(x)+{\tilde \psi}^{(q)\dagger}_{Ri}(x)T^A_s
{\tilde \psi}^{(q)}_{Ri}(x)=\nonumber \\
&=&-{\hat {\vec D}}^{(G)}_{AB}\cdot {\vec \pi}^{(G)}_B(x)+J^{{'} o}_{sA}(x)
\approx 0,\nonumber \\
&&{}\nonumber \\
\Gamma^{(\tilde A)}(x)&=&-\vec \partial \cdot {\vec {\tilde \pi}}^{(\tilde A)
{'}}(x)+ie({\vec {\tilde W}}^{'}_{+}(x)\cdot {\vec {\tilde \pi}}^{({\tilde
W}_{+}){'}}(x)-{\vec {\tilde W}}^{'}_{-}(x)\cdot {\vec {\tilde \pi}}
^{({\tilde W}_{-}){'}}(x))+\nonumber \\
&+&e \psi^{(l){'}\dagger}_{Li}(x)Q_{em}\psi^{(l){'}}_{Li}(x)+e\psi^{(l)\dagger}
_{Ri}(x)iY_w\psi^{(l)}_{Ri}(x)+\nonumber \\
&+&e\psi^{(q){'}\dagger}_{Li}(x)Q_{em}\psi^{(q){'}}_{Li}(x)+e\psi^{(q)\dagger}
_{Ri}(x)iY_w\psi^{(q)}_{Ri}(x)+\nonumber \\
&+&e{\tilde \psi}^{(q)\dagger}_{Ri}(x)iY_w{\tilde \psi}^{(q)}_{Ri}(x)=
\nonumber \\
&=&-\vec \partial \cdot {\vec {\tilde \pi}}^{(\tilde A)
{'}}(x)+ie({\vec {\tilde W}}^{'}_{+}(x)\cdot {\vec {\tilde \pi}}^{({\tilde
W}_{+}){'}}(x)-{\vec {\tilde W}}^{'}_{-}(x)\cdot {\vec {\tilde \pi}}
^{({\tilde W}_{-}){'}}(x))+e {\tilde j}^o_{(em)}(x)\approx 0,\nonumber \\
&&{}\nonumber \\
\zeta^{(\tilde Z)}(x)&=&m_Z^2 (1+{e\over {sin\, 2\theta_w\, m_Z}}H(x))^2
{\tilde Z}^{{'} o}(x)+{\tilde \zeta}^{(\tilde Z)}(x)=\nonumber \\
&=&m_Z^2 (1+{e\over {sin\, 2\theta_w\, m_Z}}H(x))^2
{\tilde Z}^{{'} o}(x)-\nonumber \\
&-&\vec \partial \cdot {\vec {\tilde \pi}}^{(\tilde
Z){'}}(x)+ie cot\, \theta_w ({\vec {\tilde W}}^{'}_{+}(x)\cdot {\vec {\tilde 
\pi}}^{({\tilde W}_{+}){'}}(x)-{\vec {\tilde W}}^{'}_{-}(x)\cdot {\vec {\tilde 
\pi}}^{({\tilde W}_{-}){'}}(x))+\nonumber \\
&+&e \psi^{(l){'}\dagger}_{Li}(x)Q_Z\psi^{(l){'}}_{Li}(x)-etg\, \theta_w
\psi^{(l)\dagger}_{Ri}(x)iY_w\psi^{(l)}_{Ri}(x)+\nonumber \\
&+&e\psi^{(q){'}\dagger}_{Li}(x)Q_Z\psi^{(q){'}}_{Li}(x)-etg\, \theta_w
\psi^{(q)\dagger}_{Ri}(x)iY_w\psi^{(q)}_{Ri}(x)-\nonumber \\
&-&etg\, \theta_w{\tilde \psi}^{(q)\dagger}_{Ri}(x)iY_w
{\tilde \psi}^{(q)}_{Ri}(x)=\nonumber \\
&=&m_Z^2 (1+{e\over {sin\, 2\theta_w\, m_Z}}H(x))^2
{\tilde Z}^{{'} o}(x)-\nonumber \\
&-&\vec \partial \cdot {\vec {\tilde \pi}}^{(\tilde
Z){'}}(x)+ie cot\, \theta_w ({\vec {\tilde W}}^{'}_{+}(x)\cdot {\vec {\tilde 
\pi}}^{({\tilde W}_{+}){'}}(x)-{\vec {\tilde W}}^{'}_{-}(x)\cdot {\vec {\tilde 
\pi}}^{({\tilde W}_{-}){'}}(x))+e {\tilde j}^o_{(NC)}(x)\approx 0,\nonumber \\
&&{}\nonumber \\
\zeta^{({\tilde W}_{\pm})}(x)&=&m_W^2(1+{e\over {sin\, 2\theta_w\,
m_Z}}H(x))^2 {\tilde W}^{{'} o}_{\mp}(x)+{\tilde \zeta}^{({\tilde W}_{\pm})}
(x)=\nonumber \\
&=&m_W^2(1+{e\over {sin\, 2\theta_w\,
m_Z}}H(x))^2 {\tilde W}^{{'} o}_{\mp}(x)-\vec \partial \cdot {\vec {\tilde \pi}}
^{({\tilde W}_{\pm}){'}}(x)\pm \nonumber \\
&\pm& ie({\vec {\tilde W}}^{'}_{\mp}(x)\cdot ({\vec {\tilde \pi}}^{(\tilde A){'}}(x)+
cot\, \theta_w {\vec {\tilde \pi}}^{(\tilde
Z){'}}(x))-({\vec {\tilde A}}^{'}(x)+cot\, \theta_w {\vec {\tilde Z}}^{'}(x))
\cdot {\vec {\tilde \pi}}^{({\tilde W}_{\pm}){'}}(x))+\nonumber \\
&+&{{ie}\over {sin\, \theta_w}}(\psi^{(l){'}\dagger}_{Li}(x)T^{\mp}_w\psi^{(l)
{'}}_{Li}(x)+\psi^{(q){'}\dagger}_{Li}(x)T^{\mp}_w\psi^{(q){'}}_{Li}(x))=
\nonumber \\
&=&m_W^2(1+{e\over {sin\, 2\theta_w\,
m_Z}}H(x))^2 {\tilde W}^{{'} o}_{\mp}(x)-\vec \partial \cdot {\vec {\tilde \pi}}
^{({\tilde W}_{\pm}){'}}(x)\pm \nonumber \\
&\pm& ie({\vec {\tilde W}}^{'}_{\mp}(x)\cdot ({\vec {\tilde \pi}}^{(\tilde A){'}}(x)+
cot\, \theta_w {\vec {\tilde \pi}}^{(\tilde
Z){'}}(x))-({\vec {\tilde A}}^{'}(x)+cot\, \theta_w {\vec {\tilde Z}}^{'}(x))
\cdot {\vec {\tilde \pi}}^{({\tilde W}_{\pm}){'}}(x))+\nonumber \\
&+&{e\over { sin\, \theta_w}} {\tilde j}^o_{(CC)\, \mp}(x)\approx 0,
\label{47}
\end{eqnarray}

\noindent where Eqs.(\ref{26}) have to be used for the fermionic
currents.

The constraints  $\Gamma^{(G)}_A(x)\approx 0$, $\Gamma^{(\tilde A)}(x)
\approx 0$, are constant of the motion and first class: therefore,
$G^o_A(x)$ and ${\tilde A}^{{'} o}(x)$ are gauge variables. Instead the
time constancy of $\zeta^{(\tilde Z)}(x)\approx 0$, $\zeta^{({\tilde W}_{\pm})}
(x)\approx 0$, determines the Dirac multipliers $\lambda_o^{(\tilde Z)}(x),
\lambda_o^{(\pm )}(x)$ (they vanish), 
so that these constraints form pairs of second class
constraints with their primaries ${\tilde \pi}^{(\tilde Z){'} o}(x)\approx
0, {\tilde \pi}^{({\tilde W}_{\pm}){'} o}(x)\approx 0$ and determine
their conjugate variables ${\tilde Z}^{{'} o}(x), {\tilde W}^{{'} o}_{\pm}(x)$,
which can be eliminated by going to Dirac brackets.

\section
{Electromagnetic and  color  Dirac observables}

We shall use the results of Ref.\cite{lusa} to find a canonical basis of
electromagnetic and color Dirac's observables, having vanishing Poisson bracket 
with all the constraints and gauge variables, and we shall use the equations 
$\zeta^{(\tilde Z)}(x)=0$, $\zeta^{({\tilde W}_{\pm})}(x)=0$, together with 
the associated Dirac brackets (still denoted as Poisson brackets), to eliminate
${\tilde Z}^{{'} o}(x)$, ${\tilde W}^{{'} o}_{\pm}(x)$.

In the electromagnetic case, we have the following decomposition (Hodge
decomposition of one-forms, when the first homotopy group of the manifold
vanish as it happen for $R^3$) of ${\vec {\tilde A}}^{'}(x)$, ${\vec {\tilde
\pi}}^{(\tilde A){'}}(x)$ [see Eqs.(2-10) and (2-9) of the second paper in 
Ref.\cite{lusa}; $\triangle =-{\vec \partial}^2$]

\begin{eqnarray}
{\vec {\tilde A}}^{'}(x)&=& \vec \partial {\tilde \eta}_{em}(x) + {\check {\vec
{\tilde A}}}_{\perp}(x),\nonumber \\
{\vec {\tilde \pi}}^{(\tilde A) {'}}(x)&=&{\check {\vec {\tilde \pi}}}_{\perp}
(x) +\nonumber \\
&+& {{\vec \partial}\over {\triangle}} [\Gamma^{(\tilde A)}(x)-ie
({\vec {\tilde W}}^{'}_{+}(x)\cdot {\vec {\tilde \pi}}^{({\tilde W}_{+}){'}}(x)
-{\vec {\tilde W}}^{'}_{-}(x)\cdot {\vec {\tilde \pi}}^{({\tilde W}_{-}){'}}
(x))-e{\tilde j}^o_{(em)}(x)],
\label{48}
\end{eqnarray}

\noindent in terms of the Dirac observables ${\check {\tilde A}}^i_{\perp}(x)=
P^{ij}_{\perp}(x) {\tilde A}^{{'}\, j}(x)$, ${\check {\tilde \pi}}^{(\tilde A)
i}_{\perp}(x)=P^{ij}_{\perp}(x) {\tilde \pi}^{(\tilde A){'} i}(x)$ [$P^{ij}
_{\perp}(x)=\delta^{ij}+{{\partial^i_x\partial^j_x}\over {\triangle_x}}$,
$\vec \partial \cdot {\check {\vec {\tilde A}}}_{\perp}(x)=\vec \partial \cdot
{\check {\vec {\tilde \pi}}}_{\perp}(x)=0$], and of

\begin{eqnarray}
{\tilde \eta}_{em}(x)&=&-{1\over {\triangle}} \vec \partial \cdot {\vec {\tilde 
A}}^{'}(x)=-\int d^3y\, \vec c (\vec x-\vec y) \cdot {\vec {\tilde A}}^{'}
(\vec y,x^o)=\nonumber \\
&=&-\int d^3y\, c(\vec x-\vec y) {\vec \partial}_y\cdot {\vec {\tilde A}}^{'}
(\vec y,x^o),\nonumber \\
&&c(\vec x-\vec y) ={1\over {\triangle}} \delta^3(\vec x-\vec y) ={{-1}\over 
{4\pi | \vec x-\vec y |}},\quad\quad \triangle c(\vec x-\vec y) =\delta^3(\vec 
x-\vec y),\nonumber \\
&&\vec c (\vec x-\vec y) ={\vec \partial}_x c(\vec x-\vec y) ={{{\vec 
\partial}_x}\over {\triangle}}\delta^3(\vec x-\vec y) ={{\vec x-\vec y}\over
{4\pi {| \vec x-\vec y |}^3}},\nonumber \\
&&{}\nonumber \\
&&\lbrace {\tilde \eta}_{em}(\vec x,x^o),\Gamma^{(\tilde A)}(\vec y,x^o)
\rbrace =-\delta^3(\vec x-\vec y),\nonumber \\
&&{}\nonumber \\
&&\lbrace {\check {\tilde A}}^i_{\perp}(\vec x,x^o),{\check {\tilde \pi}}
^{(\tilde A) j}_{\perp}(\vec y,x^o)\rbrace =-P^{ij}_{\perp}(x)\delta^3(\vec x
-\vec y).
\label{49}
\end{eqnarray}

Since we have

\begin{eqnarray}
&&\lbrace {\tilde W}^{{'} i}_{\pm}(\vec x,x^o),\Gamma^{(\tilde A)}(\vec y,x^o)
\rbrace =\pm ie {\tilde W}^{{'} i}_{\pm}(x) \delta^3(\vec x-\vec y),\nonumber \\
&&\lbrace {\tilde \pi}^{({\tilde W}_{\pm}){'} i}(\vec x,x^o),\Gamma^{(\tilde 
A)}(\vec y,x^o)\rbrace =\mp ie {\tilde \pi}^{({\tilde W}_{\pm}){'} i}(x) \delta
^3(\vec x-\vec y),\nonumber \\
&&{}\nonumber \\
&&\lbrace \psi^{(l){'}}_{Lia\alpha}(\vec x,x^o),\Gamma^{(\tilde A)}(\vec y,x^o)
\rbrace = -ieQ_{em} \psi^{(l){'}}_{Lia\alpha}(x) \delta^3(\vec x-\vec y),
\nonumber \\
&&\lbrace \psi^{(l)}_{Ri\alpha}(\vec x,x^o),\Gamma^{(\tilde A)}(\vec y,x^o)
\rbrace = eY_w \psi^{(l)}_{Ri\alpha}(x) \delta^3(\vec x-\vec y),\nonumber \\
&&\lbrace \psi^{(q){'}}_{LiAa\alpha}(\vec x,x^o),\Gamma^{(\tilde A)}(\vec y,x^o)
\rbrace = -ieQ_{em} \psi^{(q){'}}_{LiAa\alpha}(x) \delta^3(\vec x-\vec y),
\nonumber \\
&&\lbrace \psi^{(q)}_{RiA\alpha}(\vec x,x^o),\Gamma^{(\tilde A)}(\vec y,x^o)
\rbrace = eY_w \psi^{(q)}_{RiA\alpha}(x) \delta^3(\vec x-\vec y),\nonumber \\
&&\lbrace {\tilde \psi}^{(q)}_{RiA\alpha}(\vec x,x^o),\Gamma^{(\tilde A)}(\vec 
y,x^o)\rbrace = eY_w {\tilde \psi}^{(q)}_{RiA\alpha}(x) \delta^3(\vec x-\vec y),
\label{50}
\end{eqnarray}

\noindent the electromagnetic Dirac observables [having vanishing Poisson 
brackets with $\Gamma^{(\tilde A)}(x)$ and ${\tilde \eta}_{em}(x)$] are the 
following quantities, each one dressed with its Coulomb cloud

\begin{eqnarray}
&&{\check {\vec {\tilde W}}}_{\pm}(x)=e^{\pm ie {\tilde \eta}_{em}(x)} 
{\vec {\tilde W}}^{'}_{\pm}(x),\quad\quad {\check {\vec {\tilde \pi}}}
^{({\tilde W}_{\pm})}(x)= e^{\mp ie {\tilde \eta}_{em}(x)} {\vec {\tilde \pi}}
^{({\tilde W}_{\pm}){'}}(x),
\nonumber \\
&&{}\nonumber \\
&&e^{-ieQ_{em}{\tilde \eta}_{em}(x)} \psi^{(l){'}}_{Li}(x),\quad\quad
e^{eY_w{\tilde \eta}_{em}(x)} \psi^{(l)}_{Ri}(x),\nonumber \\
&&e^{-ieQ_{em}{\tilde \eta}_{em}(x)} \psi^{(q){'}}_{Li}(x),\quad\quad
e^{eY_w {\tilde \eta}_{em}(x)} \psi^{(q)}_{Ri}(x),\quad\quad
e^{eY_w {\tilde \eta}_{em}(x)} {\tilde \psi}^{(q)}_{Ri}(x).
\label{51}
\end{eqnarray}

The analogous decompositions in the non-Abelian SU(3) case can again be
obtained from the second paper in Ref.\cite{lusa}.
For the vector potential we use Eqs.(4-13), (4-16), (4-26), (4-29), (4-30),
(4-31), (4-33), (5-21), (5-24), of that paper to get

\begin{eqnarray}
{\vec G}_A(x)&=&A_{AB}(\eta^{(G)}(x)) \vec \partial \eta^{(G)}_B(x) +(P\, 
e^{\Omega_s^{(\hat \gamma )}(\eta^{(G)}(x))} )_{AB} {\check {\vec G}}_{B\perp}
(x),\quad\quad \vec \partial \cdot {\check {\vec G}}_{A\perp}(x)=0,\nonumber \\
&&{}\nonumber \\
&&{\hat T}^A_s A_{AB}(\eta^{(G)}(x)) \vec \partial \eta^{(G)}_B(x)\cdot d\vec x
=H_B(\eta^{(G)}(x)) \vec \partial \eta^{(G)}_B(x)\cdot d\vec x=\nonumber \\
&=&\Theta_A(\eta^{(G)}(x), \vec \partial \eta^{(G)}(x)) {\hat  T}^A_s= 
d_{(\hat \gamma )} \Omega_s^{(\hat \gamma )}(\eta^{(G)}(x)),\nonumber \\
&&{}\nonumber \\
&&\Omega_s^{(\hat \gamma )}(\eta^{(G)}(x))=\Omega_{sA}^{(\hat \gamma )}(\eta
^{(G)}(x)) {\hat T}^A_s = {}_{(\hat \gamma )}\int_0^{\eta^{(G)}(x,s)} H_B
(\eta^{(G)}(x;s)) {\cal D}\eta_B^{(G)}(x;s).
\label{52}
\end{eqnarray}

If $\eta_A$ are coordinates in a chart of the group manifold of SU(3), the
matrices $A_{AB}(\eta )$ satisfy the Maurer-Cartan equations, which can be
written in the zero curvature form ${{\partial H_A(\eta )}\over {\partial
\eta_B}}-{{\partial H_B(\eta )}\over {\partial \eta_A}}+[H_A(\eta ),H_B
(\eta )]=0$. We shall use only canonical coordinates of the first kind, defined
by $A_{AB}(\eta )\eta_B=\eta_A$ [ so that $A(\eta )={{e^{T\eta}-1}\over 
{T\eta}}$ with $(T\eta )_{AB}=({\hat T}^C_s)_{AB}\eta_C=c_{ABC}\eta_C$]. If 
$\theta_A=A_{AB}(\eta ) d\eta_B$ are the left-invariant (or Maurer-Cartan) 
one-forms on SU(3), the abstract Maurer-Cartan equations are $d\theta_A=-
{1\over 2}c_{ABC} \theta_B\wedge \theta_C$; then, by using the preferred line 
$\gamma_{\eta}(s)$ (s is the parameter along it) defining the canonical
coordinates of the first kind in a neighbourhood of the identity I of SU(3),
one can define $d_{(\gamma_{\eta})} \omega_A^{(\gamma_{\eta})}(\eta (s))=
\theta_A(\eta (s))$ [$d_{(\gamma_{\eta})}$ is the exterior derivative along
$\gamma_{\eta}$] with $\omega^{(\gamma_{\eta})}(\eta (s))=\omega_A^{(\gamma
_{\eta})}(\eta (s)) {\hat T}^A_s={}_{(\gamma_{\eta})}\int_0^{\eta (s)}{\hat T}
^A_s A_{AB}(\bar \eta ) d{\bar \eta}_B ={}_{(\gamma_{\eta})}\int_I^{\gamma_{eta}
(s)} \theta_A{|}_{\gamma_{\eta}} {\hat T}^A_s ={}_{(\gamma_{\eta})}\int_I
^{\gamma_{\eta}(s)} \omega_G{|}_{\gamma_{\eta}}$, where $\omega_G=\theta_A{\hat
T}^A_s$ is the canonical one-form on SU(3) in the adjoint representation.
In our case of a trivial principal SU(3)-bundle $P(R^3,SU(3))$ over $R^3$
[fixed $x^o$ slice of Minkowski spacetime], it is shown in Ref.\cite{lusa}
that $\Theta_A(\eta^{(G)}(x), \vec \partial \eta^{(G)}(x))$ and $\Omega_s
^{(\hat \gamma )}(\eta^{(G)}(x))$ are just the extension of these SU(3)
objects: in the second paper of Ref.\cite{lusa}, 
a connection-dependent coordinatization 
$(\vec x, x^o; \eta^{(G)}(\vec x,x^o))$ of the principal bundle is given with
the SU(3) fibers parametrized with parallelly transported (with respect to the
given connection) canonical coordinates of the first kind from a reference
fiber over an arbitrarily chosen origin in $R^3$. The functions $\eta^{(G)}_A
(\vec x,x^o)$ and their gradients $\vec \partial \eta_A^{(G)}(\vec x,x^o)$
vanish on the identity cross section $\sigma_I$ of the trivial principal
bundle. The path $\hat \gamma$ is a surface (in the total bundle space) of
preferred paths, associated with these generalized canonical coordinates of
the first kind, starting from the identity cross section $\sigma_I$ till
a cross section parametrized by the parameter s, in a tubolar neighbourhood of
$\sigma_I$. The operator $d_{(\hat \gamma )}$ is the exterior derivative on the 
principal SU(3)-bundle total space restricted to $\hat \gamma$; it can be
identified with the vertical derivative on the principal bundle and with the
Hamiltonian BRST operator. With these conventions, one has $\lbrace .,\Gamma
^{(G)}_A(x)\rbrace \equiv \lbrace .,
-{\hat {\vec D}}_{AB}\cdot {\vec {\tilde \pi}}^{(G){'}}_B
(x) \rbrace =- B_{BA}(\eta^{(G)}(x)) {{\tilde \delta}\over {\delta \eta^{(G)}
_B(x)}}$ [$B(\eta )=A^{-1}(\eta )$] with the functional derivative to be
interpreted as a directional derivative along the 
surface of paths $\hat \gamma$.
The longitudinal gauge variables (the non-Abelian counterpart of ${\tilde \eta}
_{em}(x)$) have a complicated formal implicit expression given in Eq.(4-49) of
the second paper of Ref.\cite{lusa} and satisfy $\lbrace \eta^{(G)}_A(\vec x,
x^o),{\tilde \Gamma}^{(G)}_B(\vec y,x^o)\rbrace =-\delta_{AB}\delta^3(\vec x
-\vec y)$, where ${\tilde \Gamma}^{(G)}_A(x)=\Gamma^{(G)}_B(x) A_{BA}(\eta^{(G)}
(x))$ are the Abelianized Gauss laws [$\lbrace {\tilde \Gamma}^{(G)}_A(\vec x,
x^o), {\tilde \Gamma}^{(G)}_B(\vec y,x^o)\rbrace =0$]. In Eq.(\ref{52}),
$A_{AB}(\eta^{(G)}(x)) \vec \partial \eta^{(G)}_B(x)$ is the pure gauge part
(saturated with $d\vec x$ it is the BRST ghost) of the vector potential 
${\vec G}_A(x)$: the magnetic field ${\vec B}^{(G)}_A(x)$ is generated only by 
the second term of Eq.(\ref{52}). In this sense, $\eta_A^{(G)}(x)=0$ is the true
generalized non-Abelian Coulomb gauge with all the same properties of the
Abelian Coulomb gauge. In suitable weighted Sobolev spaces\cite{can}, 
as discussed in
Ref.\cite{lusa}, this gauge-fixing is well defined, since all the connections 
over the principal SU(3)-bundle are completely irreducible [their holonomy
bundles (i.e. the set of points of $P(R^3,SU(3))$ which can be joined by
horizontal curves) coincide with the principal bundle itself] and there is no
form of Gribov ambiguity (i.e. of stability subgroups of the group of gauge
transformations for special connections and/or field strengths). In these
spaces, the covariant divergence is an elliptic operator without zero modes
\cite{mon}
and its Green function ${\vec \zeta}^{(G)}_{AB}(\vec x,\vec y;x^o)$ is
globally defined

\begin{eqnarray}
&&{\hat {\vec D}}^{(G)}_{AB}(x)\cdot {\vec \zeta}^{(G)}_{BC}(\vec x,\vec y;x^o)
=- \delta_{AC} \delta^3(\vec x-\vec y),\nonumber \\
&&{}\nonumber \\
&&{\vec \zeta}_{AB}^{(G)}(\vec x,\vec y;x^o)=\vec c (\vec x-\vec y) \zeta^{(G)}
_{AB}(\vec x,\vec y;x^o)=\vec c (\vec x-\vec y) (P\, e^{\int_y^x d\vec z\,
\cdot {\vec G}_C(\vec z,x^o) {\hat T}^C_s}\, )_{AB}.
\label{53}
\end{eqnarray}

\noindent The path ordering is along the straightline (flat geodesic) joining
$\vec y$ and $\vec x$.

Therefore, we have

\begin{eqnarray}
{\vec \pi}^{(G)}_A(x)&=& -{{\vec \partial}\over {\triangle}} \vec \partial 
\cdot {\vec \pi}^{(G)}_A(x) + {\vec \pi}^{(G)}_{A\perp}(x)=\nonumber \\
&=& {\vec \pi}^{(G)}_{A,D\perp}(x)+ \int d^3y\, {\vec \zeta}^{(G)}_{AB}
(\vec x,\vec y;x^o) [\Gamma^{(G)}_B(y)-J^{{'} o}_{sB}(y)],\nonumber \\
&&{}\nonumber \\
&&\vec \partial \cdot {\vec \pi}^{(G)}_{A\perp}(x) = {\hat {\vec D}}^{(G)}_{AB}
(x)\cdot {\vec \pi}^{(G)}_{B,D\perp}(x) = 0.
\label{54}
\end{eqnarray}

It is shown in Eqs. (5-7), (5-8), (5-10), of the second paper in Ref.\cite{lusa}
that we have

\begin{eqnarray}
\vec \partial \cdot {\vec \pi}^{(G)}_A(x) &=& \int d^3y\, {\vec \zeta}^{(G)}
_{AB}(\vec x,\vec y;x^o)\nonumber \\
&&[c_{BEF}{\vec G}_E(y)\cdot {\vec \pi}^{(G)}_{F\perp}(y)+ \Gamma^{(G)}_B(y) 
- J^{{'} o}_{sB}(y)],\nonumber \\
\pi^{(G) i}_{A,D\perp}(x)&=& \int d^3y\, [\delta^{ij} \delta_{AB} \delta^3
(\vec x-\vec y)-\nonumber \\
&-&{{\partial^i_x}\over {\triangle_x}} {\vec \partial}_x\cdot {\vec \zeta}
^{(G)}_{AC}(\vec x,\vec y;x^o) c_{CEB} G^j_E(y)] \pi^{(G) j}_{B\perp}(y),
\nonumber \\
&&{}\nonumber \\
&\Rightarrow& \pi^{(G) i}_{A\perp}(x)= P^{ij}_{\perp}(x) \pi^{(G) j}_{A,D\perp}
(x).
\label{55}
\end{eqnarray}

Moreover, Eqs.(5-21) and (5-25) of that paper give

\begin{eqnarray}
{\vec \pi}^{(G)}_{A,D\perp}(x)&=& (P\, e^{\Omega_s^{(\hat \gamma )}(\eta^{(G)}
(x))} )_{AB} {\check {\vec \pi}}^{(G)}_{B,D\perp}(x)=\nonumber \\
&=&(P\, e^{\Omega_s^{(\hat \gamma )}(\eta^{(G)}(x))} )_{AB} [{\check {\vec
\pi}}^{(G)}_{B\perp}(x) - {{\vec \partial}\over {\triangle}} \vec \partial 
\cdot {\check {\vec \pi}}^{(G)}_{B,D\perp}(x)],\nonumber \\
&&{}\nonumber \\
{\check \pi}^{(G) i}_{A\perp}(x)&=& P^{ij}_{\perp}(x) {\check \pi}^{(G) j}
_{A,D\perp}(x),\nonumber \\
&&\lbrace {\check {\vec \pi}}^{(G)}_{A,D\perp}(\vec x,x^o), \Gamma^{(G)}_B
(\vec y,x^o)\rbrace =0.
\label{56}
\end{eqnarray}

Therefore, the color SU(3) canonical pairs of Dirac's observables turn out to 
be ${\check {\vec G}}_{A\perp}(x)$, ${\check {\vec \pi}}^{(G)}_{A\perp}(x)$. 
They satisfy the Poisson brackets

\begin{equation}
\lbrace {\check G}^i_{A\perp}(\vec x,x^o), {\check \pi}^{(G) j}_{B\perp}(\vec
y,x^o)\rbrace = -\delta_{AB} P^{ij}_{\perp}(x) \delta^3(\vec x-\vec y).
\label{57}
\end{equation}

The original momenta can be written in the form

\begin{eqnarray}
\pi^{(G) i}_A(x)&=& \pi^{(G) i}_{A\perp}(x) -{{\partial^i}\over {\triangle}}
\vec \partial \cdot {\vec \pi}^{(G)}_A(x)=\nonumber \\
&=& P^{ij}_{\perp}(x) (P\, e^{\Omega_s^{(\hat \gamma )}(\eta^{(G)}(x))} )_{AB}
[{\check \pi}^{(G) j}_{B\perp}(x)-{{\partial^j}\over {\triangle}} \vec \partial
\cdot {\check {\vec \pi}}^{(G)}_{B,D\perp}(x)]-\nonumber \\
&-& {{\partial^i}\over {\triangle}} \int d^3y\, {\vec \partial}_x\cdot {\vec
\zeta}^{(G)}_{AB}(\vec x,\vec y;x^o) [\Gamma^{(G)}_B(y)-J^{{'} o}_{sB}(y)+
\nonumber \\
&+&c_{BCD} (\Theta^h_C(\eta^{(G)}(y)) +(P\, e^{\Omega_s^{(\hat \gamma )}
(\eta^{(G)}(x))} )_{CE} {\check G}^h_{E\perp}(y))\cdot \nonumber \\
&\cdot& P^{hk}_{\perp}(y) (P\, e^{\Omega_s^{(\hat \gamma )}(\eta^{(G)}(y))} 
)_{DF} ({\check \pi}^{(G) k}_{F\perp}(y) -{{\partial^k_y}\over {\triangle_y}}
{\vec \partial}_y\cdot {\check {\vec \pi}}^{(G)}_{F,D\perp}(y))],
\label{58}
\end{eqnarray}

\noindent with

\begin{eqnarray}
{\vec \zeta}^{(G)}&_{AB}&(\vec x,\vec y;x^o) = \vec c (\vec x-\vec y)
\nonumber \\
&&(P\, exp\lbrace \int_y^x d\vec z \cdot [{\vec \Theta}_C(\eta^{(G)}(\vec z,
x^o))+ (P\, e^{\Omega_s^{(\hat \gamma )}(\eta^{(G)}(\vec z,x^o))} )_{CD}
{\check {\vec G}}_{D\perp}(\vec z,x^o)]{\hat T}^C_s\rbrace )_{AB},
\label{59}
\end{eqnarray}

\noindent and with $\vec \partial \cdot {\check {\vec \pi}}^{(G)}_{A,D\perp}
(x)$ solution of the equation [implied by ${\hat {\vec D}}^{(G)}_{AB}(x)
\cdot {\vec \pi}^{(G)}_{B,D\perp}(x)=0$]

\begin{eqnarray}
[\delta_{AB}&-& (P\, e^{-\Omega_s^{(\hat \gamma )}(\eta^{(G)}(x))} \vec \partial
P\, e^{\Omega_s^{(\hat \gamma )}(\eta^{(G)}(x))} )_{AB} \cdot {{\vec \partial}
\over {\triangle}}-\nonumber \\
&-&(P\, e^{-\Omega_s^{(\hat \gamma )}(\eta^{(G)}(x))} )_{AC} c_{CEF} ({\vec
\Theta}_E(\eta^{(G)}(x)) + (P\, e^{\Omega_s^{(\hat \gamma )}(\eta^{(G)}(x))}
0_{ER} {\check {\vec G}}_{R\perp}(x))\cdot \nonumber \\
&\cdot& (P\, e^{\Omega_s^{(\hat \gamma )}(\eta^{(G)}(x))} )_{FB} {{\vec
\partial}\over {\triangle}} ] \vec \partial \cdot {\check {\vec \pi}}^{(G)}
_{B,D\perp}(x)=\nonumber \\
&=&-(P\, e^{-\Omega_s^{(\hat \gamma )}(\eta^{(G)}(x))} \vec \partial P\,
e^{\Omega_s^{(\hat \gamma )}(\eta^{(G)}(x))} )_{AB} \cdot {\check {\vec
\pi}}^{(G)}_{B\perp}(x)-\nonumber \\
&-&(P\, e^{\Omega_s^{(\hat \gamma )}(\eta^{(G)}(x))} )_{AB} c_{BEF} ({\vec
\Theta}_E(\eta^{(G)}(x))+ (P\, e^{\Omega_s^{(\hat \gamma )}(\eta^{(G)}(x))}
)_{FR} {\check {\vec G}}_{R\perp}(x))\cdot \nonumber \\
&\cdot& (P\, e^{\Omega_s^{(\hat \gamma )}(\eta^{(G)}(x))})_{FC} {\check {\vec
\pi}}^{(G)}_{C\perp}(x).
\label{60}
\end{eqnarray}

It is not necessary to solve this equation, since for $\Gamma^{(G)}_A(x)=0$ and
$\eta^{(G)}_A(x)=0$ [so that also $\vec \partial \eta^{(G)}_A(x)=0$, 
$\Omega_s^{(\hat \gamma )}(\eta^{(G)}(x))=0$] we get

\begin{eqnarray}
{\vec \pi}^{(G)}_A(x) &\rightarrow& {\hat {\vec \pi}}^{(G)}_A(x) = {\check 
{\vec \pi}}^{(G)}_{A\perp}(x) -\nonumber \\
&-&{{\vec \partial}\over {\triangle}} \int d^3y\, {\vec \partial}_x\cdot {\vec
\zeta}^{({\check G}_{\perp})}_{AB}(\vec x,\vec y;x^o) [c_{BCE} {\check G}^h
_{C\perp}(y){\check \pi}^{(G) h}_{E\perp}(y)-J^{{;} o}_{sB}(y)],\nonumber \\
&&{}\nonumber \\
&&{\vec \zeta}^{({\check G}_{\perp})}_{AB}(\vec x,\vec y;x^o) = \vec c (\vec x
-\vec y) (P\, e^{\int_y^x d\vec z\cdot {\check {\vec G}}_{C\perp}(\vec z,x^o)
{\hat T}^C_s} )_{AB}.
\label{61}
\end{eqnarray}

While in the electromagnetic case it is possible to get the physical Hamiltonian
without imposing the Coulomb gauge-fixing ${\tilde \eta}_{em}(x)\approx 0$
(namely it is obtained by a canonical decoupling of the gauge degrees of freedom
), this is too difficult in the non-Abelian case. 
Therefore, we shall evaluate the
physical quantities by imposing the generalized Coulomb gauge-fixings
$\eta^{(G)}_A(x)\approx 0$. Conceptually, the canonical decoupling of the gauge
degrees of freedom gives the same results for the physical quantities.

Since we have

\begin{eqnarray}
&&\lbrace \psi^{(q){'}}_{LiAa\alpha}(\vec x,x^o), \Gamma^{(G)}_B(\vec y,x^o)
\rbrace = (T^B_s)_{AC} \psi^{(q){'}}_{LiCa\alpha}(x) \delta^3(\vec x-\vec y),
\nonumber \\
&&\lbrace \psi^{(q)}_{RiA\alpha}(\vec x,x^o), \Gamma^{(G)}_B(\vec y,x^o)
\rbrace = (T^B_s)_{AC} \psi^{(q)}_{RiC\alpha}(x) \delta^3(\vec x-\vec y),
\nonumber \\
&&\lbrace {\tilde \psi}^{(q)}_{RiA\alpha}(\vec x,x^o), \Gamma^{(G)}_B(\vec y,
x^o)\rbrace = (T^B_s)_{AC} {\tilde \psi}^{(q)}_{RiC\alpha}(x) \delta^3(\vec x-
\vec y),
\label{62}
\end{eqnarray}

\noindent the fermionic Dirac observables (with vanishing Poisson brackets
with the color Gauss laws and with $\eta^{(G)}(x)$), dressed with gluon
clouds, are

\begin{equation}
(P\, e^{\Omega_s^{(\hat \gamma )}(\eta^{(G)}(x))} )_{AB} \psi^{(q){'}}
_{LiBa\alpha}(x),\quad (P\, e^{\Omega_s^{(\hat \gamma )}(\eta^{(G)}(x))} )_{AB}
 \psi^{(q)}_{RiB\alpha}(x),\quad (P\, e^{\Omega_s^{(\hat \gamma )}(\eta^{(G)}
(x))} )_{AB} {\tilde \psi}^{(q)}_{RiB\alpha}(x),
\label{63}
\end{equation}

\noindent and, by using Eqs.(5-31)-(5-33) of the second paper in Ref.\cite{lusa}
, we have for the quark fields

\begin{equation}
i\psi^{(q)\dagger}(x) \vec \alpha \cdot [\vec \partial +{\vec G}_A(x)T^A_s] 
\psi^{(q)}(x) {\rightarrow}_{\eta^{(G)}\rightarrow 0} \,\,
i{\check \psi}^{(q)\dagger}(x) \vec \alpha \cdot [\vec \partial + {\check
{\vec G}}_{A\perp}(x)T^A_s] {\check \psi}^{(q)}(x),
\label{64}
\end{equation}

\noindent where ${\check \psi}^{(q)}(x)$ are the color Dirac observables.
Analogously, in the electromagnetic case, one has for the electrically charged
fermions

\begin{eqnarray}
i\psi^{\dagger}&(x)& \vec \alpha \cdot [\vec \partial +{e\over {sin\,
\theta_w}} ({\vec {\tilde W}}^{'}_{+}(x)T^{-}_w+{\vec {\tilde W}}^{'}_{-}(x)
T^{+}_w)-ieQ_{em}{\vec {\tilde A}}^{'}(x)]\psi (x) {\rightarrow}_{{\tilde \eta}
_{em}\rightarrow 0}\nonumber \\
&&i{\check \psi}^{\dagger}(x) \vec \alpha \cdot [\vec \partial +{e\over {sin\,
\theta_w}}({\check {\vec {\tilde W}}}_{+}(x)T^{-}_w+{\check {\vec {\tilde W}}}
_{-}(x)T^{+}_w)-ieQ_{em}{\check {\vec {\tilde A}}}_{\perp}(x)] {\check \psi}
(x),
\label{65}
\end{eqnarray}

As shown in Refs.\cite{lusa,lv1}, the Noether identities implied by the second
Noether theorem, applied to the color SU(3) gauge group, give the following
result for the weak improper conserved non-Abelian Noether charges $Q^{(G)}_A$ 
and for the strong improper conserved ones $Q_{(s)A}^{(G)}$

\begin{eqnarray}
Q^{(G)}_A&=& g_s^{-2} c_{ABC} \int d^3x\, G^{ok}_B(\vec x,x^o)G^k_C(\vec x,
x^o) +\int d^3x\, J^o_{sA}(\vec x,x^o) {\buildrel \circ \over =}\nonumber \\
&{\buildrel \circ \over=}& Q_{(s)A}^{(G)}=\int d^2\vec \Sigma \cdot {\vec E}
^{(G)}_A(\vec x,x^o),
\label{66}
\end{eqnarray}

\noindent As shown in Ref.\cite{lusa}, one gets $s^{\mu}_{sA}=-g^{-2}_s\partial
_{\nu}G^{\nu\mu}_A\, {\buildrel \circ \over =}\, v^{\mu}_{sA}=-g^{-2}_sc_{ABC}
G^{\mu\nu}_BG_{C\nu}+J^{\mu}_{sA}$ [for $\mu =0$ one gets $\Gamma^{(G)}_A
\approx 0$], where $s^{\mu}_{sA}$ [$\partial_{\mu}s^{\mu}_{sA}\equiv 0$] is the
strong improper conserved current and $v^{\mu}_{sA}$ [$\partial_{\mu}v^{\mu}
_{sA}\, {\buildrel \circ \over =}\, 0$] the weak improper Noether conserved
current. One has $Q^{(G)}_{(s)A}=\int d^3x s^o_{sA}\, {\buildrel \circ \over
=}\, Q^{(G)}_A=\int d^3x v^o_{sA}$.

Then, we get [see Eqs.(6-33)-(6-35) in the second paper of 
Ref.\cite{lusa}] the following Dirac's observables

\begin{eqnarray}
Q^{(G)}_A&{\rightarrow}&_{\eta^{(G)}\rightarrow 0} {\check Q}_A^{(G)}=\int d^3x
[c_{ABC}{\check {\vec G}}_{B\perp}(\vec x,x^o)\cdot {\check {\vec \pi}}^{(G)}
_{C\perp}(\vec x,x^o)+{\check J}^o_{sA}(\vec x,x^o)],\nonumber \\
&&{}\nonumber \\
&&\lbrace {\check Q}^{(G)}_A,{\check Q}^{(G)}_B\rbrace =c_{ABC}{\check Q}
^{(G)}_C,\nonumber \\
&&\lbrace {\check {\vec G}}_{A\perp}(\vec x,x^o),{\check Q}^{(G)}_B\rbrace =
c_{ABC}{\check {\vec G}}_{C\perp}(\vec x,x^o),\nonumber \\
&&\lbrace {\check {\vec \pi}}^{(G)}_{A\perp}(\vec x,x^o),{\check Q}^{(G)}_B
\rbrace =c_{ABC} {\check {\vec \pi}}^{(G)}_{C\perp}(\vec x,x^o),\nonumber \\
&&\lbrace {\check \psi}^{(q)}_A(\vec x,x^o),{\check Q}^{(G)}_B\rbrace =
[T^B_s {\check \psi}^{(q)}(\vec x,x^o)]_A.
\label{67}
\end{eqnarray}

Instead, as shown in I and II, the original SU(2)xU(1) gauge invariance is
broken in the Higgs phase and there is no Gauss theorem associated with it due 
to the existence of the mass for the Z and $W_{\pm}$ bosons (whose electric
fields go to zero at space infinity). 
Let us remark that the second class constraints $\zeta^{(\tilde Z)}(x)\approx 0$
, $\zeta^{({\tilde W}_{\pm})}(x)\approx 0$, of Eqs.(\ref{47}) can be rewritten
as

\begin{eqnarray}
\zeta^{(\tilde Z)}(x)&=&e{\tilde {\cal Q}}_{(NC)}(x)+m_Z^2 (1+{e\over {sin\, 
2\theta_w\, m_Z}}H(x))^2{\tilde Z}^{{'} o}(x)\approx 0,\nonumber \\
\zeta^{({\tilde W}_{\pm})}(x)&=&{{e}\over {sin\, \theta_w}}{\tilde {\cal Q}}
_{(CC)\pm}(x)+m_W^2(1+{e\over {sin\, 2\theta_w\,
m_Z}}H(x))^2 {\tilde W}^{{'} o}_{\mp}(x)\approx 0.
\label{hh1}
\end{eqnarray}

The three charges

\begin{eqnarray}
{\tilde Q}_{(NC)}&=& \int d^3x {\tilde {\cal Q}}_{(NC)}(\vec x,x^o)=\nonumber \\
&=&\int d^3x [-e^{-1}\vec \partial \cdot {\vec {\tilde \pi}}^{(\tilde
Z){'}}+i cot\, \theta_w ({\vec {\tilde W}}^{'}_{+}\cdot {\vec {\tilde 
\pi}}^{({\tilde W}_{+}){'}}-{\vec {\tilde W}}^{'}_{-}\cdot {\vec {\tilde 
\pi}}^{({\tilde W}_{-}){'}}(x))+ {\tilde j}^o_{(NC)}](\vec x,x^o)\approx
\nonumber \\
&\approx& -m_Z^2 \int d^3x  (1+{e\over {sin\, 2\theta_w\, m_Z}}H(\vec x,x^o))
^2{\tilde Z}^{{'} o}(\vec x,x^o),\nonumber \\
{\tilde Q}_{(CC)\pm}&=& \int d^3x {\tilde {\cal Q}}_{(CC)\pm}(\vec x,x^o)=
\nonumber \\
&=&  sin\, \theta_w\int d^3x [-e^{-1}\vec \partial \cdot {\vec {\tilde \pi}}
^{({\tilde W}_{\pm}){'}}\pm i({\vec {\tilde W}}^{'}_{\mp}\cdot ({\vec {\tilde 
\pi}}^{(\tilde A){'}}+cot\, \theta_w {\vec {\tilde \pi}}^{(\tilde
Z){'}})-\nonumber \\
&-&({\vec {\tilde A}}^{'}+cot\, \theta_w {\vec {\tilde Z}}^{'})
\cdot {\vec {\tilde \pi}}^{({\tilde W}_{\pm}){'}})+
{1\over { sin\, \theta_w}} {\tilde j}^o_{(CC)\, \mp}](\vec x,x^o)
\approx \nonumber \\
&\approx& -m_W^2  \int d^3x (1+{e\over {sin\, 2\theta_w\, m_Z}}H(\vec x,x^o))
^2 {\tilde W}^{{'} o}_{\mp}(\vec x,x^o).
\label{hh2}
\end{eqnarray}

\noindent are not constants of the motion as can be checked by evaluating
their Poisson brackets with the Hamiltonian resulting from Eq.(\ref{46}) after 
the elimination of the gauge variables ${\tilde Z}^{{'} o}$, ${\tilde W}^{{'} 
o}_{\pm}$, by using Eq.(\ref{hh1}) and with $\lambda_o^{(\tilde Z)}(x)=\lambda
_o^{(\pm )}(x)=0$. Here, there is a difference with the result in II: in the
SU(2) Higgs model the 3 second class constraints are a vector under SU(2) and 
generate global SU(2) transformations under which the Hamiltonian (a SU(2) 
scalar) is invariant, so that there are 3 conserved charges. Here, the mixing
with the Weinberg angle implies that the 3 second class constraints (\ref{hh1})
generate global transformations under which the Hamiltonian is not a scalar
due to $m_Z\not= m_W$ and due to the mass terms of the fermions. It is not
clear, at the mathematical level, which is the distinction between  second 
class constraints generating conserved quantities like in II [it would
correspond to the first Noether theorem hidden in the second 
Noether theorem describing local gauge transformations \cite{ll}, extended to 
include the local Noether pseudogauge transformations generated by the second 
class under which the Lagrangian is invariant modulo 
those acceleration-independent combinations of its Euler-Lagrange equations 
corresponding to these secondary constraints\cite{ll}] and second class 
constraints not generating constants of motion like in this case.

By adding to these non conserved charges
the constant of the motion corresponding to the electromagnetic charge
[the second Noether theorem\cite{ll}  implies that the strong improper conserved
charge $Q_{(s)(em)}=\int d^2\vec \Sigma \cdot {\vec \pi}^{(A)}$ is connected
through the Gauss law $\Gamma^{(\tilde A)}(x)\approx 0$ of Eqs.(\ref{47})
to the weak improper
conserved charge $Q_{(em)}=\int d^3x [{\tilde j}^o_{(em)}+i({\vec W}_{+}\cdot
{\vec \pi}^{(W_{+})}-{\vec W}_{-}\cdot {\vec \pi}^{(W_{-})})]\, {\buildrel
\circ \over =}\, Q_{(s)(em)}$]

\begin{equation}
{\tilde Q}_{(em)}=\int d^3x [i({\vec {\tilde W}}^{'}_{+}\cdot 
{\vec {\tilde \pi}}^{({\tilde W}_{+}){'}}-
{\vec {\tilde W}}^{'}_{-}\cdot {\vec {\tilde \pi}}
^{({\tilde W}_{-}){'}})+ {\tilde j}^o_{(em)}](\vec x,x^o),
\label{hh3}
\end{equation}

\noindent one finds that these four  charges satisfy the algebra

\begin{eqnarray}
&&{\tilde Q}^1_w={1\over {\sqrt{2}}} ({\tilde Q}_{(CC)+}+{\tilde Q}_{(CC)-}),
\nonumber \\
&&{\tilde Q}^2_w=-{i\over {\sqrt{2}}} ({\tilde Q}_{(CC)+}-{\tilde Q}_{(CC)-}),
\nonumber \\
&&{\tilde Q}^3_w=sin\, \theta_w (sin\, \theta_w {\tilde Q}_{(em)}+cos\, \theta_w
{\tilde Q}_{(NC)}),\nonumber \\
&&{\tilde Q}_{Y_w}=cos\, \theta_w (cos\, \theta_w {\tilde Q}_{(em)}-sin\,
\theta_w {\tilde Q}_{(NC)}),\nonumber \\
&&{}\nonumber \\
&&\{ {\tilde Q}^i_w,{\tilde Q}^j_w \} = \epsilon_{ijk} {\tilde Q}^k_w,\quad\quad
\{ {\tilde Q}^i_w,{\tilde Q}_{Y_w} \} =0,\nonumber \\
&&{}\nonumber \\
&&\{ {\tilde Q}_{(em)},{\tilde Q}_{(NC)} \}=0,\quad\quad \{ {\tilde Q}_{(em)},
{\tilde Q}_{(CC)\pm} \} = \mp i {\tilde Q}_{(CC)\pm},\nonumber \\
&&\{ {\tilde Q}_{(NC)},{\tilde Q}_{(CC)\pm} \} = \mp i cot\, \theta_w {\tilde Q}
_{(CC)\pm},\nonumber \\
&& \{ {\tilde Q}_{(CC)+},{\tilde Q}_{(CC)-} \} =-i (sin^2\, \theta_w {\tilde Q}
_{(em)} + sin\, \theta_w cos\, \theta_w {\tilde Q}_{(NC)}.
\label{hh4}
\end{eqnarray}

Therefore, the charges satisfy a global $su(2)\times u(1)$ algebra, even if 
only the electromagnetic charge is conserved (custodial symmetry).

However, if we go to the electromagnetic Coulomb gauge by adding the
gauge-fixing ${\tilde \eta}_{em}(x)\approx 0$ [$\partial^o{\tilde \eta}_{em}(x)
\approx 0$ implies $\lambda_o^{(\tilde A)}(x)=0$ in Eq.(\ref{46})] and we
define the associated Dirac brackets $\{ \alpha (\vec x,x^o),\beta (\vec y,x^o)
 \}^{*} =\{ \alpha (\vec x,x^o),\beta (\vec y,x^o) \} + \int d^3z [\{ \alpha 
(\vec x,x^o),\Gamma^{(\tilde A)}(\vec z,x^o) \} \, 
\{ \eta_{em}(\vec z,x^o),\beta 
(\vec y,x^o) \} - \{ \alpha (\vec x,x^o),\eta_{em}(\vec z,x^o) \} \, \{ \Gamma
^{(\tilde A)}(\vec z,x^o),\beta (\vec y,x^o) \} ]$,
we find that the charges have their algebra  modified.
In particular, since $\{ {\tilde \eta}_{em}(\vec x,x^o),{\tilde Q}_{(CC)\pm} \}=
\mp i \int d^3y {{(\vec \partial \cdot {\vec W}_{\mp})(\vec y,x^o)}\over {4\pi
|\vec x-\vec y|}}$, $\{ \Gamma^{(\tilde A)}(\vec x,x^o),{\tilde Q}_{(CC)\pm} 
\}=-ie {\tilde {\cal Q}}_{(CC)\pm}(\vec x,x^o)$,
we get $\{ {\tilde Q}_{(CC)+},{\tilde Q}_{(CC)-} \}^{*}=
\{ {\tilde Q}_{(CC)+},{\tilde Q}_{(CC)-} \} +e \int d^3xd^3y {1\over {4\pi
|\vec x-\vec y|}} ({\tilde {\cal Q}}_{(CC)+}(\vec x,x^o) \vec \partial \cdot 
{\vec W}_{+}(\vec y,x^o)+{\tilde {\cal Q}}_{(CC)-}(\vec x,x^o) \vec \partial 
\cdot {\vec W}_{-}(\vec y,x^o) )$. To recover the algebra of charges
in the Coulomb gauge, one should redefine the charges.

This means that, due to the Weinberg rotation, ${\tilde \eta}_{em}(x)$ is not
the natural variable conjugate with $\Gamma^{(\tilde A)}(x)$, if we want
to preserve the charge algebra. It turns out that 
if we define the canonical transformation

\begin{eqnarray}
{\tilde \eta}_{em}(x) &&\mapsto {\tilde \eta}^{'}_{em}(x)={\tilde \eta}_{em}(x)
+tg\, \theta_w {1\over {\triangle}} \vec \partial \cdot {\vec {\tilde Z}}^{'} 
(x)=\nonumber \\
&&=-{1\over {\triangle}} \vec \partial \cdot ({\vec {\tilde A}}^{'}(x)-tg\,
\theta_w {\vec {\tilde Z}}^{'}(x))=-{1\over {\triangle}} \vec \partial \cdot
{\vec {\tilde V}}^{'}(x),\nonumber \\
\Gamma^{(\tilde A)}(x) &&\mapsto \Gamma^{(\tilde A)}(x),\nonumber \\
{\vec {\tilde Z}}^{'}(x) &&\mapsto {\vec {\tilde Z}}^{'}(x),\nonumber \\
{\vec \pi}^{(\tilde Z){'}}(x) &&\mapsto {\vec \pi}^{(\tilde Z) {"}}(x)=
{\vec \pi}^{(\tilde Z) {'}}(x)-tg\, \theta_w {{\vec \partial}\over {\triangle}}
\Gamma^{(\tilde A)}(x)\approx {\vec \pi}^{(\tilde Z) {'}}(x),
\label{hh5}
\end{eqnarray}

\noindent and we impose the gauge-fixing ${\tilde \eta}^{'}_{em}(x)\approx 0$
[this means that the electromagnetic Coulomb cloud is replaced by a
hypercharge cloud, associated with an effective hypercharge field ${\vec 
{\tilde V}}^{'}(x)$], 
then, since $\{ {\tilde \eta}^{'}_{em}(\vec x,x^o),{\tilde 
Q}_{(CC)\pm} \} =0$, the $su(2)\times u(1)$ algebra of charges is preserved at 
the level of Dirac brackets.

The new variable ${\tilde \eta}^{'}(x)$ implies that the first of Eqs.(\ref{48})
is modified to

\begin{eqnarray}
{\vec {\tilde A}}^{'}(x)&=&{\check {\vec {\tilde A}}}_{\perp}(x)+\vec \partial
[{\tilde \eta}^{'}_{em}(x)-tg\, \theta_w {1\over {\triangle}}\vec \partial
\cdot {\vec {\tilde Z}}^{'}(x) ]\nonumber \\
&&\rightarrow_{{\tilde \eta}^{'}_{em}(x)\rightarrow 0}\, 
{\check {\vec {\tilde A}}}_{\perp}(x)-tg\, \theta_w {{\vec \partial}\over
{\triangle}} \vec \partial \cdot {\vec {\tilde Z}}^{'}(x),
\label{hh6}
\end{eqnarray}

\noindent and that, to get the generalized Coulomb gauge, we have to add the
gauge-fixings ${\tilde \eta}^{'}_{em}(x)\approx 0$, $\eta^{(G)}_A(x)\approx 0$,
whose time constancy implies $\lambda_{Ao}(x)=\lambda_o^{(\tilde A)}(x)=0$
in Eq.(\ref{46}).

\section
{Dirac's observables for the standard model}

In conclusion a canonical basis of gauge variables for the standard model is

\begin{eqnarray}
&&G_A^o(x),\quad \pi^{(G) o}_A(x)\quad\quad \eta^{(G)}_A(x),\quad 
{\tilde \Gamma}^{(G)}_A(x),\nonumber \\
&&{\tilde A}^{'}_o(x),\quad {\tilde \pi}^{(\tilde A){'} o}(x),\quad\quad 
{\tilde \eta}^{'}_{em}(x),\quad \Gamma^{(\tilde A)}(x), \nonumber \\
&&{\tilde Z}^{{'} o}(x),\quad {\tilde \pi}^{(\tilde Z){'} o}(x),
\nonumber \\
&& {\tilde W}^{{'}o}_{\pm}(x),\quad {\tilde \pi}^{({\tilde W}_{\pm}){'} o}(x).
\label{68}
\end{eqnarray}

The associated canonical basis of Dirac's observables is [on the
SU(2)-singlets one has $Q_{em}=iY_w$]

\begin{eqnarray}
&&{\check {\vec G}}_{\perp A}(x),\quad {\check {\vec \pi}}^{(G)}_{A\perp}(x),
\nonumber \\
&&{\check {\vec {\tilde A}}}_{\perp}(x),\quad {\check {\vec {\tilde \pi}}}
_{\perp}(x),\nonumber \\
&&{\check {\vec {\tilde Z}}}(x)={\vec {\tilde Z}}^{'}(x),\quad {\check {\vec 
{\tilde \pi}}}^{(\tilde Z)}(x)={\vec {\tilde \pi}}^{(\tilde Z){'}}(x),
\nonumber \\
&&{\check {\vec {\tilde W}}}_{\pm}(x)=e^{\pm ie{\tilde \eta}^{'}_{em}(x)}{\vec 
{\tilde W}}^{'}_{\pm}(x),\quad {\check {\vec {\tilde \pi}}}^{({\tilde W}_{\pm})
}(x)=e^{\mp ie{\tilde \eta}^{'}_{em}(x)}{\vec {\tilde \pi}}^{({\tilde W}
_{\pm}){'}}(x),\nonumber \\
&&H(x),\quad \pi_H(x),\nonumber \\
&&{\check \psi}^{(l)}_{Li}(x)=e^{-ie{\tilde \eta}^{'}_{em}(x)Q_{em}} 
e^{-\theta_a(x)T^a_w}\psi^{(l)}_{Li}(x),\nonumber \\
&&{\check \psi}^{(l)}_{Ri}(x)=e^{e{\tilde \eta}^{'}_{em}(x)Y_w}\psi^{(l)}_{Ri}
(x),\nonumber \\
&&{\check \psi}^{(q)}_{Li}(x)=e^{-ie{\tilde \eta}^{'}_{em}(x)Q_{em}} 
e^{-\theta_a(x)T^a_w}[P\, e^{\Omega^{(\hat \gamma )}_s(\eta^{(G)}(x))}] 
\psi^{(q)}_{Li}(x),\nonumber \\
&&{\check \psi}^{(q)}_{Ri}(x)=e^{e{\tilde \eta}^{'}_{em}(x)Y_w} [P\, e^{\Omega
^{(\hat \gamma )}_s(\eta^{(G)}(x))}] \psi^{(q)}_{Ri}(x),\nonumber \\
&&{\check {\tilde \psi}}^{(q)}_{Ri}(x)=e^{e{\tilde \eta}^{'}_{em}(x)Y_w} 
[P\, e^{\Omega^{(\hat \gamma )}_s(\eta^{(G)}(x))}] {\tilde \psi}^{(q)}_{Ri}(x).
\label{69}
\end{eqnarray}

Note that the fermion fields ${\check \psi}^{(l)}_{Li}$, ${\check \psi}
^{(q)}_{Li}$, are dressed with clouds of would-be Goldstone bosons
$\theta_a(x)$, which also contribute to the definition of the fields
${\check {\vec {\tilde A}}}_{\perp}(x)$, ${\check {\vec {\tilde Z}}}(x)$,
${\check {\vec {\tilde W}}}_{\pm}(x)$, through Eqs.(\ref{8}) and the first
of Eqs.(\ref{39}).

The Dirac's observables for the fermionic
charge densities are [one has $iT^3_w=sin\,
\theta_w (sin\, \theta_w Q_{em} + cos\, \theta_w Q_Z)$, $iY_w = cos\, \theta_w
(cos\, \theta_w Q_{em} - sin\, \theta_w Q_Z)$]

\begin{eqnarray}
{\check {\tilde j}}^o_{(em)}(x)&=&{\check \psi}^{(l)\dagger}_{Li}(x)Q_{em}
{\check \psi}^{(l)}_{Li}(x)+{\check \psi}^{(l)\dagger}_{Ri}(x)iY_w{\check \psi}
^{(l)}_{Ri}(x)+\nonumber \\
&+&{\check \psi}^{(q)\dagger}_{Li}(x)Q_{em}{\check \psi}^{(q)}_{Li}(x)+
{\check \psi}^{(q)\dagger}_{Ri}(x)iY_w{\check \psi}^{(q)}_{Ri}(x)+
{\check {\tilde \psi}}^{(q)\dagger}_{Ri}(x)iY_w{\check {\tilde \psi}}^{(q)}
_{Ri}(x),\nonumber \\
&&{}\nonumber \\
{\check {\tilde j}}^o_{(NC)}(x)&=&{\check \psi}^{(l)\dagger}_{Li}(x)Q_Z{\check 
\psi}^{(l)}_{Li}(x)-tg\, \theta_w {\check \psi}^{(l)\dagger}_{Ri}(x)iY_w
{\check \psi}^{(l)}_{Ri}(x)+\nonumber \\
&+&{\check \psi}^{(q)\dagger}_{Li}(x)Q_Z{\check \psi}^{(q)}_{Li}(x)-tg\,
\theta_w {\check \psi}^{(q)\dagger}_{Ri}(x)iY_w{\check \psi}^{(q)}_{Ri}(x)-
tg\, \theta_w {\check {\tilde \psi}}^{(q)\dagger}_{Ri}(x)iY_w{\check {\tilde 
\psi}}^{(q)}_{Ri}(x),\nonumber \\
&&{}\nonumber \\
{\check {\tilde j}}^o_{(CC)\, \pm}(x)&=&{\check \psi}^{(l)\dagger}_{Li}(x)
iT^{\pm}_w{\check \psi}^{(l)}_{Li}(x)+{\check \psi}^{(q)\dagger}_{Li}(x)
iT^{\pm}_w{\check \psi}^{(q)}_{Li}(x),\nonumber \\
&&{}\nonumber \\
{\check {\tilde j}}^o_{(CC)\, 3}(x)&=&{\check \psi}^{(l)\dagger}_{Li}(x)iT^3_w
{\check \psi}^{(l)}_{Li}(x)+{\check \psi}^{(q)\dagger}_{Li}(x)iT^3_w
{\check \psi}^{(q)}_{Li}(x)=\nonumber \\
&=& sin\, \theta_w [sin\, \theta_w {\check {\tilde j}}^o_{(em)}(x)+cos\,
\theta_w {\check {\tilde j}}^o_{(NC)}(x)],\nonumber \\
&&{}\nonumber \\
{\check {\tilde j}}^o_{Y_w}(x)&=&{\check \psi}^{(l)\dagger}_{Li}(x)iY_w{\check 
\psi}^{(l)}_{Li}(x)+{\check \psi}^{(l)\dagger}_{Ri}(x)iY_w{\check \psi}^{(l)}
_{Ri}(x)+\nonumber \\
&+&{\check \psi}^{(q)\dagger}_{Li}(x)iY_w{\check \psi}^{(q)}_{Li}(x)+{\check 
\psi}^{(q)\dagger}_{Ri}(x)iY_w{\check \psi}^{(q)}_{Ri}(x)+{\check {\tilde \psi}}
^{(q)\dagger}_{Ri}(x)iY_w{\check {\tilde \psi}}^{(q)}_{Ri}(x)=\nonumber \\
&=& cos\, \theta_w [cos\, \theta_w {\check {\tilde j}}^o_{(em)}(x) -sin\,
\theta_w {\check {\tilde j}}^o_{(NC)}(x)],\nonumber \\
&&{}\nonumber \\
{\check J}^{(G) o}_{sA}(x)&=&
{\check \psi}^{(q)\dagger}_{Li}(x)iT_s^A{\check \psi}
^{(q)}_{Li}(x)+{\check \psi}^{(q)\dagger}_{Ri}(x)iT_s^A{\check \psi}^{(q)}_{Ri}
(x)+{\check {\tilde \psi}}^{(q)\dagger}_{Ri}(x)iT^A_s{\check {\tilde \psi}}
^{(q)}_{Ri}(x),
\label{70}
\end{eqnarray}

\noindent whose expression in terms of the physical mass eigenstates
${\check \nu}_e, ..., \check e,...$ coincide with Eqs.(\ref{26}).

By collecting all the previous results, by using Section 6 of the second paper
in Ref.\cite{lusa} [especially Eq.(6-27)] and by using the following
notation for the non-Abelian counterpart of $c(\vec x-\vec y)={1\over
{\triangle}} \delta^3(\vec x-\vec y)$ [see Eq.(3-25) of that paper; one has
${\hat {\vec D}}^{(G)}_{AB}(x)\cdot {\hat {\vec D}}^{(G)}_{BC}(x) C^{(G)}
_{\triangle ,CD}(\vec x,\vec y;x^o)=-\delta_{AD}\delta^3(\vec x-\vec y)$ and,
if one puts equal to zero the structure constants $c_{ABC}$, this equation 
becomes $\triangle C^{(G)(o)}_{\triangle ,AB}(\vec x,\vec y;x^o)=\delta_{AB}
\delta^3(\vec x-\vec y)$ so that $C^{(G)(o)}_{\triangle ,AB}(\vec x,\vec y;x^o)
=\delta_{AB} c(\vec x-\vec y)$]

\begin{eqnarray}
c(\vec x-\vec y)&=&{1\over {\triangle}}\delta^3(\vec x-\vec y)=-{1\over 
{4\pi |\vec x-\vec y |}},\nonumber \\
C^{({\check {\vec G}}_{\perp})}_{\triangle ,AB}({\vec y}_1,{\vec y}_2; x^o)&=&
\delta_{AB}\, c({\vec y}_1-{\vec y}_2)-\nonumber \\
&-&2\int d^3z\,  c({\vec y}_1-\vec z)\, [{\vec \partial}_zc(\vec z-{\vec y}_2)]
\, c_{AUV}\, {\check {\vec G}}_{V\perp}(\vec z,x^o)\, \zeta_{UB}^{({\check 
{\vec G}}_{\perp})}(\vec z,{\vec y}_2;x^o)+\nonumber \\
&+&\int d^3z_1d^3z_2\, c({\vec y}_1-{\vec z}_1)\, [\partial^h_{z_1}c({\vec z}_1-
{\vec z}_2)]\, [\partial^k_{z_2}c({\vec z}_2-{\vec y}_2)]\nonumber \\
&&c_{AUR}\, {\check G}^h_{R\perp}({\vec z}_1,x^o)\, \zeta_{UV}^{({\check {\vec 
G}}_{\perp})}({\vec z}_1,{\vec z}_2;x^o)\, c_{VTS}\, {\check G}^k_{S\perp}
({\vec z}_2,x^o)\, \zeta_{TB}^{({\check {\vec G}}_{\perp})}({\vec z}_2,{\vec y}
_2;x^o).
\label{71}
\end{eqnarray}

\noindent we obtain the physical Hamiltonian density in the generalized
Coulomb gauge ${\tilde \eta}^{'}_{em}(x)=\eta^{(G)}_A(x)=0$
[we also rescale the
SU(3) vector potential ${\vec G}_{A\perp}(x) =g_s {\vec {\tilde G}}_{A\perp}
(x)$ so that $g_s$ now appears in the field strength]. It consists of 
four pieces

\begin{equation}
{\cal H}_{phys}(x)={\cal H}_o(x)+{\cal H}_{magn}(x)+{\cal H}_{self}(x)+
\theta {\check Q}_{top}(x)
\label{72}
\end{equation}

\noindent i) The field kinetic terms and the linear couplings of the bosonic
fields to the fermions

\begin{eqnarray}
{\cal H}_o(x)&=&{1\over 2}\sum_A{\check {\vec {\tilde \pi}}}^{({\check {\tilde
G}})2}_{A\perp}(x)+{1\over 2}{\check {\vec {\tilde \pi}}}^{({\check {\tilde 
A}})2}_{\perp}(x)+{1\over 2}{\check {\vec {\tilde \pi}}}^{({\check {\tilde 
Z}})2}(x)+{\check {\vec {\tilde \pi}}}^{({\check {\tilde W}}_{+})}(x)\cdot 
{\check {\vec {\tilde \pi}}}^{({\check {\tilde W}}_{-})}(x)+\nonumber \\
&+&m^2_W(1+{{|e|}\over {sin\, 2\theta_wm_Z}}H(x))^2{\check {\vec {\tilde W}}}
_{+}(x)\cdot {\check {\vec {\tilde W}}}_{-}(x)+\nonumber \\
&+&{1\over 2}m^2_Z(1+{{|e|}\over 
{sin\, 2\theta_wm_Z}}H(x))^2 {\check {\vec {\tilde Z}}}^2(x)+\nonumber \\
&+&{1\over 2}[\pi_H^2(x)+(\vec \partial H(x))^2]+{1\over 2}m^2_H\, H^2(x)
(1+{{|e|}\over {2sin\, 2\theta_wm_Z}}H(x))^2+\nonumber \\
&+&{\check \psi}^{(l)\dagger}_{Li}(x)[i\vec \alpha \cdot (\vec \partial +
{{e}\over {sin\, \theta_w}}({\check {\vec {\tilde W}}}_{+}(x)T^{-}_w+{\check 
{\vec {\tilde W}}}_{-}(x)T^{+}_w)-\nonumber \\
&-&ieQ_{em}({\check {\vec {\tilde A}}}_{\perp}(x)-tg\, \theta_w {{\vec \partial}
\over {\triangle}} \vec \partial \cdot {\check {\vec {\tilde Z}}}(x))
-ieQ_Z{\check {\vec {\tilde 
Z}}}(x))]{\check \psi}^{(l)}_{Li}(x)+\nonumber \\
&+&{\check \psi}^{(l)\dagger}_{Ri}(x)[i\vec \alpha \cdot (\vec \partial +
{e\over {cos\, \theta_w}}({\check {\vec {\tilde A}}}_{\perp}(x)
-tg\, \theta_w {{\vec \partial}
\over {\triangle}} \vec \partial \cdot {\check {\vec {\tilde Z}}}(x)-
\nonumber \\
&-&tg\, \theta_w
{\check {\vec {\tilde Z}}}(x))Y_w)]{\check \psi}^{(l)}_{Ri}(x)-\nonumber \\
&-&(1+{{|e|}\over {sin\, 2\theta_wm_Z}} H(x))\nonumber \\
&&[{\check {\bar \psi}}^{(l)}_{Li}(x)\cdot \left( \begin{array}{l} 0\\ 1
\end{array} \right) M^{(l)}_{ij} {\check \psi}^{(l)}_{Rj}(x)+{\check {\bar
\psi}}^{(l)}_{Ri}(x) M^{(l)\dagger}_{ij} \left( \begin{array}{ll} 0& 1
\end{array} \right) \cdot {\check \psi}^{(l)}_{Lj}(x)]+\nonumber \\
&+&{\check \psi}^{(q)\dagger}_{Li}(x)[i\vec \alpha \cdot (\vec \partial +g_s
{\check {\vec {\tilde G}}}_{A\perp}(x)T_s^A+{{e}\over {sin\, \theta_w}}
({\check {\vec {\tilde W}}}_{+}(x)T^{-}_w+{\check {\vec {\tilde W}}}_{-}(x)
T^{+}_w)-\nonumber \\
&-&ieQ_{em}({\check {\vec {\tilde A}}}_{\perp}(x)
-tg\, \theta_w {{\vec \partial}
\over {\triangle}} \vec \partial \cdot {\check {\vec {\tilde Z}}}(x))
-ieQ_Z{\check {\vec {\tilde 
Z}}}(x))]{\check \psi}^{(q)}_{Li}(x)+\nonumber \\
&+&{\check \psi}^{(q)\dagger}_{Ri}(x)[i\vec \alpha \cdot (\vec \partial +g_s
{\check {\vec {\tilde G}}}_{A\perp}(x)T_s^A+\nonumber \\
&+&{e\over {cos\, \theta_w}}
({\check {\vec {\tilde A}}}_{\perp}(x)
-tg\, \theta_w {{\vec \partial}
\over {\triangle}} \vec \partial \cdot {\check {\vec {\tilde Z}}}(x)
-tg\, \theta_w{\check {\vec {\tilde Z}}}
(x))Y_w)]{\check \psi}^{(q)}_{Ri}(x)+\nonumber \\
&+&{\check {\tilde \psi}}^{(q)\dagger}_{Ri}(x)[i\vec \alpha \cdot (\vec
\partial +g_s{\check {\vec {\tilde G}}}_{A\perp}(x)T_s^A+\nonumber \\
&+&{e\over {cos\,
\theta_w}}({\check {\vec {\tilde A}}}_{\perp}(x)
-tg\, \theta_w {{\vec \partial}
\over {\triangle}} \vec \partial \cdot {\check {\vec {\tilde Z}}}(x)
- tg\, \theta_w {\check {\vec 
{\tilde Z}}}(x))Y_w)]{\check {\tilde \psi}}^{(q)}_{Ri}(x)-\nonumber \\
&-&(1+{{|e|}\over {sin\, 2\theta_wm_Z}}H(x))\cdot \nonumber \\
&&[{\check {\bar \psi}}^{(q)}_{Li}(x)\cdot \left( \begin{array}{l}0\\ 1
\end{array} \right) M^{(q)}_{ij} {\check \psi}^{(q)}_{Rj}(x)+{\check {\bar
\psi}}^{(q)}_{Ri}(x) M^{(q)\dagger}_{ij} \left( \begin{array}{ll} 0& 1
\end{array} \right) \cdot {\check \psi}^{(q)}_{Lj}(x)+\nonumber \\
&+&{\check {\bar \psi}}^{(q)}_{Li}(x)\cdot \left( \begin{array}{l}1\\
0\end{array} \right) {\tilde M}^{(q)}_{ij} {\check {\tilde \psi}}^{(q)}_{Rj}(x)
+{\check {\bar {\tilde \psi}}}^{(q)}_{Ri}(x) {\tilde M}^{(q)\dagger}_{ij}
\left( \begin{array}{ll} 1& 0\end{array} \right) \cdot {\check \psi}^{(q)}_{Lj}
(x)]=\nonumber \\
&&{}\nonumber \\
&=&{1\over 2}\sum_A{\check {\vec {\tilde \pi}}}^{({\check {\tilde
G}})2}_{A\perp}(x)+{1\over 2}{\check {\vec {\tilde \pi}}}^{({\check {\tilde 
A}})2}_{\perp}(x)+{1\over 2}{\check {\vec {\tilde \pi}}}^{({\check {\tilde 
Z}})2}(x)+{\check {\vec {\tilde \pi}}}^{({\check {\tilde W}}_{+})}(x)\cdot 
{\check {\vec {\tilde \pi}}}^{({\check {\tilde W}}_{-})}(x)+\nonumber \\
&+&m^2_W(1+{{|e|}\over {sin\, 2\theta_wm_Z}}H(x))^2{\check {\vec {\tilde W}}}
_{+}(x)\cdot {\check {\vec {\tilde W}}}_{-}(x)+\nonumber \\
&+&{1\over 2}m^2_Z(1+{{|e|}\over 
{sin\, 2\theta_wm_Z}}H(x))^2 {\check {\vec {\tilde Z}}}^2(x)+\nonumber \\
&+&{1\over 2}[\pi_H^2(x)+(\vec \partial H(x))^2]+{1\over 2}m^2_H\, H^2(x)
(1+{{|e|}\over {2sin\, 2\theta_wm_Z}}H(x))^2+\nonumber \\
&+& \left( \begin{array}{ccc} {\check \nu}^{(m)\dagger}_e(x) & {\check 
\nu}^{(m)\dagger}_{\mu}(x)& {\check \nu}^{(m)\dagger}_{\tau}(x) \end{array} 
\right) i\vec \alpha \cdot \vec \partial {1\over 2}(1-\gamma_5)
\left( \begin{array}{c} {\check \nu}^{(m)}_e(x)\\ {\check \nu}^{(m)}_{\mu}(x)\\ 
{\check \nu}^{(m)}_{\tau}(x) \end{array} \right) +\nonumber \\
&+& \left( \begin{array}{ccc} {\check e}^{(m)\dagger}(x) & {\check \mu}
^{(m)\dagger}(x) & {\check \tau}^{(m)\dagger}(x) \end{array} \right) 
\nonumber \\
&&[i\vec \alpha \cdot \vec \partial +\beta (1+ {{|e|}\over
{sin\, 2\theta_w m_Z}} H(x)) \left( \begin{array}{ccc} 
m_e & 0 & 0\\ 0 & m_{\mu}
& 0\\ 0 & 0 & m_{\tau} \end{array} \right) ] \left( \begin{array}{c} {\check e}
^{(m)}(x) \\ {\check \mu}^{(m)}(x) \\ {\check \tau}^{(m)}(x) \end{array} 
\right) +\nonumber \\
&+&\left( \begin{array}{ccc} {\check u}^{(m)\dagger}(x) & {\check c}
^{(m)\dagger}(x) & {\check t}^{(m)\dagger}(x) \end{array} \right) \nonumber \\
&&[i\vec \alpha \cdot \vec \partial +\beta (1+ {{|e|}\over
{sin\, 2\theta_w m_Z}} H(x)) \left( \begin{array}{ccc} m_u & 0 & 0\\ 0 & m_c
& 0\\ 0 & 0 & m_t \end{array} \right) ] \left( \begin{array}{c} {\check u}^{(m)}
(x) \\ {\check c}^{(m)}(x) \\ {\check t}^{(m)}(x) \end{array} \right) +
\nonumber \\
&+&\left( \begin{array}{ccc} {\check d}^{(m)\dagger}(x) & {\check s}
^{(m)\dagger}(x) & {\check b}^{(m)\dagger}(x) \end{array} \right) \nonumber \\
&&[i\vec \alpha \cdot \vec \partial +\beta (1+ {{|e|}\over
{sin\, 2\theta_w m_Z}} H(x)) \left( \begin{array}{ccc} m_d & 0 & 0\\ 0 & m_s
& 0\\ 0 & 0 & m_b \end{array} \right) ] \left( \begin{array}{c} {\check d}^{(m)}
(x) \\ {\check s}^{(m)}(x) \\ {\check b}^{(m)}(x) \end{array} \right) +
\nonumber \\
&+&g_s{\check {\vec {\tilde G}}}_{A\perp}(x)\cdot {\check {\vec {\tilde J}}}
_{sA}(x)+e({\check {\vec {\tilde A}}}_{\perp}(x)
-tg\, \theta_w {{\vec \partial}
\over {\triangle}} \vec \partial \cdot {\check {\vec {\tilde Z}}}(x))
\cdot {\check {\vec {\tilde j}}}
_{(em)}(x)+\nonumber \\
&+&e{\check {\vec {\tilde Z}}}(x) {\check {\vec {\tilde j}}}_{(NC)}(x)+
{e\over {sin\, \theta_w}} ({\check {\vec {\tilde W}}}
_{+}(x)\cdot {\check {\vec{\tilde j}}}_{(CC)\, -}(x)+ {\check {\vec {\tilde 
W}}}_{-}(x) \cdot {\check {\vec {\tilde j}}}_{(CC)\, +}(x).
\label{73}
\end{eqnarray}

Let us remark that the right fermions couple to ${\check {\vec {\tilde Z}}}
_{\perp}(x)={\check {\vec {\tilde Z}}}(x)+{{\vec \partial}\over {\triangle}}
\vec \partial \cdot {\check {\vec {\tilde Z}}}(x)$ in this generalized
[${\tilde \eta}^{'}_{em}(x)=0$] gauge.

\noindent ii) The magnetic bilinear, trilinear and quadrilinear terms 
[${\check B}^{({\check {\tilde A}}) i} =-\epsilon^{ijk}\partial^j{\check 
{\tilde A}}^k_{\perp}$, ${\check B}^{({\check {\tilde Z}}) i} =-\epsilon^{ijk} 
\partial^j{\check {\tilde Z}}^k$, ${\check B}^{({\check {\tilde W}}_{\pm}) i} 
=-\epsilon^{ijk} \partial^j{\check {\tilde W}}^k_{\pm}=-{1\over 2}\epsilon
^{ijk}{\cal W}_{\pm}^{jk}$ are the ``Abelian" 
magnetic fields, while for the color non-Abelian magnetic field we have
${1\over 2}\sum_A {\check {\vec B}}_A^{({\check {\tilde G}}_{\perp})\, 2}(x)=
\partial^i{\check {\tilde G}}^j_{A\perp}(x)\partial^i
{\check {\tilde G}}^j_{A\perp}(x)+2g_sc_{ABC}\partial^i{\check {\tilde G}}^j_
{A\perp}(x){\check {\tilde G}}^i_{B\perp}(x){\check {\tilde G}}^j_{C\perp}(x)+
{1\over 2}g^2_sc_{ABC}c_{AUV}{\check {\tilde G}}^i_{B\perp}(x){\check 
{\tilde G}}^j_{C\perp}(x){\check {\tilde G}}^i_{U\perp}(x){\check {\tilde G}}
^j_{V\perp}(x)$]

\begin{eqnarray}
{\cal H}_{magn}(x)&=&{1\over 2}\sum_A {\check {\vec B}}_A^{({\check {\tilde 
G}}_{\perp})\, 2}(x)+\nonumber \\
&+& {1\over 2}{\check {\vec B}}^{({\check {\tilde A}}) 2}(x) + {1\over 2} 
{\check {\vec B}}^{({\check {\tilde Z}}) 2}(x) + {\check {\vec B}}^{({\check 
{\tilde W}}_{+})}(x)\cdot {\check {\vec B}}^{({\check {\tilde W}}_{-})}(x)+
\nonumber \\
&+&ie [{\check {\vec {\tilde W}}}_{+}(x) \times {\check {\vec {\tilde W}}}_{-}
(x) \cdot ({\check {\vec B}}^{({\check {\tilde A}})}(x)+ cot\, \theta_w {\check 
{\vec B}}^{({\check {\tilde Z}})}(x))]+\nonumber \\
&+& ie({\check {\vec {\tilde A}}}_{\perp}(x)
-tg\, \theta_w {{\vec \partial}
\over {\triangle}} \vec \partial \cdot {\check {\vec {\tilde Z}}}(x)
+cot\, \theta_w {\check {\vec {\tilde Z}}}(x)
) \cdot \nonumber \\
&&[\times {\check {\vec {\tilde W}}}_{-}(x) \cdot 
{\check {\vec B}}^{({\check {\tilde 
W}}_{+})}(x)-\times {\check {\vec {\tilde W}}}_{+}(x) \cdot 
{\check {\vec B}}^{({\check
{\tilde W}}_{-})}(x)) ]-\nonumber \\
&-&{{e^2}\over {2sin^2\, \theta_w}} [ {\check {\vec {\tilde W}}}^{2}_{+}(x) 
{\check {\vec {\tilde W}}}^{2}_{-}(x) -( {\check {\vec {\tilde W}}}_{+}(x) 
\cdot {\check {\vec {\tilde W}}}_{-}(x) )^2]+\nonumber \\
&+& e^2 [{\check {\vec {\tilde W}}}_{+}(x) \cdot {\check {\vec {\tilde W}}}_{-}
(x)({\check {\vec {\tilde A}}}_{\perp}(x)
-tg\, \theta_w {{\vec \partial}
\over {\triangle}} \vec \partial \cdot {\check {\vec {\tilde Z}}}(x)
+cot\, \theta_w {\check {\vec {\tilde 
Z}}}(x))^2 -\nonumber \\
&-&{\check {\vec {\tilde W}}}_{+}(x) \cdot ({\check {\vec {\tilde A}}}_{\perp}
(x)
-tg\, \theta_w {{\vec \partial}
\over {\triangle}} \vec \partial \cdot {\check {\vec {\tilde Z}}}(x)
+cot\, \theta_w{\check {\vec {\tilde Z}}}(x)) \cdot \nonumber \\
&&{\check {\vec {\tilde W}}}_{-}
(x) \cdot ({\check {\vec {\tilde A}}}_{\perp}(x)
-tg\, \theta_w {{\vec \partial}
\over {\triangle}} \vec \partial \cdot {\check {\vec {\tilde Z}}}(x)
+cot\, \theta_w {\check {\vec 
{\tilde Z}}}(x))].
\label{74}
\end{eqnarray}

\noindent iii) The nonperturbative terms with the self-interactions ($y^o=x^o$)

\begin{eqnarray}
\int d^3x\, {\cal H}_{self}(x)&=&-{1\over 2}g^2_s\int d^3xd^3y\nonumber \\ 
&&[c_{ARS}{\check {\vec {\tilde \pi}}}^{({\check {\tilde G}})}_{R\perp}(x)\cdot 
{\check {\vec {\tilde G}}}_{S\perp}(x)+{\check J}^o_{sA}(x)]\nonumber \\
&&C^{({\check {\tilde G}}_{\perp})}_{\triangle ,AB}(\vec x,\vec y;x^o)\cdot
\nonumber \\
&&[c_{BUV}{\check {\vec {\tilde \pi}}}^{({\check {\tilde G}})}_{U\perp}(y)
\cdot {\check {\vec {\tilde G}}}_{V\perp}(y)+{\check J}^o_{sB}(y)]+
\nonumber \\
&&{}\nonumber \\
&+&{{e^2}\over 2}\int d^3xd^3y\nonumber \\ 
&&[i({\check {\vec {\tilde W}}}_{+}(x)\cdot 
{\check {\vec {\tilde \pi}}}^{({\check {\tilde W}}_{-})}(x)-{\check {\vec 
{\tilde W}}}_{-}(x)\cdot {\check {\vec {\tilde \pi}}}^{({\check {\tilde W}}
_{+})}(x))+{\check {\tilde j}}^o_{(em)}(x)]\nonumber \\
&&{1\over {4\pi |\vec x-\vec y|}}\nonumber \\
&&[i({\check {\vec {\tilde W}}}_{+}(y)\cdot {\check {\vec {\tilde \pi}}}
^{({\check {\tilde W}}_{-})}(y)-{\check {\vec {\tilde W}}}_{-}(y)\cdot 
{\check {\vec {\tilde \pi}}}^{({\check {\tilde W}}_{+})}(x))+{\check {\tilde
j}}^o_{(em)}(y)]+\nonumber \\
&&{}\nonumber \\
&+&{1\over 2} \int d^3xd^3y\nonumber \\
&&[\vec \partial \cdot {\check {\vec {\tilde \pi}}}^{({\check {\tilde Z}})}(x)
-iecotg\, \theta_w({\check {\vec {\tilde W}}}_{+}(x)\cdot {\check {\vec {\tilde 
\pi}}}^{({\check {\tilde W}}_{-})}(x)-{\check {\vec {\tilde W}}}_{-}(x)
{\check {\vec {\tilde \pi}}}^{({\check {\tilde W}}_{+})}(x))-\nonumber \\
&-&e{\check {\tilde j}}^o_{(NC)}(x)]\nonumber \\
&&{1\over {m_Z^2(1+{{|e|}\over {sin\, 2\theta_wm_Z}}H(x))^2 }}
\delta^3(\vec x-\vec y)\nonumber \\
&&[\vec \partial \cdot {\check {\vec {\tilde \pi}}}^{({\check {\tilde Z}})}(y)
-iecotg\, \theta_w({\check {\vec {\tilde W}}}_{+}(y)\cdot {\check {\vec {\tilde 
\pi}}}^{({\check {\tilde W}}_{-})}(y)-{\check {\vec {\tilde W}}}_{-}(y)
{\check {\vec {\tilde \pi}}}^{({\check {\tilde W}}_{+})}(y))-\nonumber \\
&-&e{\check {\tilde j}}^o_{(NC)}(y)]+\nonumber \\
&&{}\nonumber \\
&+&\int d^3xd^3y\nonumber \\
&&\lbrace \vec \partial \cdot {\check {\vec {\tilde \pi}}}^{({\check {\tilde W}}
_{+})}(x) +ie [{\check {\vec {\tilde W}}}_{+}(x)\cdot ({\check {\vec {\tilde
\pi}}}_{\perp}^{({\check{\tilde A}})}(x)-\nonumber \\
&-& e{{\vec \partial}\over {\triangle}}
[i({\check {\vec {\tilde W}}}_{+}(x)\cdot {\check {\vec {\tilde \pi}}}
^{({\check {\tilde W}}_{-})}(x)-{\check {\vec {\tilde W}}}_{-}(x)\cdot
{\check {\vec {\tilde \pi}}}^{({\check {\tilde W}}_{+})}(x))+{\check {\tilde
j}}^o_{(em)}(x)]+\nonumber \\
&+&cot\, \theta_w {\check {\vec {\tilde \pi}}}^{({\check {\tilde Z}})}(x))-
({\check {\vec {\tilde A}}}_{\perp}(x)
-tg\, \theta_w {{\vec \partial}
\over {\triangle}} \vec \partial \cdot {\check {\vec {\tilde Z}}}(x)+
\nonumber \\
&+&cot\, \theta_w {\check {\vec {\tilde
Z}}}(x)) \cdot {\check {\vec {\tilde \pi}}}^{({\check {\tilde W}}_{-})}(x)]
-\nonumber \\
&-&{e\over {sin\, \theta_w}} {\check {\tilde j}}^o_{(CC) -}(x)
\rbrace \nonumber \\
&&{1\over {m_W^2(1+{{|e|}\over {sin\, 2\theta_wm_Z}}H(x))^2}} 
\delta^3(\vec x-\vec y)\nonumber \\
&&\lbrace \vec \partial \cdot {\check {\vec {\tilde \pi}}}^{({\check {\tilde W}}
_{-})}(y) -ie [{\check {\vec {\tilde W}}}_{-}(y)\cdot ({\check {\vec {\tilde
\pi}}}_{\perp}^{({\check{\tilde A}})}(y)-\nonumber \\
&-&e{{{\vec \partial}_y}\over {\triangle_y}}
[i({\check {\vec {\tilde W}}}_{+}(y)\cdot {\check {\vec {\tilde \pi}}}
^{({\check {\tilde W}}_{-})}(y)-{\check {\vec {\tilde W}}}_{-}(y)\cdot
{\check {\vec {\tilde \pi}}}^{({\check {\tilde W}}_{=})}(y))+{\check {\tilde
j}}^o_{(em)}(y)]+\nonumber \\
&+&cot\, \theta_w {\check {\vec {\tilde \pi}}}^{({\check {\tilde Z}})}(y))-
({\check {\vec {\tilde A}}}_{\perp}(y)
-tg\, \theta_w {{\vec \partial}
\over {\triangle}} \vec \partial \cdot {\check {\vec {\tilde Z}}}(x)+
\nonumber \\
&+&cot\, \theta_w {\check {\vec {\tilde
Z}}}(y)) \cdot {\check {\vec {\tilde \pi}}}^{({\check {\tilde W}}_{+})}(y)]
-\nonumber \\
&-&{e\over {sin\, \theta_w}} {\check {\tilde j}}^o_{(CC) +}(y)
\rbrace .
\label{75}
\end{eqnarray}

\noindent iv) the topological term 

\begin{eqnarray}
{\check Q}_{top}&=& -g^2_s \int d^3x {\vec B}^{({\check {\tilde G}})}_A(\vec x,
x^o)\cdot [{\check {\vec {\tilde \pi}}}^{({\check {\tilde G}})}_{A\perp}
(\vec x,x^o)+\nonumber \\
&+&c_{ABC}{\check {\vec {\tilde G}}}_{C\perp}(\vec x,x^o) {1\over {\triangle}}
\int d^3y\, {\vec \partial}_x\cdot {\vec \zeta}^{({\check {\tilde G}})}_{BD}
(\vec x,\vec y;x^o){\check J}_{sD}^o(\vec y,x^o)].
\label{76}
\end{eqnarray}

There is a relevant asymmetry between the electromagnetic and color (massless
fields) nonperturbative self-energies, which are ``nonlocal" (more exactly
bilocal), and the weak (massive fields) ones, which are ``local": at the
classical nonperturbative level, the Fermi 4-fermion interactions ${\check 
{\tilde j}}^o_{(NC)}(x) {\check {\tilde j}}^o_{(NC)}(x)$ and ${\check {\tilde
j}}^o_{(CC)+}(x) {\check {\tilde j}}^o_{(CC)-}(x)$ reappear, notwithstanding
that they had been eliminated from the ``tree level"

If one would add the not-Lorentz-invariant terms
$-{1\over 2}\vec \partial {\tilde Z}^{{'} o}(x) \cdot \vec \partial {\tilde Z}
^{{'} o}(x)-\vec \partial {\tilde W}_{+}^{{'} o}(x) \cdot \vec \partial 
{\tilde W}_{-}^{{'} o}(x)$ to the Hamiltonian density (\ref{46}) [the same terms
with the opposite sign to the unitary gauge Lagrangian density (\ref{21})], 
then in Eqs.(\ref{47}) the coefficients of ${\tilde Z}^{{'}o}(x)$ and of 
${\tilde W}^{{'}o}_{\pm}(x)$ would become $\triangle +m^2_Z (1+{e\over {sin\, 
2\theta_w m_Z}} H(x))^2$ and $\triangle +m_W^2 (1+{e\over {sin\, 2\theta_w
m_Z}} H(x))^2$ respectively, and the last two terms in Eq.(\ref{75}) would 
become bilocal (like the non-weak ones) with the expected massive Yukawa
Green functions for the weak self-energies in analogy with
 the electromagnetic and color massless ones. Once
the model will be reformulated in a covariant way on spacelike hypersurfaces,
this modification can be done in a covariant way.

Let us remark that if, in analogy to I and II, we add the holonomic constraint 
$H(x)\approx 0$ to the physical Lagrangian with a multiplier $\lambda (x)$,
then its time constancy would imply the constraint $\pi_H(x)\approx 0$ and
$\partial^o \pi_H(x)\approx 0$ would determine $\lambda (x)$. Therefore, it is
consistent to put $H(x) = \pi_H(x) =0$ in the physical Lagrangian: we would
get the Hamiltonian corresponding to treat the fields $Z^{\mu}(x)$ and 
$W^{\mu}_{\pm}(x)$ as massive vector fields (Proca field theory) with masses
$m_Z$ and $m_W=cos\, \theta_w m_Z$ respectively [see the discussion in I and
II].

The elimination of $H(x)$ can also be thought
as a limiting classical result of the so-called ``triviality problem"
[triviality of the $\lambda \phi^4$ theory \cite{tri}], which however would 
imply a quantization (but how?) of the
Higgs phase alone without the residual Higgs field, so that also its
quantum fluctuations would be absent. Instead these fluctuations are the main 
left quantum effect in the limit $m_H\rightarrow \infty$, which is known to
produce\cite{ab}, in the non-Abelian case, a gauge theory coupled to a
nonlinear $SU(2)_L\times SU(2)_R$ 
$\sigma$-model, equivalent\cite{bs} to a massive Yang-Mills theory.

Finally, let us note that the coupling to the Higgs field $H(x)$ is always
proportional to the charge-mass ratio $|e|/ sin\, 2\theta_w m_Z$.

\section
{Physical Hamilton equations.}

Instead of evaluating the physical Lagrangian [it can be made with the
inverse Legendre transformation like in Ref.\cite{lusa}], 
which is not particularly illuminating
being even more nonlocal of the physical Hamiltonian, we will present the
physical Hamilton equations

However, before doing that, let us introduce the following notations: i) the
charge densities appearing in the self-energies (\ref{75}) are those of Eqs.
(\ref{hh1}), (\ref{hh3}), specialized to the generalized Coulomb gauge,
and satisfy the $su(3)\times su(2)\times u(1)$ algebra [see Eqs.(\ref{67}) and
(\ref{hh4})]. They will be denoted as
[$(\vec \partial /\triangle )\cdot \vec f(\vec x)=\int d^3y \vec c(\vec x-
\vec y)\cdot \vec f(\vec y)$, $\vec c(\vec x)=\vec x / 4\pi |\vec x|^3$]

\begin{eqnarray}
{\check {\cal Q}}_{sA}(x)&=&{\check J}^o_{sA}(x)+
c_{ARS}{\check {\vec {\tilde \pi}}}^{({\check {\tilde G}})}_{R\perp}(x)\cdot 
{\check {\vec {\tilde G}}}_{S\perp}(x),\nonumber \\
&&{}\nonumber \\
{\check {\cal Q}}_{(em)}(x)&=&{\check {\tilde j}}^o_{(em)}(x)+
i[{\check {\vec {\tilde 
W}}}_{+}(x)\cdot {\check {\vec {\tilde \pi}}}^{({\check {\tilde W}}_{+})}(x)-
{\check {\vec {\tilde W}}}_{-}(x)\cdot {\check {\vec {\tilde \pi}}}
^{({\check {\tilde W}}_{-})}(x)],\nonumber \\
&&{}\nonumber \\
{\check {\cal Q}}_{(NC)}(x)&=&{\check {\tilde j}}^o_{(NC)}(x)-
e^{-1}\vec \partial \cdot
{\check {\vec {\tilde \pi}}}^{({\check {\tilde Z}})}(x)+\nonumber \\
&+&i cot\, \theta_w [{\check {\vec {\tilde 
W}}}_{+}(x)\cdot {\check {\vec {\tilde \pi}}}^{({\check {\tilde W}}_{+})}(x)-
{\check {\vec {\tilde W}}}_{-}(x)\cdot {\check {\vec {\tilde \pi}}}
^{({\check {\tilde W}}_{-})}(x)],\nonumber \\
&&{}\nonumber \\
{\check {\cal Q}}_{(CC)\pm}(\vec x,x^o)&=&
{\check {\tilde j}}^o_{(CC)\mp}(\vec x,x^o)-e^{-1}sin\, 
\theta_w\vec \partial \cdot {\check {\vec {\tilde \pi}}}^{({\check {\tilde W}}
_{\pm})}(\vec x,x^o) \pm \nonumber \\
&\pm& isin\, \theta_w 
[{\check {\vec {\tilde W}}}_{\mp}\cdot ({\check {\vec
{\tilde \pi}}}_{\perp}^{({\check {\tilde A}})}+cot\, \theta_w\,
{\check {\vec {\tilde \pi}}}^{({\check {\tilde Z}})})-\nonumber \\
&-&({\check {\vec {\tilde A}}}_{\perp}
-tg\, \theta_w {{\vec \partial}\over {\triangle}}\vec \partial \cdot {\check
{\vec {\tilde Z}}}
+cot\, \theta_w {\check {\vec {\tilde
Z}}})\cdot {\check {\vec {\tilde \pi}}}^{({\check {\tilde W}}_{\pm})}](\vec x,
x^o)\mp \nonumber \\
&\mp& ie\, sin\, \theta_w {\check {\vec {\tilde W}}}_{\mp}(\vec x,x^o)
\cdot \int d^3z {{\vec x-
\vec z}\over {4\pi {|\vec x-\vec z|}^3}}
{\check {\cal Q}}_{(em)}(\vec z,x^o),\nonumber \\
&&{}\nonumber \\
&&{\check Q}^{(G)}_A=\int d^3x {\check {\cal Q}}_{sA}(\vec x,x^o),\nonumber \\
&&{\check Q}_{(em)}=\int d^3x{\check {\cal Q}}_{(em)}(\vec x,x^o),\nonumber \\
&&{\check Q}_{(NC)} = \int d^3x {\check {\cal Q}}_{(NC)}(\vec x,x^o),
\nonumber \\
&&{\check Q}_{(CC)\pm} = \int d^3x {\check {\cal Q}}_{(CC)\pm}(\vec x,x^o),
\nonumber \\
&&{}\nonumber \\
&&{\check Q}^1_w={1\over {\sqrt{2}}} ({\check Q}_{(CC)+}+{\check Q}_{(CC)-}),
\nonumber \\
&&{\check Q}^2_w=-{i\over {\sqrt{2}}} ({\check Q}_{(CC)+}-{\check Q}_{(CC)-}),
\nonumber \\
&&{\check Q}^3_w=sin\, \theta_w (sin\, \theta_w {\check Q}_{(em)}+cos\, \theta_w
{\check Q}_{(NC)}),\nonumber \\
&&{\check Q}_{Y_w}=cos\, \theta_w (cos\, \theta_w {\check Q}_{(em)}-sin\,
\theta_w {\check Q}_{(NC)}),
\label{80}
\end{eqnarray}

\noindent so that Eq.(\ref{75}) may be rewritten in the form

\begin{eqnarray}
H_{self}&=& \int d^3x {\cal H}_{self}(\vec x,x^o) = \nonumber \\ 
&=&-{1\over 2}g^2_s\int d^3xd^3y
{\check {\cal Q}}_{sA}(\vec x,x^o)\, C^{({\check {\tilde G}}_{\perp})}
_{\triangle ,AB}(\vec x,\vec y;x^o)\, {\check {\cal Q}}_{sB}(\vec y,x^o)+
\nonumber \\
&+&{1\over 2}e^2 \int d^3xd^3y {\check {\cal Q}}_{(em)}(\vec x,x^o)\, {1\over 
{4\pi |\vec x-\vec y|}}\, {\check {\cal Q}}_{(em)}(\vec y,x^o)+\nonumber \\
&+&{1\over 2}e^2 \int d^3xd^3y {\check {\cal Q}}_{(NC)}(\vec x,x^o)\, 
{{\delta^3(\vec x-\vec y)}\over {m^2_Z\, (1+{{|e|}\over {sin\, 2\theta_w m_Z}}
H(\vec x,x^o))^2}}\, {\check {\cal Q}}_{(NC)}(\vec y,x^o)+\nonumber \\
&+&{{e^2}\over {sin^2\, \theta_w}}
\int d^3xd^3y {\check {\cal Q}}_{(CC)+}(\vec x,x^o)\, {{\delta^3(\vec x-
\vec y)}\over {m^2_W\, (1+{{|e|}\over {sin\, 2\theta_w m_Z}}H(\vec x,x^o))^2}}
\, {\check {\cal Q}}_{(CC)-}(\vec y,x^o);
\label{81}
\end{eqnarray}

\noindent Therefore, the weak self-energies contain the densities of the
neutral and charged charges ${\check Q}_{(NC)}$, ${\check Q}_{(CC)\pm}$: even 
if these charges are not constants of motion, the associated charges ${\check 
Q}^a_w$, ${\check Q}_{Y_w}$, still satisfy a su(2)xu(1) algebra in this
generalized [${\tilde \eta}^{'}_{em}(x)=0$] gauge [this is the reason why we
choose this gauge rather then the standard Coulomb one ${\tilde \eta}_{em}(x)=0$
].

The derivatives of $H_{self}$ with respect to the charge densities
, which will be left implicit in what follows, are

\begin{eqnarray}
{{\delta H_{self}}\over {\delta {\check {\cal Q}}_{sA}(x)}}&=&-g^2_s \int d^3y
C^{({\check {\tilde G}}_{\perp})}_{\triangle ,AB}(\vec x,\vec y;x^o)\,
{\check {\cal Q}}_{sB}(\vec y,x^o),\nonumber \\
{{\delta H_{self}}\over {\delta {\check {\cal Q}}_{(em)}(x)}}&=&e^2 \int d^3y 
{{ {\check {\cal Q}}_{(em)}(\vec y,x^o)}\over {4\pi |\vec x-\vec y|}}+
\nonumber \\
&+&i{{e^3}\over {sin\, \theta_w}} \int d^3y {{\vec x-\vec y}\over 
{4\pi {|\vec x-\vec y|}^3}} \cdot \nonumber \\
&\cdot& {{ {\check {\vec {\tilde W}}}_{-}(\vec y,x^o){\check {\cal Q}}_{(CC)-}
(\vec y,x^o)-{\check {\vec {\tilde W}}}_{+}(\vec y,x^o){\check {\cal Q}}_{(CC)+}
(\vec y,x^o)}\over
{m^2_Z\, (1+{{|e|}\over {sin\, 2\theta_w m_Z}}H(\vec y,x^o))^2}},\nonumber \\
{{\delta H_{self}}\over {\delta {\check {\cal Q}}_{(NC)}(x)}}&=&
{{ e^2 {\check {\cal Q}}_{(NC)}(x)}
\over {m^2_Z\, (1+{{|e|}\over {sin\, 2\theta_w m_Z}}H(x))^2}},\nonumber \\
{{\delta H_{self}}\over {\delta {\check {\cal Q}}_{(CC)\pm}(x)}}&=&
{{e^2}\over {sin^2\, \theta_w}} {{ {\check {\cal Q}}_{(CC)\mp}
(x)}\over {m^2_W\, (1+{{|e|}\over {sin\, 2\theta_w m_Z}}H(x))^2}}.
\label{82}
\end{eqnarray}

The Hamilton equations are [$H_{phys}=\int d^3x {\cal H}_{phys}(\vec x,x^o)=
H_o+H_{magn}+H_{self}+\theta H_{top}$; let us remark that $H_{top}=\int d^3x
{\check Q}_{top}(\vec x,x^o)$ contributes to the reduced equations of motion
for ${\check {\vec {\tilde G}}}_A(x)$, ${\check {\vec {\tilde \pi}}}_A^{(
{\check {\tilde G}})}(x)$, since the reduced topological charge is no more a
surface term: the phenomenological result $\theta \approx 0$ may be rephrased
as the absence of these terms in the equations of motion]

\begin{eqnarray}
\partial^o&& {\check {\tilde G}}^i_{A\perp}(\vec x,x^o)\, {\buildrel \circ 
\over =}\, \lbrace {\check {\tilde G}}^i_{A\perp}(\vec x,x^o),H_{phys}\rbrace =
-{{\delta H_{phys}}\over {{\check {\tilde \pi}}^{({\check {\tilde G}})\, i}
_{A\perp}(\vec x,x^o)}}=\nonumber \\
&=&-{\check {\tilde \pi}}_{A\perp}^{({\check {\tilde G}})\, i}(\vec x,x^o)-
\nonumber \\
&-&g^2_s c_{ABC} P^{ij}_{\perp}(\vec x) [{\check {\tilde \pi}}^{({\check {\tilde
G}})\, j}_{B\perp}(\vec x,x^o) \int d^3y C^{({\check {\tilde G}}_{\perp})}
_{\triangle ,CD}(\vec x,\vec y;x^o){\check {\cal Q}}_{sD}(\vec y,x^o)]-
\nonumber \\
&-&\theta g^2_s P^{ij}_{\perp}(\vec x) B^{({\check {\tilde G}})\, j}_A
(\vec x,x^o),\nonumber \\
\partial^o&& {\check {\tilde \pi}}^{({\check {\tilde G}})\, i}_{A\perp}(\vec x,
x^o)\, {\buildrel \circ \over =}\, \lbrace {\check {\tilde \pi}}^{({\check 
{\tilde G}})\, i}_{A\perp}(\vec x,x^o),H_{phys}\rbrace = {{\delta H_{phys}}
\over {\delta {\check {\tilde G}}^i_{A\perp}(\vec x,x^o)}}=\nonumber \\
&=&P^{ij}_{\perp}(\vec x) (\, {\hat {\vec D}}^{({\check {\tilde G}}_{\perp})}
_{AB}(\vec x,x^o) \times [({\hat {\vec D}}_{BC}^{({\check {\tilde G}}_{\perp})}
(\vec x,x^o)-{1\over 2}c_{BDC}{\check {\vec {\tilde G}}}_{D\perp}(\vec x,x^o))
\times {\check {\vec {\tilde G}}}_{C\perp}(\vec x,x^o)]\, )^j+\nonumber \\
&+&g_s P^{ij}_{\perp}(\vec x) {\check J}^j_{sA}(\vec x,x^o)-\nonumber \\
&-&g^2_s P^{ij}_{\perp}(\vec x) c_{ABC}{\check {\tilde \pi}}_{B\perp}^{({\check
{\tilde G}})\, j}(\vec x,x^o) \int d^3y C^{({\check {\tilde G}}_{\perp})}
_{\triangle ,CD}(\vec x,\vec y;x^o) {\check {\cal Q}}_{sD}(\vec y,x^o)-
\nonumber \\
&-&{1\over 2}g^2_s \int d^3zd^3y {\check {\cal Q}}_{sB}(\vec z,x^o) 
{{\delta C_{\triangle
,BC}^{({\check {\tilde G}}_{\perp})}(\vec z,\vec y;x^o)}\over {\delta {\check 
{\tilde G}}^i_{A\perp}(\vec x,x^o)}} {\check {\cal Q}}_{sC}(\vec y,x^o)+
\nonumber \\
&+& \theta g^2_s [\, (\vec \partial \times {\check {\vec {\tilde \pi}}}_{A\perp}
^{({\check {\tilde G}})}(\vec x,x^o)\, )^i-\nonumber \\
&-&c_{ACB} P^{ij}_{\perp}(\vec x){\check B}_C^{({\check {\tilde G}})\, j}(\vec 
x,x^o) {1\over {\triangle_x}} \int d^3y {\vec \partial}_x\cdot {\vec \zeta}
_{BD}^{({\check {\tilde G}}_{\perp})}(\vec x,\vec y;x^o){\check J}^o_{sD}(\vec 
y,x^o)-\nonumber \\
&-&c_{CBE}\int d^3z {\check {\vec B}}_C^{({\check {\tilde G}})}(\vec z,x^o)\cdot
{\check {\vec {\tilde G}}}_{E\perp}(\vec z,x^o){1\over {\triangle_z}} \int d^3y
{\vec \partial}_z\cdot {{\delta {\vec \zeta}_{BD}^{({\check {\tilde G}}
_{\perp})}(\vec z,\vec y;x^o)}\over {\delta {\check {\tilde G}}^i_{A\perp}(\vec 
x,x^o)}}{\check J}^o_{sD}(\vec y,x^o)\, ],\nonumber \\
&&{}\nonumber \\
\partial^o&& {\check {\tilde A}}^i_{\perp}(\vec x,x^o)\,{\buildrel \circ 
\over =}\, \lbrace {\check {\tilde A}}^i_{\perp}(\vec x,x^o),H_{phys}\rbrace =
-{{\delta H_{phys}}\over {{\check {\tilde \pi}}^{({\check {\tilde A}})\,i}
_{A\perp}(\vec x,x^o)}}=\nonumber \\
&=&-{\check {\tilde \pi}}_{\perp}^{({\check {\tilde A}})\, i}(\vec x,x^o)-
\nonumber \\
&-&i{{e^2}\over {sin\, \theta_w}} 
P_{\perp}^{ij}(\vec x) {{ {\check {\tilde W}}^j_{-}(\vec x,x^o){\check {\cal Q}}
_{(CC)-}(\vec x,x^o)-{\check {\tilde W}}^j_{+}(\vec x,x^o){\check {\cal Q}}
_{(CC)+}(\vec x,x^o)}\over {m^2_W\, (1+{{|e|}\over {sin\, 2\theta_w m_Z}}
H(\vec x,x^o))^2}},\nonumber \\
\partial^o&& {\check {\tilde \pi}}_{\perp}^{({\check {\tilde A}})\, i}(\vec x,
x^o)\, {\buildrel \circ \over =}\, \lbrace {\check {\tilde \pi}}^{({\check 
{\tilde A}})\, i}_{\perp}(\vec x,x^o),H_{phys}\rbrace = {{\delta H_{phys}}
\over {\delta {\check {\tilde A}}^i_{\perp}(\vec x,x^o)}}=\nonumber \\
&=&\triangle {\check {\tilde A}}_{\perp}^i(\vec x,x^o)+ e P^{ij}_{\perp}(\vec x)
{\check {\tilde j}}^j_{(em)}(\vec x,x^o)-\nonumber \\
&-&P^{ij}_{\perp}(\vec x) [\, ie ({\check {\vec {\tilde W}}}_{-}\times {\check 
{\vec B}}^{({\check {\tilde W}}_{+})}-{\check {\vec {\tilde W}}}_{+}\times
{\check {\vec B}}^{({\check {\tilde W}}_{-})}\, )^j(\vec x,x^o)-ie (\, \vec
\partial \times [{\check {\vec {\tilde W}}}_{+}\times {\check {\vec {\tilde W}}}
_{-}]\, )^j(\vec x,x^o)+\nonumber \\
&+&e^2 [2({\check {\tilde A}}^j_{\perp}
-tg\, \theta_w {{\vec \partial}\over {\triangle}}\vec \partial \cdot {\check
{\vec {\tilde Z}}}(\vec x,x^o)
+cot\, \theta_w{\check {\tilde Z}}^j)
{\check {\vec {\tilde W}}}_{+}\cdot {\check {\vec {\tilde W}}}_{-}-\nonumber \\
&-&({\check {\tilde W}}^j_{+}{\check {\vec {\tilde W}}}_{-}+
{\check {\tilde W}}^j_{-}{\check 
{\vec {\tilde W}}}_{+})\cdot ({\check {\vec {\tilde A}}}_{\perp}
-tg\, \theta_w {{\vec \partial}\over {\triangle}}\vec \partial \cdot {\check
{\vec {\tilde Z}}}(\vec x,x^o)
+cot\, \theta_w
{\check {\vec {\tilde Z}}})](\vec x,x^o)-\nonumber \\
&-&{{e^2}\over {sin\, \theta_w}} { {i{\check {\tilde \pi}}^{({\check {\tilde 
W}}_{+})\, j}{\check {\cal Q}}_{(CC)-}-i{\check {\tilde \pi}}^{({\check {\tilde 
W}}_{-})\, j}{\check {\cal Q}}_{(CC)+}}\over {m^2_W (1+{{|e|}\over 
{sin\, 2\theta
_w m_Z}}H)^2} }(\vec x,x^o)\, ],\nonumber \\
&&{}\nonumber \\
\partial^o&& {\check {\tilde Z}}^i(\vec x,x^o)\, {\buildrel \circ 
\over =}\, \lbrace {\check {\tilde Z}}^i(\vec x,x^o),H_{phys}\rbrace =
-{{\delta H_{phys}}\over {{\check {\tilde \pi}}^{({\check {\tilde Z}})\, i}
(\vec x,x^o)}}=\nonumber \\
&=&-{\check {\tilde \pi}}^{({\check {\tilde Z}})\, i}(\vec x,x^o)
-\partial^i \cdot {{e {\check {\cal Q}}_{(NC)}(\vec x,x^o)}\over
{m^2_Z\, (1+{{|e|}\over {sin\, 2\theta_w m_Z}}H(\vec x,x^o))^2}}-\nonumber \\
&-&i{{e^2 cos\, \theta_w}\over {sin^2\, \theta_w}} 
{{ {\check {\tilde W}}^i_{-}(\vec x,x^o){\check {\cal Q}}_{(CC)-}
(\vec x,x^o)-{\check {\tilde W}}^i_{+}(\vec x,x^o){\check {\cal Q}}_{(CC)+}
(\vec 
x,x^o)}\over {m^2_W\, (1+{{|e|}\over {sin\, 2\theta_w m_Z}}H(\vec x,x^o))^2}},
\nonumber \\
\partial^o&& {\check {\tilde \pi}}^{({\check {\tilde Z}})}(\vec x,x^o)\,
{\buildrel \circ \over =}\, \lbrace {\check {\tilde \pi}}^{({\check 
{\tilde Z}})\, i}(\vec x,x^o),H_{phys}\rbrace = {{\delta H_{phys}}
\over {\delta {\check {\tilde Z}}^i(\vec x,x^o)}}=\nonumber \\
&=&\triangle {\check {\tilde Z}}^i(\vec x,x^o)+\partial^i\vec \partial \cdot 
{\check {\vec {\tilde Z}}}(\vec x,x^o)+e{\check {\tilde j}}^i_{(NC)}(\vec 
x,x^o)+\nonumber \\
&+&m^2_Z (1+{{|e|}\over {sin\, 2\theta_w m_Z}}H(\vec x,x^o))^2 {\check {\tilde 
Z}}^i(\vec x,x^o)+\nonumber \\
&+&ie cot\, \theta_w [{\check {\vec {\tilde W}}}_{-}\times {\check {\vec B}}
^{({\check {\tilde W}}_{+})}-{\check {\vec {\tilde W}}}){+}\times {\check {\vec 
B}}^{({\check {\tilde W}}_{-})}]^i(\vec x,x^o)+\nonumber \\
&+&e^2cot\, \theta_w[2({\check {\tilde A}}^i_{\perp}
-tg\, \theta_w {{\vec \partial}\over {\triangle}}\vec \partial \cdot {\check
{\vec {\tilde Z}}}(\vec x,x^o)
+cot\, \theta_w{\check 
{\tilde Z}}^i){\check {\vec {\tilde W}}}_{+}\cdot {\check {\vec {\tilde W}}}
_{-}-\nonumber \\
&-&({\check {\tilde W}}^i_{+}{\check {\vec {\tilde W}}}_{-}+{\check {\tilde 
W}}^i_{-}{\check {\vec {\tilde W}}}_{+})\cdot ({\check {\vec {\tilde A}}}
_{\perp}
-tg\, \theta_w {{\vec \partial}\over {\triangle}}\vec \partial \cdot {\check
{\vec {\tilde Z}}}(\vec x,x^o)
+cot\, \theta_w{\check {\vec {\tilde Z}}})](\vec x,x^o)-\nonumber \\
&-&{{e^2 cos\, \theta_w}\over {sin^2\, \theta_w}}{ {i{\check {\tilde \pi}}
^{({\check {\tilde W}}_{+})\, i}{\check {\cal Q}}_{(CC)-}-i{\check {\tilde \pi}}
^{({\check {\tilde W}}_{-})\, i}{\check {\cal Q}}_{(CC)+} }\over 
{m^2_W (1+{{|e|}
\over {sin\, 2\theta_w m_Z}}H)^2} }(\vec x,x^o)+\nonumber \\
&+&tg\, \theta_w \int d^3y {{\partial^i \vec \partial \cdot \vec F(\vec y,x^o)}
\over {4\pi |\vec x-\vec y|}},\nonumber \\
with&&\nonumber \\
&&\vec F(\vec x,x^o)=e {\check {\tilde j}}^i_{(em)}(\vec x,x^o)-\nonumber \\
&-&[\, ie ({\check {\vec {\tilde W}}}_{-}\times {\check 
{\vec B}}^{({\check {\tilde W}}_{+})}-{\check {\vec {\tilde W}}}_{+}\times
{\check {\vec B}}^{({\check {\tilde W}}_{-})}\, )^j(\vec x,x^o)-ie (\, \vec
\partial \times [{\check {\vec {\tilde W}}}_{+}\times {\check {\vec {\tilde W}}}
_{-}]\, )^j(\vec x,x^o)+\nonumber \\
&+&e^2 [2({\check {\tilde A}}^j_{\perp}
-tg\, \theta_w {{\vec \partial}\over {\triangle}}\vec \partial \cdot {\check
{\vec {\tilde Z}}}(\vec x,x^o)
+cot\, \theta_w{\check {\tilde Z}}^j)
{\check {\vec {\tilde W}}}_{+}\cdot {\check {\vec {\tilde W}}}_{-}-\nonumber \\
&-&({\check 
{\tilde W}}^j_{+}{\check {\vec {\tilde W}}}+{\check {\tilde W}}^j_{-}{\check 
{\vec {\tilde W}}}_{+})\cdot ({\check {\vec {\tilde A}}}_{\perp}
-tg\, \theta_w {{\vec \partial}\over {\triangle}}\vec \partial \cdot {\check
{\vec {\tilde Z}}}(\vec x,x^o)
+cot\, \theta_w
{\check {\vec {\tilde Z}}})](\vec x,x^o)-\nonumber \\
&-&{{e^2}\over {sin\, \theta_w}} { {i{\check {\tilde \pi}}^{({\check {\tilde 
W}}_{+})\, j}{\check {\cal Q}}_{(CC)-}-i{\check {\tilde \pi}}^{({\check {\tilde 
W}}_{-})\, j}{\check {\cal Q}}_{(CC)+}}\over 
{m^2_W (1+{{|e|}\over {sin\, 2\theta
_w m_Z}}H)^2} }(\vec x,x^o)\, ],\nonumber \\
&&{}\nonumber \\
\partial^o&& {\check {\tilde W}}_{\pm}^i(\vec x,x^o)\, {\buildrel \circ 
\over =}\, \lbrace {\check {\tilde W}}^i_{\pm}(\vec x,x^o),H_{phys}\rbrace =
-{{\delta H_{phys}}\over {{\check {\tilde \pi}}^{({\check {\tilde W}}_{\pm})\, 
i}(\vec x,x^o)}}=\nonumber \\
&=&-{\check {\tilde \pi}}^{({\check {\tilde W}}_{\mp})\, i}(\vec x,x^o)\pm
i {\check {\tilde W}}^i_{\mp}(\vec x,x^o) \int d^3y {{ {\check {\cal Q}}
_{(em)}(\vec y,
x^o)}\over {4\pi |\vec x-\vec y|}}\pm \nonumber \\
&\pm& ie^2 cot\, \theta_w {{ {\check {\cal Q}}_{(NC)}(\vec x,x^o)}\over 
{m^2_Z\, (1+
{{|e|}\over {sin\, 2\theta_wm_Z}}H(\vec x,x^o))^2}}-\partial^i {{ 
e sin^{-1}\, \theta_w {\check {\cal Q}}_{(CC)
\mp}(\vec x,x^o)}\over {m^2_W\, (1+{{|e|}\over {sin\, 2\theta_wm_Z}}H(\vec x,
x^o))^2}}+\nonumber \\
&+&i{{e^3}\over {sin\, \theta_w}} 
{\check {\tilde W}}^i_{\mp}(\vec x,x^o) \int d^3y {{(\vec x-\vec y)
\cdot {\check {\vec {\tilde W}}}_{\mp}(\vec y,x^o)}\over {4\pi |\vec x-\vec y|
^3}} {{ {\check {\cal Q}}_{(CC)\mp}(\vec y,x^o)}\over {m^2_W\, (1+{{|e|}\over 
{sin\, 2\theta_wm_Z}}H(\vec y,x^o))^2}},\nonumber \\
\partial^o&& {\check {\tilde \pi}}^{({\check {\tilde W}}_{\pm})\, i}(\vec x,x^o)
\, {\buildrel \circ \over =}\, \lbrace {\check {\tilde \pi}}^{({\check 
{\tilde W}}_{\pm})\, i}(\vec x,x^o),H_{phys}\rbrace = {{\delta H_{phys}}
\over {\delta {\check {\tilde W}}_{\pm}^i(\vec x,x^o)}}=\nonumber \\
&=&\triangle {\check {\tilde W}}^i_{\mp}(\vec x,x^o)+\partial^i\vec \partial 
\cdot {\check {\vec {\tilde W}}}_{\mp}(\vec x,x^o)+{e\over {sin\, \theta_w}}
{\check {\tilde j}}^i_{(CC)\mp}(\vec x,x^o)+\nonumber \\
&+&m^2_W (1+{{|e|}\over {sin\, 2\theta_w m_Z}}H(\vec x,x^o))^2{\check {\tilde
W}}^i_{\mp}(\vec x,x^o)\pm \nonumber \\
&\pm& ie [{\check {\vec {\tilde W}}}_{\mp}\times ({\check {\vec B}}^{({\check 
{\tilde A}})}+cot\, \theta_w {\check {\vec B}}^{({\check {\tilde Z}})})]
^i(\vec x,x^o)\pm \nonumber \\
&\pm& ie [2({\check {\vec {\tilde A}}}_{\perp}
-tg\, \theta_w {{\vec \partial}\over {\triangle}}\vec \partial \cdot {\check
{\vec {\tilde Z}}}(\vec x,x^o)
+cot\, \theta_w{\check {\vec 
{\tilde Z}}})\times {\check {\vec B}}^{({\check {\tilde W}}_{\mp})})^i+
\nonumber \\
&+&{\check {\tilde W}}^i_{\mp}(cot\, \theta_w\vec \partial \cdot {\check {\vec 
{\tilde Z}}})-
{\check {\vec {\tilde W}}}_{\mp}\cdot \partial^i({\check {\vec {\tilde
A}}}_{\perp}
-tg\, \theta_w {{\vec \partial}\over {\triangle}}\vec \partial \cdot {\check
{\vec {\tilde Z}}}(\vec x,x^o)
+cot\, \theta_w{\check {\vec {\tilde Z}}})](\vec x,x^o)-\nonumber \\
&-&{{e^2}\over {sin^2\, \theta_w}} [{\check {\tilde W}}^i_{\pm}{\check {\vec 
{\tilde W}}}^2_{\mp}-{\check {\tilde W}}^i_{\mp}{\check {\vec {\tilde W}}}_{+}
\cdot {\check {\vec {\tilde W}}}_{-}](\vec x,x^o)+\nonumber \\
&+&e^2[{\check {\tilde W}}^i_{\mp}({\check {\vec {\tilde A}}}_{\perp}
-tg\, \theta_w {{\vec \partial}\over {\triangle}}\vec \partial \cdot {\check
{\vec {\tilde Z}}}(\vec x,x^o)
+cot\, \theta_w{\check {\vec {\tilde Z}}})^2-\nonumber \\
&-&({\check {\tilde A}}^i_{\perp}
-tg\, \theta_w {{\vec \partial}\over {\triangle}}\vec \partial \cdot {\check
{\vec {\tilde Z}}}(\vec x,x^o)
+cot\, 
\theta_w{\check {\tilde Z}}^i)\nonumber \\
&&{\check {\vec {\tilde W}}}_{\mp}\cdot ({\check
{\vec {\tilde A}}}_{\perp}
-tg\, \theta_w {{\vec \partial}\over {\triangle}}\vec \partial \cdot {\check
{\vec {\tilde Z}}}(\vec x,x^o)
+cot\, \theta_w{\check {\vec {\tilde Z}}})](\vec 
x,x^o)\pm \nonumber \\
&\pm& ie^2 \int d^3y {{ {\check {\cal Q}}_{(em)}(\vec y,x^o)}\over 
{4\pi |\vec x-
\vec y|}} {\check {\tilde \pi}}^{({\check {\tilde W}}_{\mp})\, i}(\vec x,x^o)
\pm \nonumber \\
&\pm& ie^2 cot\, \theta_w {{ {\check {\cal Q}}_{(NC)}(\vec x,x^o)}\over 
{m^2_Z (1+{{|e|}
\over {sin\, 2\theta_w m_Z}}H(\vec x,x^o))^2}} {\check {\tilde \pi}}^{({\check 
{\tilde W}}_{\mp})\, i}(\vec x,x^o)\mp \nonumber \\
&\mp& i {{e^2}\over {sin^2\, \theta_w}} {{ {\check {\cal Q}}_{(CC)\pm}(\vec x,
x^o)}\over {m^2_W (1+{{|e|}\over {sin\, 2\theta_w m_Z}}H(\vec x,x^o))^2}}\cdot 
\nonumber \\
&\cdot& [{\check {\tilde \pi}}_{\perp}^{({\check {\tilde A}})\, i}(\vec x,x^o)-
e \int d^3z {{x^i-z^i}\over {4\pi |\vec x-\vec z|^3}}{\check {\cal Q}}
_{(em)}(\vec 
z,x^o)+cot\, \theta_w {\check {\tilde \pi}}^{({\check {\tilde Z}})\, i}(\vec 
x,x^o)]-\nonumber \\
&-&{{e^3}\over {sin\, \theta_w}} \int d^3y {{(\vec x-\vec y)\cdot {\check 
{\vec {\tilde W}}}_{\mp}(\vec y,x^o)\, {\check {\cal Q}}_{(CC)\mp}(\vec y,x^o)}
\over {4\pi |\vec x-\vec y|^3\,\, m^2_W (1+{{|e|}\over {sin\, 2\theta_w m_Z}}
H(\vec x,x^o))^2}} {\check {\tilde \pi}}^{({\check {\tilde W}}_{\mp})\, i}(\vec
x,x^o),\nonumber \\
&&{}\nonumber \\
\partial^o&& H(\vec x,x^o)\,{\buildrel \circ \over =}\, \lbrace H(\vec x,x^o),
H_{phys}\rbrace = {{\delta H_{phys}}\over {\delta \pi_H(\vec x,x^o)}}=
\pi_H(\vec x,x^o),\nonumber \\
\partial^o&& \pi_H(\vec x,x^o)\, {\buildrel \circ \over =}\, \lbrace \pi_H
(\vec x,x^o),H_{phys}\rbrace = -{{\delta H_{phys}}
\over {\delta H(\vec x,x^o)}}=\nonumber \\
&=&\triangle_x H(\vec x,x^o)-m^2_HH(\vec x,x^o) (1+{{|e|}\over {2sin\, 
2\theta_w m_Z}}H(\vec x,x^o))^2-\nonumber \\
&-& {{|e| m^2_H}\over {2sin\, 2\theta_w m_Z}} H^2(\vec x,x^o) (1+{{|e|}\over 
{2sin\, 2\theta_w m_Z}}H(\vec x,x^o))-\nonumber \\
&-&{{|e|}\over {sin\, 2\theta_w m_Z}}(1+{{|e|}\over {sin\, 2\theta_w m_Z}}
H(\vec x,x^o)) \cdot \nonumber \\
&&[m^2_Z {\check {\vec {\tilde Z}}}^2(\vec x,x^o)+2m^2_W {\check 
{\vec {\tilde W}}}_{+}(\vec x,x^o)\cdot {\check {\vec {\tilde W}}}_{-}(\vec
x,x^o)]+\nonumber \\
&+&{{|e|e^2}\over {4sin\, 2\theta_w m_Z}} {1\over {(1+{{|e|}\over {sin\, 
2\theta_w m_Z}}H(\vec x,x^o))^3}}\cdot \nonumber \\
&\cdot& [{1\over {m^2_Z}}{\check {\cal Q}}^2_{(NC)}(\vec x,x^o)+
{1\over {m^2_W sin^2\,
\theta_w}}{\check {\cal Q}}_{(CC)+}(\vec x,x^o){\check {\cal Q}}_{(CC)-}(\vec x,
x^o)]-\nonumber \\
&-&{{|e|}\over {sin\, 2\theta_w m_Z}} [\, 
\left( \begin{array}{ccc} {\bar {\check e}}^{(m)}(x) & {\bar {\check \mu}}
^{(m)}(x) & {\bar {\check \tau}}^{(m)}(x) \end{array} \right) 
\beta \left( \begin{array}{ccc} m_e & 0 & 0\\ 0 & m_{\mu}
& 0\\ 0 & 0 & m_{\tau} \end{array} \right)  \left( \begin{array}{c} {\check e}
^{(m)}(x) \\ {\check \mu}^{(m)}(x) \\ {\check \tau}^{(m)}(x) \end{array} 
\right) +\nonumber \\
&+&\left( \begin{array}{ccc} {\bar {\check u}}^{(m)}(x) & {\bar {\check c}}
^{(m)}(x) & {\bar {\check t}}^{(m)}(x) \end{array} \right) 
\beta \left( \begin{array}{ccc} m_u & 0 & 0\\ 0 & m_c
& 0\\ 0 & 0 & m_t \end{array} \right)  \left( \begin{array}{c} {\check u}^{(m)}
(x) \\ {\check c}^{(m)}(x) \\ {\check t}^{(m)}(x) \end{array} \right) +
\nonumber \\
&+&\left( \begin{array}{ccc} {\bar {\check d}}^{(m)}(x) & {\bar {\check s}}
^{(m)}(x) & {\bar {\check b}}^{(m)}(x) \end{array} \right) 
\beta \left( \begin{array}{ccc} m_d & 0 & 0\\ 0 & m_s
& 0\\ 0 & 0 & m_b \end{array} \right)  \left( \begin{array}{c} {\check d}^{(m)}
(x) \\ {\check s}^{(m)}(x) \\ {\check b}^{(m)}(x) \end{array} \right) \, ],
\nonumber \\
&&{}\nonumber \\
i\partial^o&& {\check \psi}^{(l)}_{Li}(\vec x,x^o)\, {\buildrel \circ \over
=}\, \lbrace {\check \psi}^{(l)}_{Li}(\vec x,x^o),H_{phys}\rbrace =\nonumber \\
&=&i\vec \alpha \cdot [\vec \partial +{e\over {sin\, \theta_w}}({\check {\vec
{\tilde W}}}_{+}(\vec x,x^o)T^{-}_w+{\check {\vec {\tilde W}}}_{-}(\vec x,x^o)
T^{+}_w)-\nonumber \\
&-&ieQ_{em}{\check {\vec {\tilde A}}}_{\perp}(\vec x,x^o)
-tg\, \theta_w {{\vec \partial}\over {\triangle}}\vec \partial \cdot {\check
{\vec {\tilde Z}}}(\vec x,x^o)-\nonumber \\
&-&ieQ_Z{\check {\vec
{\tilde Z}}}(\vec x,x^o)] {\check \psi}^{(l)}_{Li}(\vec x,x^o)-\nonumber \\
&-&(1+{{|e|}\over {sin\, 2\theta_w m_Z}}H(\vec x,x^o)) \beta
\left( \begin{array}{l} 0\\ 1\end{array} \right) M^{(l)}_{ij} {\check \psi}
^{(l)}_{Rj}(\vec x,x^o)+\nonumber \\
&+&e^2 \int d^3y {{ {\check {\cal Q}}_{(em)}(\vec y,x^o)}\over 
{4\pi |\vec x-\vec y|}}
Q_{em}{\check \psi}^{(l)}_{Li}(\vec x,x^o)+\nonumber \\
&+&e^2 {{ {\check {\cal Q}}_{(NC)}(\vec x,x^o)}\over 
{m^2_Z (1+{{|e|}\over {sin\, 
2\theta_w m_Z}}H(\vec x,x^o))^2}} Q_Z{\check \psi}^{(l)}_{Li}(\vec x,x^o)+
\nonumber \\
&+&{{e^2}\over {sin^2\, \theta_w}} {{ {\check {\cal Q}}_{(CC)-}iT_w^{+}+{\check 
{\cal Q}}_{(CC)-} iT_w^{-}}\over {m^2_W (1+{{|e|}\over {sin\, 2\theta_w m_Z}}H)
^2}}(\vec x,x^o) {\check \psi}^{(l)}_{Li}(\vec x,x^o)+\nonumber \\
&+&{{e^3}\over {sin\, \theta_w}}\int d^3y {{ (\vec x-\vec y)\cdot [i{\check
{\vec {\tilde W}}}_{-}{\check {\cal Q}}_{(CC)-}-i{\check {\vec {\tilde W}}}_{+}
{\check {\cal Q}}_{(CC)+}](\vec y,x^o)}\over {4\pi |\vec x-\vec y|^3\,\, m^2_W 
(1+{{|e|}\over {sin\, 2\theta_w m_Z}}H(\vec y,x^o))^2}} Q_{em}{\check \psi}
^{(l)}_{Li}(\vec x,x^o),\nonumber \\
&&{}\nonumber \\
i\partial^o&& {\check \psi}^{(l)}_{Ri}(\vec x,x^o)\, {\buildrel \circ \over
=}\, \lbrace {\check \psi}^{(l)}_{Ri}(\vec x,x^o),H_{phys}\rbrace =\nonumber \\
&=&i\vec \alpha \cdot [\vec \partial +{e\over {cos\, \theta_w}}({\check {\vec
{\tilde A}}}_{\perp}(\vec x,x^o)
-tg\, \theta_w {{\vec \partial}\over {\triangle}}\vec \partial \cdot {\check
{\vec {\tilde Z}}}(\vec x,x^o)-\nonumber \\
&-&tg\, \theta_w{\check {\vec {\tilde Z}}}(\vec
x,x^o))Y_w]{\check \psi}^{(l)}_{Ri}(\vec x,x^o)-\nonumber \\
&-&(1+{{|e|}\over {sin\, 2\theta_w m_Z}}H(\vec x,x^o)) \beta
M^{(l)\dagger}_{ij} \left( \begin{array}{ll} 0 & 1\end{array} \right) \cdot 
{\check \psi}^{(l){'}}_{Lj}(\vec x,x^o)+\nonumber \\
&+&e^2 \int d^3y {{ {\check {\cal Q}}_{(em)}(\vec y,x^o)}\over 
{4\pi |\vec x-\vec y|}} 
iY_w {\check \psi}^{(l)}_{Ri}(\vec x,x^o)-\nonumber \\
&-&e^2 tg\, \theta_w {{ {\check {\cal Q}}_{(NC)}(\vec x,x^o)}\over {m^2_Z
(1+{{|e|}\over {sin\, 2\theta_w m_Z}}H(\vec x,x^o))^2}} iY_w {\check \psi}^{(l)}
_{Ri}(\vec x,x^o)+\nonumber \\
&+&{{e^3}\over {sin\, \theta_w}}\int d^3y {{(\vec x-\vec y)\cdot [i{\check
{\vec {\tilde W}}}_{-}{\check {\cal Q}}_{(CC)-}-i{\check {\vec {\tilde W}}}_{+}
{\check {\cal Q}}_{(CC)+}](\vec y,x^o)}\over {4\pi |\vec x-\vec y|^3\,\, m^2_W
(1+{{|e|}\over {sin\, 2\theta_w m_Z}}H(\vec x,x^o))^2}} iY_w {\check \psi}^{(l)}
_{Ri}(\vec x,x^o),\nonumber \\
&&{}\nonumber \\
i\partial^o&& {\check \psi}^{(q)}_{Li}(\vec x,x^o)\, {\buildrel \circ \over
=}\, \lbrace {\check \psi}^{(q)}_{Li}(\vec x,x^o),H_{phys}\rbrace =\nonumber \\
&=&i\vec \alpha \cdot [\vec \partial +g_s{\check {\vec {\tilde G}}}_{A\perp}
(\vec x,x^o)T^A_s+{e\over {sin\, \theta_w}}({\check {\vec {\tilde W}}}_{+}
(\vec x,x^o)T^{-}_w+\nonumber \\
&+&{\check {\vec {\tilde W}}}_{-}(\vec x,x^o)T^{+}_w)-ieQ_{em}{\check {\vec
{\tilde A}}}_{\perp}(\vec x,x^o)
-tg\, \theta_w {{\vec \partial}\over {\triangle}}\vec \partial \cdot {\check
{\vec {\tilde Z}}}(\vec x,x^o)-\nonumber \\
&-&ieQ_Z{\check {\vec {\tilde Z}}}(\vec x,x^o)]
{\check \psi}^{(q)}_{Li}(\vec x,x^o)-\nonumber \\
&-&(1+{{|e|}\over {sin\, 2\theta_w m_Z}}H(\vec x,x^o)) \beta
[\left( \begin{array}{l} 0 \\ 1 \end{array} \right) M^{(q)}_{ij}{\check \psi}
^{(q)}_{Rj}(\vec x,x^o)+\left( \begin{array}{l} 1 \\ 0 \end{array}
\right) {\tilde M}^{(q)}_{ij}{\check {\tilde \psi}}^{(q)}_{Rj}(\vec x,x^o)]-
\nonumber \\
&-&g^2_s\int d^3y iT^A_s C^{({\check {\tilde G}}_{\perp})}_{\triangle ,AB}
(\vec x,\vec y;x^o){\check {\cal Q}}_{sB}(\vec y,x^o) 
{\check \psi}^{(q)}_{Li}(\vec 
x,x^o)+\nonumber \\
&+&e^2 \int d^3y {{ {\check {\cal Q}}_{(em)}(\vec y,x^o)}\over 
{4\pi |\vec x-\vec y|}} 
Q_{em} {\check \psi}^{(q)}_{Li}(\vec x,x^o)-\nonumber \\
&-&e^2 tg\, \theta_w {{ {\check {\cal Q}}_{(NC)}(\vec x,x^o)}\over {m^2_Z
(1+{{|e|}\over {sin\, 2\theta_w m_Z}}H(\vec x,x^o))^2}} Q_Z {\check \psi}
^{(q)}_{Li}(\vec x,x^o)+\nonumber \\
&+&{{e^2}\over {sin^2\, \theta_w}} {{ {\check {\cal Q}}_{(CC)-}(\vec x,x^o)iT_w
^{+}+{\check {\cal Q}}_{(CC)+}(\vec x,x^o)iT_w^{-}}\over {m^2_W
(1+{{|e|}\over {sin\, 2\theta_w m_Z}}H(\vec x,x^o))^2}} {\check \psi}^{(q)}
_{Li}(\vec x,x^o)+\nonumber \\
&+&{{e^3}\over {sin\, \theta_w}}\int d^3y {{(\vec x-\vec y)\cdot [i{\check 
{\vec {\tilde W}}}_{-}{\check {\cal Q}}_{(CC)-}-i{\check {\vec {\tilde W}}}_{+}
{\check {\cal Q}}_{(CC)+}](\vec y,x^o)}\over {4\pi |\vec x-\vec y|^3\,\, m^2_W
(1+{{|e|}\over {sin\, 2\theta_w m_Z}}H(\vec y,x^o))^2}} Q_{em} {\check \psi}
^{(q)}_{Li}(\vec x,x^o),\nonumber \\
&&{}\nonumber \\
i\partial^o&& {\check \psi}^{(q)}_{Ri}(\vec x,x^o)\, {\buildrel \circ \over
=}\, \lbrace {\check \psi}^{(q)}_{Ri}(\vec x,x^o),H_{phys}\rbrace =\nonumber \\
&=&i\vec \alpha \cdot [\vec \partial +g_s{\check {\vec {\tilde G}}}_{A\perp}
(\vec x,x^o)T^A_s+\nonumber \\
&+&{e\over {cos\, \theta_w}}({\check {\vec {\tilde A}}}_{\perp}(\vec x,x^o)
-tg\, \theta_w {{\vec \partial}\over {\triangle}}\vec \partial \cdot {\check
{\vec {\tilde Z}}}(\vec x,x^o)-\nonumber \\
&-&
tg\, \theta_w{\check {\vec {\tilde Z}}}(\vec x,x^o)) Y_w] {\check \psi}^{(q)}
_{Ri}(\vec x,x^o)-\nonumber \\
&-&(1+{{|e|}\over {sin\, 2\theta_w m_Z}}H(\vec x,x^o)) \beta
M_{ij}^{(q)\dagger} \left( \begin{array}{ll}0 & 1 \end{array} \right) \cdot 
{\check \psi}^{(q)}_{Lj}(\vec x,x^o)+\nonumber \\
&+&e^2 \int d^3y {{ {\check {\cal Q}}_{(em)}(\vec y,x^o)}\over 
{4\pi |\vec x-\vec y|}} 
iY_w {\check \psi}^{(q)}_{Ri}(\vec x,x^o)-\nonumber \\
&-&e^2 tg\, \theta_w {{ {\check {\cal Q}}_{(NC)}(\vec x,x^o)}\over {m^2_Z
(1+{{|e|}\over {sin\, 2\theta_w m_Z}}H(\vec x,x^o))^2}} iY_w {\check \psi}
^{(q)}_{Ri}(\vec x,x^o)+\nonumber \\
&+&{{e^3}\over {sin\, \theta_w}}\int d^3y {{(\vec x-\vec y)\cdot [i{\check 
{\vec {\tilde W}}}_{-}{\check {\cal Q}}_{(CC)-}-i{\check {\vec {\tilde W}}}_{+}
{\check {\cal Q}}_{(CC)+}](\vec y,x^o)}\over {4\pi |\vec x-\vec y|^3\,\, m^2_W
(1+{{|e|}\over {sin\, 2\theta_w m_Z}}H(\vec y,x^o))^2}} iY_w {\check \psi}
^{(q)}_{Ri}(\vec x,x^o),\nonumber \\
&&{}\nonumber \\
i\partial^o&& {\check {\tilde \psi}}^{(q)}_{Ri}(\vec x,x^o)\, {\buildrel \circ 
\over =}\, \lbrace {\check {\tilde \psi}}^{(q)}_{Ri}(\vec x,x^o),H_{phys}
\rbrace =\nonumber \\
&=&i\vec \alpha \cdot [\vec \partial +g_s{\check {\vec {\tilde G}}}_{A\perp}
(\vec x,x^o)T^A_s+\nonumber \\
&+&{e\over {cos\, \theta_w}}({\check {\vec {\tilde A}}}_{\perp}(\vec x,x^o)
-tg\, \theta_w {{\vec \partial}\over {\triangle}}\vec \partial \cdot {\check
{\vec {\tilde Z}}}(\vec x,x^o)-\nonumber \\
&-&
tg\, \theta_w{\check {\vec {\tilde Z}}}(\vec x,x^o)) Y_w] {\check {\tilde 
\psi}}^{(q)}_{Ri}(\vec x,x^o)-\nonumber \\
&-&(1+{{|e|}\over {sin\, 2\theta_w m_Z}}H(\vec x,x^o)) \beta
{\tilde M}_{ij}^{(q)\dagger} \left( \begin{array}{ll} 1 & 0 
\end{array} \right) \cdot {\check \psi}^{(q)}_{Lj}(\vec x,x^o)+\nonumber \\
&+&e^2 \int d^3y {{ {\check {\cal Q}}_{(em)}(\vec y,x^o)}\over 
{4\pi |\vec x-\vec y|}} 
iY_w {\check {\tilde \psi}}^{(q)}_{Ri}(\vec x,x^o)-\nonumber \\
&-&e^2 tg\, \theta_w {{ {\check {\cal Q}}_{(NC)}(\vec x,x^o)}\over {m^2_Z
(1+{{|e|}\over {sin\, 2\theta_w m_Z}}H(\vec x,x^o))^2}} iY_w {\check {\tilde
\psi}}^{(q)}_{Ri}(\vec x,x^o)+\nonumber \\
&+&{{e^3}\over {sin\, \theta_w}}\int d^3y {{(\vec x-\vec y)\cdot [i{\check 
{\vec {\tilde W}}}_{-}{\check {\cal Q}}_{(CC)-}-i{\check {\vec {\tilde W}}}_{+}
{\check {\cal Q}}_{(CC)+}](\vec y,x^o)}\over {4\pi |\vec x-\vec y|^3\,\, m^2_W
(1+{{|e|}\over {sin\, 2\theta_w m_Z}}H(\vec y,x^o))^2}} iY_w {\check {\tilde
\psi}}^{(q)}_{Ri}(\vec x,x^o),\nonumber \\
&&{}\nonumber \\
&&\Downarrow \nonumber \\
&&{}\nonumber \\
i\partial^o&& \left( \begin{array}{c} {\check \nu}^{(m)}_e(\vec x,x^o)\\ {\check
\nu}^{(m)}_{\mu}(\vec x,x^o)\\ {\check \nu}^{(m)}_{\tau}(\vec x,x^o) 
\end{array} \right) \, {\buildrel \circ \over =}\, [i\vec \alpha \cdot (\vec
\partial -{{ie}\over {sin\, 2\theta_w}}{\check {\vec {\tilde Z}}}(\vec x,x^o))+
\nonumber \\
&+&{{e^2}\over {sin\, 2\theta_w}} {{ {\check {\cal Q}}_{(NC)}(\vec x,x^o)}\over 
{m^2_Z
(1+{{|e|}\over {sin\, 2\theta_w m_Z}}H(\vec x,x^o))^2}} {1\over 2}(1-\gamma_5)
\left( \begin{array}{c} {\check \nu}^{(m)}_e(\vec x,x^o)\\ {\check
\nu}^{(m)}_{\mu}(\vec x,x^o)\\ {\check \nu}^{(m)}_{\tau}(\vec x,x^o) 
\end{array} \right) +\nonumber \\
&+&{e\over {sin\, \theta_w}}[\vec \alpha \cdot {\check {\vec {\tilde W}}}
_{-}(\vec x,x^o)+{e\over {2cos\, \theta_w}}{{ {\check {\cal Q}}_{(NC)}
(\vec x,x^o)}\over
{m^2_Z (1+{{|e|}\over {sin\, 2\theta_w m_Z}}H(\vec x,x^o))^2}}]\cdot
\nonumber \\
&& {1\over 2}(1-
\gamma_5)\left( \begin{array}{c} {\check e}^{(m)}(\vec x,x^o) \\ {\check \mu}
^{(m)}(\vec x,x^o) \\ {\check \tau}^{(m)}(\vec x,x^o) \end{array} \right),
\nonumber \\
&&{}\nonumber \\
i\partial^o&& \left( \begin{array}{c} {\check e}^{(m)}(\vec x,x^o) \\ {\check 
\mu}^{(m)}(\vec x,x^o) \\ {\check \tau}^{(m)}(\vec x,x^o) \end{array} \right)\,
{\buildrel \circ \over =}\, [i\vec \alpha \cdot (\vec \partial -ie{\check {\vec
{\tilde A}}}_{\perp}(\vec x,x^o)
-tg\, \theta_w {{\vec \partial}\over {\triangle}}\vec \partial \cdot {\check
{\vec {\tilde Z}}}(\vec x,x^o)-\nonumber \\
&-&ie{{2sin^2\, \theta_w-{1\over 2}(1-\gamma_5)}
\over {sin\, 2\theta_w}}{\check {\vec {\tilde Z}}}(\vec x,x^o))+\nonumber \\
&+&\beta (1+{{|e|}\over {sin\, 2\theta_w m_Z}}H(\vec x,x^o)) \left( 
\begin{array}{ccc} m_e & 0 & 0\\ 0 & m_{\mu} & 0\\ 0 & 0 & m_{\tau} 
\end{array} \right) +e^2\int d^3y {{ {\check {\cal Q}}_{(em)}(\vec y,x^o)}
\over {4\pi
|\vec x-\vec y|}}+\nonumber \\
&+&e^2 {{ {\check {\cal Q}}_{(NC)}(\vec x,x^o)}\over 
{m^2_Z (1+{{|e|}\over {sin\, 
2\theta_w m_Z}}H(\vec x,x^o))^2}} {{2sin^2\, \theta_w-{1\over 2}(1-\gamma_5)}
\over {sin\, 2\theta_w}}+\nonumber \\
&+&{{e^3}\over {sin\, \theta_w}}\int d^3y {{(\vec x-\vec y)\cdot [i{\check
{\vec {\tilde W}}}_{-}{\check {\cal Q}}_{(CC)-}-i{\check {\vec {\tilde W}}}_{+}
{\check {\cal Q}}_{(CC)+}](\vec y,x^o)}\over {4\pi |\vec x-\vec y|^3\,\, m^2_W
(1+{{|e|}\over {sin\, 2\theta_w m_Z}}H(\vec x,x^o))^2}}\,\, ]
\left( \begin{array}{c} {\check e}^{(m)}(\vec x,x^o) \\ {\check 
\mu}^{(m)}(\vec x,x^o) \\ {\check \tau}^{(m)}(\vec x,x^o) \end{array} \right)
+\nonumber \\
&+&{{e^2}\over {sin^2\, \theta_w}} {{ {\check {\cal Q}}_{(CC)-}(\vec x,x^o)iT^{+}
_w}\over {m^2_W (1+{{|e|}\over {sin\, 2\theta_w m_Z}}H(\vec x,x^o))^2}} 
{1\over 2}(1-\gamma_5)
\left( \begin{array}{c} {\check \nu}^{(m)}_e(\vec x,x^o)\\ {\check
\nu}^{(m)}_{\mu}(\vec x,x^o)\\ {\check \nu}^{(m)}_{\tau}(\vec x,x^o) 
\end{array} \right),\nonumber \\
&&{}\nonumber \\
i\partial^o&& \left( \begin{array}{c} {\check u}^{(m)}(\vec x,x^o) \\ {\check 
c}^{(m)}(\vec x,x^o) \\ {\check t}^{(m)}(\vec x,x^o) \end{array} \right) \,
{\buildrel \circ \over =}\, [i\vec \alpha \cdot (\vec \partial +g_s{\check {\vec
{\tilde G}}}_{A\perp}(\vec x,x^o)T^A_s+\nonumber \\
&+&i{2\over 3}e{\check {\vec {\tilde A}}}
_{\perp}(\vec x,x^o)
-tg\, \theta_w {{\vec \partial}\over {\triangle}}\vec \partial \cdot {\check
{\vec {\tilde Z}}}(\vec x,x^o)-\nonumber \\
&-&ie{\check {\vec {\tilde Z}}}(\vec x,x^o){{{1\over 2}(1-\gamma_5)-{4\over 3}
sin^2\, \theta_w}\over {sin\, 2\theta_w}}
+\beta (1+{{|e|}\over {sin\, 2\theta_w m_Z}}H(\vec x,x^o)) 
\left( \begin{array}{ccc} m_u & 0 & 0\\ 0 & m_c
& 0\\ 0 & 0 & m_t \end{array} \right) -\nonumber \\
&-&g^2_s\int d^3y iT^A_s C^{({\check {\tilde G}}_{\perp})}_{\triangle ,AB}(\vec
x,\vec y;x^o) {\check {\cal Q}}_{sB}(\vec y,x^o)-\nonumber \\
&-&{2\over 3}e^2\int d^3y {{ {\check {\cal Q}}_{(em)}(\vec y,x^o)}\over 
{4\pi |\vec x-
\vec y|}}+{{e^2 {\check {\cal Q}}_{(NC)}(\vec x,x^o)}\over {m^2_Z
(1+{{|e|}\over {sin\, 2\theta_w m_Z}}H(\vec x,x^o))^2}} {{{1\over 2}(1-
\gamma_5)-{4\over 3}sin^2\, \theta_w}\over {sin\, 2\theta_w}}-\nonumber \\
&-&{{2e^2}\over {3sin\, \theta_w}}\int d^3y {{(\vec x-\vec y)\cdot [i{\check
{\vec {\tilde W}}}_{-}{\check {\cal Q}}_{(CC)-}-i{\check {\vec {\tilde W}}}_{+}
{\check {\cal Q}}_{(CC)+}](\vec y,x^o)}\over {4\pi |\vec x-\vec y|^3\,\, m^2_W
(1+{{|e|}\over {sin\, 2\theta_w m_Z}}H(\vec y,x^o))^2}}-\nonumber \\
&-&\theta g^2_s\int d^3y c_{ABC}{\check {\vec B}}^{({\check {\tilde G}})}_A(\vec
y,x^o)\cdot {\check {\vec {\tilde G}}}_{C\perp}(\vec y,x^o) {{{\vec \partial}
_y}\over {\triangle_y}}\cdot {\vec \zeta}^{({\check {\tilde G}}_{\perp})}
_{BD}(\vec y,\vec x;x^o) iT^D_s\, ]
\left( \begin{array}{c} {\check u}^{(m)}(\vec x,x^o) \\ {\check 
c}^{(m)}(\vec x,x^o) \\ {\check t}^{(m)}(\vec x,x^o) \end{array} \right)
+\nonumber \\
&+&{e\over {sin\, \theta_w}}[\vec \alpha \cdot {\check {\vec {\tilde W}}}
_{+}(\vec x,x^o)+{e\over {sin\, \theta_w}}{{ {\check {\cal Q}}_{(CC)+}(\vec x,
x^o) T^{-}_w}\over {m^2_W (1+{{|e|}\over {sin\, 2\theta_w m_Z}}H(\vec x,x^o))
^2}}]\cdot \nonumber \\
&&{1\over 2}(1-\gamma_5) V_{CKM}
\left( \begin{array}{c} {\check d}^{(m)}(\vec x,x^o) \\ {\check
s}^{(m)}(\vec x,x^o) \\ {\check b}^{(m)}(\vec x,x^o) \end{array} \right),
\nonumber \\
&&{}\nonumber \\
i\partial^o&& \left( \begin{array}{c} {\check d}^{(m)}(\vec x,x^o) \\ {\check
s}^{(m)}(\vec x,x^o) \\ {\check b}^{(m)}(\vec x,x^o) \end{array} \right) \,
{\buildrel \circ \over =}\, [i\vec \alpha \cdot (\vec \partial +g_s{\check {\vec
{\tilde G}}}_{A\perp}(\vec x,x^o)T^A_s-\nonumber \\
&-&i{1\over 3}e{\check {\vec {\tilde A}}}
_{\perp}(\vec x,x^o)
-tg\, \theta_w {{\vec \partial}\over {\triangle}}\vec \partial \cdot {\check
{\vec {\tilde Z}}}(\vec x,x^o)-\nonumber \\
&-&ie{\check {\vec {\tilde Z}}}(\vec x,x^o){{{2\over 3}
sin^2\, \theta_w-{1\over 2}(1-\gamma_5)}\over {sin\, 2\theta_w}}
+\beta (1+{{|e|}\over {sin\, 2\theta_w m_Z}}H(\vec x,x^o))
\left( \begin{array}{ccc} m_d & 0 & 0\\ 0 & m_s
& 0\\ 0 & 0 & m_b \end{array} \right) -\nonumber \\
&-&g^2_s\int d^3y iT^A_s C^{({\check {\tilde G}}_{\perp})}_{\triangle ,AB}(\vec
x,\vec y;x^o) {\check {\cal Q}}_{sB}(\vec y,x^o)+\nonumber \\
&+&{1\over 3}e^2\int d^3y {{ {\check {\cal Q}}_{(em)}(\vec y,x^o)}\over 
{4\pi |\vec x-
\vec y|}}+{{e^2 {\check {\cal Q}}_{(NC)}(\vec x,x^o)}\over {m^2_Z
(1+{{|e|}\over {sin\, 2\theta_w m_Z}}H(\vec x,x^o))^2}} {{{2\over 3}sin^2\, 
\theta_w-{1\over 2}(1-\gamma_5)}\over {sin\, 2\theta_w}}+\nonumber \\
&+&{{e^2}\over {3sin\, \theta_w}}\int d^3y {{(\vec x-\vec y)\cdot [i{\check
{\vec {\tilde W}}}_{-}{\check {\cal Q}}_{(CC)-}-i{\check {\vec {\tilde W}}}_{+}
{\check {\cal Q}}_{(CC)+}](\vec y,x^o)}\over {4\pi |\vec x-\vec y|^3\,\, m^2_W
(1+{{|e|}\over {sin\, 2\theta_w m_Z}}H(\vec y,x^o))^2}}-\nonumber \\
&-&\theta g^2_s\int d^3y c_{ABC}{\check {\vec B}}^{({\check {\tilde G}})}_A(\vec
y,x^o)\cdot {\check {\vec {\tilde G}}}_{C\perp}(\vec y,x^o) {{{\vec \partial}
_y}\over {\triangle_y}}\cdot {\vec \zeta}^{({\check {\tilde G}}_{\perp})}
_{BD}(\vec y,\vec x;x^o) iT^D_s\, ]
\left( \begin{array}{c} {\check d}^{(m)}(\vec x,x^o) \\ {\check
s}^{(m)}(\vec x,x^o) \\ {\check b}^{(m)}(\vec x,x^o) \end{array} \right)
+\nonumber \\
&+&{e\over {sin\, \theta_w}}[\vec \alpha \cdot {\check {\vec {\tilde W}}}
_{-}(\vec x,x^o)+{e\over {sin\, \theta_w}}{{ {\check {\cal Q}}_{(CC)-}(\vec x,
x^o) T^{+}_w}\over {m^2_W (1+{{|e|}\over {sin\, 2\theta_w m_Z}}H(\vec x,x^o))
^2}}]\cdot \nonumber \\
&&{1\over 2}(1-\gamma_5) V_{CKM}^{\dagger}
\left( \begin{array}{c} {\check u}^{(m)}(\vec x,x^o) \\ {\check 
c}^{(m)}(\vec x,x^o) \\ {\check t}^{(m)}(\vec x,x^o) \end{array} \right) .
\label{83}
\end{eqnarray}

One could deduce the Euler-Lagrange equations from here without making the 
inverse Legendre transformation to find the physical Lagrangian. See Ref.
\cite{alba} for the form of the reduced second order equation for the transverse
Yang-Mills field (formulated on spacelike hypersurfaces) when only the color
SU(3) field is present.

Let us remark that, like in papers I and II, if we assume that the Higgs field 
H(x) is a weak nearly constant field [$H(x)\approx 0$, $\partial^oH(x)\approx 
0$], from its equation of motion we get the following restriction on the
bosonic field ${\check {\tilde Z}}(x)$, ${\check {\tilde W}}_{\pm}(x)$"

\begin{eqnarray}
&&[m^2_Z {\check {\vec {\tilde Z}}}^2 +2 m^2_W {\check {\vec {\tilde W}}}_{+}
\cdot {\check {\vec {\tilde W}}}_{-}](x)=\nonumber \\
&&=m^2_Z [{\check {\vec {\tilde Z}}}^2 +2 cos^2\, \theta_w {\check {\vec 
{\tilde W}}}_{+}\cdot {\check {\vec {\tilde W}}}_{-}](x)\approx \nonumber \\
&&\approx {{e^2}\over {4m^2_Z}} [{\check {\cal Q}}^2_{(NC)}+{4\over {sin^2\,
2\theta_w}}{\check {\cal Q}}_{(CC)+}{\check {\cal Q}}_{(CC)-}](x)-\nonumber \\
&-&[\, 
\left( \begin{array}{ccc} {\bar {\check e}}^{(m)}(x) & {\bar {\check \mu}}
^{(m)}(x) & {\bar {\check \tau}}^{(m)}(x) \end{array} \right) 
\beta \left( \begin{array}{ccc} m_e & 0 & 0\\ 0 & m_{\mu}
& 0\\ 0 & 0 & m_{\tau} \end{array} \right)  \left( \begin{array}{c} {\check e}
^{(m)}(x) \\ {\check \mu}^{(m)}(x) \\ {\check \tau}^{(m)}(x) \end{array} 
\right) +\nonumber \\
&+&\left( \begin{array}{ccc} {\bar {\check u}}^{(m)}(x) & {\bar {\check c}}
^{(m)}(x) & {\bar {\check t}}^{(m)}(x) \end{array} \right) 
\beta \left( \begin{array}{ccc} m_u & 0 & 0\\ 0 & m_c
& 0\\ 0 & 0 & m_t \end{array} \right)  \left( \begin{array}{c} {\check u}^{(m)}
(x) \\ {\check c}^{(m)}(x) \\ {\check t}^{(m)}(x) \end{array} \right) +
\nonumber \\
&+&\left( \begin{array}{ccc} {\bar {\check d}}^{(m)}(x) & {\bar {\check s}}
^{(m)}(x) & {\bar {\check b}}^{(m)}(x) \end{array} \right) 
\beta \left( \begin{array}{ccc} m_d & 0 & 0\\ 0 & m_s
& 0\\ 0 & 0 & m_b \end{array} \right)  \left( \begin{array}{c} {\check d}^{(m)}
(x) \\ {\check s}^{(m)}(x) \\ {\check b}^{(m)}(x) \end{array} \right) \, ].
\label{84}
\end{eqnarray}

Finally, one should check that the dressed (Dirac observable) charges ${\check 
Q}_V$, ${\check Q}_A$, ${\check Q}_{V,\bar A}$, ${\check Q}_{A,\bar A}$, 
corresponding to Eqs.(\ref{a2}), are constants of the motion in the limit 
$m_u=m_d=m_s=0$, $\theta_{13}=\theta_{23}=\delta_{13}=0$, and that the strong 
and weak chiral symmetries are verified in the appropriate limits.

\section {Conclusions.}

We have given the missing complete Hamiltonian treatment of the standard
model of elementary particles in the Higgs phase. A canonical basis of Dirac's
observables for the electromagnetic, strong and weak interactions has been 
found and the noncovariant canonical reduction (the generalized Coulomb gauge)
has been done. We have evaluated the physical noncovariant, nonlocal and 
nonpolynomial Hamiltonian. An unexpected result is that the self-energy terms 
of the weak interactions, associated with the Z and $W_{\pm}$ bosons, are
``local". Therefore, the Fermi 4-fermion interaction reappears at the
nonperturbative level after the solution of the Gauss laws in the Higgs phase
and the elimination of the unphysical degrees of freedom. It is interesting
to note that, even if only the electromagnetic charge is conserved (custodial
symmetry in the electroweak sector, at each instant there is a global su(2)xu(1)
algebra of non conserved charges.

This physical Hamiltonian appears as the final stage of the reduction of the
non-renormalizable unitary gauge. To go to the quantum level, one has to learn 
how to quantize this nonlocal and nonpolynomial field theory. Since the
Hamiltonian is bilinear in the momenta, with a nonlocal and nonpolynomial
coordinate-dependent metric connecting them, the natural technology to apply 
for the canonical quantization seems to be the one used for field theory in 
curved spacetimes. Moreover, as said in the Introduction, one now has an 
intrinsic classical unit of lenght (the M$\o$ller radius) to be used as an
intrinsic physical ultraviolet cutoff in the spirit of Dirac and Yukawa.

However, before trying to quantize, we have to covariantize the generalized
Coulomb gauge (see the Introduction) and to unify the standard model with
tetrad gravity at the classical level [see Ref.\cite{russo}]: since in the
asymptotically flat case one can define the same classical unit of lenght in 
terms of the asymptotic Poincar\'e charges, one would have a unified 
description of the four interactions with a universal ultraviolet cutoff and
a physical nonlocal Hamiltonian bilinear in the momenta.

As in the cases of papers I and II, the covariant R-gauge-fixings\cite{rgau}, 
of the type $\partial^{\mu}U_{a\mu}(x)+\xi \theta_a(x)\approx 0$,
used in the proof of renormalizability and in the evaluation
of radiative corrections, are ambiguous like the Gauss laws: they can be solved 
either  in the Higgs fields (would-be Goldstone bosons) $\theta_a(x)$
[Higgs phase] or in $U_{ao}(x)$ [unbroken  phase] or in a mixed way [the other
mixed phases]. It turns out that in the proofs of renormalizability
one is mixing all the existing disjoint phases (the only physical ones are
the Higgs one and, maybe, the unbroken phase, which  could be relevant in
cosmology; all the mixed non-covariant phases are unphysical), and only at the
end, in the limit $\xi \rightarrow \infty$, one is recovering the Higgs
phase.

As said in the Introduction, the covariantization of these results can be
done by reformulating the theory on spacelike hypersurfaces in Minkowski
spacetime. However, before getting it, one has to end the study of Dirac and 
chiral fermion fields on spacelike hypersurfaces and to understand whether 
there is a classical background for the chiral anomaly. In the covariantized
theory there will be the possibility to avoid the Fermi 4-fermion interaction
in a covariant way as said in Section VII.

\vfill\eject


\begin{references}

\bibitem{lv1}L.Lusanna and P.Valtancoli, ``Dirac's Observables for the Higgs
model: I) the Abelian Case", to appear in Int.J.Mod.Phys. A (HEP-TH/9606078).
\bibitem{lv2}L.Lusanna and P.Valtancoli, ``Dirac's Observables for the Higgs
model: II) the non-Abelian SU(2) Case", to appear in Int.J.Mod.Phys. A
(HEP-TH/9606079)>
\bibitem{lusa}L.Lusanna, Int.J.Mod.Phys. {\bf A10}, 3531 and 3675 (1995).
\bibitem{lus1}L.Lusanna, Int.J.Mod.Phys. {\bf A12}, 645 (1997).
\bibitem{albad}D.Alba and L.Lusanna, ``The Lienard-Wiechert Potential of 
Charged Scalar Particles and their Relation to Scalar Electrodynamics in the
Rest-Frame Instant Form", Firenze Univ. preprint 1997 (HEP-TH/9705155).
\bibitem{alba}D.Alba and L.Lusanna, ``The Classical Relativistic Quark Model
in the Rest-Frame Wigner-Covariant Coulomb Gauge", Firenze Univ.preprint 1997
(HEP-TH/9705156).
\bibitem{dep}F.Bigazzi, R.DePietri and L.Lusanna, ``Fermion Fields on 
Spacelike Hypersurfaces", in preparation.
\bibitem{hen}J.Geheniau and M.Henneaux, Gen.Rel.Grav. {\bf 8}, 611 (1977).
M.Henneaux, Gen.Rel.Grav> {\bf 9}, 1031 (1978).
\bibitem{re}``Solving Gauss' Laws and Searching
Dirac Observables for the Four Interactions", talk at the ``Second Conf. on
Constrained Dynamics and Quantum Gravity", S.Margherita Ligure 1996 
(HEP-TH/9702114). ``Unified Description and Canonical Reduction to Dirac's
Observables of the Four Interactions", talk at the Int.Workshop ``New non
Perturbative Methods and Quantization on the Light Cone", Les Houches 1997
(HEP-TH/9705154).
\bibitem{mon}V.Moncrief, J.Math.Phys. {\bf 20}, 579 (1979).
\bibitem{can}M.Cantor, Bull.Am.Math.Soc. {\bf 5}, 235 (1981).
\bibitem{sm}K.Huang, ``Quarks, Leptons and Gauge Fields" (World Scientific,
Singapore, 1982). 
\bibitem{dgh}J.F.Donoghue, E.Golowich and B.R.Holstein, ``Dynamics of the
Standard Model" (Cambridge Univ.Press, Cambridge, 1992).
\bibitem{daw}S.Dawson, ``Introduction to the Physics of Higgs Bosons",
Lectures given at the 1994 Theoretical Advanced Study Institute, Boulder CO,
BNL-61012 (October 1994) preprint.
\bibitem{her}
M.J.Herrero, ``Introduction to the Symmetry Breaking Sector", Lectures given 
at the XXIII Int.Meeting on Fundamental Physics, Santander (Spain) 1995, 
preprint FTUAM Jan/96/1, hep-ph/9601, of the Universidad Autonoma de Madrid.
\bibitem{pich}A.Pich, ``The Standard Model of Electroweak Interactions", 
Lectures at 
the XXII Int. Meeting on Fundamental Physics ``The Standard Model and Beyond",
Jaca(Spain) 1994, and CERN Academic Training, Gen\'eva 1993.
``Quantum Chromodynamics", Lectures at the 1994 European School of High
Energy Physics", Sorrento (Italy) 1994.
``Flavourdynamics", Lectures at the XXIII Int. Meeting on Fundamental
Interactions, Comillas (Spain) 1995.
\bibitem{ll}L.Lusanna, Int.J.Mod.Phys. {\bf A8}, 4193 (1993);
Phys.Rep. {\bf 185}, 1 (1990); Riv. Nuovo Cimento {\bf 14}, n.3, 1 (1991);
J.Math.Phys. {\bf 31}, 2126 (1990); J.Math.Phys. 
{\bf 31}, 428 (1990).
\bibitem{tri}K.Wilson, Phys.Rev. {\bf B4}, 3184 (1971).
\bibitem{dob}A.Dobado and M.T.Urdiales, `Determination of the Electroweak 
Chiral-Lagrangian Parameters at the LHC', hep-th/9502255.
\bibitem{heavy}T.Mannel, `Review of Heavy Quark Effective Theory', talk at the
Workshop `Heavy Quarks at Fixed Target', Rheinfals Castle 1996, hep-ph/9611411.
\bibitem{chai}M.Chaichian, D.Louis Martinez and L.Lusanna, Ann.Phys.(N.Y.)
{\bf 232}, 40 (1994).
K.Wilson and J.Kogut, Phys.Rep. {\bf 12}, 75 (1974).
J.Fr\"olich, in ``Progress in Gauge Field Theory", Carg\`ese 1983, eds.
G.'t Hooft, A.Jaffe, H.Lehmann, P.K.Mitter, I.M.Singer and R.Stora,
NATO ASI B115 (Plenum, New York, 1984).
M.A.B.B\'eg and R.C.Furlong, Phys.Rev. {\bf D31}, 1370 (1985). 
\bibitem{ab}T.Appelquist and C.Bernard, Phys.Rev. {\bf D22}, 200 (1980).
\bibitem{bs}W.A.Bardeen and K.Shizuya, Phys.Rev. {\bf D18}, 1969 (1978).
\bibitem{rgau}B.W.Lee and J.Zinn-Justin, Phys.Rev. {\bf D5}, 3121, 3137 and 
3155 (1972).
A.Salam and J.Strathdee, Nuovo Cim. {\bf 11A}, 397 (1972).
K.Fujikawa, B.W.Lee and A.I.Sanda, Phys.Rev. {\bf D6}, 2923 (1972).
Y.P.Yao, Phys.Rev. {\bf D7}, 1647 (1973).
E.Abers and B.Lee, Phys.Rep. {\bf 9}, 1 (1975).
\bibitem{russo}L.Lusanna and S.Russo, ``Dirac's Observables for Tetrad Gravity",
in preparation.

\end{references}
\end{document}